\documentclass[aps,prd,onecolumn,showpacs,superscriptaddress,nofootinbib,preprintnumbers]{revtex4-2}

\usepackage{graphicx}
\usepackage{comment}
\usepackage{color}
\usepackage{amsmath}
\usepackage{amssymb}
\usepackage{enumerate}
\usepackage{subcaption}   
\usepackage{caption}
\usepackage[colorlinks=true,citecolor=blue,urlcolor=blue]{hyperref}

\graphicspath{ {Figures/} }
\newcommand{\CL}{\mathcal{L}}

\newcommand{\CO}{\mathcal{O}}

\begin{document}

\title{Projected Sensitivity of Paleo-Detectors \\ to Dark Matter Effective Interactions with Nuclei}

\author{Dionysios~P.~Theodosopoulos}
\email{d.theodosopoulos@utexas.edu}
\affiliation{Department of Physics, The University of Texas at Austin, Austin, TX 78712, USA}

\author{Katherine~Freese}%
 \email{ktfreese@utexas.edu}
\affiliation{Department of Physics, The University of Texas at Austin, Austin, TX 78712, USA}
\affiliation{The Oskar Klein Centre, Department of Physics, Stockholm University, AlbaNova, SE-106 91 Stockholm, Sweden}
\affiliation{Nordita, Stockholm University and KTH Royal Institute of Technology, Hannes Alfvéns väg 12, SE-106 91 Stockholm, Sweden}

\author{Chris~Kelso}
\email{ckelso@unf.edu}
\affiliation{
 Department of Physics and Astronomy, University of North Florida, 1 UNF Dr, Jacksonville, FL 32224, USA
}%

\author{Patrick~Stengel}
\email{patrick.stengel@ijs.si}
\affiliation{
 Jo\v{z}ef Stefan Institute, Jamova 39, 1000 Ljubljana, Slovenia
}%

\begin{abstract}
Paleo-detectors are a proposed experimental technique for direct detection of dark matter (DM) via the read-out of DM-induced nuclear recoil tracks in natural minerals. The large detector mass required for the sensitivity of conventional direct detection experiments to rare events is replaced by the exposure of paleo-detectors to DM-induced nuclear recoils over geological timescales. In this paper, we generalize the previous theoretical predictions that focused on canonical spin-independent coherent and spin-dependent scattering (proportional to $A^2$ and the spin of the nucleus, respectively). We estimate the sensitivity of paleo-detectors to interactions between weakly interacting massive particle (WIMP) DM and nuclei within the framework of a Non-Relativistic Effective Field Theory (NREFT), considering isoscalar couplings to nucleons for both elastic and inelastic scattering. Taking into account cosmogenic, astrophysical and radiogenic backgrounds, we project the 90\% confidence-level upper limits on the isoscalar NREFT coupling constants for both scattering types. We consider representative read-out scenarios and examine a wide variety of target minerals. The projected sensitivities of paleo-detectors are compared with the 90$\%$ confidence-level limits from the XENON100, LUX–ZEPLIN, and PandaX–II experiments, as well as with the 95$\%$ Bayesian credible region of the two-dimensional marginalized posterior distribution from SuperCDMS. For DM masses from $1\, \mathrm{GeV}/c^2 - 10\, \mathrm{GeV}/c^2$, we find that paleo-detectors are projected to have sensitivity superior to that of conventional experiments for WIMP–nucleus interactions mediated by all NREFT operators, largely independent of read-out scenario or target mineral. For heavier DM masses in the range $10\, \mathrm{GeV}/c^2 - 5\, \mathrm{TeV}/c^2$, we find that the sensitivity of paleo-detectors is projected to be comparable to or better than that of conventional experiments for WIMP–nucleus interactions mediated by several NREFT operators, depending on the read-out scenario and target mineral.  
\end{abstract}

\maketitle

\section{Introduction}

Although cosmological observations indicate that non-baryonic cold dark matter constitutes more than 80\% of matter in the Universe, its nature remains a mystery. The Weakly Interacting Massive Particle (WIMP) is one of the most well-studied dark matter candidates and is theoretically motivated by a variety of Beyond the Standard Model physics scenarios. WIMP searches are pursued via several approaches: their production in colliders leading to signatures including missing energy; indirect detection searches for WIMP annihilation products; discovery of Dark Stars; and, as is discussed in this paper, direct detection experiments which are designed to detect the scattering of WIMP dark matter off atomic nuclei.

Conventional direct detection experiments measure the energy deposited by WIMP interactions with nuclei via some combination of scintillation light, ionization, and phonons produced by WIMP-induced nuclear recoils in the detectors~\cite{Drukier:1984vhf,Goodman:1984dc,Drukier:1986tm}. To avoid being overwhelmed by cosmic ray backgrounds on the Earth's surface, WIMP detectors must be deep underground. WIMPs distributed in the Milky Way dark matter halo are predicted to have low rates of interactions with target nuclei contained within the instrumented volume of a direct detection experiment. Thus, it is essential for the exposure of direct detection experiments, defined as the product of the target mass and the exposure time of the experiment, to be maximized in order to probe rare interactions. The traditional approach to obtaining the required large exposure is to build detectors of ever larger mass (and volume). For example, liquid noble gas based detectors have now achieved several tonnes in target mass, as such detectors become more and more sensitive to interactions between WIMPs and nuclei~\cite{LZ:2022lsv,DarkSide-50:2022qzh,XENON:2023cxc,PandaX:2024qfu}.

Paleo-detectors have been proposed as a novel alternative to conventional direct detection searches for signatures of nuclear recoils from WIMP interactions~\cite{PhysRevLett.74.4133,Collar:1995aw,Snowden-Ifft:1996dug,Baum:2018tfw,Drukier:2018pdy,Edwards:2018hcf,SinghSidhu:2019znk,Ebadi:2021cte,Acevedo:2021tbl,Baum:2021jak,Fung:2025cub}. Rather than requiring increasingly large instrumented volumes of target material to probe even more rare interactions, paleo-detectors can achieve much larger exposures by extracting ancient minerals from deep underground that have been collecting damage features caused by WIMP interactions for geological timescales, as long as $\sim 1 \,$Gyr. Depending on the read-out technique, the exposures envisaged for paleo-detectors can be up to $\sim 10^3 - 10^4$ times larger than conventional direct detection experiments. In other words, the enormous detector mass of conventional experiments can be replaced by the long exposure time of paleo-detectors to potentially achieve better sensitivity to WIMP scattering. Whereas a conventional direct experiment expects to see only a handful of WIMP-induced nuclear recoils at the threshold of sensitivity necessary for a discovery, the equivalent number of signal events expected in paleo-detectors can be very large, even reaching into the millions, due to the long exposure times. However, paleo-detectors are also exposed to potentially much more numerous backgrounds which are more easily suppressed in a controlled laboratory experiment. The key to the sensitivity of a future paleo-detector experiment is, thus, the development of efficient microscopy techniques to read out the nano- and micro-scale damage features induced in natural minerals with enough precision to measure a statically significant excess of WIMP events over the relevant backgrounds.

Originally proposed as detectors for dark matter in the galactic halo, subsequent work has shown that paleo-detectors can be used for a variety of other studies as well. Paleo-detectors can be used to search for neutrinos; by digging up rocks of different ages, it is possible to measure the evolution of supernova~\cite{Baum:2019fqm,Baum:2022wfc}, solar~\cite{Tapia-Arellano:2021cml}, and atmospheric~\cite{Jordan:2020gxx} neutrino fluxes over geological timescales.  For example, paleo-detectors could trace the number of core collapse supernova explosions in the Milky Way over the last $\sim 1 \,$Gyr.  Paleo-detectors have also been proposed as a probe of dark matter substructure in the Milky Way~\cite{Baltz:1997dw,Baum:2021chx,Bramante:2021dyx,Zhang:2025xzi}. The exposure of paleo-detectors over geological timescales could also be leveraged to trace the history of the galactic cosmic ray flux~\cite{Caccianiga:2024otm,Galelli:2025gss} or to probe exotica such as magnetic monopoles~\cite{Price:1986ky}, charged black hole remnants~\cite{Lehmann:2019zgt} and proton decay~\cite{Baum:2024sst}.

Paleo-detectors are reaching an exciting new stage of research and development. Groups from around the world are currently experimenting with advanced microscopy techniques to efficiently read out damage to the crystalline structure of natural minerals induced by a variety of interesting potential signals in paleo-detectors~\cite{Baum:2023cct,Baum:2024eyr,Hirose:2025jht}. Hence further theoretical studies of the capabilities of these detectors have become essential.  Our focus is to study the particle physics aspects of the WIMP--nucleus interaction in making predictions for upcoming experiments in a wide variety of target minerals. Previous analyses of the sensitivity of paleo-detectors to WIMPs have only examined the two most commonly studied types of WIMP--nucleus interactions: canonical spin-independent (SI) and spin-dependent (SD) elastic scattering, both assuming interactions which are independent of the WIMP velocity or momentum transferred to the target nucleus. SI interactions yield coherent scattering between the WIMP and the nucleus, leading to $A^2$ enhancement of the cross section, where $A$ is the atomic mass of the target nucleus. SD interactions, on the other hand, depend on the spin of the target nucleus and have no coherent enhancement of the cross section. 

In this work, we extend the analysis of paleo-detector sensitivity to additional interactions of WIMPs with nuclei, beyond the canonical SI and SD interactions. Using a Non-Relativistic Effective Field Theory (NREFT), for a variety of mineral targets and read-out scenarios, we project the sensitivity of paleo-detectors to WIMP scattering via a set of effective operators which can represent all possible WIMP--nucleus interactions in the non-relativistic limit~\cite{Fan:2010gt,Fitzpatrick:2012ix}. We begin with studies of elastic WIMP--nucleus scattering. Within the same NREFT framework, we also consider the sensitivity of paleo-detectors to inelastic WIMP--nucleus interactions in which the WIMP upscatters to a heavier state~\cite{Tucker-Smith:2001myb}. Similar to previous studies, throughout the paper we consider the track length predicted by the stopping power of a given nucleus recoiling in the target mineral to be a proxy for nuclear recoil energy.\footnote{Previous phenomenological studies have also considered the detection of nuclear recoils in synthetic crystals~\cite{Cogswell:2021qlq} or natural minerals~\cite{Baum:2024sst} via the creation of color center defects read out by optical microscopy techniques~\cite{BUDNIK2018242,Rajendran2017,ebadi_directional_2022,Araujo:2025rhr}.}

The sensitivity of paleo-detectors can complement conventional direct detection searches for WIMP interactions with nuclei, with the sensitivity of paleo-detectors projected to be comparable to or better than that of conventional experiments across a range of WIMP masses. Specifically, conventional experiments reach optimal sensitivity for WIMPs in the $10 - 100\, \mathrm{GeV}/c^2$ mass range. However, we project that, both at lower and higher WIMP masses, the sensitivity of paleo-detectors to the couplings of several NREFT operators can outperform conventional experiments, due to the combination of low effective nuclear recoil energy thresholds and large exposures possible for paleo-detectors. In comparison to constraints from current experiments on elastic scattering, paleo-detectors are projected to increase the sensitivity to WIMPs with $\mathcal{O}(\mathrm{GeV}/ c^2)$ masses by up to five orders of magnitude in the squared coupling of the operator (proportional to the scattering cross section for canonical WIMP-nucleus interactions). The projected sensitivity of paleo-detectors to WIMPs with $\mathcal{O}(\mathrm{TeV}/ c^2)$ masses is improved by up to an order of magnitude compared to current constraints in the squared coupling of the operator for elastic WIMP-nucleus scattering and inelastic scattering scenarios with dark matter mass splittings $\lesssim 50  \, \mathrm{keV} /c^2$. 

The rest of this paper is outlined as follows. In Section~\ref{nreft}, we describe the the WIMP–nucleon NREFT framework used throughout our analysis. Nuclear recoils signatures from WIMP signals and various background sources are described in Sections~\ref{DMpaleo} and~\ref{background}, respectively. In Section~\ref{sensitivity}, we describe the statistical approach used to project the sensitivity of paleo-detectors and present sensitivity projections for the full set of effective operators compared to current constraints from conventional direct detection experiments. We conclude in Section~\ref{conclusion} with a summary and discussion of potential avenues for future work. In the main body of the text, we discuss the nuclear recoil track length spectra for the possible spin-independent operators \(\mathcal{O}_{1}^{s}\), \(\mathcal{O}_{5}^{s}\), \(\mathcal{O}_{8}^{s}\), and \(\mathcal{O}_{11}^{s}\), compared to background spectra in paleo-detectors, and relegate the discussion of the track length spectra for NREFT operators with spin-dependence to the Appendix (in the interest of brevity of the main text). In the Appendix, we also include the track length spectra for additional target minerals and the corresponding paleo-detector sensitivity projections. 

\section{Non-Relativistic Effective Field Theory} \label{nreft}

A comprehensive formulation of the Non-Relativistic Effective Field Theory (NREFT) for WIMP–nucleon elastic interactions, encompassing all interactions relevant at non-relativistic energies, was developed in Refs.~\cite{Fan:2010gt,Fitzpatrick:2012ix}. An effective field theory for direct detection experiments describes the interactions of a Dark Matter (DM) particle $\chi$ with a nucleon $N$ through a set of four-fermion operators in the non-relativistic limit. In this section, we first review the operators governing elastic WIMP--nucleon scattering and subsequently extend the discussion to the inelastic case.

The NREFT framework explored in \cite{Fitzpatrick:2012ix} provides a powerful, model-independent method for interpreting experimental data. Moreover, its power-counting scheme \cite{Buchalla:2013eza} enables a systematic and controlled expansion of the scattering cross-section in terms of the momentum transfer, which is essential for evaluating the sensitivity of direct detection experiments to WIMP DM.\footnote{In the NREFT framework, the effective Lagrangian is expanded in powers of the momentum transferred during the WIMP--nucleus interaction. The cutoff scale of this theory is set by the maximum momentum transfer, roughly given by $q_{\max} \sim 2 \mu v$ $\sim$ 200 MeV, where $\mu$ is the reduced mass of the WIMP--nucleon system and $v$ is the relative WIMP--nucleon speed~\cite{Fitzpatrick:2012ix}. The pion degrees of freedom are integrated out of this EFT, restricting its validity to energies below the pion mass. Whereas this NREFT provides a reliable description of all possible elastic interactions at these energies, matching it to an ultraviolet theory of dark matter requires reinstating the QCD dynamics neglected in constructing the NREFT. Illustrations of such matching procedures can be found in \cite{Bishara:2017pfq,Hoferichter:2018acd}. Experimental constraints on WIMP–pion interactions are reported in \cite{XENON:2018clg}, and upper limits for a chiral effective field theory and a model of inelastic DM are given in \cite{XENON:2022avm}.} The derivation of the NREFT introduces both momentum- and velocity-dependent operators to capture all possible WIMP–nucleon elastic interactions. By enforcing momentum conservation and Galilean invariance, these operators can be reduced to a basis of four Hermitian quantities
\begin{equation}
     i\frac{\vec{q}}{m_N},\quad \vec{v}^{\perp} \equiv \vec{v} + \frac{\vec{q}}{2\mu},\quad \vec{S}_{\chi},\quad \vec{S}_{N},\label{basis}
\end{equation}
where $\vec{q}$ is the momentum transferred from the WIMP to the nucleon, $m_{N}$ is the mass of the nucleon, $\vec{v}^{\perp}$ is the component of the WIMP–nucleon relative velocity ($\vec{v}$) perpendicular to the momentum transfer, $\vec{S}_{\chi}$ is the spin of the WIMP, and $\vec{S}_{N}$ is the spin of the nucleon. By forming linear combinations of these four Galilean invariant and Hermitian operators and retaining terms up to second order in $\vec{q}$, the general Lagrangian for all possible non-relativistic interactions between WIMPs $\chi$ and 
nucleons $N$ has been shown to be
\begin{equation}
\CL_{\rm int} = \sum_{N =n,p} \sum_i c_i^{(N)}  \CO_i \chi^+ \chi^- N^+ N^-,
\end{equation}
where $\CO_i$ denotes one of the linearly independent and dimensionless NREFT operators \cite{Fitzpatrick:2012ix,Anand:2013yka}. The NREFT operators for elastic interactions are listed in Table \ref{tab:NREFT_operators}. Additionally, $c_i^{p}$ and $c_i^{n}$ are the NREFT coupling constants for the WIMP–proton and WIMP–neutron interactions, respectively. The superscripts ($-,+$) label the incoming and outgoing particles. For elastic scattering, $\chi^+ \equiv \chi^-$, while for inelastic scattering the incoming and outgoing dark matter states have different masses. Throughout this work, for both elastic and inelastic scattering, we assume that the nucleon state remains unchanged by the interaction, i.e.\ $N^+ \equiv N^-$. We further note that the operators $\mathcal{O}_1$ and $\mathcal{O}_4$ correspond to the canonical spin-independent (SI) and spin-dependent (SD) interactions, respectively, which have been the primary focus of previous studies of paleo-detector sensitivity. From our discussion we will exclude the operator $\mathcal{O}_{2}=\left( v^{\perp} \right)^2$, as this operator cannot be obtained from the leading-order non-relativistic reduction of a manifestly relativistic operator \cite{Anand:2013yka}. Moreover, since we focus on non-relativistic interactions, any operator with a quadratic or higher dependence on $v^{\perp}$ is consider negligible. 

Thus, the NREFT interaction considered in this paper reads
\begin{equation}
\sum_{N = n,p} \sum_{i=1}^{15} c_i^{(N)} \mathcal{O}_i^{(N)}, 
\qquad c_2^{(N)} \equiv 0~.
\end{equation}
We work in the isoscalar basis, where $c_i^{p}=c_i^{n} \equiv c_i$, considering that isospin is a good quantum number in nuclear systems. We denote the operators in this basis as $\mathcal{O}^{s}_{i}$. Moreover, working in the isoscalar basis enables direct comparison with the limits on the NREFT coupling constants reported by the XENON100 \cite{XENON:2017fdd}, LUX--ZEPLIN \cite{LZ:2023lvz}, and PandaX-II \cite{PandaX-II:2018woa} experiments.

\begin{table}[t]
\centering
\captionsetup{justification=raggedright,singlelinecheck=false}
\begin{tabular}{l l}
\hline\hline
\multicolumn{2}{c}{NREFT operators relevant for elastic WIMP--nucleon scattering} \\
\hline
$\mathcal{O}_1 = 1_\chi 1_N$ 
& $\mathcal{O}_9 = i\, \vec{S}_\chi \cdot \left( \vec{S}_N \times \dfrac{\vec{q}}{m_N} \right)$ \\[0.8em]

$\mathcal{O}_3 = i\, \vec{S}_N \cdot \left( \dfrac{\vec{q}}{m_N} \times \vec{v}^{\perp} \right)$ 
& $\mathcal{O}_{10} = i\, \vec{S}_N \cdot \dfrac{\vec{q}}{m_N}$ \\[0.8em]

$\mathcal{O}_4 = \vec{S}_\chi \cdot \vec{S}_N$ 
& $\mathcal{O}_{11} = i\, \vec{S}_\chi \cdot \dfrac{\vec{q}}{m_N}$ \\[0.8em]

$\mathcal{O}_5 = i\, \vec{S}_\chi \cdot \left( \dfrac{\vec{q}}{m_N} \times \vec{v}^{\perp} \right)$ 
& $\mathcal{O}_{12} = \vec{S}_\chi \cdot \left( \vec{S}_N \times \vec{v}^{\perp} \right)$ \\[0.8em]

$\mathcal{O}_6 = \left( \vec{S}_\chi \cdot \dfrac{\vec{q}}{m_N} \right)
                  \left( \vec{S}_N \cdot \dfrac{\vec{q}}{m_N} \right)$
& $\mathcal{O}_{13} = i \left( \vec{S}_\chi \cdot \vec{v}^{\perp} \right)
                       \left( \vec{S}_N \cdot \dfrac{\vec{q}}{m_N} \right)$ \\[0.8em]

$\mathcal{O}_7 = \vec{S}_N \cdot \vec{v}^{\perp}$ 
& $\mathcal{O}_{14} = i \left( \vec{S}_\chi \cdot \dfrac{\vec{q}}{m_N} \right)
                       \left( \vec{S}_N \cdot \vec{v}^{\perp} \right)$ \\[0.8em]

$\mathcal{O}_8 = \vec{S}_\chi \cdot \vec{v}^{\perp}$ 
& $\mathcal{O}_{15} = - \left( \vec{S}_\chi \cdot \dfrac{\vec{q}}{m_N} \right)
                       \left[ \left( \vec{S}_N \times \vec{v}^{\perp} \right)
                       \cdot \dfrac{\vec{q}}{m_N} \right]$ \\[0.8em]
\hline\hline
\end{tabular}
\caption{Quantum mechanical operators defining the non-relativistic effective theory of WIMP--nucleon elastic interactions~\cite{Fan:2010gt,Fitzpatrick:2012ix}. The operators $\mathcal{O}_{1}$ and $\mathcal{O}_{4}$ correspond to canonical spin-independent (SI) and spin-dependent (SD) interactions, respectively. Operator $\mathcal{O}_{2}$, which is quadratic in $\vec{v}^{\perp}$, and operator $\mathcal{O}_{16}$, which is a linear combination of $\mathcal{O}_{12}$ and $\mathcal{O}_{15}$, are not considered here~\cite{Anand:2013yka}.}
\label{tab:NREFT_operators}
\end{table}

We will also examine inelastic scattering processes, in which the masses of the incoming and outgoing DM particles differ. Such interactions allow a WIMP to transition to a heavier state during scattering~\cite{Tucker-Smith:2001myb}. The mass splitting, denoted by $\delta_m \equiv m_{\chi,\text{out}} - m_{\chi,\text{in}}$, imposes a minimum recoil energy for an inelastic event to occur. Consequently, the recoil rate at low energies is suppressed and the signal is shifted toward higher recoil energies. In scenarios where elastic scattering is strongly suppressed, inelastic transitions between WIMPs and nucleons can dominate~\cite{Han:1997wn,Hall:1997ah}. 

To account for inelastic WIMP--nucleon interactions, a slight modification of the standard Hermitian basis vectors is required. In elastic scattering, energy conservation implies $\vec{v}^{\perp}\!\cdot \vec{q}=0$. For inelastic scattering with nonzero $\delta_m$, energy conservation instead requires~\cite{Barello:2014uda}
\begin{equation}
  \delta_m + \vec{v}\!\cdot \vec{q} + \frac{|\vec{q}|^2}{2\mu_N} = 0,
\end{equation}
where $\mu_N$ is the reduced mass of the WIMP--nucleon system. This condition is implemented in the NREFT formalism by modifying the Hermitian basis, specifically the perpendicular velocity given in Eq.~(\ref{basis}), to incorporate the nonzero mass splitting
\begin{equation}
\vec v_{\text{inel}}^{\perp} \equiv \vec v + \frac{\vec q}{2\mu_N} + \frac{\delta_m}{|\vec q|^2}\vec q
= \vec v^{\perp} + \frac{\delta_m}{|\vec q|^2}\vec q ~.\label{u_inelastic}
\end{equation}
The operators describing inelastic interactions are then obtained from the operators $\mathcal{O}_i$ listed in Table~\ref{tab:NREFT_operators} by replacing $\vec v^{\perp}$ with $\vec v_{\text{inel}}^{\perp}$. In this paper, we focus on non-relativistic WIMP--nucleon elastic scattering produced via the operators in Table~\ref{tab:NREFT_operators}, with additional discussion of inelastic scattering.

\section{Dark Matter Signals in Paleo-Detectors}\label{DMpaleo}

We can now make predictions for dark matter signals in paleo-detectors arising from WIMP--nucleus interactions via the NREFT operators introduced in the previous section. Our primary goal in this section is to compute the predicted track length spectra, i.e.\ the number of events as a function of track length, for a variety of target minerals. In this section, we focus on the spin-independent operators $\mathcal{O}_{1}^{s}$, $\mathcal{O}_{5}^{s}$, $\mathcal{O}_{8}^{s}$, and $\mathcal{O}_{11}^{s}$. Among these, $\mathcal{O}_{1}^{s}$ corresponds to the canonical spin-independent interaction commonly used to present mass--cross-section constraints from direct detection experiments. Track length spectra for the remaining NREFT operators are deferred to the Appendix, in order to keep the number of figures in the main text manageable. The Appendix also includes results for target minerals different from those considered here.

The signature of WIMP DM in paleo-detectors is the spectrum of damage-track lengths produced by nuclei recoiling in the mineral after WIMP–nucleus scattering. The ancient minerals have been collecting damage tracks over geological timescales, with track lengths determined by the stopping power of a given nucleus in a given material; hence we assume the track length as a proxy for the nuclear recoil energy. For a WIMP of mass $m_{\chi}$ scattering off a given target nucleus $(Z,A)$ of mass $m_{T}$, the differential nuclear-recoil rate in the NREFT framework reads \cite{Fitzpatrick:2012ix,Anand:2013yka}
\begin{equation}
    \left(\frac{dR}{dE_R}\right)_{(Z,A)} = N_T \frac{\rho_\chi m_T}{32 \pi m_\chi^3 m_N^2} \left\langle \frac{1}{v} \sum_{i j} \sum_{N,N'=p,n} c_i^{(N)} c_j^{(N')} F_{ij,\;(Z,A)}^{(N,N')}\!\left(v^2, q^2\right) \right\rangle~,\label{recoilrate}
\end{equation}
where $R$ is the number of recoiling nuclei per unit exposure,\footnote{In this work, we assume a one-to-one correspondence between the number of tracks and the number of recoiling nuclei; under this assumption, the rate $R$ in Eq.~\eqref{recoilrate} is consistent with the interpretation as the number of tracks per unit exposure.} $E_{R}$ is the nuclear recoil energy, $N_T$ is the number of target nuclei per detector mass, $m_{N}$ is the nucleon mass, $\rho_{\chi}$ is the local DM density, $v$ is the WIMP speed following the Standard Halo Model velocity distribution in the laboratory frame, $F_{ij,\;(Z,A)}^{(N, N')}$ are the form factors defined in \cite{Fitzpatrick:2012ix,Anand:2013yka}, and $\langle...\rangle$ indicates average over the DM halo velocity distribution. Following the conventions in \cite{Baxter:2021pqo}, the local DM density reads $\rho_{\chi}=0.3~\text{GeV}/\text{cm}^{3}$ \cite{Read:2014qva}, and the Standard Halo Model is used to describe the WIMP velocity distribution, with $\vec{v}_{\odot}=(11.1, 12.24,7.25)~\text{km}/\text{s}$ (solar peculiar velocity) \cite{10.1111/j.1365-2966.2010.16253.x}, $\vec{v}_{0}=(0,238,0)~\text{km}/\text{s}$ (local standard of rest velocity) \cite{Sch_nrich_2012,Bland_Hawthorn_2016,2021}, and $v_{\text{esc}}=544~\text{km}/\text{s}$ (galactic escape speed) \cite{10.1111/j.1365-2966.2007.11964.x}. For the calculation of the nuclear-recoil rate, we used the open-source codes \textit{WimPyDD} \cite{Jeong:2021bpl} and \textit{dmscatter} \cite{Gorton:2022eed}, with results verified by comparison to those of the Mathematica package \textit{DMFormFactor} \cite{Anand:2013yka}. 

To compute the form factors $F_{ij,\;(Z,A)}^{(N, N')}$, nuclear-structure input is required in the form of one-body density matrices between many-body eigenstates. Accordingly, we used \textit{WimPyDD} for isotopes with nuclear response functions provided in Ref.~\cite{Catena:2015uha}. For the remaining isotopes, we employed \textit{BIGSTICK} \cite{Johnson:2018hrx}, an open-source shell-model code, to compute the one-body density matrices, which were then incorporated into \textit{dmscatter} for the final calculation of the form factors and the nuclear-recoil rate. Specifically, to calculate the one-body density matrix we used the USDB interaction \cite{Brown:2006gx} for $^{35}\mathrm{Cl}$, an effective $pf$-shell interaction \cite{Honma:2004xk} for $^{55}\mathrm{Mn}$, and the SDPF-NR interaction \cite{Retamosa:1996rz} for $^{37}\mathrm{Cl}$ and $^{41}\mathrm{K}$. 

According to the formalism introduced in Refs.~\cite{Fitzpatrick:2012ix,Anand:2013yka}, each NREFT operator corresponds to a distinct combination of six nuclear response operators: $M$, $\Sigma''$, $\Sigma'$, $\Phi''$, $\tilde{\Phi}'$, and $\Delta$. The standard spin-independent $M$ response contributes to the spin-independent operators $\mathcal{O}_{1}$, $\mathcal{O}_{5}$, $\mathcal{O}_{8}$, and $\mathcal{O}_{11}$, while $\Delta$ contributes only to $\mathcal{O}_{5}$ and $\mathcal{O}_{8}$. The $\Delta$ response at zero momentum transfer probes the angular momentum content of the nucleus. Consequently, nuclei with an unpaired nucleon (proton or neutron) in a non-$s$-shell orbital exhibit stronger $\Delta$ responses. For example, $^{23}\mathrm{Na}$, which has an unpaired proton outside the $s$-shell, shows an enhanced response, whereas $^{19}\mathrm{F}$, with an unpaired proton occupying the $2s_{1/2}$ orbital, is comparatively disfavored. At finite momentum transfer $\vec{q}$, the $\Delta$ response acquires a kinematic enhancement proportional to the mass number $A$, which can compete with the coherent enhancement characteristic of the standard SI response $M$. As a result, contributions from operators such as $\mathcal{O}_5$ and $\mathcal{O}_8$ can become significant, and even dominant, for nuclei containing unpaired nucleons in high-angular-momentum orbitals.  

Operators $\mathcal{O}_3$, $\mathcal{O}_{12}$, $\mathcal{O}_{13}$, and $\mathcal{O}_{15}$ would normally appear in the conventional framework as interactions coupling through $\Sigma'$ and $\Sigma''$, and would therefore contribute only for target nuclei with total spin $j_{(Z,A)} \geq 1/2$. However, $\mathcal{O}_3$, $\mathcal{O}_{12}$, and $\mathcal{O}_{15}$ possess dominant scalar couplings through $\Phi''$, which is proportional to the spin–orbit interaction $\vec{\sigma}(i)\!\cdot\!\vec{\ell}(i)$. Consequently, these operators can yield nonzero contributions even for target nuclei with spin-zero ground states, favoring isotopes with partially filled angular momentum orbitals. The leading contribution from $\mathcal{O}_{13}$ instead arises from the tensor operator $\tilde{\Phi}'_2$, which is proportional to the interaction $\vec{\sigma}(i)\!\times\!\vec{\ell}(i)$ and requires $j_{(Z,A)} \geq 1$, implying a vanishing spectrum for nuclei with a spin-zero ground state. 

The range $x_{T}$ of a recoiling nucleus $(Z,A)$ with recoil energy $E_{R}$ in a target material reads
\begin{equation}
    x_T(E_{R})=\int_{0}^{E_{R}}dE~\left| \frac{dE}{dx} \right|_{(Z,A)}^{-1}~, \label{length}
\end{equation}
where $(dE/dx)_{(Z,A)}$ denotes the stopping power of nucleus $(Z,A)$ in the target material, with $dE$ the energy lost by a nucleus moving a distance $dx$ in the target material. Note that $x_T$ depends on the type of nucleus $(Z,A)$, but for brevity we have dropped the subscript $(Z,A)$ for this quantity. The stopping power is calculated using \textit{SRIM}, a software package for the stopping and range of ions in matter \cite{ZIEGLER20101818}. The observed length of a damage track in a given target material may differ from the range as computed above. For example, the nucleus might produce a permanent damage track only along part of the distance it travels through the material, or its trajectory might deviate significantly from a straight line. Nevertheless, previous studies indicate that such effects can be ignored for phenomenological studies estimating the sensitivity of paleo-detectors \cite{Drukier:2018pdy}, making Eq.~(\ref{length}) a valid approximation for the track length for scattering on one nucleus. Thermal annealing could also alter track lengths over geological timescales; however, any such modifications would be similar for both signal and background recoils. For this reason, throughout this work we use Eq.~(\ref{length}) to define the track length for scattering off a given nuclear species $(Z,A)$, while future detailed experimental studies for each target material will aim to quantify and correct for these possible effects. In our analysis, the tracks of light ions with charge $Z \leq 2$ are excluded, as either the tracks are not sufficiently long-lived or the recoiling nuclei are not sufficiently ionizing to form tracks \cite{Drukier:2018pdy}.

In a paleo-detector, the target material is a mineral crystal composed of several different nuclear species (and isotopes) $(Z,A)$, each contributing a mass fraction $\xi_{(Z,A)}$, defined as the nuclear mass per unit target mass. Accordingly, the total number of events in an experiment is given by the sum of scattering events over all nuclear species present in the target mineral. A track of length $x_T$ may therefore originate from scattering off any one of these nuclei and corresponds to the track length computed using Eq.~\eqref{length}.

In the following, we present predictions for track length spectra in various target minerals, with chemical compositions and fiducial ${}^{238}$U concentrations listed in Table~\ref{tab:U238_concentration}. In order to suppress backgrounds induced by radioactive contaminants (see Sec.~\ref{background}), we propose to use minerals found in Marine Evaporites (MEs) or in Ultra-Basic Rocks (UBRs). Such minerals have significantly lower concentrations of radioactive contaminants, e.g. uranium, than typical minerals found in the Earth's crust. In the main text, we focus on four representative target minerals commonly found in nature: two in UBRs, olivine and muscovite---the latter containing hydrogen---and two in MEs, halite and gypsum, with gypsum also containing hydrogen. Both the ${}^{238}$U concentration and the presence of hydrogen in the target material significantly affect the expected background rates in paleo-detector experiments, as discussed in Section~\ref{background}. For searches for WIMPs with masses $m_\chi > 10 \, \mathrm{GeV/c^2}$, where the background is dominated by neutrons from radioactive processes, target minerals containing hydrogen are particularly useful. This is because hydrogen is an effective moderator of fast neutrons, reducing the number of neutron-induced nuclear recoils. In the Appendix, we present track length spectra and projected sensitivities for several additional minerals, namely sinjarite, epsomite, phlogopite, and nchwaningite, with chemical compositions and fiducial ${}^{238}$U concentrations also listed in Table~\ref{tab:U238_concentration}.
\begin{table}[ht]
\captionsetup{justification=raggedright,singlelinecheck=false}
\centering
\begin{tabular}{l l c}
\hline
\textbf{Mineral} & \textbf{Composition} & \textbf{Fiducial ${}^{238}$U Concentration [per Weight, g/g]} \\
\hline
Gypsum        & Ca(SO$_4$)$\cdot$2(H$_2$O)     & $10^{-11}$ \\
Halite        & NaCl                          & $10^{-11}$ \\
Sinjarite     & CaCl$_2$$\cdot$2(H$_2$O)     & $10^{-11}$ \\
Epsomite     & Mg(SO$_4$)$\cdot$7(H$_2$O)     & $10^{-11}$ \\
Olivine       & Mg$_{1.6}$Fe$^{2+}_{0.4}$(SiO$_4$) & $10^{-10}$ \\
Muscovite    & KAl$_2$(AlSi$_3$O$_{10}$)(F,OH)$_{2}$   & $10^{-10}$ \\
Phlogopite    & KMg$_3$AlSi$_3$O$_{10}$F(OH)   & $10^{-10}$ \\
Nchwaningite  & Mn$^{2+}_2$SiO$_3$(OH)$_2\cdot$(H$_2$O) & $10^{-10}$ \\
\hline
\end{tabular}
\caption{Minerals considered in this work, together with their chemical compositions and the fiducial ${}^{238}$U concentrations $(C^{238})$ assumed for radiopure samples.}
\label{tab:U238_concentration}
\end{table}

\begin{figure}
    \captionsetup{justification=raggedright,singlelinecheck=false}
    \centering
    \begin{subfigure}{0.33\textwidth}
    \includegraphics[width=\linewidth]{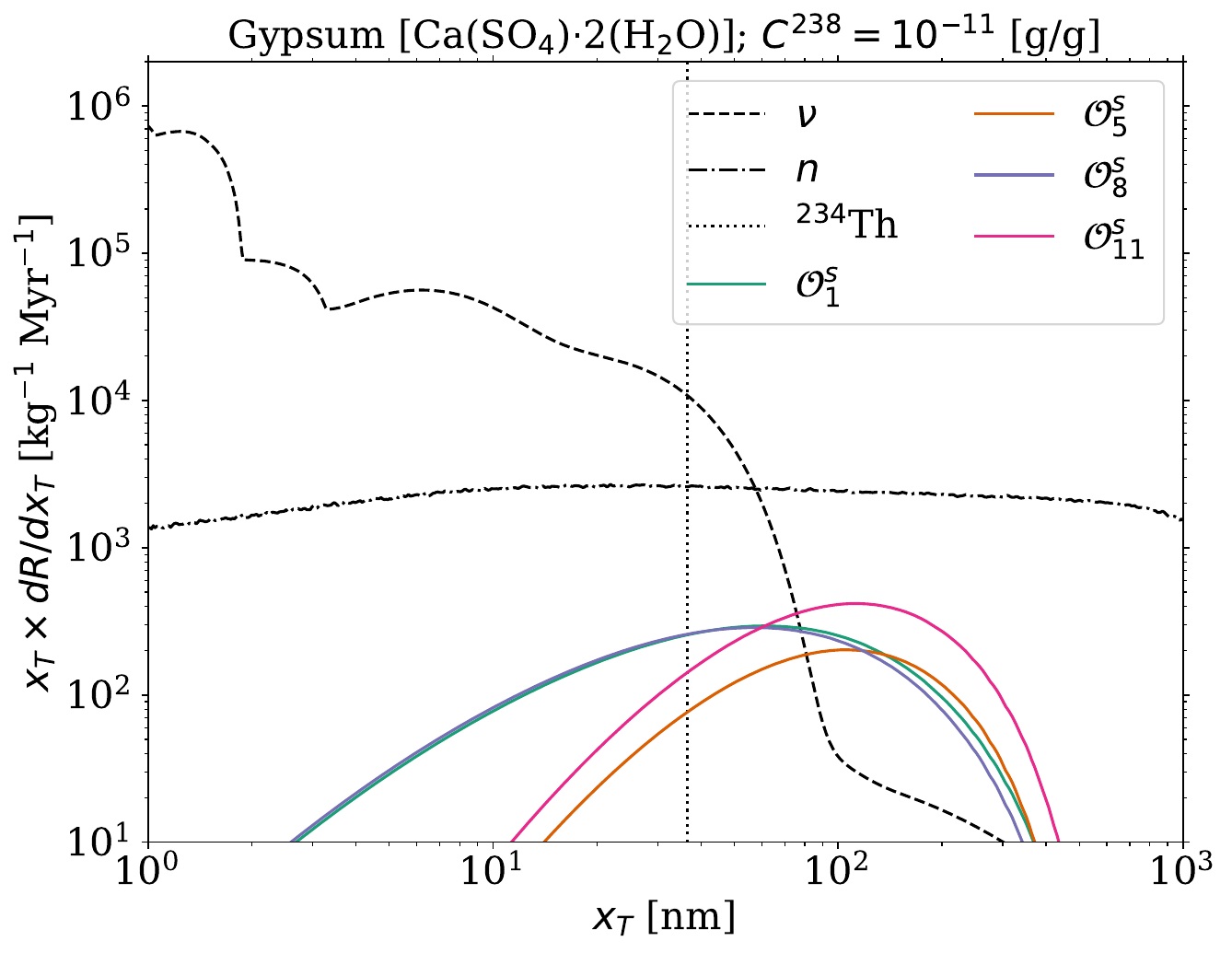}
    \end{subfigure}
    \begin{subfigure}{0.33\textwidth}
    \includegraphics[width=\linewidth]{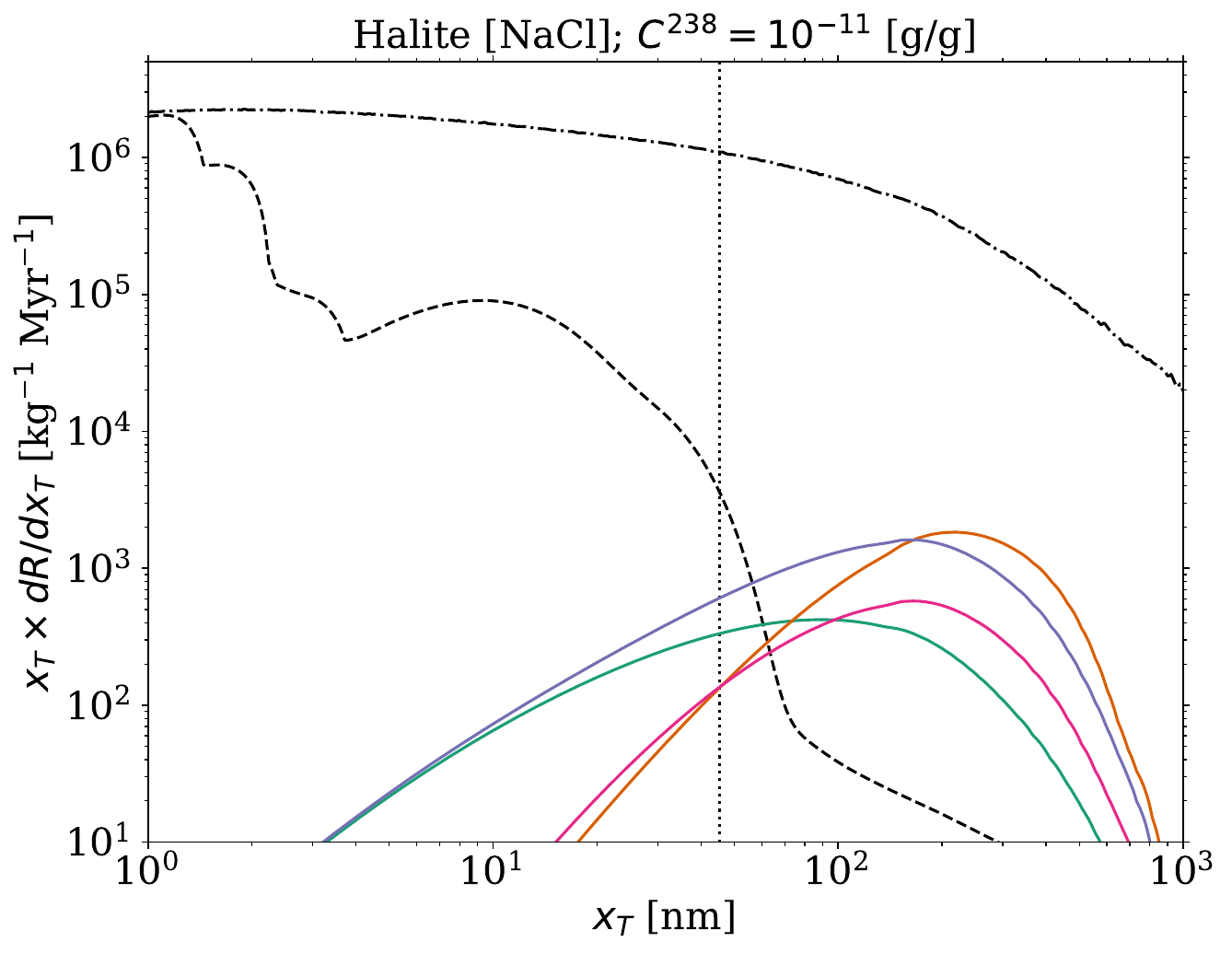}
    \end{subfigure}
    
    \vspace{0.1ex}

    \begin{subfigure}{0.33\textwidth}
    \includegraphics[width=\linewidth]{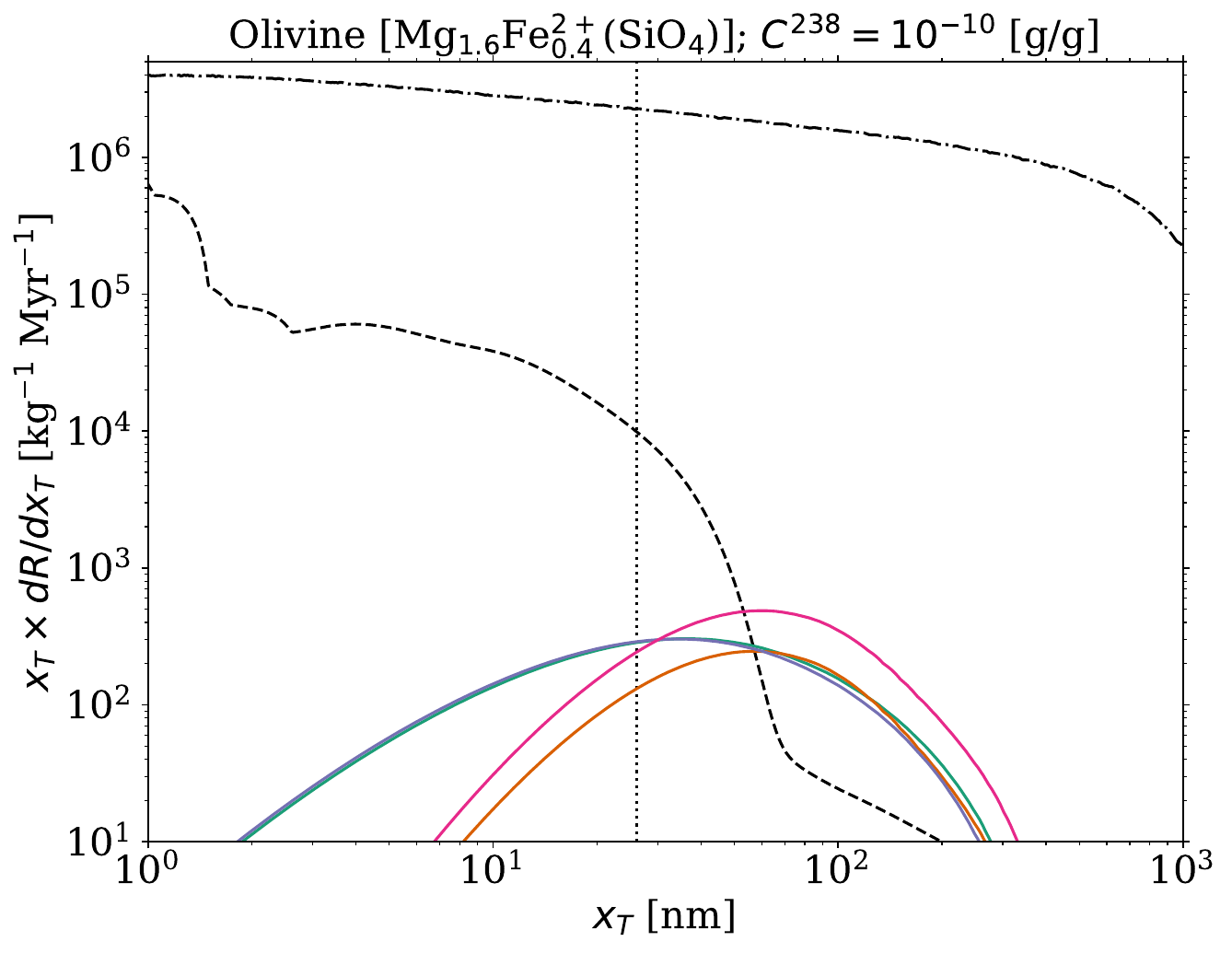}
    \end{subfigure}
    \begin{subfigure}{0.33\textwidth}
    \includegraphics[width=\linewidth]{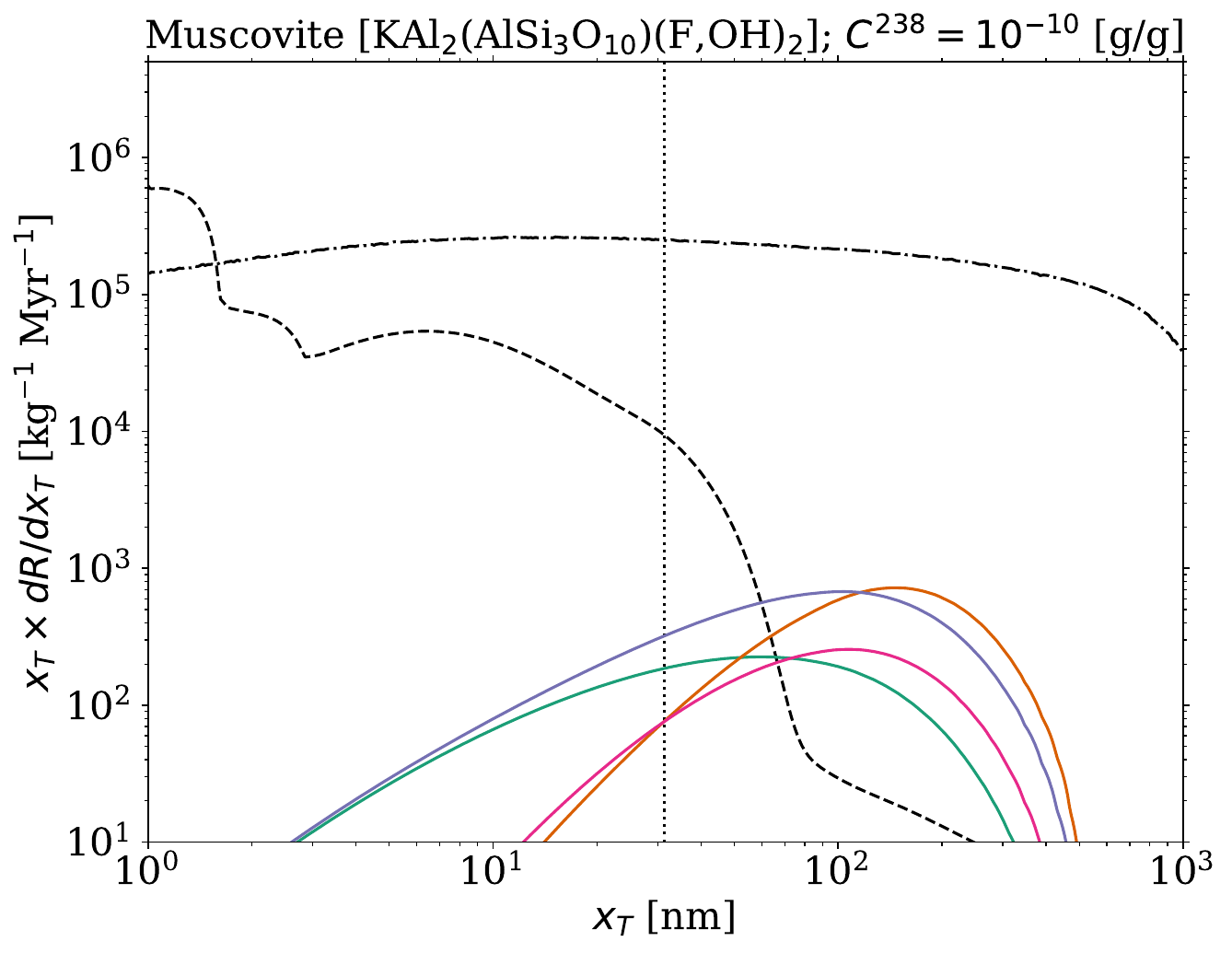}
    \end{subfigure}
    \caption{Differential track production rate $dR/d(\ln x_T)$ for elastic scattering, given per unit exposure and per logarithmic interval in track length, shown as a function of the track length $x_T$ for the spin-independent NREFT operators $\mathcal{O}_{1}^{s}$, $\mathcal{O}_{5}^{s}$, $\mathcal{O}_{8}^{s}$, and $\mathcal{O}_{11}^{s}$ and for a variety of target minerals. We note that $\mathcal{O}^{s}_{1}$ is the standard spin-independent case considered by many direct detection experiments. The calculations assume  isoscalar interactions ($c^{p}=c^{n}$). Results are shown for gypsum and halite with ${}^{238}$U concentrations of $10^{-11}\ \mathrm{g/g}$, and for olivine and muscovite with ${}^{238}$U concentrations of $10^{-10}\ \mathrm{g/g}$. The DM mass is fixed at $500\ \mathrm{GeV}/c^{2}$. For comparison, we also show background spectra induced by neutrinos ($\nu$), radiogenic neutrons ($n$), and ${}^{238}\text{U}\to{}^{234}\text{Th}+\alpha$ recoils (${}^{234}\text{Th}$); see Sec.~\ref{background}. For the normalization of the DM track production spectra, we set the NREFT coupling constants to $(c^{s}_{1}\ m_{v}^{2})^{2}=4\cdot10^{-9}$, $(c^{s}_{5}\ m_{v}^{2})^{2}=2.5$, $(c^{s}_{8}\ m_{v}^{2})^{2}=1.5\cdot10^{-2}$, and $(c^{s}_{11}\ m_{v}^{2})^{2}=5\cdot10^{-6}$, where $m_{v} = 246.2\ \mathrm{GeV}$ denotes the electroweak mass scale. These values are chosen to be compatible with the upper limits from the LUX–ZEPLIN experiment \cite{LZ:2023lvz}.}
    \label{fig:Spectrum_SI}
\end{figure}

\begin{figure}
    \captionsetup{justification=raggedright,singlelinecheck=false}
    \centering
    \includegraphics[width=0.35\textwidth]{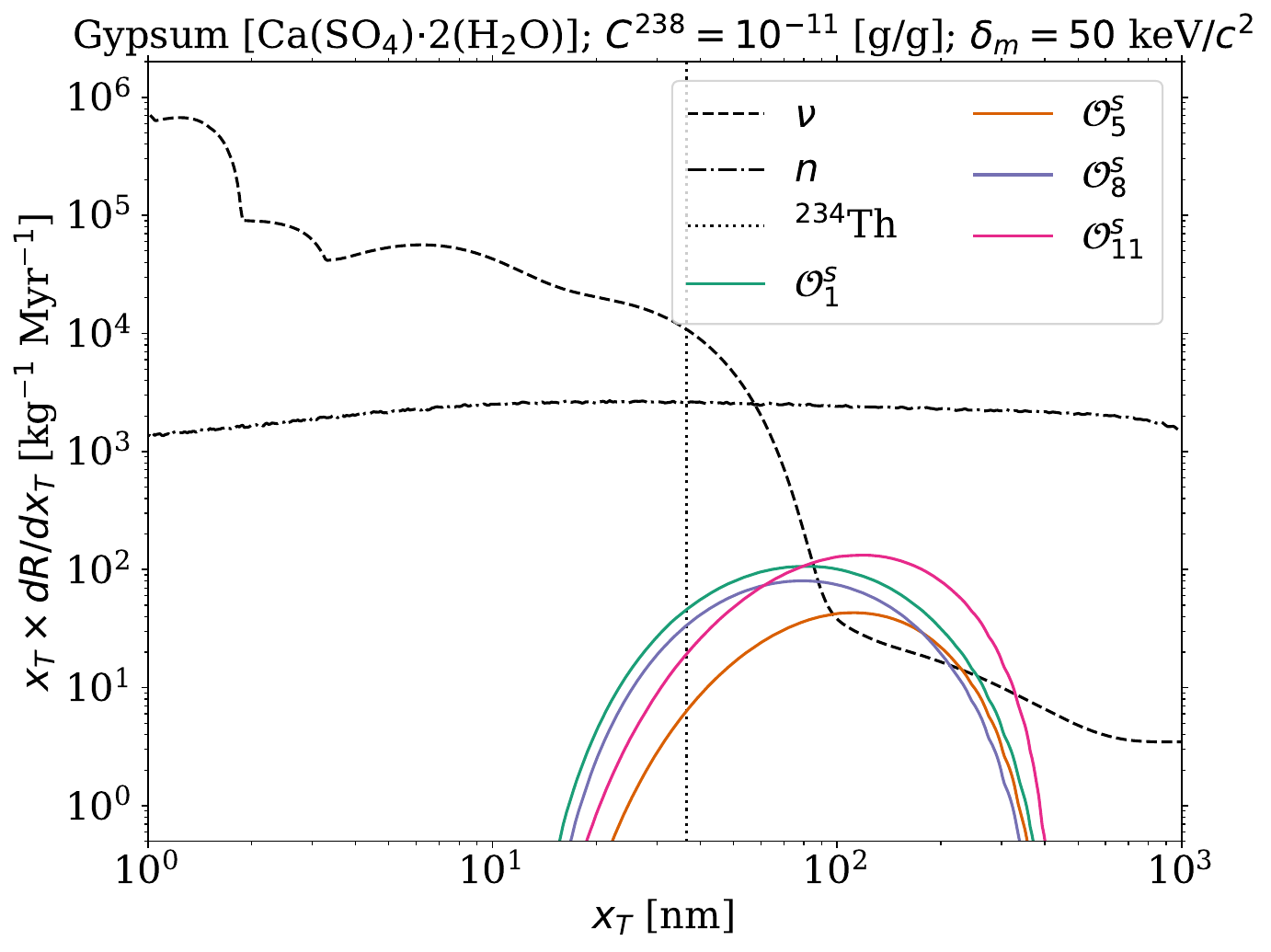}
    \includegraphics[width=0.356\textwidth]{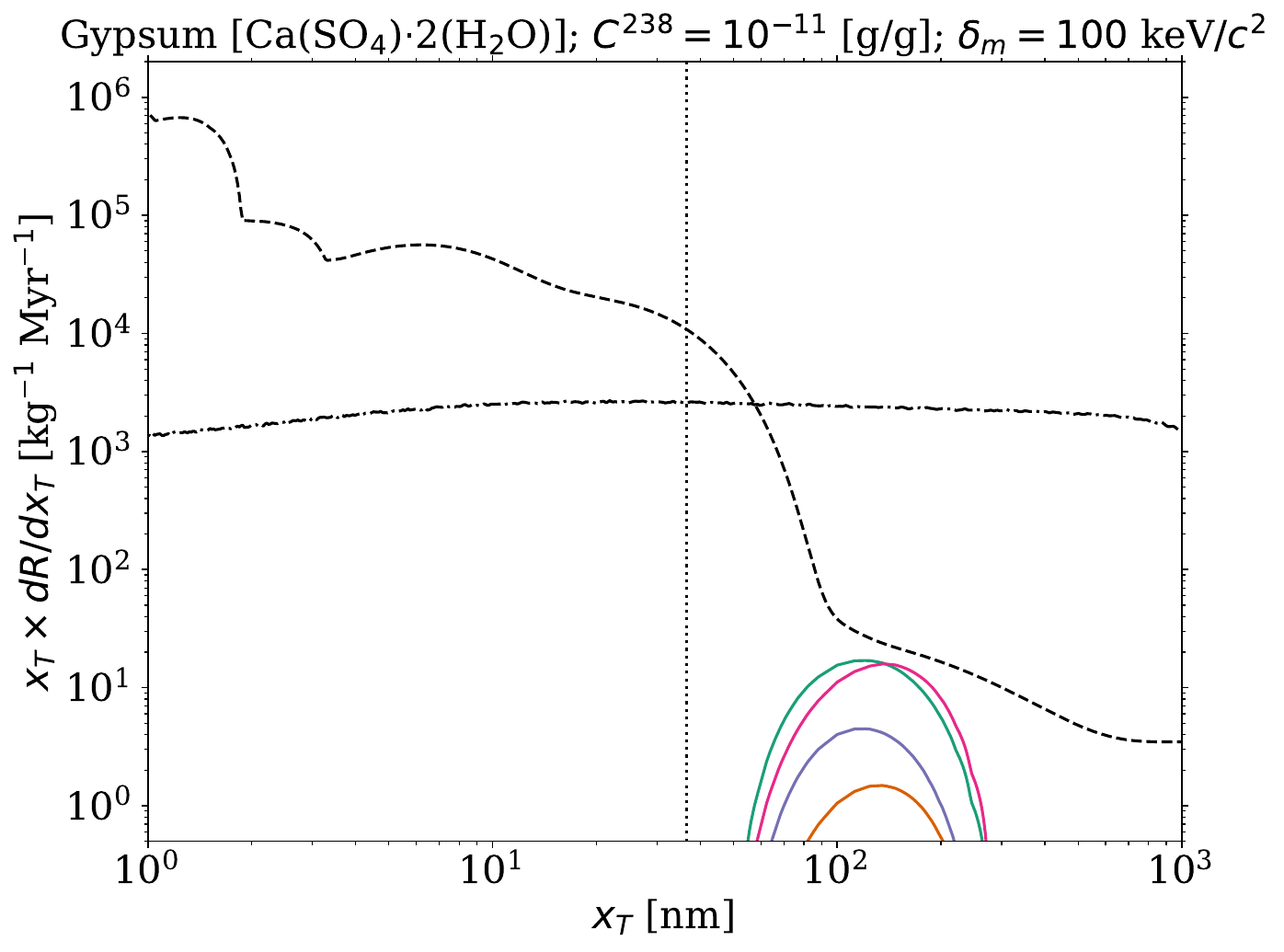}
    \caption{Differential track production rate $dR/d(\ln x_T)$ for inelastic scattering, given per unit exposure and per logarithmic interval in track length, shown as a function of the track length $x_T$ for the spin-independent NREFT operators $\mathcal{O}_{1}^{s}$, $\mathcal{O}_{5}^{s}$, $\mathcal{O}_{8}^{s}$, and $\mathcal{O}_{11}^{s}$ and for a variety of target minerals. The calculations assume isoscalar interactions ($c^{p}=c^{n}$). The DM mass is fixed at $500\ \mathrm{GeV}/c^{2}$. In the left panel, the DM mass splitting is $\delta_{m}=50\ \mathrm{keV}/c^{2}$. The NREFT coupling constants are set to $(c^{s}_{1} m_{v}^{2})^{2}=1.5\cdot10^{-8}$, $(c^{s}_{5} m_{v}^{2})^{2}=5$, $(c^{s}_{8} m_{v}^{2})^{2}=5.6\cdot10^{-2}$, and $(c^{s}_{11} m_{v}^{2})^{2}=1.2\cdot10^{-5}$, where $m_{v}=246.2\ \mathrm{GeV}$ denotes the electroweak mass scale. In the right panel, the DM mass splitting is $\delta_{m}=100\ \mathrm{keV}/c^{2}$. The coupling constants are set to $(c^{s}_{1} m_{v}^{2})^{2}=1.3\cdot10^{-7}$, $(c^{s}_{5} m_{v}^{2})^{2}=1.4\cdot10^{1}$, $(c^{s}_{8} m_{v}^{2})^{2}=3.7\cdot10^{-1}$, and $(c^{s}_{11} m_{v}^{2})^{2}=5.4\cdot10^{-5}$. In both panels, the values for the coupling constants are chosen to be compatible with the upper limits for inelastic WIMP--nucleon scattering from the LUX--ZEPLIN experiment \cite{LZ:2023lvz}. For comparison, background spectra induced by neutrinos ($\nu$), radiogenic neutrons ($n$), and ${}^{238}\text{U}\to{}^{234}\text{Th}+\alpha$ recoils (${}^{234}\text{Th}$) are also shown; see Sec.~\ref{background}. Results are presented for gypsum with a ${}^{238}$U concentration of $10^{-11}\,\mathrm{g/g}$.}
    \label{fig:Spectrum_SI_inelastic}
\end{figure}

Ultimately, the quantity of interest is the track length spectrum, which plays a role analogous to the recoil energy spectrum in conventional direct detection experiments. We first compute predictions for the differential event rate $dR/dx_T$ with respect to the nuclear track length $x_T$, defined per unit time, per unit detector mass, and per unit track length, and summed over all nuclear species $(Z,A)$ present in a given mineral. The track length $x_T$ for a nuclear recoil of energy $E_R$ is given by the integral over the stopping power in Eq.~(\ref{length}). The differential event rate in paleo-detectors is given by
\begin{equation}
    \frac{dR}{dx_T}=\sum_{(Z,A)}\xi_{(Z,A)}\left(\frac{dR}{dE_{R}}\right)_{(Z,A)}\left(\frac{dE_R}{dx_T}\right)_{(Z,A)}~,\label{dRdx}
\end{equation}
where $(dR/dE_R)_{(Z,A)}$ is given by Eq.~\eqref{recoilrate}, and the final factor $(dE_R/dx)_{(Z,A)}$ accounts for the conversion from recoil energy to track length and is evaluated using stopping power calculations from \textit{SRIM}~\cite{ZIEGLER20101818}. Figs.~\ref{fig:Spectrum_SI} and~\ref{fig:Spectrum_SI_inelastic} illustrate our results for the differential event rate for WIMP--nucleon interactions via the spin-independent NREFT operators \(\mathcal{O}_{1}^{s}\), \(\mathcal{O}_{5}^{s}\), \(\mathcal{O}_{8}^{s}\), and \(\mathcal{O}_{11}^{s}\), assuming a DM mass of \(500~\mathrm{GeV}/c^{2}\). 
Specifically, we plot $x_T\frac{dR}{dx_T}$, the differential track production rate per unit exposure and per logarithmic track length, as a function of track length $x_T$. In Fig.~\ref{fig:Spectrum_SI}, we show this quantity for elastic, isoscalar ($c^{p}=c^{n}$) interactions, for four target minerals—gypsum, halite, olivine, and muscovite--together with the background spectra discussed in Section~\ref{background}. For the normalization of the DM track production spectra, the NREFT coupling constants are chosen to be consistent with the upper limits set by the LUX–ZEPLIN experiment~\cite{LZ:2023lvz}. In Fig.~\ref{fig:Spectrum_SI_inelastic}, we present the corresponding spectra for inelastic, isoscalar interactions described by the same spin-independent NREFT operators, for a DM mass of \(500~\mathrm{GeV}/c^{2}\) and mass splittings of \(\delta_{m}=50~\mathrm{keV}/c^{2}\) (left panel) and \(\delta_{m}=100~\mathrm{keV}/c^{2}\) (right panel). Here, the NREFT coupling constants are set to values consistent with the upper limits for inelastic WIMP–nucleon scattering from the LUX–ZEPLIN experiment~\cite{LZ:2023lvz}. The DM track production spectra for the remaining NREFT operators, for both elastic and inelastic scattering, and for additional minerals, are discussed in the Appendix.


Nuclear recoil tracks in a paleo-detector sample can be read out using a variety of techniques \cite{Drukier:2018pdy}. In this work we consider two benchmark scenarios: 

\bigskip

\textbullet $\,\,\,\,\,$ {\it{High-Resolution Read-out Scenario}}~(HR): Here we assume that $10~\mathrm{mg}$ of material can be analyzed with a track length resolution of $\sigma_x = 1~\mathrm{nm}$. Such performance may be achievable with Helium Ion Beam Microscopy combined with pulsed-laser and fast-ion-beam ablation techniques \cite{HILL201265,VANGASTEL20122104,Joens2013,ECHLIN20151,PFEIFENBERGER2017109,10.1116/1.5047806}. This scenario is particularly advantageous for low-mass WIMP searches ($m_\chi \lesssim 10~\mathrm{GeV}/c^{2}$)~\cite{Baum:2021jak}. 

\bigskip

\textbullet $\,\,\,\,\,$ {\it{High-Exposure Read-out Scenario}}~(HE): Here we assume that $100~\mathrm{g}$ of material can be analyzed with $\sigma_x = 15~\mathrm{nm}$. This could be realized using Small-Angle X-ray Scattering tomography at a synchrotron facility \cite{RODRIGUEZ2014150,Schaff2015,Holler2014} and is better suited for heavier WIMPs ($m_\chi \gtrsim 10~\mathrm{GeV}/c^{2}$) \cite{Baum:2021jak}. 

\bigskip

These two read-out options match those adopted in previous paleo-detector studies \cite{Drukier:2018pdy,Baum:2018tfw,Edwards:2018hcf,Baum:2019fqm,Jordan:2020gxx,Baum:2021jak}. Although technically challenging, demonstrating the feasibility and scalability of these methods to the proposed sample masses would be an important step toward a successful paleo-detector experiment. Ongoing efforts to demonstrate the read-out of few-keV recoils using various microscopy techniques are under way at various institutions across the world~\cite{Baum:2023cct,Baum:2024eyr,Hirose:2025jht}.

One of the main results of this section is the predicted binned track length spectrum, defined as the number of tracks from WIMP--nucleus interactions binned by track length. This experimental observable incorporates the finite resolution of paleo-detectors in the sensitivity calculations, as discussed in detail in Ref.~\cite{Baum:2021jak}. The number of tracks in the $k$-th bin with length $x_T\in [x_{T,k}^{\text{min}},x_{T,k}^{\text{max}}]$ in a mineral sample of mass $M$ and age $t_{\text{age}}$, reads
\begin{equation}
    \mathcal{N}_k = M \times t_{\mathrm{age}} \int dx_T' \, W\!\left(x_T'; x_{T,k}^{\min}, x_{T,k}^{\max}\right) \frac{dR}{dx_T}(x_T')~,\label{bin}
\end{equation}
where $dR/dx_T$ is given by Eq.~(\ref{dRdx}) and is assumed to be time independent. We take the probability of measuring a track length $x_T$ from a true length $x_T'$ to be Gaussian-distributed with variance $\sigma_x^2$, the square of the read-out resolution. The corresponding window function is therefore 
\begin{equation}
    W\!\left(x_T'; x_{T,k}^{\min}, x_{T,k}^{\max}\right) = \frac{1}{2} \left[ \operatorname{erf} \left( \frac{x_{T}' - x_{T,k}^{\min}}{\sqrt{2}\,\sigma_{x_T}} \right) - \operatorname{erf} \left( \frac{x_{T}' - x_{T,k}^{\max}}{\sqrt{2}\,\sigma_{x_T}} \right) \right]~.
\end{equation}
To avoid artificial sensitivity to tracks significantly shorter than the read-out's spatial resolution, we discard all tracks with true length $x'_T < \sigma_x/2$ from the spectrum $\bigl(dR/dx_T\bigr)$ before applying Eq.~(\ref{bin}). For all numerical results presented in this paper, we use 100 logarithmically spaced bins between $\sigma_x/2$ and $1000~\mathrm{nm}$. 

\begin{figure}
    \captionsetup{justification=raggedright,singlelinecheck=false}
    \centering
    \includegraphics[width=0.40\textwidth]{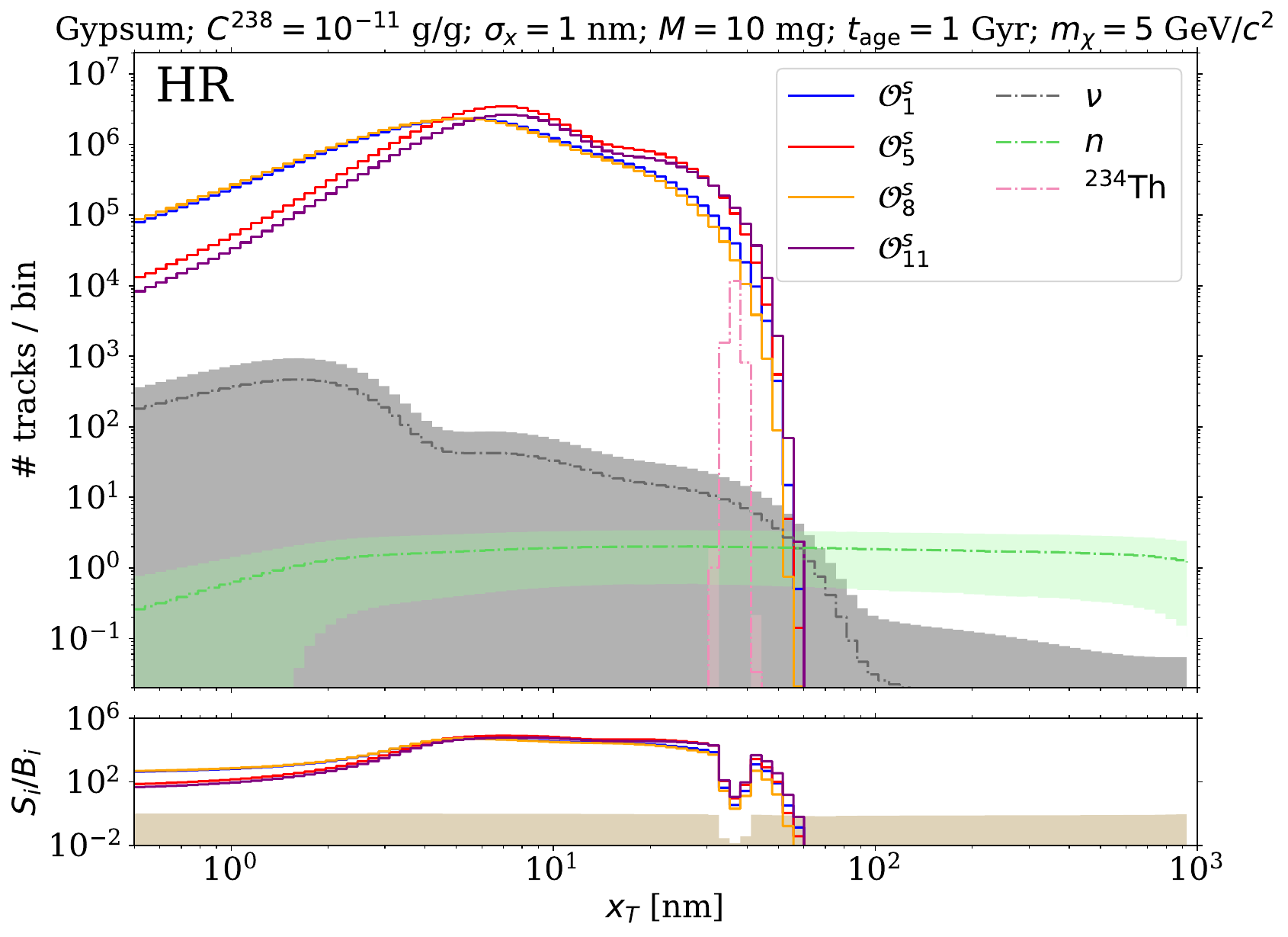}
    \includegraphics[width=0.41\textwidth]{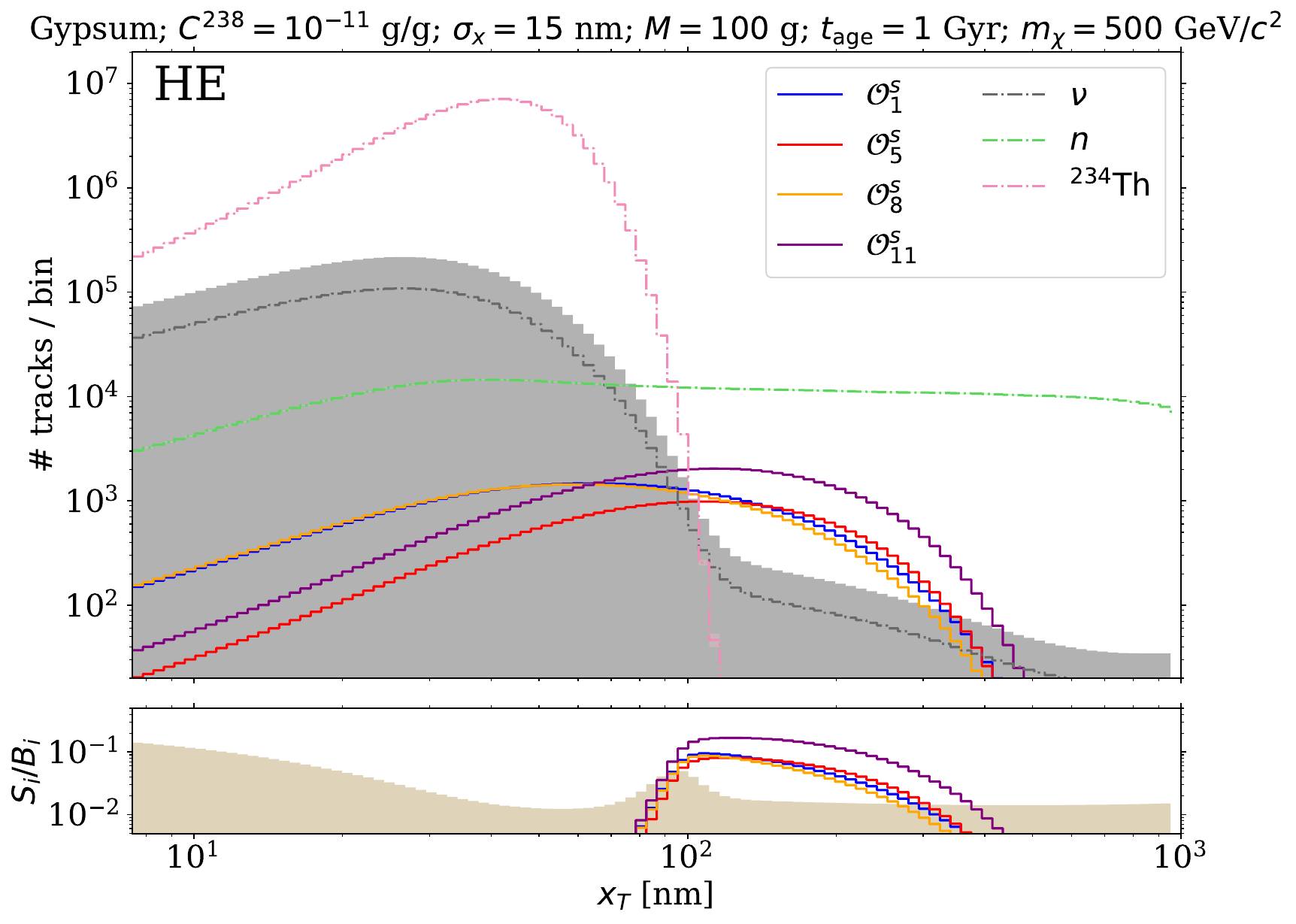}
    \caption{
    Elastic Scattering. Upper panels: track length spectra (number of tracks binned by track length)
    for the spin–independent NREFT operators $\mathcal{O}^{s}_{1}$, $\mathcal{O}^{s}_{5}$, $\mathcal{O}^{s}_{8}$, and $\mathcal{O}^{s}_{11}$ in gypsum, assuming  the case of elastic, isoscalar interactions ($c^{p}=c^{n}$). Note the DM spectra for $\mathcal{O}^{s}_{1}$ (blue line) and $\mathcal{O}^{s}_{8}$ (orange line) are nearly identical in both panels. Left panel: High Resolution (HR) scenario, with a read-out resolution of $\sigma_{x}=1\ \mathrm{nm}$ and a mineral mass of $M=10\ \mathrm{mg}$. The DM mass is set to $m_{\chi}=5\ \mathrm{GeV}/c^{2}$. For the normalization of the DM track production spectra, we set the NREFT coupling constants to $(c^{s}_{1} m_{v}^{2})^{2}=9.5\cdot10^{-3}$, $(c^{s}_{5} m_{v}^{2})^{2}=5.7\cdot10^{8}$, $(c^{s}_{8} m_{v}^{2})^{2}=3.7\cdot10^{4}$, and $(c^{s}_{11} m_{v}^{2})^{2}=4\cdot10^{2}$, where $m_{v}=246.2\ \mathrm{GeV}$ denotes the electroweak mass scale. These values are chosen to be compatible with the upper limits from the SuperCDMS experiment \cite{SuperCDMS:2022crd}. Right panel: High Exposures (HE) scenario, with $\sigma_{x}=15\ \mathrm{nm}$, $M=100\ \mathrm{g}$, $m_{\chi}=500\ \mathrm{GeV}/c^{2}$, and $(c^{s}_{1} m_{v}^{2})^{2}=4\cdot10^{-9}$, $(c^{s}_{5} m_{v}^{2})^{2}=2.5$, $(c^{s}_{8} m_{v}^{2})^{2}=1.5\cdot10^{-2}$, and $(c^{s}_{11} m_{v}^{2})^{2}=5\cdot10^{-6}$, compatible with the upper limits from the LUX--ZEPLIN experiment \cite{LZ:2023lvz}. For comparison, background spectra induced by neutrinos ($\nu$), radiogenic neutrons ($n$), and ${}^{238}\text{U}\to{}^{234}\text{Th}+\alpha$ recoils (${}^{234}\text{Th}$) are also shown; see Sec.~\ref{background}. Here, we include shaded bands around background components, representing the combined statistical (Poisson) and systematic uncertainties in the background predictions. In the HE scenario, these uncertainties are much smaller than the background event rates, so the corresponding bands are extremely narrow and effectively invisible in the plots. Bottom panels: Colored lines show the ratio of signal events from the spin–independent NREFT operators ($S_{i}$) to background events ($B_{i}$) in each bin. The sand–colored band indicates the relative uncertainty ($\delta B_i/B_i$) for the number of background events per bin, such that the ratio of the colored lines to the upper edge of the sand-colored band gives an estimate of the signal-to-noise ratio ($S_i/\delta B_i$) in each bin. Results are presented for gypsum with a ${}^{238}$U concentration of $10^{-11}\,\mathrm{g/g}$. The ratio $S_i/B_i$ is largest at shorter (longer) track lengths for lower (higher) DM masses, where neutrino (neutron) backgrounds dominate, as shown in the left (right) panel.}
    \label{fig:Spectrum_binned_SI}
\end{figure}

\begin{figure}
    \captionsetup{justification=raggedright,singlelinecheck=false}
    \centering
    \includegraphics[width=0.40\textwidth]{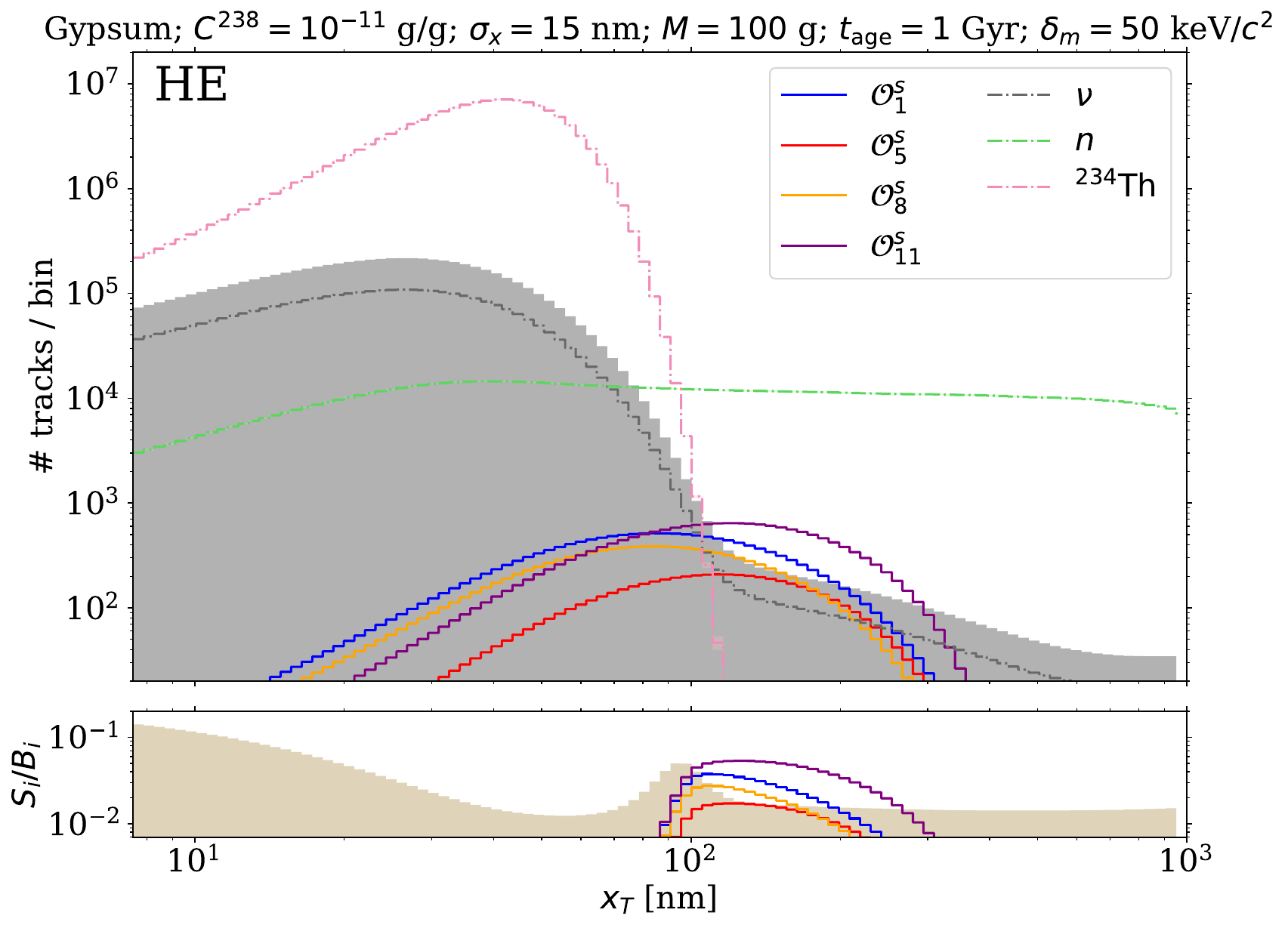}
    \includegraphics[width=0.405\textwidth]{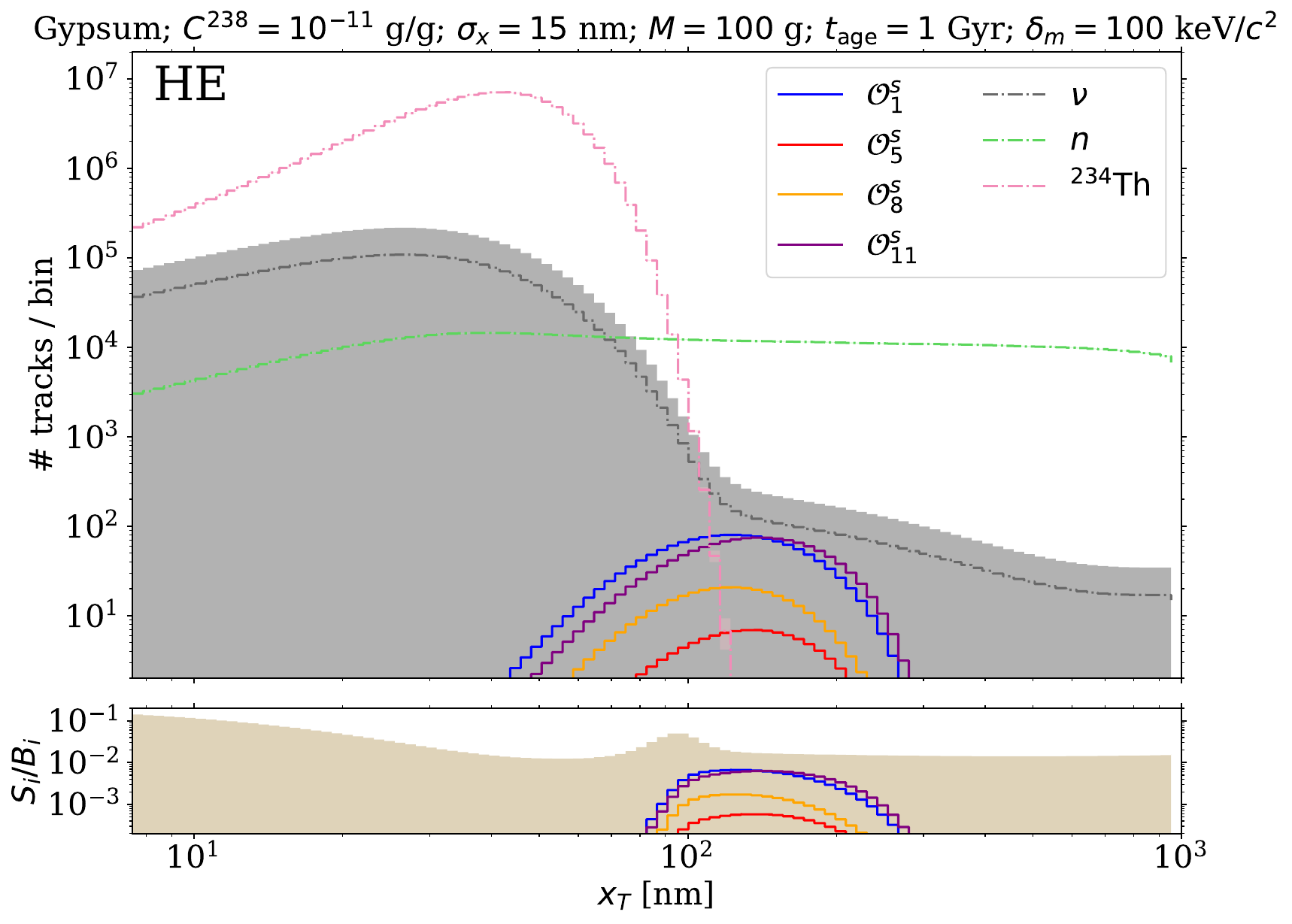}
    \caption{Inelastic Scattering. Upper panels: track length spectra (number of tracks binned by track length) for the spin–independent NREFT operators $\mathcal{O}^{s}_{1}$, $\mathcal{O}^{s}_{5}$, $\mathcal{O}^{s}_{8}$, and $\mathcal{O}^{s}_{11}$ in gypsum for the case of inelastic, isoscalar interactions ($c^{p}=c^{n}$), and a DM mass of $m_{\chi}=500\ \mathrm{GeV}/c^{2}$. The HE scenario is considered, with a read-out resolution of $\sigma_{x}=15\ \mathrm{nm}$ and a mineral mass of $M=100\ \mathrm{g}$. In the left panel, the mass splitting is $\delta_{m}=50\ \mathrm{keV}/c^{2}$. The NREFT coupling constants are set to $(c^{s}_{1} m_{v}^{2})^{2}=1.5\cdot10^{-8}$, $(c^{s}_{5} m_{v}^{2})^{2}=5$, $(c^{s}_{8} m_{v}^{2})^{2}=5.6\cdot10^{-2}$, and $(c^{s}_{11} m_{v}^{2})^{2}=1.2\cdot10^{-5}$, where $m_{v}=246.2\ \mathrm{GeV}$ denotes the electroweak mass scale. In the right panel, the mass splitting is $\delta_{m}=100\ \mathrm{keV}/c^{2}$. The coupling constants are set to $(c^{s}_{1} m_{v}^{2})^{2}=1.3\cdot10^{-7}$, $(c^{s}_{5} m_{v}^{2})^{2}=1.4\cdot10^{1}$, $(c^{s}_{8} m_{v}^{2})^{2}=3.7\cdot10^{-1}$, and $(c^{s}_{11} m_{v}^{2})^{2}=5.4\cdot10^{-5}$. In both panels, the values for the coupling constants are chosen to be compatible with the upper limits for inelastic WIMP--nucleon scattering from the LUX--ZEPLIN experiment \cite{LZ:2023lvz}. For comparison, background spectra induced by neutrinos ($\nu$), radiogenic neutrons ($n$), and ${}^{238}\text{U}\to{}^{234}\text{Th}+\alpha$ recoils (${}^{234}\text{Th}$) are also shown; see Sec.~\ref{background}. Here, we include shaded bands around background components, representing the combined statistical (Poisson) and systematic uncertainties in the background predictions. Bottom panels: Ratio of signal events from the spin–independent NREFT operators ($S_{i}$) to background events ($B_{i}$) in each bin (colored lines). The sand–colored band indicates the relative uncertainty ($\delta B_i/B_i$) for the number of background events per bin, such that the ratio of the colored lines to the upper edge of the sand-colored band gives an estimate of the signal-to-noise ratio ($S_i/\delta B_i$) in each bin. Results are presented for gypsum with a ${}^{238}$U concentration of $10^{-11}\,\mathrm{g/g}$. The ratio $S_i/B_i$ peaks at track lengths $x_T \sim 100~\mathrm{nm}$ for both mass splittings, $\delta_m = 50~\mathrm{keV}/c^{2}$ (left panel) and $\delta_m = 100~\mathrm{keV}/c^{2}$ (right panel). The signal-to-noise ratio is larger for $\delta_m = 50~\mathrm{keV}/c^{2}$, indicating improved sensitivity relative to current limits in this case.} 
    \label{fig:Spectrum_binned_SI_inelastic}
\end{figure}

Figs.~\ref{fig:Spectrum_binned_SI} and~\ref{fig:Spectrum_binned_SI_inelastic} show track length spectra defined in Eq.~(\ref{bin}) for the spin-independent operators $\mathcal{O}^{s}_{1}$, $\mathcal{O}^{s}_{5}$, $\mathcal{O}^{s}_{8}$, and $\mathcal{O}^{s}_{11}$ for the case of elastic and inelastic WIMP--nucleon scattering, respectively.  In both figures, the target mineral is taken to be  gypsum. Fig.~\ref{fig:Spectrum_binned_SI}  illustrates the binned spectra for the HR (left panel) and HE (right panel) scenarios. As noted in Ref.~\cite{Baum:2021jak}, the HR and HE scenarios are particularly sensitive to low–mass and high–mass WIMPs, respectively. Accordingly, the left panel of Fig.~\ref{fig:Spectrum_binned_SI} presents the binned spectra together with the relevant background contributions (see Sec.~\ref{background}) for a DM mass of \(5~\mathrm{GeV}/c^{2}\), using NREFT coupling constants consistent with the 95\% Bayesian credible region of the two–dimensional marginalized posterior distribution reported by SuperCDMS~\cite{SuperCDMS:2022crd}. The right panel shows the binned version of the spectra from the top-left panel of Fig.~\ref{fig:Spectrum_SI}, corresponding to a DM mass of 500~$\mathrm{GeV}$. 
In the bottom panels of Fig.~\ref{fig:Spectrum_binned_SI}, we plot with colored lines the ratio of signal events, $S_{i}$, to background events, $B_{i}$, per bin. Lighter WIMPs produce softer nuclear recoils, leading on average to shorter track lengths, as seen in the left panel of Fig.~\ref{fig:Spectrum_binned_SI}. Hence, achieving high track length resolution is essential for resolving the signal and background features in the region where the signal–to–background ratio is largest for low-mass WIMPs. Conversely, heavier WIMPs generate longer tracks on average, as shown in the right panel of Fig.~\ref{fig:Spectrum_binned_SI}. Current constraints from direct detection experiments already exclude much smaller NREFT coupling constants for \(m_{\chi} \gtrsim 10~\mathrm{GeV}/c^{2}\). Consequently, probing WIMPs in this mass range with coupling constants below existing limits requires very large exposures. In the top panels, we also include shaded bands around the background components representing the combined statistical (Poisson) and systematic uncertainties in the background predictions. In the bottom panels, we instead show a sand-colored band indicating the relative uncertainty, $\delta B_i/B_i$, in the number of background events per bin. The ratio of the colored lines to the upper edge of the sand-colored band provides an estimate of the signal-to-noise ratio, $S_i/\delta B_i$, in each bin. The signal-to-noise ratio offers a qualitative measure of the sensitivity of paleo-detectors. Further details on the background sources are given in Section~\ref{background}, while the interpretation of the sensitivity in terms of the signal-to-noise ratio is discussed in Section~\ref{sensitivity}.

Fig.~\ref{fig:Spectrum_binned_SI_inelastic} shows the binned versions of the track length spectra displayed in Fig.~\ref{fig:Spectrum_SI_inelastic}, obtained using Eq.~(\ref{bin}), for inelastic scattering in gypsum. We consider mass splittings of $\delta_m = 50~\mathrm{keV}/c^{2}$ (left panel) and $\delta_m = 100~\mathrm{keV}/c^{2}$ (right panel), assuming the HE scenario with $m_\chi = 500~\mathrm{GeV}/c^{2}$. Similar to Fig.~\ref{fig:Spectrum_binned_SI}, the bottom panels in Fig.~\ref{fig:Spectrum_binned_SI_inelastic} show, with colored curves, the ratio of signal events to background events per bin. The sand-colored bands indicate the relative uncertainty in the number of background events per bin.

In this section, we have presented predictions for two key quantities relevant to WIMP--nucleus scattering in paleo-detectors: (i) the differential event rate $dR/dx_T$, corresponding to the track production rate per unit time, per unit detector mass, and per unit track length (see Figs.~\ref{fig:Spectrum_SI} and~\ref{fig:Spectrum_SI_inelastic}), and (ii) the binned track length spectrum, which constitutes the experimentally observable quantity (see Figs.~\ref{fig:Spectrum_binned_SI} and~\ref{fig:Spectrum_binned_SI_inelastic}). Results are shown for the spin-independent NREFT operators $\mathcal{O}_{1}^{s}$, $\mathcal{O}_{5}^{s}$, $\mathcal{O}_{8}^{s}$, and $\mathcal{O}_{11}^{s}$, considering both elastic and inelastic scattering. Similar results for the differential event rate $dR/dx_T$ for the remaining NREFT operators can be found in the Appendix.

In Section~\ref{background}, we discuss the background sources relevant for paleo-detector experiments. The corresponding background predictions have already been incorporated into the figures in this section in order to facilitate a direct visual comparison between signal and background.


\section{Backgrounds in Paleo-Detectors}\label{background}

The backgrounds in DM searches with paleo-detectors are the same types as in conventional direct detection experiments, but their relative importance differs.  Paleo-detectors employ much smaller target masses ($\lesssim 0.1~\mathrm{kg}$) and integrate over far longer timescales ($t_{\mathrm{age}}\sim0.1$--$1~\mathrm{Gyr}$) without timing information for individual events. The observables in paleo-detectors are nuclear damage tracks, providing essentially perfect rejection of electronic recoils. Natural mineral defects are either single-site or span the entire crystalline volume and do not mimic nuclear tracks. Thus, only nuclear recoils constitute a relevant background. Below we briefly discuss the contributions from three background sources of nuclear recoils: cosmogenic muons, astrophysical neutrinos, and radiogenics; for detailed discussions see Refs.~\cite{Drukier:2018pdy,Baum:2019fqm}.

Cosmogenics muon backgrounds are suppressed by using minerals shielded by a large overburden while recording nuclear damage tracks, as in conventional deep-underground DM experiments. Since for a paleo-detector only a few g of target material are needed, samples can be taken from boreholes at depths greater than typical underground laboratories. At about $5~\mathrm{km}$ depth the cosmogenic-muon-induced neutron flux is $\mathcal{O}(10^{2})~\mathrm{cm}^{-2}\,\mathrm{Gyr}^{-1}$ \cite{Mei:2005gm}, making this background negligible for DM searches with paleo-detectors. At depths $\gtrsim6~\mathrm{km}$, neutron production from atmospheric neutrinos becomes comparable to the backgrounds from cosmogenic muons \cite{SNO:2009oor}. After extraction, samples can be stored near the surface; even at $\sim50~\mathrm{m}$ depth the muon-induced neutron flux remains $\lesssim0.2~\mathrm{cm}^{-2}\,\mathrm{yr}^{-1}$, leading to negligible background contributions accrued during the timescale of performing the experiment.

Astrophysical neutrinos induce nuclear damage tracks in paleo-detector samples. As noted in the introduction, neutrino-induced nuclear recoils can constitute an interesting signal for paleo-detectors~\cite{Baum:2019fqm,Jordan:2020gxx,Tapia-Arellano:2021cml}; however, in this paper  we focus on the DM detection capabilities of paleo-detectors and treat neutrinos as a background.  In our discussion, we include neutrinos from the Sun, from core-collapse supernovae, and from cosmic-ray interactions in the Earth's atmosphere \cite{OHare:2020lva}. Exposed to these neutrino fluxes for up to $\sim1~\mathrm{Gyr}$, paleo-detectors are sensitive not only to the Diffuse Supernova Neutrino Background (DSNB) from distant galaxies but also to neutrinos from local core-collapse supernovae; the Milky Way supernova rate is estimated at $2$--$3$ per century~\cite{Cappellaro:2003eg,Diehl:2006cf,Strumia:2006db,Leaman_2011,Botticella_2012,Adams:2013ana}. The DSNB spectrum and the Galactic supernova contribution are computed following Ref.~\cite{Baum:2019fqm} (see also Refs.~\cite{Adams:2013ana,Santos:2025plx,Billard:2013qya,Madau:2014bja,Strolger:2015kra}). At short track lengths ($x_T \lesssim 100~\mathrm{nm}$), the neutrino–induced background is dominated by solar neutrinos. The small variations observed in this track length region arise from the combined contributions of different solar neutrino components, including $pp$, $pep$, $hep$, $^{7}\mathrm{Be}$, $^{8}\mathrm{B}$, $^{13}\mathrm{N}$, $^{15}\mathrm{O}$, and $^{17}\mathrm{F}$ neutrinos~\cite{Baum:2023cct,Vahsen:2020pzb}. At intermediate track lengths of a few hundred nm, the dominant contribution comes from supernova neutrinos, whereas at even longer track lengths the neutrino background is primarily due to atmospheric neutrinos. 

Radiogenic backgrounds arise from trace amounts of radioactive material present in any natural mineral used as a paleo-detector. The dominant radioactive isotope for paleo-detectors is $^{238}\mathrm{U}$. Typical minerals formed in the Earth's crust contain $^{238}\mathrm{U}$ with concentrations $C^{238}\sim10^{-6}\,\mathrm{g/g}$, which would generate significant backgrounds for DM searches. To limit these backgrounds, we propose to use minerals with lower concentrations of $^{238}\mathrm{U}$, such as Ultra-Basic Rocks (UBRs), derived from Earth's mantle, and Marine Evaporites (MEs), formed from evaporated sea water ~\cite{THOMSON1954169,doi:10.1126/science.165.3888.57,ADAMS1959298,SEITZ197397,10.2110/scn.78.01.0086,1998488,article} (see also Ref.~\cite{Drukier:2018pdy} and the appendix of Ref.~\cite{Baum:2019fqm}). Following previous paleo-detector studies~\cite{Baum:2018tfw,Drukier:2018pdy,Edwards:2018hcf,Baum:2019fqm,Jordan:2020gxx,Baum:2021jak}, we adopt benchmark concentrations of $C^{238}=10^{-10}\,\mathrm{g/g}$ for UBRs and $C^{238}=10^{-11}\,\mathrm{g/g}$ for MEs. The most relevant radiogenic backgrounds arise from two processes: $\alpha$ decays and spontaneous fission, which we model following Refs.~\cite{Drukier:2018pdy,Baum:2021jak}. Regarding $\alpha$ decays, because heavy nuclei other than ${}^{238}\mathrm{U}$ have relatively short half-lives, the vast majority of $^{238}\mathrm{U}$ decays lead to the full chain of $\alpha$ decays, which are easily rejected as background with relatively large clusters of tracks leaving much more damage to the crystal than isolated nuclear recoils~\cite{PhysRevLett.74.4133}. The first background contribution due to isolated nuclear recoils comes from $\alpha$ decays of $^{238}\mathrm{U}$ which only undergo the initial decay $^{238}\mathrm{U}\to ^{234}\mathrm{Th}+\alpha$, yielding a recoil of $^{234}\mathrm{Th}$ with energy of 72 keV \cite{Collar:1995aw,Snowden-Ifft:1996dug}. These single-$\alpha$ decays lead to a population of monochromatic isolated  thorium tracks with approximate number density $n(^{234}\mathrm{Th}) \simeq 10^{6}\,\mathrm{g}^{-1} \times \left(C^{238}/10^{-11}\,\mathrm{g/g}\right)$ \cite{Drukier:2018pdy}; these decays are easy to identify as background due to the thorium pile-up in a single energy (track length) bin. The background from the $\alpha$ particles arising from the initial decay $^{238}\mathrm{U}\to ^{234}\mathrm{Th}+\alpha$ can be ignored, as these low-ionizing particles do not leave lasting damage tracks in most materials and have a range of order 10 $\mu$m in natural minerals, orders of magnitude larger than the tracks from DM-induced nuclear recoils. 

The second significant radiogenic background contribution arises from the spontaneous fission of heavy isotopes, as well as from $(\alpha,n)$ reactions in which $\alpha$ particles produced by $\alpha$ decays interact with nuclei of the target mineral, generating neutrons with MeV energies. These neutrons in turn scatter off the nuclei in the paleo-detector sample, giving rise to a broad spectrum of nuclear recoil tracks. We compute the neutron spectra using SOURCES-4A \cite{sources4a1999}, and the induced nuclear recoil spectra using TENDL-2017 \cite{KONING20122841,Rochman2016,Sublet2015,Fleming2015} neutron-nucleus cross sections accessed through the JANIS4.0 \cite{SOPPERA2014294} database. Neutron-induced backgrounds are strongly reduced in hydrogen-containing target materials. Since neutrons and protons have nearly equal masses, a neutron loses a large fraction of its energy in a single collision with hydrogen, efficiently moderating radiogenic neutrons even when hydrogen makes up only a small fraction of the target atoms \cite{Baum:2018tfw,Drukier:2018pdy}.

In Figs.~\ref{fig:Spectrum_SI} and \ref{fig:Spectrum_SI_inelastic}, we show the background track length spectra induced by neutrinos, radiogenic neutrons, and recoils of $^{234}\mathrm{Th}$. The $^{234}\mathrm{Th}$ spectrum appears as a delta function, since $^{234}\mathrm{Th}$ nuclei receive a fixed recoil energy of 72~keV following the $\alpha$ decay of $^{238}\mathrm{U}$. In contrast, radiogenic neutrons produce a broad track length distribution, with a shape and normalization dependent on the hydrogen content of the target mineral. For example, the normalization of the neutron–induced spectrum in gypsum, which contains hydrogen, is significantly lower than in halite, which lacks hydrogen, despite both minerals being assumed to have the same $^{238}\mathrm{U}$ concentration, $C^{238}=10^{-11}\,\mathrm{g/g}$. A similar trend is observed between olivine (hydrogen-free) and muscovite (hydrogen-bearing), both of which assume $C^{238}=10^{-10}\,\mathrm{g/g}$. Furthermore, for minerals with otherwise similar concentrations of hydrogen, smaller $^{238}\mathrm{U}$ concentrations correspond to reduced neutron backgrounds; for instance, the neutron background in gypsum is smaller than in muscovite, while in halite it is smaller than in olivine.

In Figs.~\ref{fig:Spectrum_binned_SI} and \ref{fig:Spectrum_binned_SI_inelastic}, in addition to the binned WIMP track length spectra, we also plot the binned background spectra. A comparison with the benchmark WIMP spectra in Fig.~\ref{fig:Spectrum_binned_SI} shows that solar neutrinos dominate the background for low-mass WIMPs (a few~GeV$/c^{2}$), while radiogenic neutrons set the sensitivity limit in searches for heavier WIMPs with $m_{\chi}\gtrsim10\ \mathrm{GeV}/c^{2}$. Due to the large exposures achievable with paleo-detectors, a substantial number of background events are expected. Consequently, a DM search using paleo-detectors differs significantly from conventional direct detection experiments, where the strategy typically involves defining a signal region with minimal or no background and searching for a few ($\mathcal{O}(1)$) signal events. In contrast, paleo-detector analyses rely on accurately modeling the distribution of background events—such as in track length space—and identifying deviations from the background–only expectation. As a result, the uncertainty in the background prediction becomes more critical than the absolute number of background events.

To illustrate this, in the top panels of Figs.~\ref{fig:Spectrum_binned_SI} and \ref{fig:Spectrum_binned_SI_inelastic}, we include shaded bands around each background component, representing both statistical (Poisson) and systematic uncertainties in the background prediction (added in quadrature). The systematic uncertainties are further discussed in Section~\ref{sensitivity}. The assumed systematic uncertainties are chosen to demonstrate how the sensitivity of paleo-detectors to DM signals of various masses depends on different background sources in a cut–and–count framework (see Refs.~\cite{Baum:2018tfw,Drukier:2018pdy,Edwards:2018hcf,Baum:2019fqm,Jordan:2020gxx,Baum:2021jak}). Nevertheless, as discussed in Ref.~\cite{Baum:2021jak}, the sensitivity in a spectral analysis is largely independent of prior assumptions for the systematic uncertainties of the background normalizations. The bottom panels of Figs.~\ref{fig:Spectrum_binned_SI} and~\ref{fig:Spectrum_binned_SI_inelastic} display, with colored curves, the ratio of the number of signal events $S_i$ to the total number of background events $B_i$ in each bin. The sand-colored region represents the relative uncertainty in the background events per bin. Specifically, the upper edge of the sand-colored region corresponds to the ratio of the total statistical and systematic uncertainty in the background events, $\delta B_i$, to the expected number of background events, $B_i$. Consequently, the ratio of the colored curves ($S_i/B_i$) to the upper edge of the sand-colored region ($\delta B_i/B_i$) provides an estimate of the signal-to-noise ratio ($S_i/\delta B_i$) in each bin. This signal-to-noise ratio offers a qualitative measure of the sensitivity of paleo-detectors: when it exceeds unity, the sensitivity of paleo-detectors are expected to outperform the corresponding direct detection experiment from which current limits are used to normalize the track length spectra shown in the plots.

\section{Projected Sensitivity of Paleo-Detectors}\label{sensitivity}

In this section, we evaluate the projected sensitivity of paleo-detectors to WIMP-nucleon interactions. Because of the large exposure of paleo-detectors, any mineral sample will contain a very large number of background events.  In some cases large numbers of signal events are predicted as well, possibly even more than background events. To search for dark matter, a spectral analysis of the recoil track length distribution can be performed, testing whether it is better described by a background-only model or by a combined background-plus-signal hypothesis. Thus, a profile-likelihood–ratio approach, also used in Ref. \cite{Baum:2021jak}, is employed to project the sensitivity of paleo-detectors to a DM signal through spectral analysis.

The  binned track-length spectrum (specifically the number of tracks in the $k^{th}$ bin) in a paleo-detector, including backgrounds and DM signal, reads
\begin{equation}
\begin{aligned}
    \mathcal{N}_k(\vec{\theta}; c_j, m_{\chi}) &=
    \mathcal{N}_{\nu,k}^{\mathrm{sol}}\!\left(\Phi_{\nu}^{\mathrm{sol}}\right)
    + \mathcal{N}_{\nu,k}^{\mathrm{DSNB}}\!\left(\Phi_{\nu}^{\mathrm{DSNB}}\right)
    +\mathcal{N}_{\nu,k}^{\mathrm{GSNB}}\!\left(\Phi_{\nu}^{\mathrm{GSNB}}\right)
    + \mathcal{N}_{\nu,k}^{\mathrm{atm}}\!\left(\Phi_{\nu}^{\mathrm{atm}}\right) \\
    &\quad + \mathcal{N}_{\mathrm{rad},k}^{^{234}\mathrm{Th}}\!\left(C^{238}\right)
    + \mathcal{N}_{\mathrm{rad},k}^{n}\!\left(C^{238}\right)
    + \mathcal{N}_{\mathrm{DM},k}\!\left(c_j, m_{\chi}\right) ~,
\end{aligned}
\end{equation}
where $\mathcal{N}_{s,k}$ are the binned track length spectra given by Eq.~(\ref{bin}) for the various signal and background contributions, different $\Phi_{\nu}$ denote the various astrophysical neutrino fluxes, $C^{238}$ is the concentration of $U^{238}$, $c_j$ is the coupling constant of a NREFT interaction, $m_{\chi}$ is the DM mass, and $\vec{\theta}$ is the set of nuisance parameters
\begin{equation}
    \vec{\theta}=\left\{M, t_{\text{age}},\Phi_{\nu}^{\mathrm{sol}},\Phi_{\nu}^{\mathrm{DSNB}},\Phi_{\nu}^{\mathrm{GSNB}},\Phi_{\nu}^{\mathrm{atm}},C^{238}\right\}~,
\end{equation}
with $M$ and $t_{\text{age}}$ the mass and age of the mineral sample, respectively. The neutrino contributions are induced by solar neutrinos ($\mathcal{N}_\nu^{\mathrm{sol}}$), the DSNB ($\mathcal{N}_\nu^{\mathrm{DSNB}}$), the Galactic Supernova Neutrino Background ($N_\nu^{\mathrm{GSNB}}$), and atmospheric neutrinos ($\mathcal{N}_\nu^{\mathrm{atm}}$). The radiogenics contributions come from two sources: (i) the isolated $^{234}$Th decays ($\mathcal{N}_{\mathrm{rad}}^{^{234}\mathrm{Th}}$) from $^{238}$U nuclei which have undergone only the initial $\alpha$ decay during the time the sample has been recording nuclear damage tracks; and (ii) the radiogenic neutrons ($\mathcal{N}_{\mathrm{rad}}^{n}$). The DM signal $\mathcal{N}_{\mathrm{DM},k}\!\left(c_j, m_{\chi}\right)$ depends on the NREFT coupling constant $c_j$, and the DM mass $m_{\chi}$. In the case of $\vec{S}_{\chi}$-dependent NREFT operators, the DM signal also depends on the spin of the DM particle, which is considered to be $1/2$ in this study. All contributions are proportional to $M$, and all contributions except for $\mathcal{N}_{\mathrm{rad},k}^{^{234}\mathrm{Th}}$ are proportional to $t_{\mathrm{age}}$.

To facilitate comparison with current limits and projected sensitivities of conventional direct detection experiments, we compute the 90\% confidence level projected upper (exclusion) limits on the NREFT coupling constants as a function of $m_\chi$ for a paleo-detector. Paleo-detectors are a counting experiment, and hence we expect the data to be Poisson-distributed in each bin. In this case, the log-likelihood to observe a data set $\mathbf{D}$ given the parameter set $\left\{\vec{\theta},c_j,m_{\chi}\right\}$ is
\begin{equation}
    \ln \mathcal{L}_{\text{Poisson}}\bigl(\mathbf{D}\,|\,\vec{\theta}; c_j, m_{\chi}\bigr)
    = \sum_{k} \left[ D_{k} \ln \mathcal{N}_{k}\bigl(\vec{\theta}; c_j, m_{\chi}\bigr)
    - \mathcal{N}_{k}\bigl(\vec{\theta}; c_j, m_{\chi}\bigr) \right]~,
\end{equation}
where we ignore constant factors in the expression of the likelihood, which cancel in the likelihood ratio we are ultimately interested in. The nuisance parameters are constrained by other measurements. For instance, neutrino fluxes are determined from existing neutrino experiments and theoretical models, while the sample's mass, age, and uranium content can be directly measured. In a frequentist framework, these external constraints can be incorporated by jointly fitting the ancillary measurements together with the track length spectrum data. To represent this, we include Gaussian constraints on the nuisance parameters
\begin{equation}
\ln \mathcal{L}_{\text{ext. const.}}\left(\vec{\theta}\right)
= -\frac{1}{2}\sum_{i}
\left( \frac{\theta_{i} - \bar{\theta}_{i}}{\ell_{i}\,\bar{\theta}_{i}} \right)^{2} ~,
\end{equation}
where $\bar{\theta}_{i}$ denotes the central value of the $i$-th parameter inferred from ancillary data, and $\ell_{i}$ is its relative uncertainty. Thus, the full likelihood reads
\begin{equation}
\ln \mathcal{L}\bigl(\mathbf{D}\,|\,\vec{\theta}; c_j, m_{\chi}\bigr)
= \ln \mathcal{L}_{\text{Poisson}}\bigl(\mathbf{D}\,|\,\vec{\theta}; c_j, m_{\chi}\bigr)
+ \ln \mathcal{L}_{\text{ext. const.}}\bigl(\vec{\theta}~\bigr).
\end{equation}
To estimate the projected sensitivity of paleo-detectors, we use the maximum log-likelihood ratio~\cite{Cowan:2010js,Billard_2012,Conrad:2014nna}:
\begin{equation}
q(c_j; m_{\chi})
= -2 \ln \left[
\frac{ \mathcal{L}\left( \mathbf{D} \big| \hat{\hat{\vec{\theta}}}; c_j, m_{\chi} \right)}
{ \mathcal{L}\left( \mathbf{D} \big| \hat{\vec{\theta}}; \hat{c}_j, m_{\chi} \right)}
\right]~,
\end{equation}
where $\hat{\hat{\vec{\theta}}}$ denotes the set of nuisance parameters that maximizes the likelihood $\mathcal{L}$ in the numerator for fixed values of the NREFT coupling constant $c_j$ and the DM mass $m_{\chi}$. In the denominator, $\mathcal{L}$ is maximized over both $\vec{\theta}$ and $c_j$, for a fixed $m_{\chi}$, yielding the best-fit values $\hat{\vec{\theta}}$ and $\hat{c}_j$. To obtain the projected exclusion limit, for the case of no DM signal (all $c_j = 0$), we adopt the \emph{Asimov} data set \cite{Cowan:2010js} 
\begin{equation}
D_i = \mathcal{N}_{i}\left(\bar{\vec{\theta}}; c_j = 0 \right),
\end{equation}
where $\bar{\theta}$ represents a fiducial choice of nuisance parameters. The 90\% confidence level exclusion limit for a given $m_{\chi}$ is the smallest $c_j$ such that $q(c_j; m_{\chi}) \ge q_{\mathrm{crit}} = 2.71$. According to Wilks' theorem~\cite{Wilks:1938dza}, $q$ is asymptotically $\chi^2$ distributed; for an one-dimensional $\chi^2$ distribution, $\chi^2 = 2.71$ corresponds to a $p$-value of 0.1.

In our analysis, for the HR and HE scenarios, we adopt a sample mass of $\bar{M} = 10\,$mg and $\bar{M} = 100\,$g, respectively, and for both cases we assume $\bar{t}_{\text{age}} = 1\,$Gyr for the time the sample has been recording damage tracks. These parameters are treated as known external constraints with relative uncertainties $\ell_{M}=0.01\%$ and $\ell_{t_{\text{age}}}=5\%$. Whereas the mass of a target sample can be measured with high precision, its age can typically be determined to within a few percent using standard geological dating methods \cite{Gradstein2012,Gallagher,vandenHaute1998}. The fiducial values for the solar and atmospheric neutrino fluxes are taken from \cite{OHare:2020lva}, while the DSNB spectrum and the contribution from Galactic supernovae are calculated as in \cite{Baum:2019fqm}. The ${}^{238}\mathrm{U}$ concentration in the target minerals are summarized in Table~\ref{tab:U238_concentration}. As a benchmark, we assume the associated nuisance parameters are constrained to $\ell_{\Phi^i_\nu}=100\%$ and $\ell_{C^{238}}=1\%$, matching the systematic uncertainties of the backgrounds considered in earlier studies~\cite{Baum:2018tfw,Drukier:2018pdy,Edwards:2018hcf,Baum:2019fqm,Jordan:2020gxx}. Nevertheless, as discussed in Ref. \cite{Baum:2021jak}, the projected sensitivity of paleo-detectors to a DM signal is largely unaffected by these external constraints on the nuisance parameters. 

\begin{figure}
    \captionsetup{justification=raggedright,singlelinecheck=false}
    \centering
    \begin{subfigure}{0.30\textwidth}
    \includegraphics[width=\linewidth]{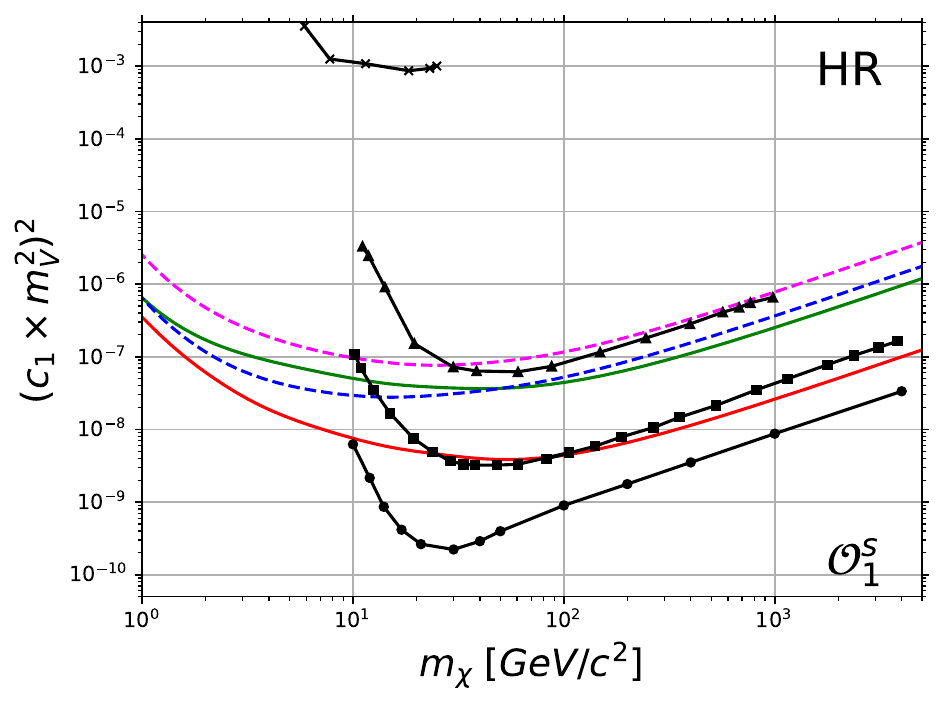}
    \end{subfigure}
    \begin{subfigure}{0.30\textwidth}
    \includegraphics[width=\linewidth]{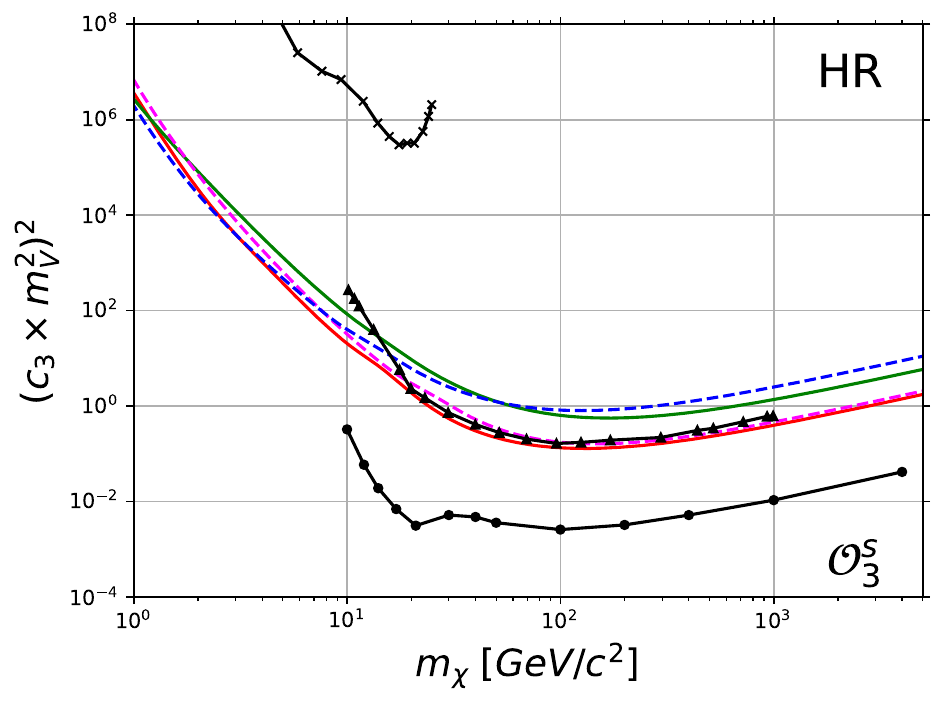}
    \end{subfigure}
    \begin{subfigure}{0.30\textwidth}
    \includegraphics[width=\linewidth]{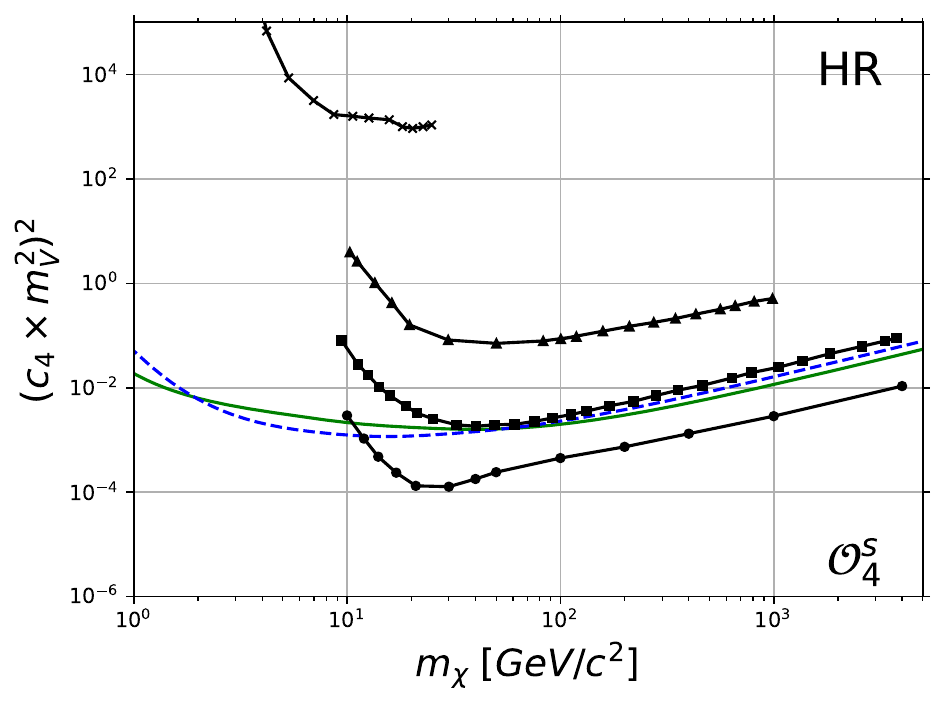}
    \end{subfigure}

    \vspace{0.1ex}

    \begin{subfigure}{0.30\textwidth}
    \includegraphics[width=\linewidth]{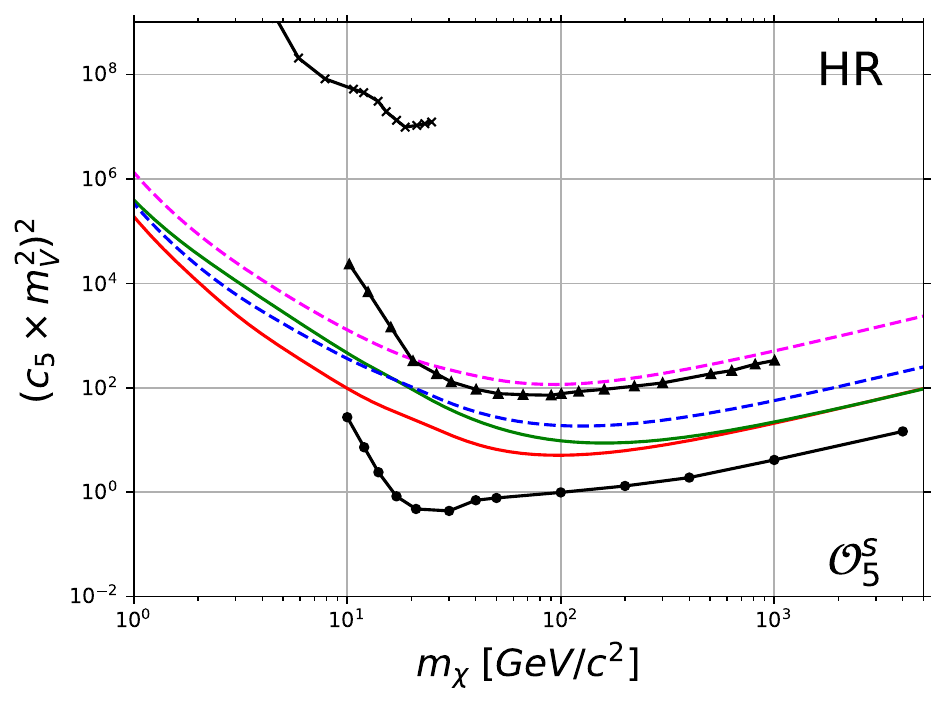}
    \end{subfigure}
    \begin{subfigure}{0.30\textwidth}
    \includegraphics[width=\linewidth]{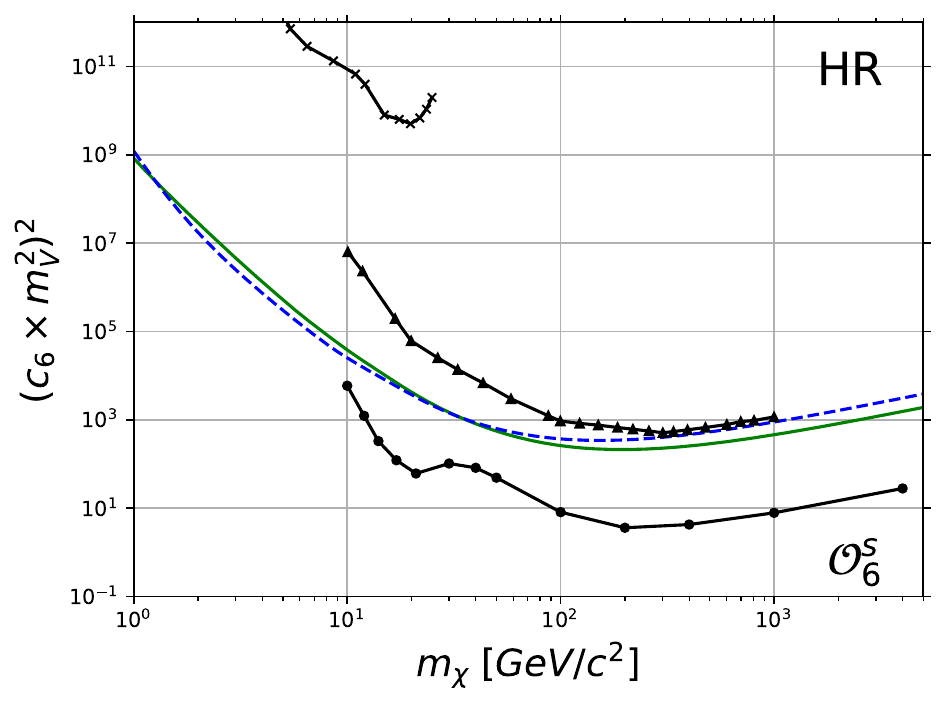}
    \end{subfigure}
    \begin{subfigure}{0.30\textwidth}
    \includegraphics[width=\linewidth]{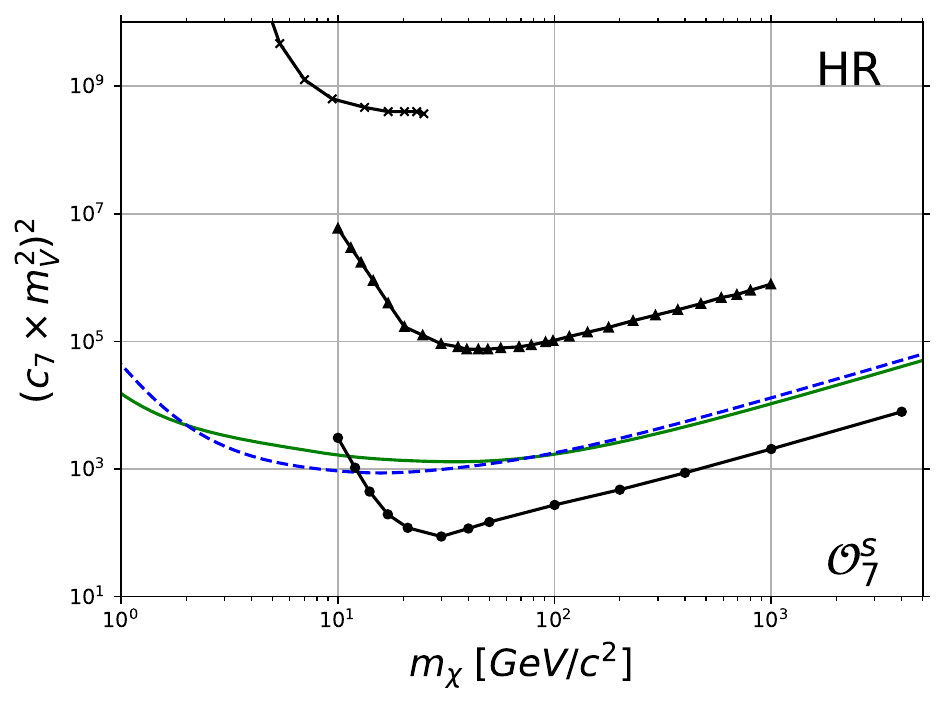}
    \end{subfigure}

    \vspace{0.1ex}

    \begin{subfigure}{0.30\textwidth}
    \includegraphics[width=\linewidth]{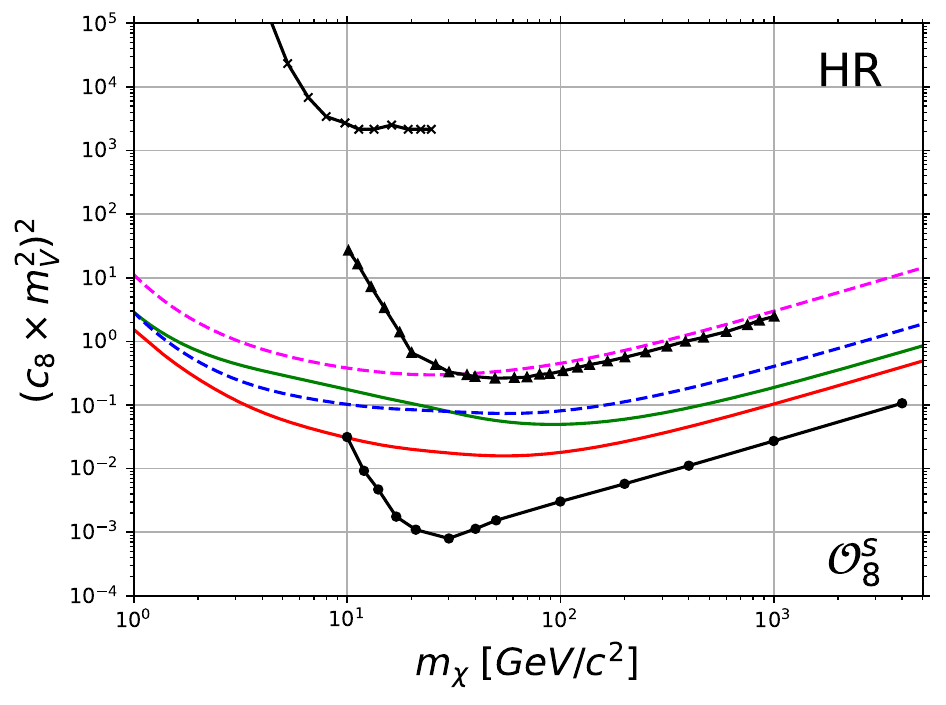}
    \end{subfigure}
    \begin{subfigure}{0.30\textwidth}
    \includegraphics[width=\linewidth]{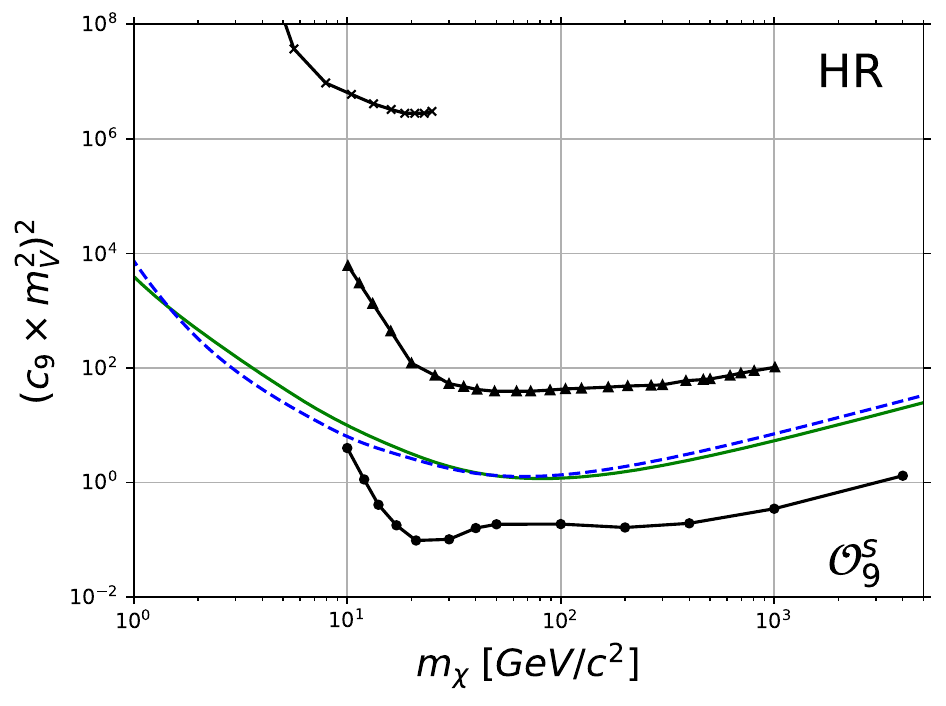}
    \end{subfigure}
    \begin{subfigure}{0.30\textwidth}
    \includegraphics[width=\linewidth]{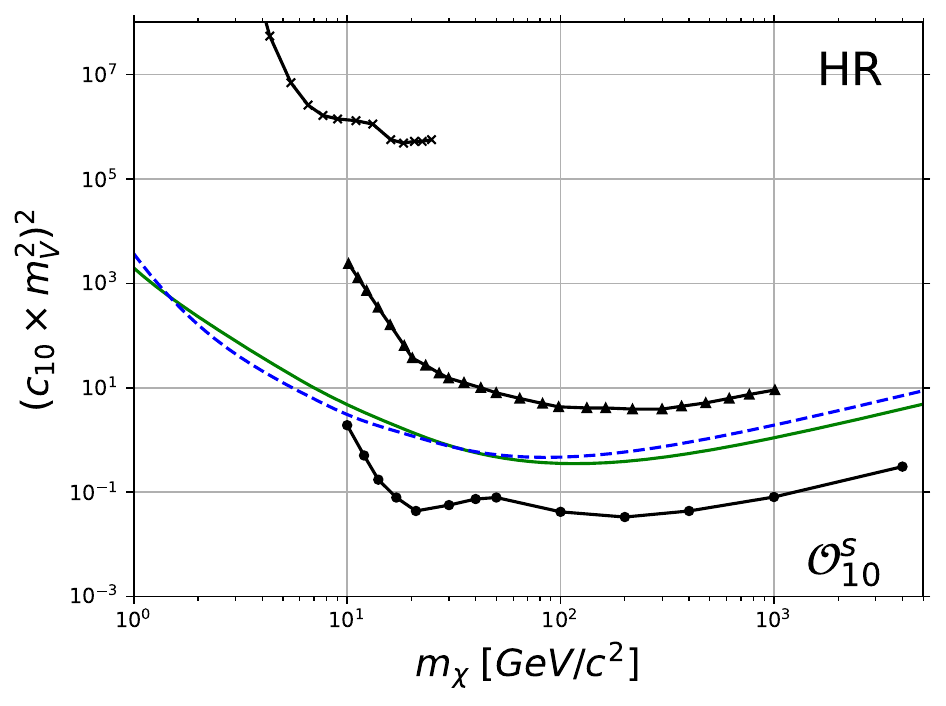}
    \end{subfigure}

    \vspace{0.1ex}

    \begin{subfigure}{0.30\textwidth}
    \includegraphics[width=\linewidth]{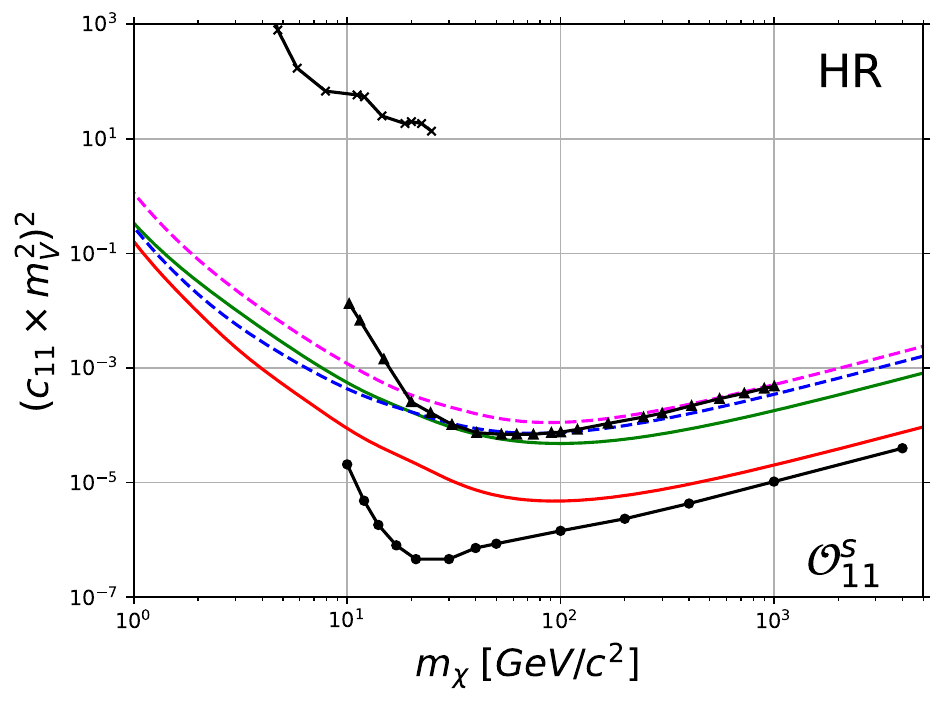}
    \end{subfigure}
    \begin{subfigure}{0.30\textwidth}
    \includegraphics[width=\linewidth]{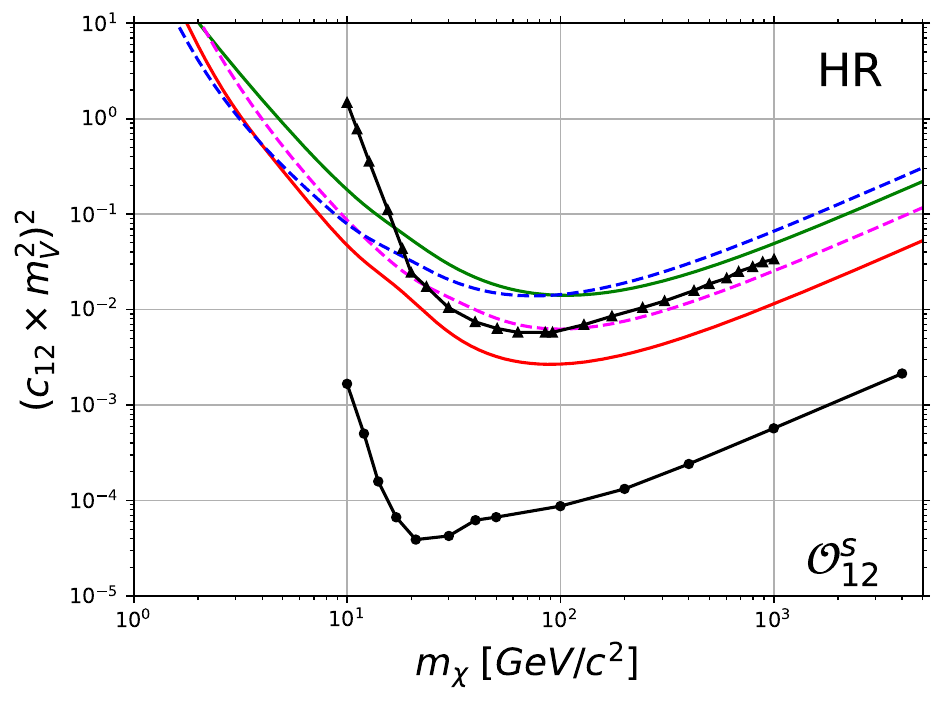}
    \end{subfigure}
    \begin{subfigure}{0.30\textwidth}
    \includegraphics[width=\linewidth]{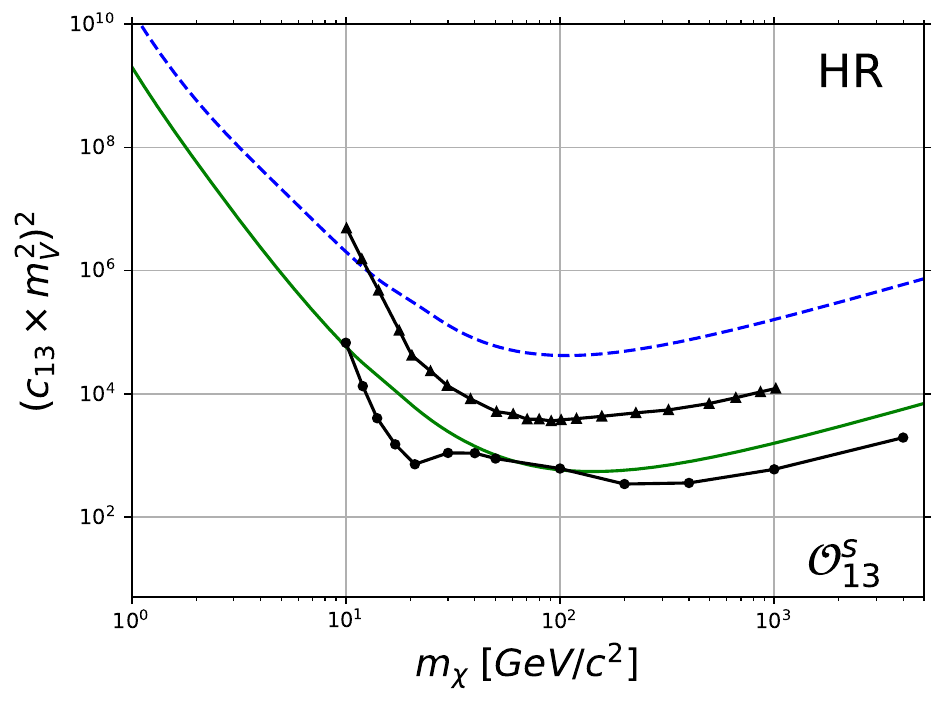}
    \end{subfigure}

    \vspace{0.1ex}

    \begin{subfigure}{0.30\textwidth}
    \includegraphics[width=\linewidth]{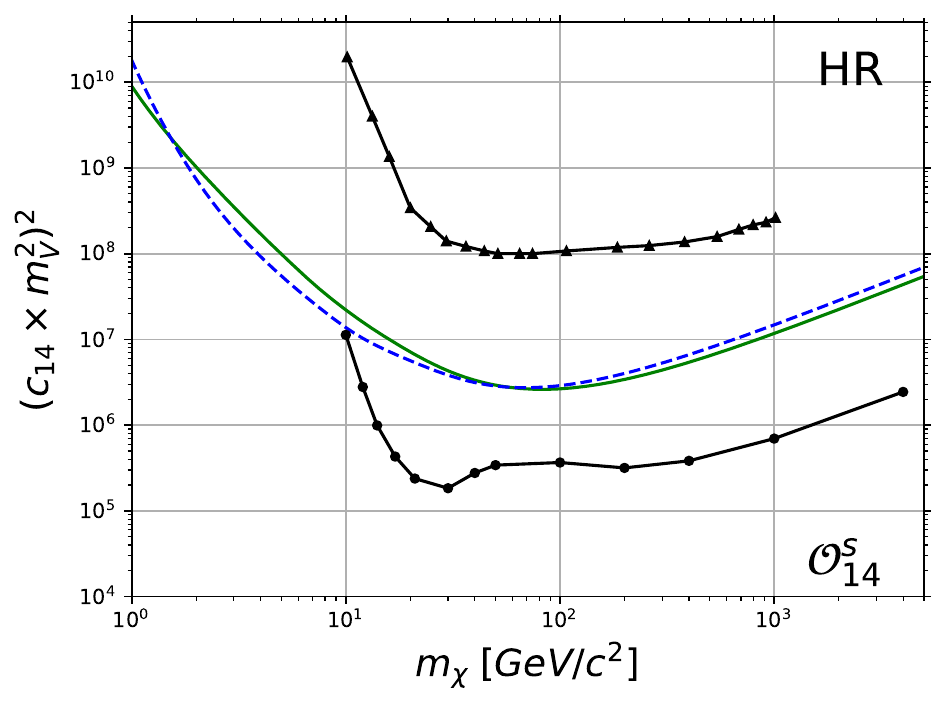}
    \end{subfigure}
    \begin{subfigure}{0.30\textwidth}
    \includegraphics[width=\linewidth]{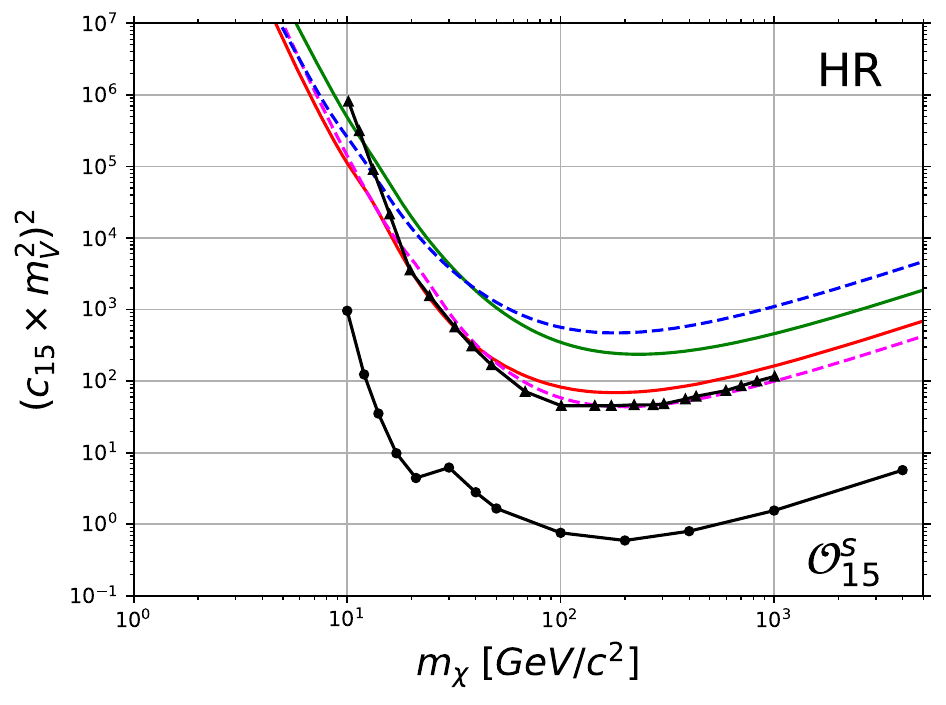}
    \end{subfigure}\hspace{4.85em}
    \begin{subfigure}{0.21\textwidth}
    \includegraphics[width=\linewidth]{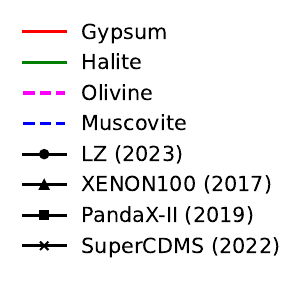} 
    \end{subfigure}
    \caption{Projected 90\% confidence level upper limits on the dimensionless isoscalar WIMP--nucleon NREFT coupling constants for elastic scattering, assuming a small exposure $M=10$ mg and $t_{ \text{age}} = 1$ Gyr, and a high read-out resolution $\sigma_{x}=1$ nm. The solid (dashed) lines indicate minerals with $C^{238}=10^{-11}$ g/g ($C^{238}=10^{-10}$ g/g), see Table \ref{tab:U238_concentration}. Black lines show the NREFT results from conventional direct detection experiments: the 90\% confidence level upper limits from XENON100 \cite{XENON:2017fdd}, LUX--ZEPLIN \cite{LZ:2023lvz}, PandaX--II \cite{PandaX-II:2018woa}, as well as the 95\% Bayesian credible region of the two-dimensional marginalized posterior distribution from SuperCDMS \cite{SuperCDMS:2022crd}. For elastic scattering, the HR scenario illustrates sensitivity of paleo-detectors to WIMP--nucleon cross sections below existing experimental direct detection bounds for $m_\chi \lesssim 10~\mathrm{GeV}/c^2$.}
    \label{fig:HR}
\end{figure}

\begin{figure}
    \captionsetup{justification=raggedright,singlelinecheck=false}
    \centering
    \begin{subfigure}{0.30\textwidth}
    \includegraphics[width=\linewidth]{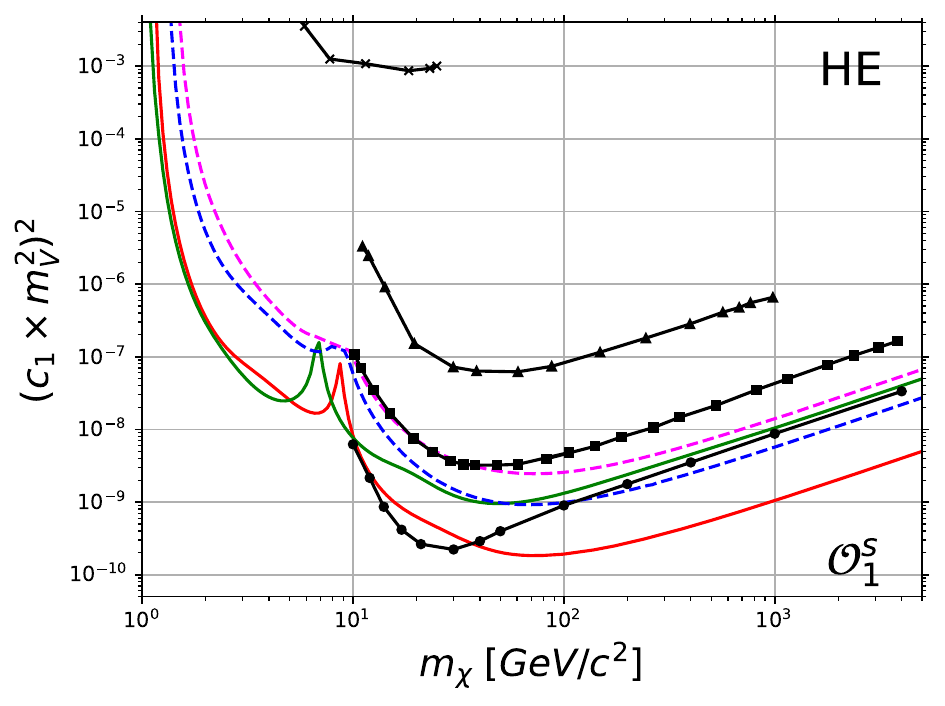}
    \end{subfigure}
    \begin{subfigure}{0.30\textwidth}
    \includegraphics[width=\linewidth]{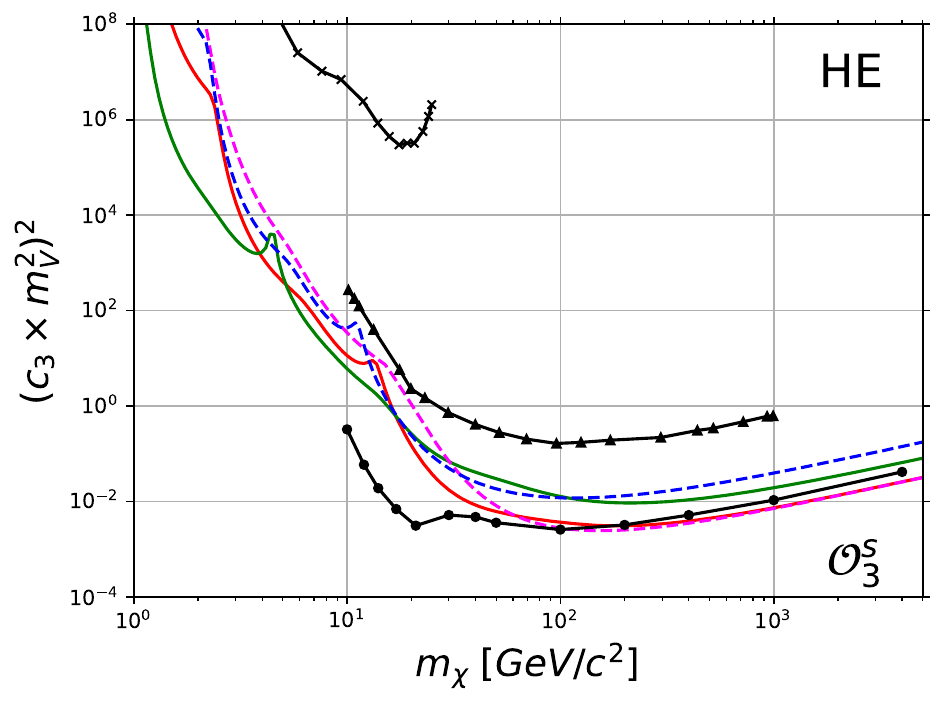}
    \end{subfigure}
    \begin{subfigure}{0.30\textwidth}
    \includegraphics[width=\linewidth]{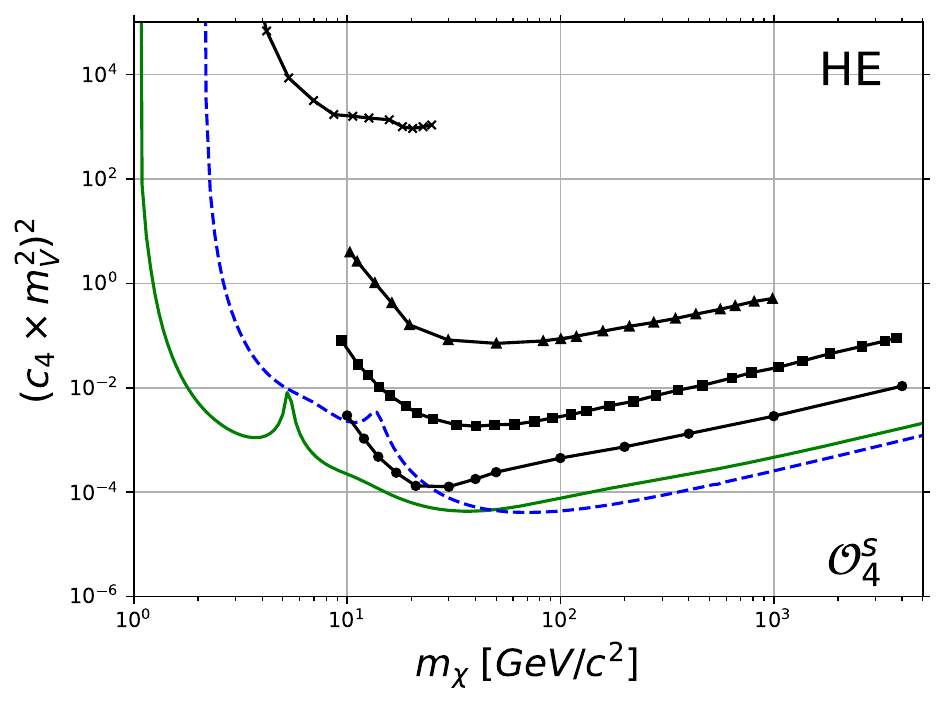}
    \end{subfigure}

    \vspace{0.1ex}

    \begin{subfigure}{0.30\textwidth}
    \includegraphics[width=\linewidth]{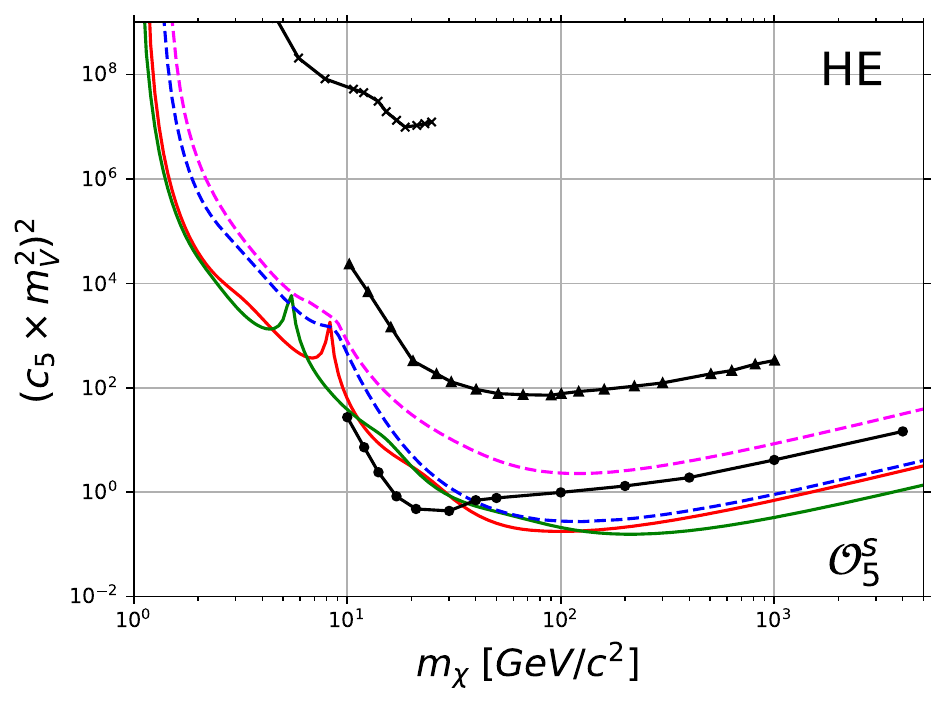}
    \end{subfigure}
    \begin{subfigure}{0.30\textwidth}
    \includegraphics[width=\linewidth]{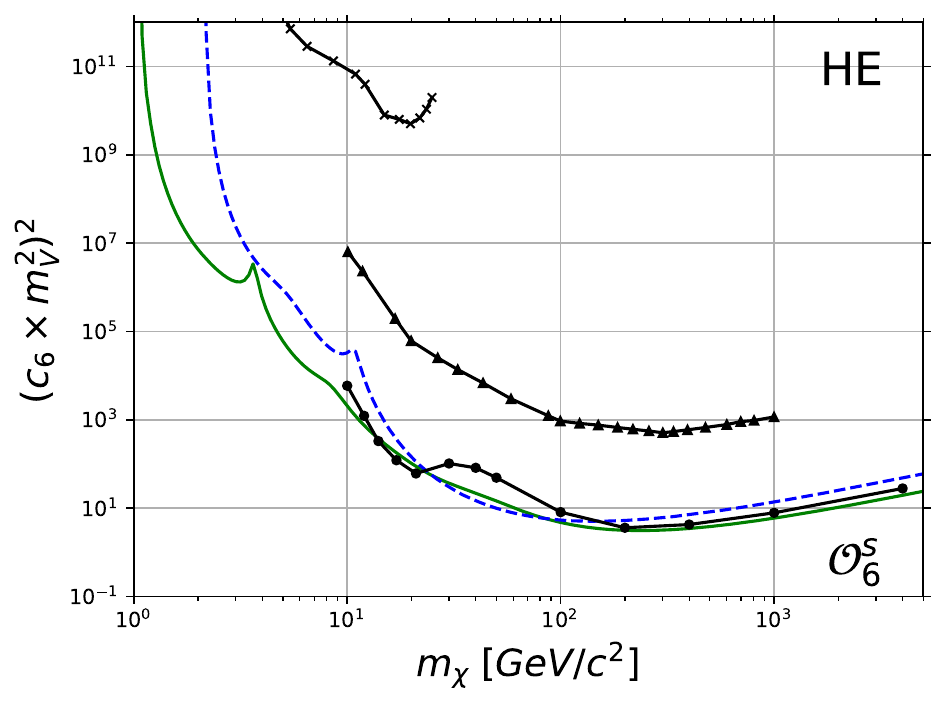}
    \end{subfigure}
    \begin{subfigure}{0.30\textwidth}
    \includegraphics[width=\linewidth]{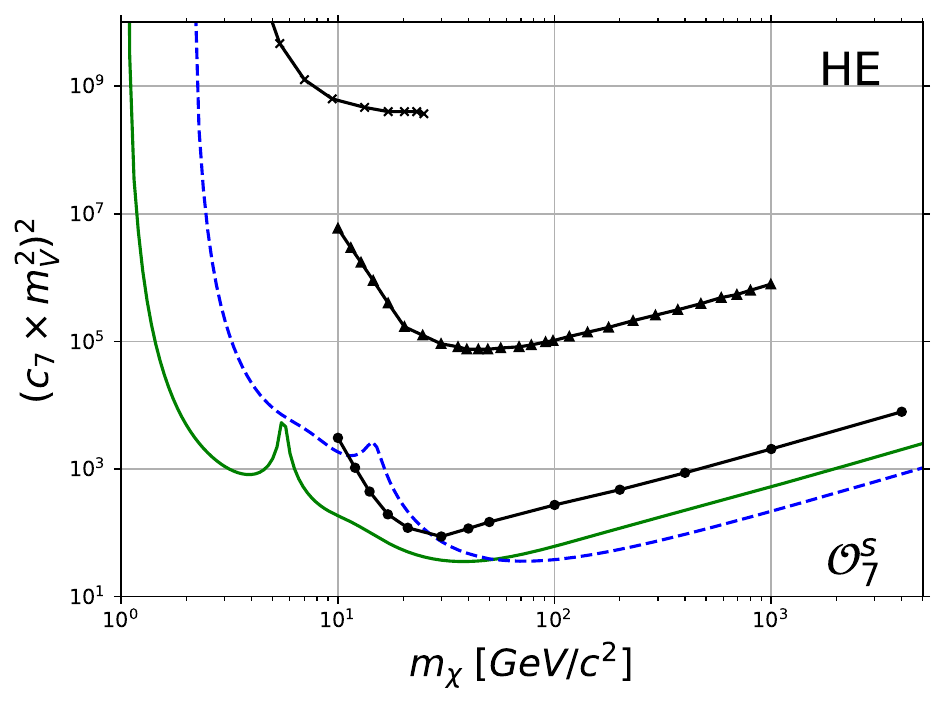}
    \end{subfigure}

    \vspace{0.1ex}

    \begin{subfigure}{0.30\textwidth}
    \includegraphics[width=\linewidth]{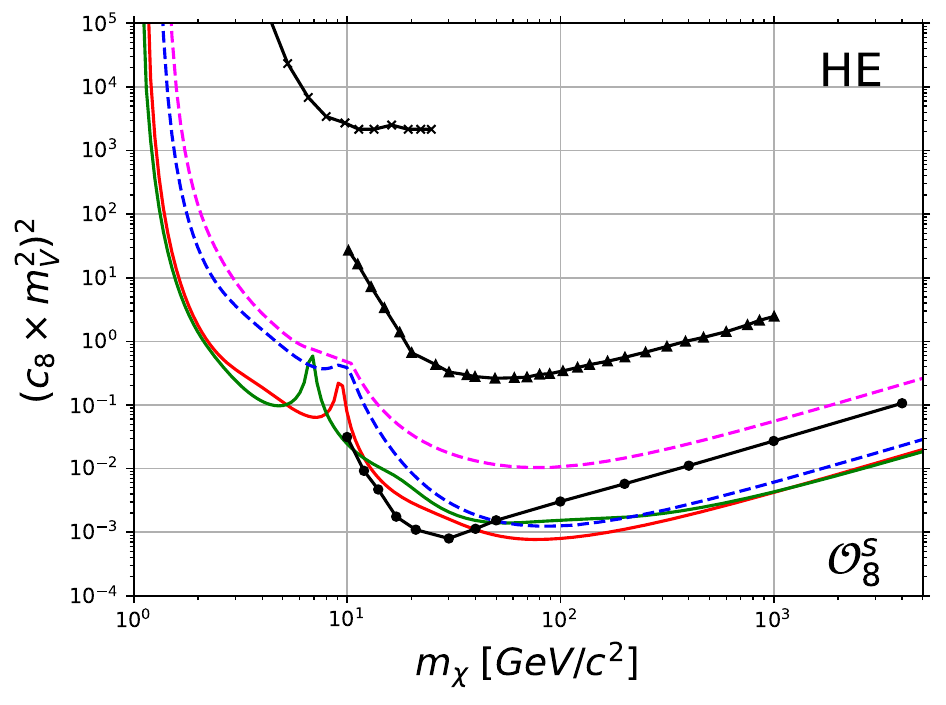}
    \end{subfigure}
    \begin{subfigure}{0.30\textwidth}
    \includegraphics[width=\linewidth]{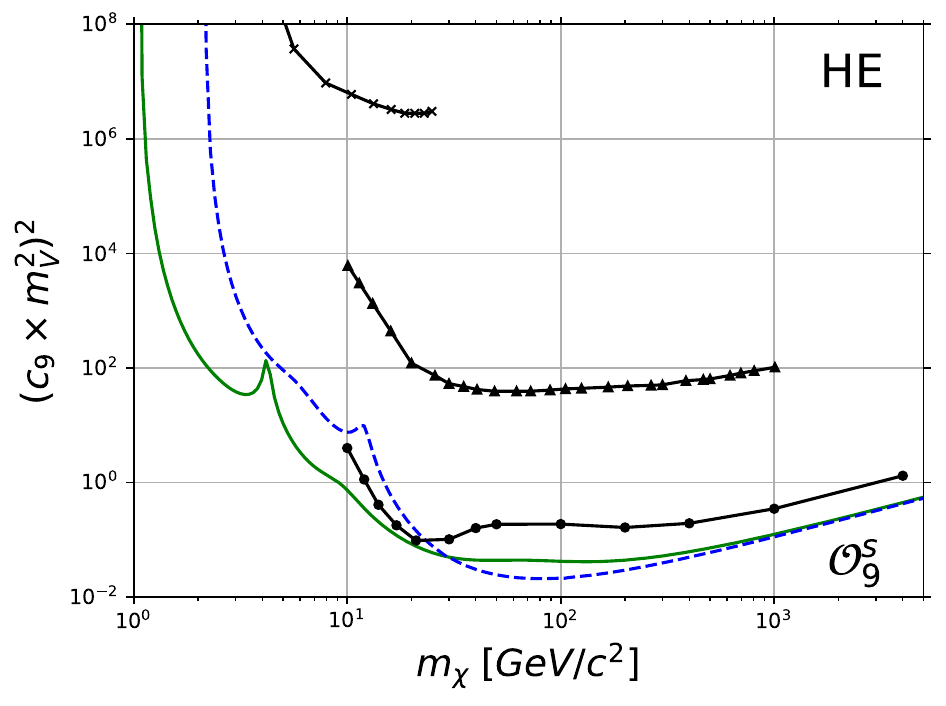}
    \end{subfigure}
    \begin{subfigure}{0.30\textwidth}
    \includegraphics[width=\linewidth]{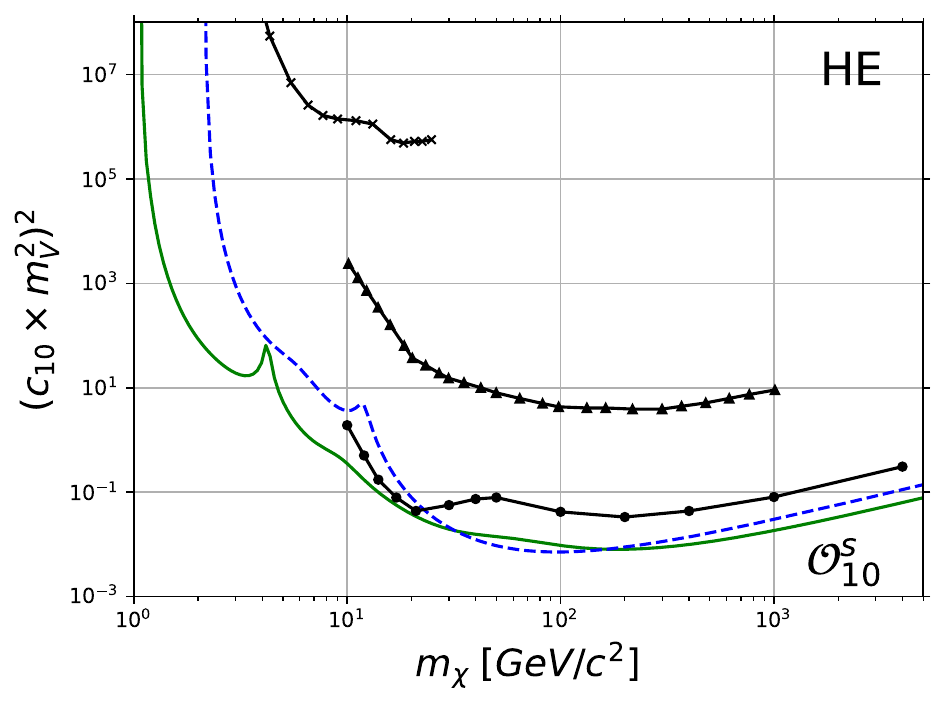}
    \end{subfigure}

    \vspace{0.1ex}

    \begin{subfigure}{0.30\textwidth}
    \includegraphics[width=\linewidth]{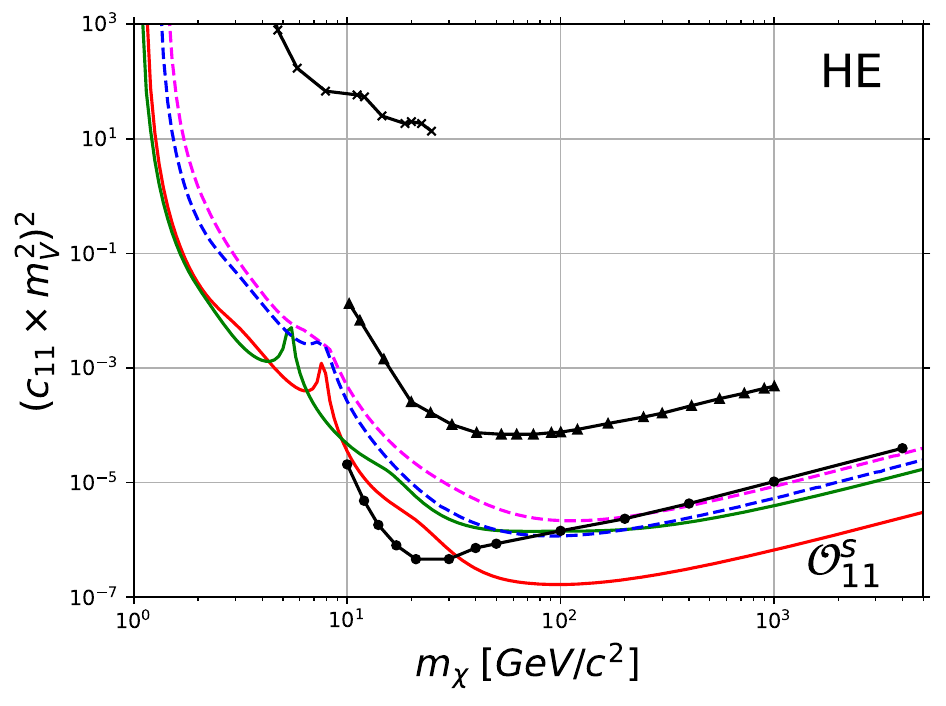}
    \end{subfigure}
    \begin{subfigure}{0.30\textwidth}
    \includegraphics[width=\linewidth]{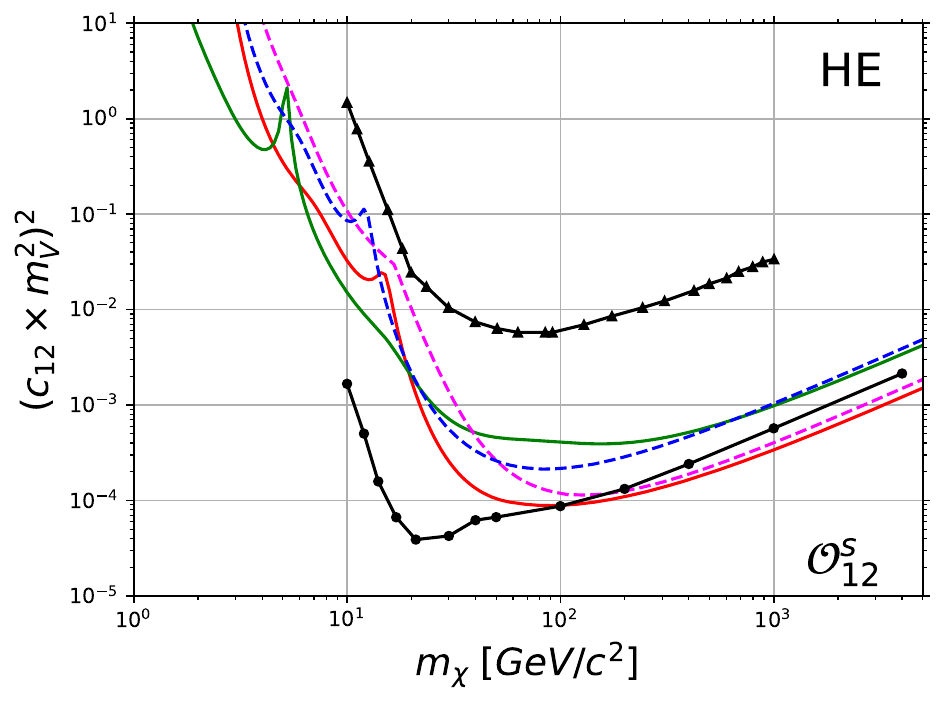}
    \end{subfigure}
    \begin{subfigure}{0.30\textwidth}
    \includegraphics[width=\linewidth]{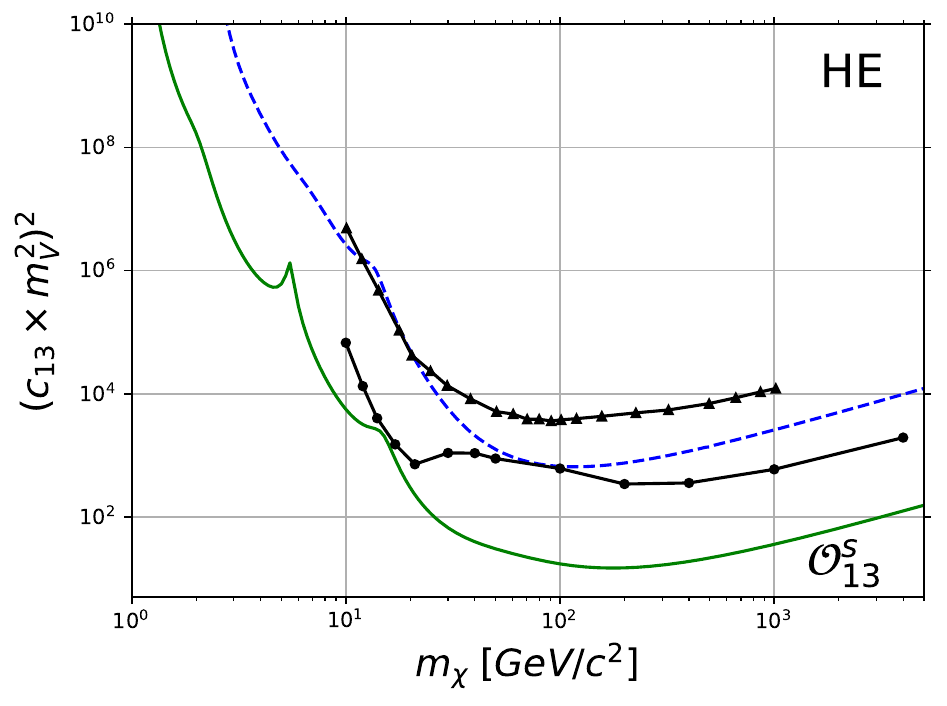}
    \end{subfigure}

    \vspace{0.1ex}

    \begin{subfigure}{0.30\textwidth}
    \includegraphics[width=\linewidth]{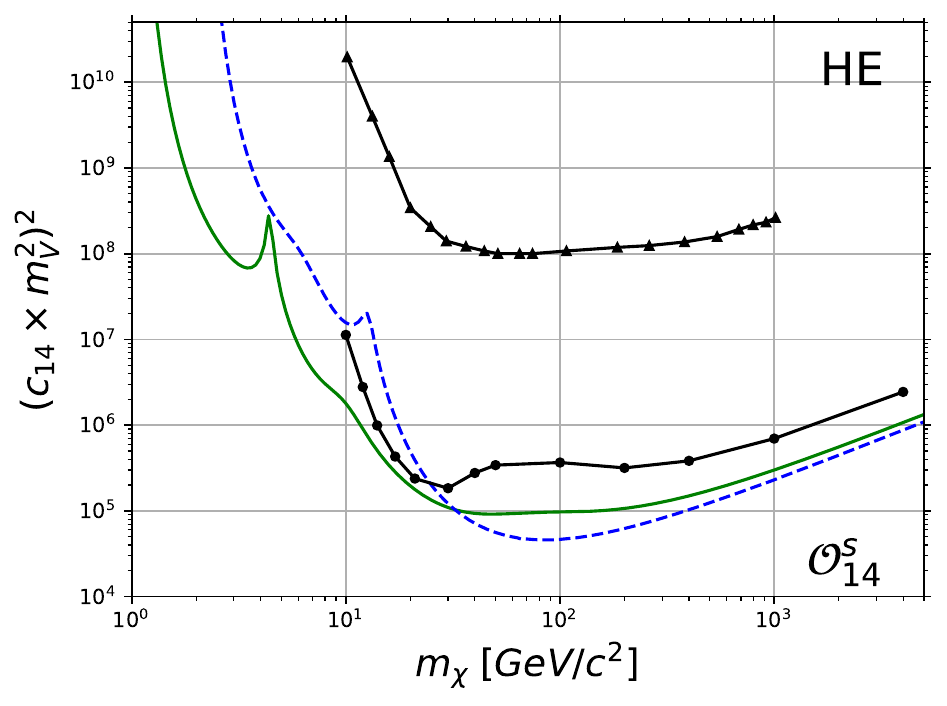}
    \end{subfigure}
    \begin{subfigure}{0.30\textwidth}
    \includegraphics[width=\linewidth]{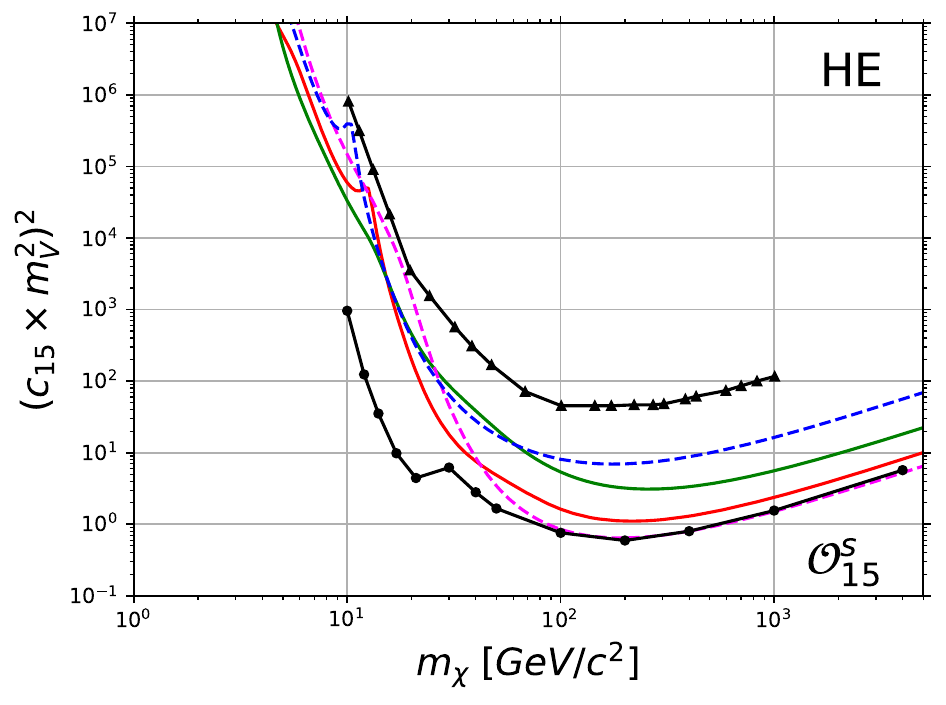}
    \end{subfigure}\hspace{4.85em}
    \begin{subfigure}{0.21\textwidth}
    \includegraphics[width=\linewidth]{figures/legend.pdf} 
    \end{subfigure}
    \caption{Projected 90\% confidence level upper limits on the dimensionless isoscalar WIMP--nucleon NREFT coupling constants for elastic scattering, assuming a large exposure $M=100$ g and $t_{ \text{age}} = 1$ Gyr, and a low read-out resolution $\sigma_{x}=15$ nm. The solid (dashed) lines indicate minerals with $C^{238}=10^{-11}$ g/g ($C^{238}=10^{-10}$ g/g), see Table \ref{tab:U238_concentration}. Black lines present the NREFT results from conventional direct detection experiments: the 90\% confidence level upper limits from XENON100 \cite{XENON:2017fdd}, LUX--ZEPLIN \cite{LZ:2023lvz}, PandaX--II \cite{PandaX-II:2018woa}, as well as the 95\% Bayesian credible region of the two-dimensional marginalized posterior distribution from SuperCDMS \cite{SuperCDMS:2022crd}. 
    For elastic scattering, the HE scenario illustrates enhanced sensitivity of paleo-detectors to WIMP--nucleon cross sections below existing experimental direct detection bounds for $m_\chi \gtrsim 100~\mathrm{GeV}/c^2$, particularly for minerals such as gypsum.}
    \label{fig:HE}
\end{figure}

In Figs.~\ref{fig:HR} and \ref{fig:HE}, we illustrate the projected 90\% confidence level exclusion limits on the isoscalar WIMP--nucleon NREFT coupling constants ($c^{p}=c^{n}$) for elastic scattering over a DM mass range from $1\ \mathrm{GeV}/c^2$ to $5000\ \mathrm{GeV}/c^2$, for the HR scenario ($M=10$~mg, $\sigma_{x}=1$~nm) and the HE scenario ($M=100$~g, $\sigma_{x}=15$~nm), respectively. The limits are expressed in weak-interaction units, where $c_j \times m_{v}^{2}$ is dimensionless, with $m_{v} = 246.2\ \mathrm{GeV}$ denoting the electroweak mass scale. We consider four target minerals common in nature (see Table~\ref{tab:U238_concentration}): two in Ultra-Basic Rocks (UBRs)—olivine and muscovite, the latter containing hydrogen—and two in Marine Evaporites (MEs), halite and gypsum, with the latter also containing hydrogen.

As discussed above, the HR scenario provides better sensitivity for WIMP masses $m_\chi \lesssim 10\ \mathrm{GeV}/c^2$, where solar neutrinos dominate the background. Thus, for relatively light DM particles, the bounds imposed on the NREFT coupling constants in the HR scenario (Fig. \ref{fig:HR}) are tighter than those imposed in the HE scenario (Fig. \ref{fig:HE}). In this mass range, excellent track length resolution is essential to distinguish the spectral shapes of signal and background events, and therefore the coupling constants remain almost unbounded in the HE scenario. On the contrary, for $m_\chi \gtrsim 10\ \mathrm{GeV}/c^2$, the HE scenario yields better sensitivity because radiogenic backgrounds become dominant and a large exposure is crucial for optimal sensitivity.

Comparing different target minerals, those containing hydrogen generally exhibit better sensitivity because hydrogen efficiently suppresses radiogenic neutron backgrounds (see Sec.~\ref{background}). Moreover, MEs typically outperform UBRs, since we assume a lower $^{238}$U concentration for MEs ($C^{238}=10^{-11},\mathrm{g/g}$) than for UBRs ($C^{238}=10^{-10},\mathrm{g/g}$). As an illustrative example, we examine gypsum and olivine (see Table~\ref{tab:U238_concentration}). As discussed in Section~\ref{DMpaleo}, nuclei with spin-zero ground states yield nonzero spectra only for the spin-independent NREFT operators $\mathcal{O}_{1}^{s}$, $\mathcal{O}_{5}^{s}$, $\mathcal{O}_{8}^{s}$, and $\mathcal{O}_{11}^{s}$, as well as for $\mathcal{O}_{3}^{s}$, $\mathcal{O}_{12}^{s}$, and $\mathcal{O}_{15}^{s}$, which exhibit dominant scalar couplings through the $\Phi''$ response. Thus, the track length spectra for gypsum and olivine, each with dominant isotopes in spin-zero ground states, are nonzero only for these NREFT operators. Additionally, the isotopes of these minerals do not possess unpaired nucleons; hence, the $\Delta$ response does not contribute to interactions via the $\mathcal{O}_{5}^{s}$ and $\mathcal{O}_{8}^{s}$ operators. Consequently, the track length spectra for spin-independent interactions in gypsum and olivine (see Fig.~\ref{fig:Spectrum_SI}) appear similar, as the dominant contribution arises from the $M$ response and the isotopes of both minerals have comparable mass numbers $A$. Nevertheless, as shown in Figs.~\ref{fig:HR} and \ref{fig:HE}, gypsum provides better sensitivity than olivine for the spin-independent NREFT operators in both read-out scenarios (HR and HE), owing to its hydrogen content and lower $^{238}$U concentration, which together lead to reduced $^{238}$U-induced backgrounds. In contrast, the spectra for the $\mathcal{O}_{3}^{s}$, $\mathcal{O}_{12}^{s}$, and $\mathcal{O}_{15}^{s}$ operators in olivine are larger than those in gypsum (see Figs.~\ref{fig:Spectrum_SDq0} and \ref{fig:Spectrum_SDq1v1} in the Appendix), since the dominant isotopes of olivine have partially filled angular-momentum orbitals, which favor $\Phi''$ responses. Thus, olivine and gypsum yield similar sensitivities for the $\mathcal{O}_{3}^{s}$, $\mathcal{O}_{12}^{s}$, and $\mathcal{O}_{15}^{s}$ operators.      

Halite and muscovite contain isotopes with nonzero-spin ground states and therefore yield non-vanishing track length spectra for all NREFT operators. When compared with gypsum, halite exhibits similar spectra for the $\mathcal{O}_{1}^{s}$ and $\mathcal{O}_{11}^{s}$ operators, as shown in Fig.~\ref{fig:Spectrum_SI}, since these interactions depend solely on the $M$ response and the relevant isotopes have comparable mass numbers. However, because halite lacks hydrogen, its radiogenic backgrounds are higher, leading to poorer sensitivity to $\mathcal{O}_{1}^{s}$ and $\mathcal{O}_{11}^{s}$, as illustrated in Figs.~\ref{fig:HR} and~\ref{fig:HE}. In contrast to gypsum, the dominant isotopes of halite possess unpaired nucleons, which induce an additional contribution from the $\Delta$ response. Consequently, the spectra for the $\mathcal{O}_{5}$ and $\mathcal{O}_{8}$ operators (Fig.~\ref{fig:Spectrum_SI}) are enhanced in halite, resulting in sensitivities to $\mathcal{O}_{5}^{s}$ and $\mathcal{O}_{8}^{s}$ that are comparable to those of gypsum. Moreover, halite and muscovite exhibit similar spectra for the spin-dependent operators and comparable radiogenic backgrounds for track lengths of a few hundred nm, as shown in Figs.~\ref{fig:Spectrum_SDq0}, \ref{fig:Spectrum_SDq1v0}, and~\ref{fig:Spectrum_SDq1v1} in the Appendix. Consequently, these minerals provide similar sensitivities to the spin-dependent operators, except for $\mathcal{O}_{13}^{s}$. The dominant contribution to the nuclear form factor for interactions with the $\mathcal{O}_{13}^{s}$ operator arises from the $\Phi'$ response. This response favors Cl in halite and K in muscovite, with the former yielding a nuclear response function nearly an order of magnitude larger than the latter. In addition, the abundance of Cl in halite is approximately one order of magnitude higher than that of K in muscovite. Together, these factors lead to a total difference of about two orders of magnitude between the signal spectra of halite and muscovite for $\mathcal{O}_{13}^{s}$, as shown in Fig.~\ref{fig:Spectrum_SDq1v1}. Hence, the sensitivity of halite to the $\mathcal{O}_{13}^{s}$ operator is roughly two orders of magnitude greater than that of muscovite in both read-out scenarios shown in Figs.~\ref{fig:HR} and~\ref{fig:HE}.

In the Appendix, we explore the sensitivity of additional minerals, including rare minerals such as sinjarite and nchwaningite (see Table~\ref{tab:U238_concentration}), and present the corresponding 90\% confidence-level exclusion limits on the isoscalar WIMP–nucleon NREFT coupling constants for elastic scattering in Figs.~\ref{fig:HR_ap} and \ref{fig:HE_ap}, for the HR and HE scenarios, respectively. Although likely too rare in nature to be considered feasible as paleo-detectors, sinjarite and nchwaningite serve as illustrative examples of how the sensitivity of paleo-detectors can depend on the mineral's chemical composition and on the presence of isotopes with non-zero–spin nuclear ground states.

Figs.~\ref{fig:HR}, \ref{fig:HE}, \ref{fig:HR_ap}, and \ref{fig:HE_ap} also display 90\% confidence level upper limits from the conventional direct detection experiments XENON100 \cite{XENON:2017fdd}, LUX--ZEPLIN \cite{LZ:2023lvz}, and PandaX--II \cite{PandaX-II:2018woa}, as well as the 95\% Bayesian credible region of the two-dimensional marginalized posterior distribution from SuperCDMS \cite{SuperCDMS:2022crd}. The latter does not differ significantly from the 90\% confidence-level upper limits, as reported in Ref.~\cite{SuperCDMS:2022crd}. Paleo-detectors with high read-out resolution can reach WIMP–nucleon cross sections far below current experimental bounds for $m_\chi \lesssim 10\ \mathrm{GeV}/c^2$. For higher masses ($m_\chi \gtrsim 10\ \mathrm{GeV}/c^2$), the high-exposure scenario provides sufficient events for paleo-detectors to impose tighter limits on the NREFT coupling constants than conventional direct detection experiments. 

For the spin-independent NREFT operators $\mathcal{O}_{1}^{s}$, $\mathcal{O}_{5}^{s}$, $\mathcal{O}_{8}^{s}$, and $\mathcal{O}_{11}^{s}$, the sensitivity projected for gypsum is approximately five orders of magnitude greater than that reported by SuperCDMS~\cite{SuperCDMS:2022crd} for dark matter with $m_{\chi} \sim 10~\mathrm{GeV}/c^{2}$, and about one order of magnitude greater than that reported by LUX--ZEPLIN~\cite{LZ:2023lvz} for $m_{\chi} \sim 1~\mathrm{TeV}/c^{2}$. This enhancement in sensitivity arises from the large exposure achievable with paleo-detectors, which leads to more projected signal events, combined with the relatively low background levels in gypsum compared with other minerals—particularly for DM masses of a few hundred~$\mathrm{GeV}/c^{2}$, where radiogenic backgrounds dominate. As shown in the left panel of Fig.~\ref{fig:Spectrum_binned_SI}, for gypsum in the HR scenario with normalization consistent with the limits reported by SuperCDMS~\cite{SuperCDMS:2022crd}, the spectra for the spin-independent NREFT operators are roughly three orders of magnitude larger than the neutrino background. In addition, the signal-to-background ratio for these operators, displayed in the bottom-left panel of Fig.~\ref{fig:Spectrum_binned_SI}, is clearly distinguishable from the relative uncertainty of the total background per bin, resulting in better projected sensitivity than that reported by SuperCDMS for spin-independent operators at small DM masses ($m_{\chi} \lesssim 30~\mathrm{GeV}/c^{2}$). Similarly, as illustrated in the right panel of Fig.~\ref{fig:Spectrum_binned_SI}, for gypsum in the HE scenario with normalization consistent with the limits reported by LUX--ZEPLIN~\cite{LZ:2023lvz}, the spectra for the spin-independent NREFT operators are about one order of magnitude larger than the neutrino background and one order of magnitude smaller than the neutron background. Nevertheless, the signal-to-background ratio for these operators, shown in the bottom-right panel of Fig.~\ref{fig:Spectrum_binned_SI}, remains clearly distinguishable from the relative background uncertainty, leading to better projected sensitivity than that reported by LUX--ZEPLIN for spin-independent operators at larger DM masses ($m_{\chi} \gtrsim 50~\mathrm{GeV}/c^{2}$). 

The four spin-independent NREFT operators can be categorized based on their dependence on the momentum transfer $\vec{q}$ and the component of the WIMP–nucleon relative velocity perpendicular to the momentum transfer, $\vec{v}^{\perp}$. Specifically, two operators, $\mathcal{O}^{s}_{1}$ and $\mathcal{O}^{s}_{8}$, are independent of $\vec{q}$, while $\mathcal{O}^{s}_{5}$ and $\mathcal{O}^{s}_{11}$ depend explicitly on it. In terms of normalization, the rates of the latter two operators carry an additional factor of $(\vec{q}/m_{v})^{2}\sim10^{-3}$. Regarding the spectral shape, the differential recoil rate ($dR/dE_{R}$), as given by Eq.~(\ref{recoilrate}), peaks at relatively small recoil energies for $\mathcal{O}^{s}_{1}$ and $\mathcal{O}^{s}_{8}$, while for $\mathcal{O}^{s}_{5}$ and $\mathcal{O}^{s}_{11}$ the peak shifts toward higher recoil energies. Similarly, $\mathcal{O}^{s}_{1}$ and $\mathcal{O}^{s}_{11}$ are independent of $\vec{v}^{\perp}$, whereas $\mathcal{O}^{s}_{5}$ and $\mathcal{O}^{s}_{8}$ depend on it. The normalization of the latter is thus suppressed by an additional factor of $(\vec{v}^{\perp})^{2}\sim10^{-6}$. This velocity dependence also affects the spectral shape, slightly broadening the peaks and shifting them toward higher recoil energies due to the contribution of the $\Delta$ response, particularly for isotopes with unpaired nucleons. The general features of $dR/dE_{R}$ are reflected in the corresponding track length spectra for the minerals shown in Fig.~\ref{fig:Spectrum_SI}. Specifically, for halite and muscovite, which possess isotopes with unpaired nucleons, the spectra associated with $\mathcal{O}_{5}^{s}$ and $\mathcal{O}_{8}^{s}$--operators with $\vec{v}^{\perp}$ dependence--extend to larger track lengths compared to those of $\mathcal{O}_{1}^{s}$ and $\mathcal{O}_{11}^{s}$. Additionally, the peaks of $\mathcal{O}_{1}^{s}$ and $\mathcal{O}_{8}^{s}$ appear at shorter track lengths relative to $\mathcal{O}_{11}^{s}$ and $\mathcal{O}_{5}^{s}$, respectively, which depend on $\vec{q}$. In gypsum and olivine, where the dominant isotopes are symmetric and do not have unpaired nucleons, the pattern expected from the $\vec{v}^{\perp}$ dependence is not observed.

For conventional direct detection experiments, which are effectively background–free and sensitive to high-mass WIMPs (e.g., $m_{\chi}\sim1\ \mathrm{TeV}/c^{2}$), the limits can be approximately scaled from $\mathcal{O}^{s}_{1}$ based on the normalization factors discussed above. According to the results from the LUX–ZEPLIN experiment~\cite{LZ:2023lvz}, the upper limits on the coupling constants are roughly $(c_{1}^{\mathrm{LZ}} m_{v}^{2})^{2}\sim10^{-8}$, $(c_{5}^{\mathrm{LZ}} m_{v}^{2})^{2}\sim10^{1}$, $(c_{8}^{\mathrm{LZ}} m_{v}^{2})^{2}\sim10^{-2}$, and $(c_{11}^{\mathrm{LZ}} m_{v}^{2})^{2}\sim10^{-5}$. These scaling relations apply to gypsum and olivine, which do not contain isotopes with unpaired nucleons and therefore contribute only through the $M$ response to the spectra of the spin-independent NREFT operators. For example, for a WIMP mass of $1~\mathrm{TeV}/c^{2}$, where the dominant background arises from radiogenic neutrons, the projected limits for gypsum in the HE scenario, shown in Fig.~\ref{fig:HE}, are about one order of magnitude stronger than those from the LUX--ZEPLIN experiment. In contrast, these scaling relations do not directly apply to halite and muscovite, since their sensitivities to the $\mathcal{O}^{s}_{1}$ and $\mathcal{O}^{s}_{11}$ operators for a WIMP mass of $1~\mathrm{TeV}/c^{2}$ are comparable to those reported by LUX--ZEPLIN, whereas their sensitivities to the $\mathcal{O}^{s}_{5}$ and $\mathcal{O}^{s}_{8}$ operators are approximately an order of magnitude greater. This enhancement arises because these minerals contain isotopes with unpaired nucleons, which lead to significant contributions from the $\Delta$ response to the spectra of $\mathcal{O}^{s}_{5}$ and $\mathcal{O}^{s}_{8}$. As reported in Ref.~\cite{Fitzpatrick:2012ix}, for Na the $\Delta$ response provides a larger contribution than the $M$ response for the $\mathcal{O}^{s}_{8}$ operator, whereas for Xe, used in the LUX--ZEPLIN experiment, the $\Delta$ contribution to $\mathcal{O}^{s}_{8}$ is negligible compared with that of the $M$ response.

Furthermore, in Figs.~\ref{fig:HE} and~\ref{fig:HE_ap}, a local bump appears for every mineral at DM masses around $10\ \mathrm{GeV}/c^2$. This occurs because, for those masses, the track length spectrum of DM particles closely resembles that produced by solar neutrinos, making it difficult to distinguish the two and thus causing the limits to weaken locally. In the case of sinjarite, which provides better sensitivity than most other minerals, two additional local bumps occur at DM masses of about $2\ \mathrm{GeV}/c^2$ and $20\ \mathrm{GeV}/c^2$. The former is again due to the solar neutrino fluxes which dominate the background spectrum for $m_\chi \lesssim 10\ \mathrm{GeV}/c^2$, while the latter arises because the DM track length spectrum is similar to those from the GSNB and from $^{234}$Th, making it difficult to separate the signal from these backgrounds and leading to a local weakening in the limits. These two local bumps are absent in other minerals with lower sensitivity.

In addition to elastic scattering, we also explore inelastic DM scattering. Paleo-detectors have a limitation on how large a mass splitting $\delta_{m}$ they can probe because of the small masses of the constituent isotopes. The minimum velocity an incoming WIMP must have in the target reference frame to deposit recoil energy $E_{R}$ while upscattering to an excited state with mass splitting $\delta_m$ is given by \cite{Tucker-Smith:2001myb,Barello:2014uda}
\begin{equation}
v_{T,\min}(E_{R}) = \frac{1}{\sqrt{2 m_{T} E_{R}}} \left| \frac{m_{T} E_{R}}{\mu_{T}} + \delta_{m}c^{2} \right|~,
\end{equation}
where $m_{T}$ is the nuclear target mass, and $\mu_{T}$ is the reduced mass of the WIMP–nucleus system. By minimizing $v_{T,\min}(E_{R})$ with respect to $E_{R}$, we obtain
\begin{equation}
    \min_{E_R} v_{T,\min} = \sqrt{\frac{2\delta_{m}c^{2}}{\mu_{ T}}}\equiv v_{\delta}~.
\end{equation}
Assuming the maximum velocity of DM particles in the target's frame is $v_{\max}\sim700\ \mathrm{km/s}$ according to the Standard Halo Mode~\cite{Baxter:2021pqo}, and imposing $v_{\max}>v_{\delta}$, we find the maximum mass splitting $\delta_{\max}$ that paleo-detectors can probe
\begin{equation}
    \delta_{\max}=\frac{\mu_{T}v_{\max}^{2}}{2c^{2}}~.
\end{equation}
For DM masses much larger than the target nucleus, $\mu_{T}\approx m_{T}$. For the paleo-detectors in Table \ref{tab:U238_concentration}, the heavier isotopes have masses around $m_{T}\approx40\ \mathrm{GeV}/c^{2}$, and thus the corresponding maximum mass splitting is $\delta_{\max}\approx100\ \mathrm{keV}/c^{2}$. 

\begin{figure}
    \captionsetup{justification=raggedright,singlelinecheck=false}
    \centering
    \begin{subfigure}{0.30\textwidth}
    \includegraphics[width=\linewidth]{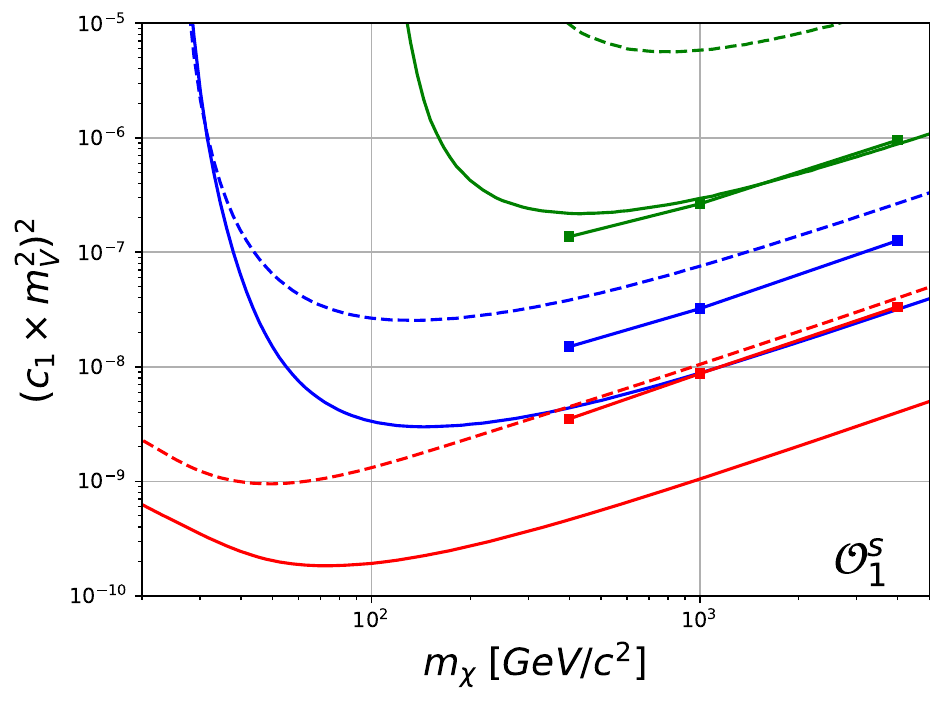}
    \end{subfigure}
    \begin{subfigure}{0.30\textwidth}
    \includegraphics[width=\linewidth]{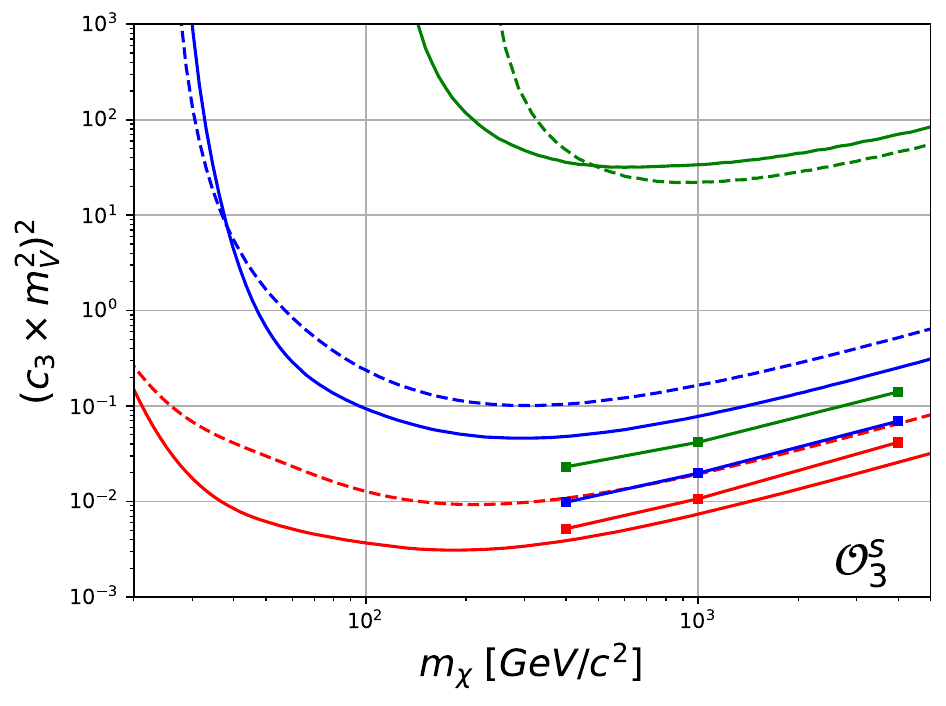}
    \end{subfigure}
    \begin{subfigure}{0.30\textwidth}
    \includegraphics[width=\linewidth]{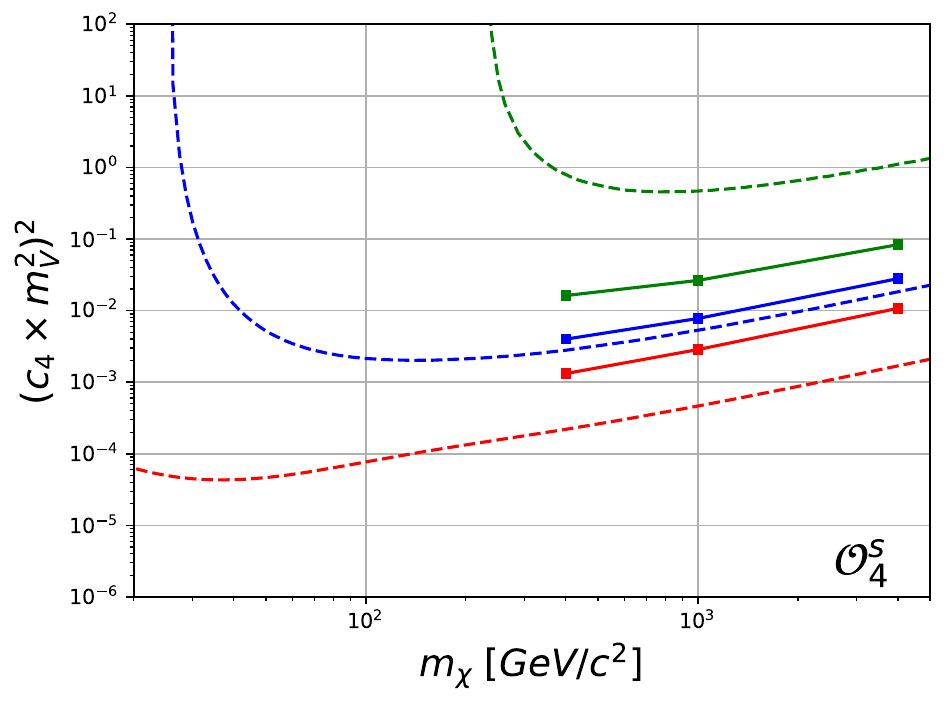}
    \end{subfigure}

    \vspace{0.1ex}

    \begin{subfigure}{0.30\textwidth}
    \includegraphics[width=\linewidth]{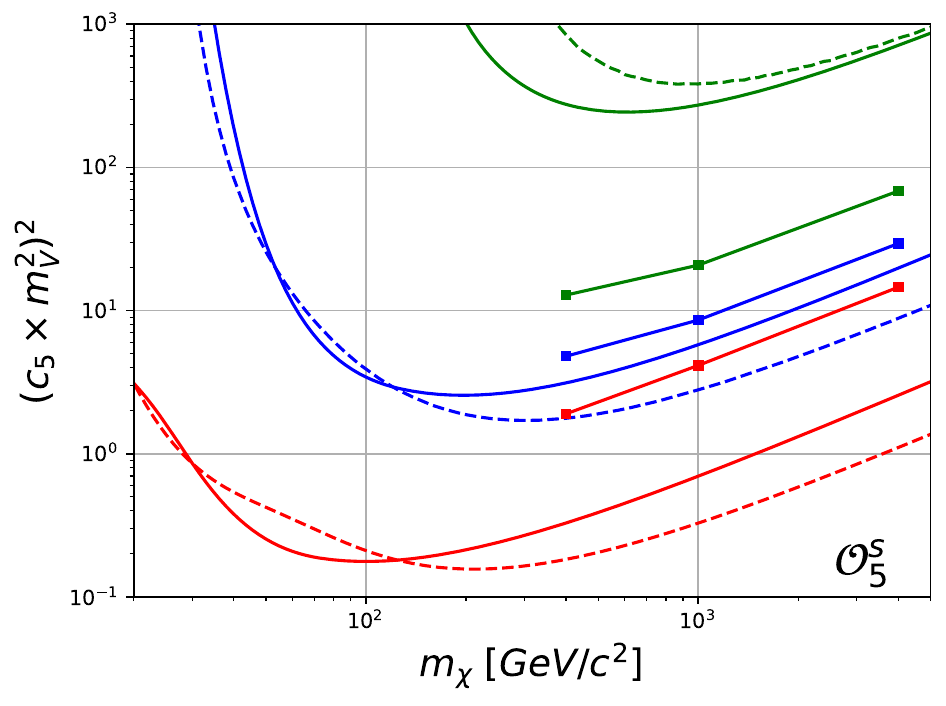}
    \end{subfigure}
    \begin{subfigure}{0.30\textwidth}
    \includegraphics[width=\linewidth]{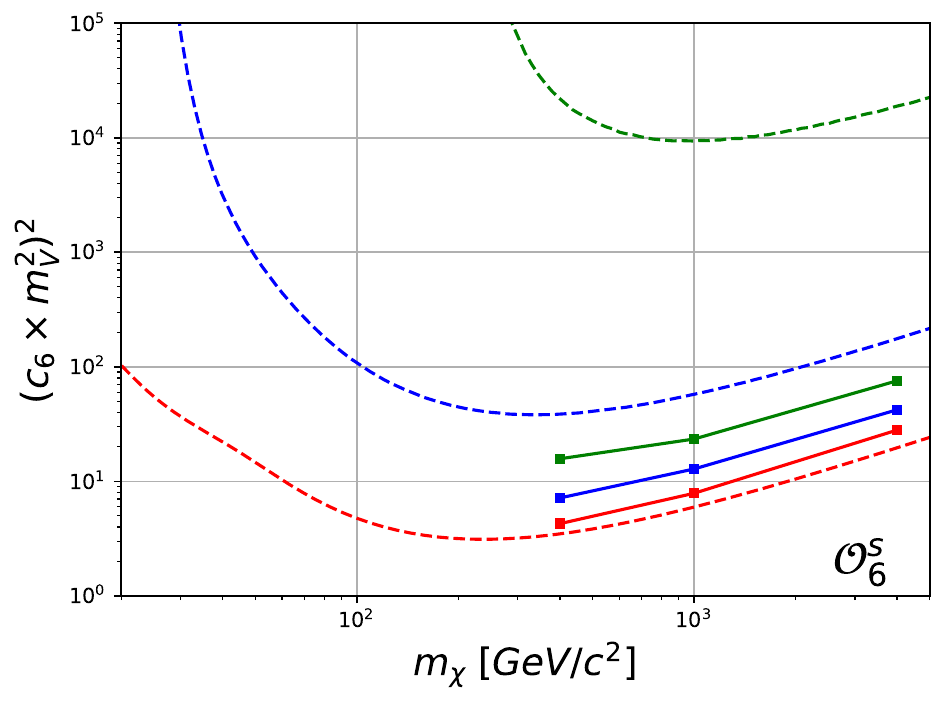}
    \end{subfigure}
    \begin{subfigure}{0.30\textwidth}
    \includegraphics[width=\linewidth]{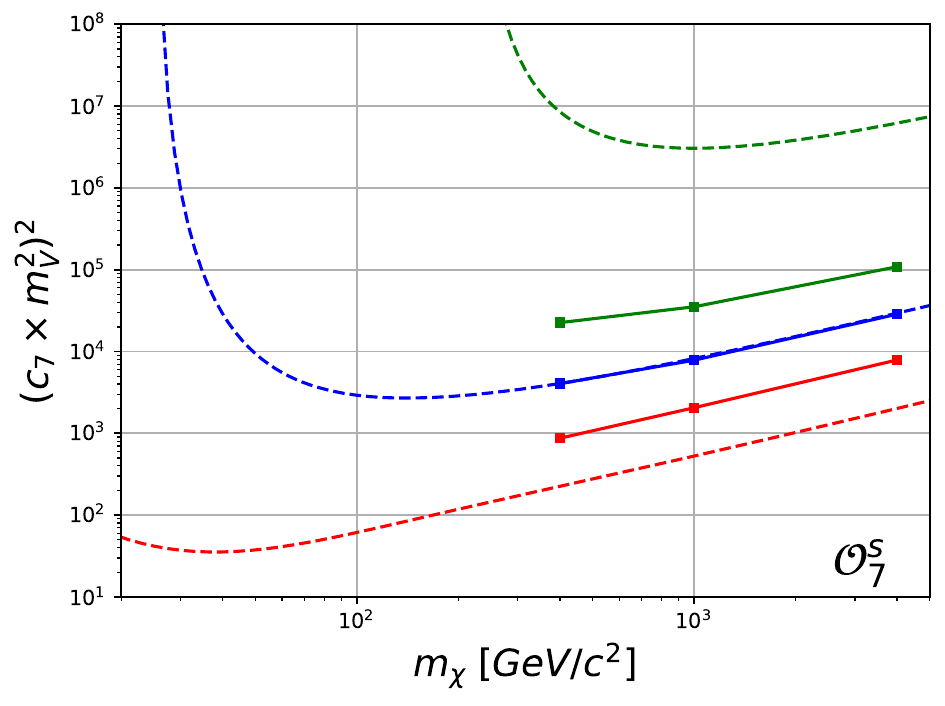}
    \end{subfigure}

    \vspace{0.1ex}

    \begin{subfigure}{0.30\textwidth}
    \includegraphics[width=\linewidth]{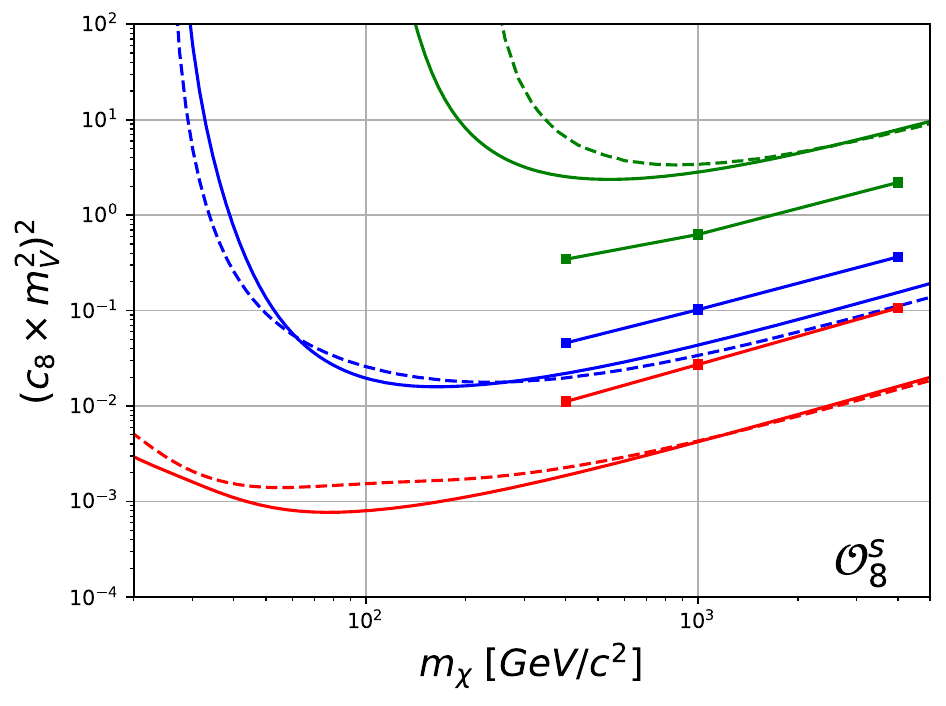}
    \end{subfigure}
    \begin{subfigure}{0.30\textwidth}
    \includegraphics[width=\linewidth]{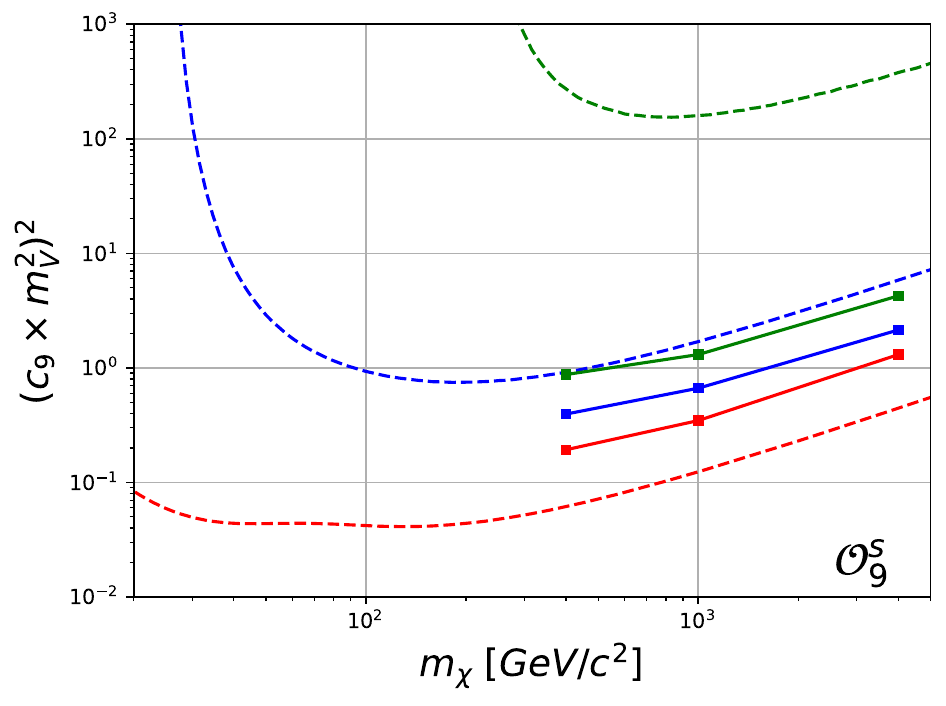}
    \end{subfigure}
    \begin{subfigure}{0.30\textwidth}
    \includegraphics[width=\linewidth]{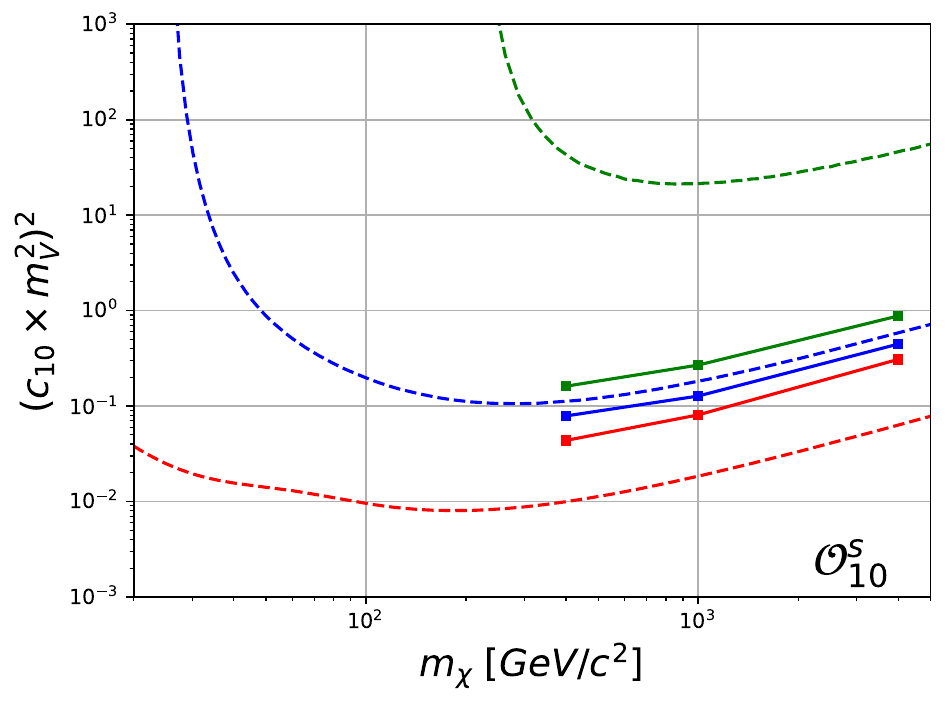}
    \end{subfigure}

    \vspace{0.1ex}

    \begin{subfigure}{0.30\textwidth}
    \includegraphics[width=\linewidth]{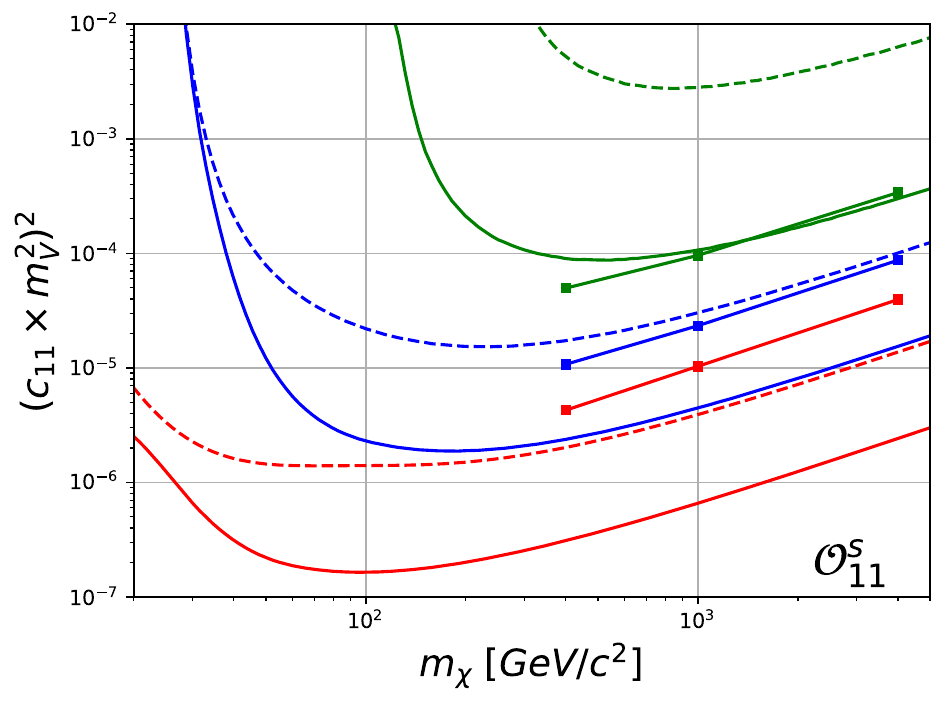}
    \end{subfigure}
    \begin{subfigure}{0.30\textwidth}
    \includegraphics[width=\linewidth]{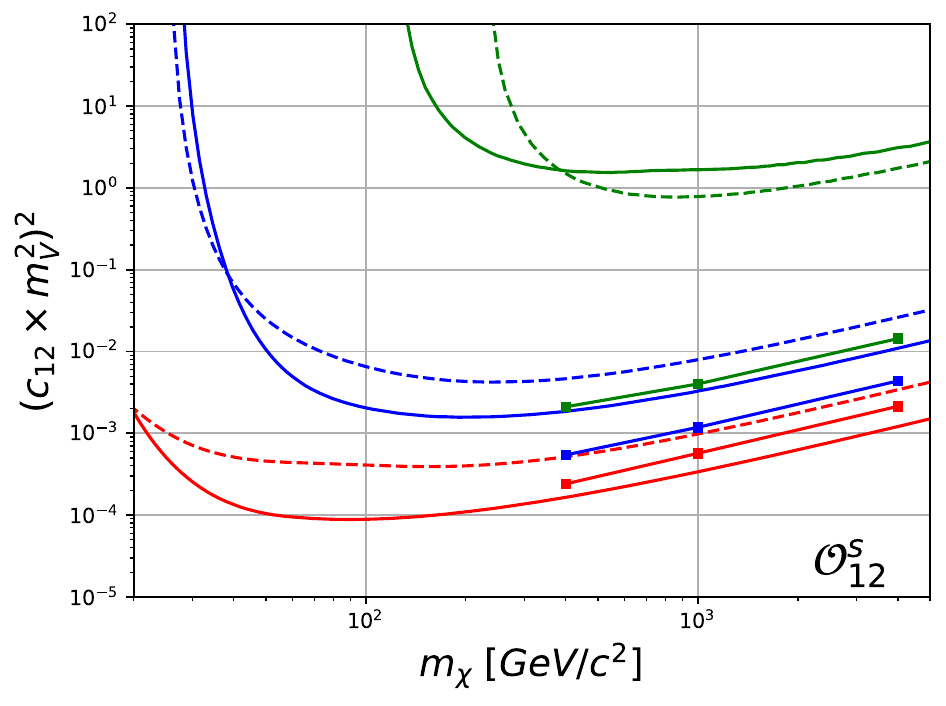}
    \end{subfigure}
    \begin{subfigure}{0.30\textwidth}
    \includegraphics[width=\linewidth]{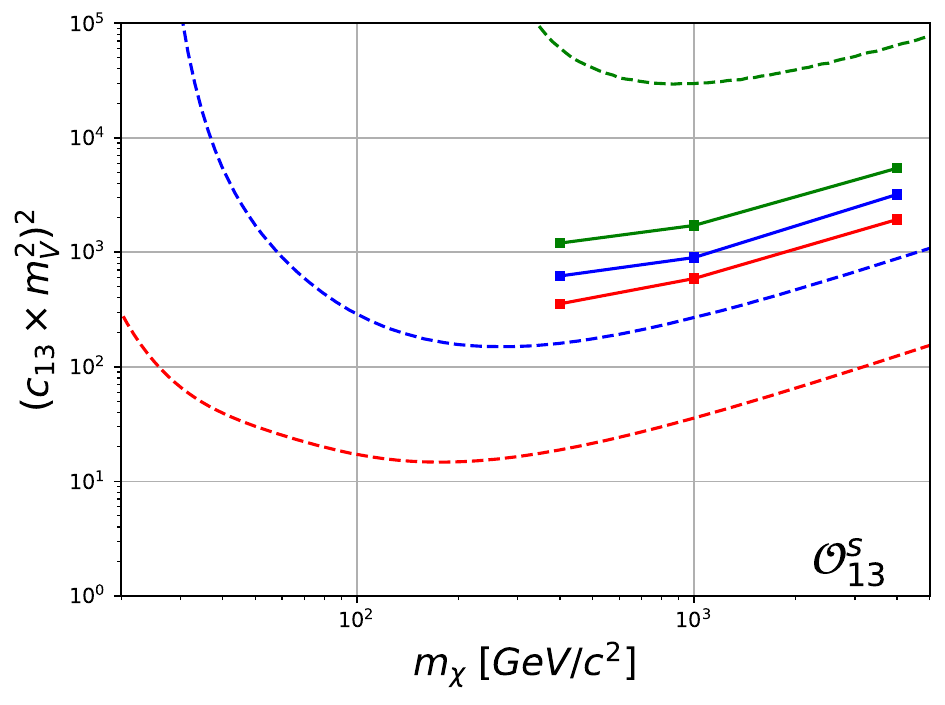}
    \end{subfigure}

    \vspace{0.1ex}

    \begin{subfigure}{0.30\textwidth}
    \includegraphics[width=\linewidth]{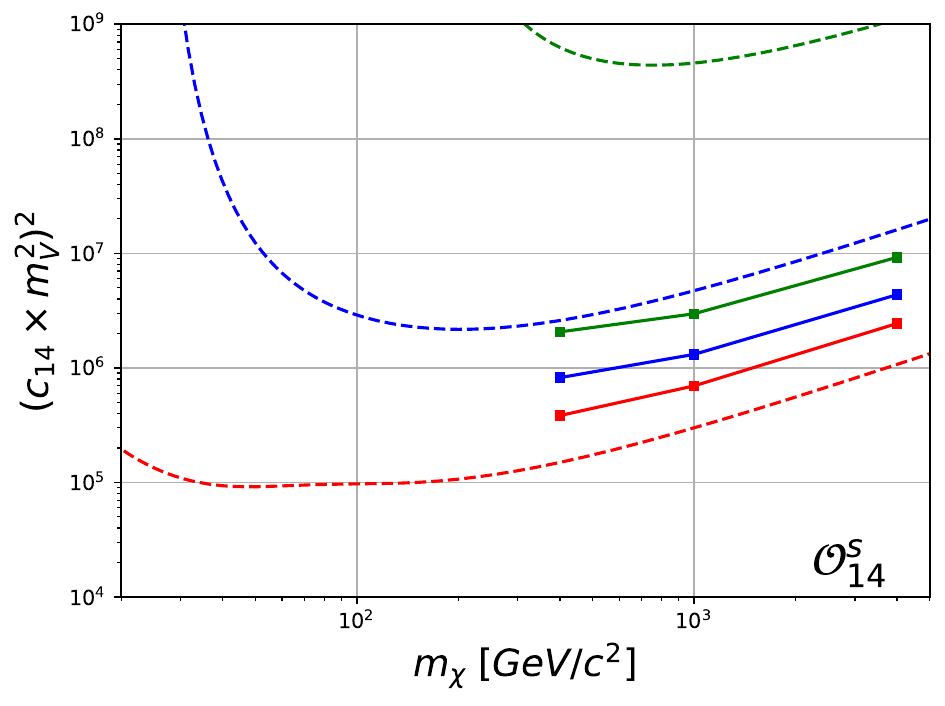}
    \end{subfigure}
    \begin{subfigure}{0.30\textwidth}
    \includegraphics[width=\linewidth]{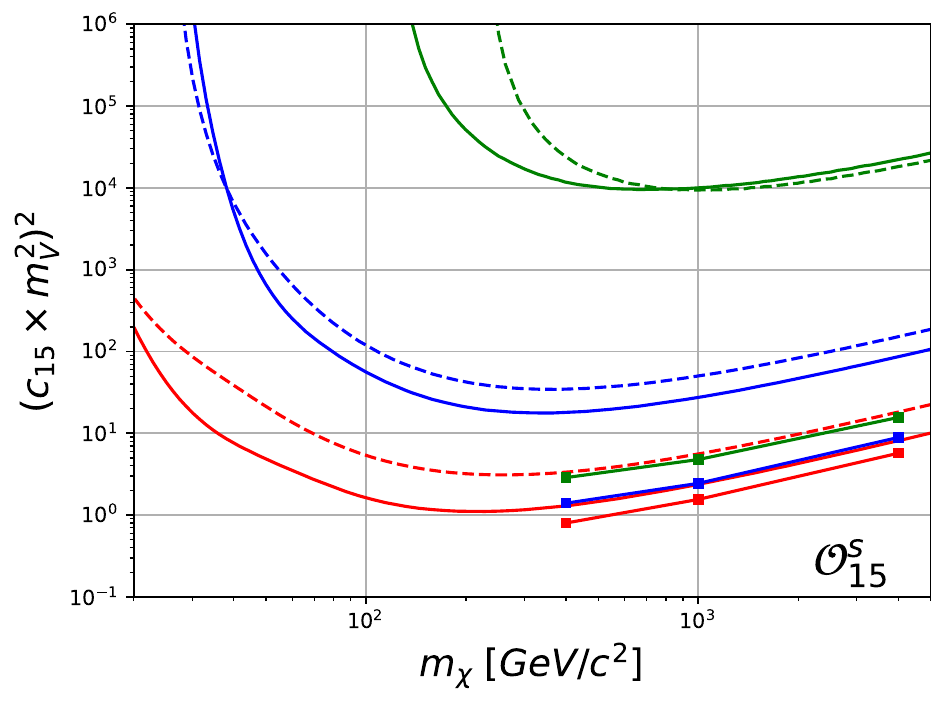}
    \end{subfigure}\hspace{4.8em}
    \begin{subfigure}{0.21\textwidth}
    \includegraphics[width=\linewidth]{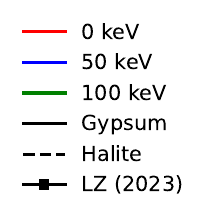} 
    \end{subfigure}
    \caption{Projected 90\% confidence-level upper limits on the dimensionless isoscalar WIMP–nucleon NREFT coupling constants for inelastic scattering, assuming a large exposure ($M = 100$\,g, $t_{\text{age}} = 1$\,Gyr) and a low read-out resolution ($\sigma_{x} = 15$\,nm). Colored lines correspond to different mass splittings $\delta_m$, while square markers indicate the NREFT limits from the LUX--ZEPLIN experiment~\cite{LZ:2023lvz}. Solid (dashed) lines show the projected upper limits for gypsum (halite) as the target mineral. For inelastic scattering, the results indicate sensitivity comparable to or exceeding that of the LZ experiment for most NREFT operators, assuming a mass splitting $\delta_m \lesssim 50~\mathrm{keV}/c^2$ and DM masses $m_\chi \gtrsim 50~\mathrm{GeV}/c^2$.}
    \label{fig:inelastic}
\end{figure}

In Fig.~\ref{fig:inelastic}, we demonstrate the projected 90\% confidence level exclusion limits on the isoscalar WIMP--nucleon NREFT coupling constants ($c^{p}=c^{n}$) for inelastic scattering, for DM masses between $20\ \mathrm{GeV}/c^2$ and $5000\ \mathrm{GeV}/c^2$, in the HE scenario ($M=100$~g, $\sigma_{x}=15$~nm), and for mass-splitting values $\delta_m$ in the range $0$–$100~\mathrm{keV}/c^{2}$---a parameter space well motivated by many WIMP models \cite{Tucker-Smith:2001myb,Barello:2014uda}. We assume gypsum and halite as the target minerals, which contain Ca and Cl with masses $\sim40\ \mathrm{GeV}/c^{2}$ and $\sim35\ \mathrm{GeV}/c^{2}$, respectively (see Table~\ref{tab:U238_concentration}). For large mass splittings, the track length spectrum is strongly suppressed unless the DM particles are very heavy. Therefore, we focus on the HE scenario for inelastic DM scattering, which favors more massive DM particles. This trend is evident in Fig. \ref{fig:inelastic}, where we have limits on the NREFT coupling constants only for very large DM masses, whereas for $m_{\chi}<100\ \mathrm{GeV}/c^{2}$ the couplings remain unconstrained when $\delta_{m}=100\ \mathrm{GeV}/c^{2}$.

Limits from the direct detection experiment LUX--ZEPLIN~\cite{LZ:2023lvz} are also shown in Fig.~\ref{fig:inelastic}. For the spin-independent NREFT operators, gypsum is projected to have more stringent constraints than LUX--ZEPLIN for a mass splitting of $50~\mathrm{keV}/c^{2}$, following trends similar to those observed in the elastic-scattering case. As shown in the top-left panel of Fig.~\ref{fig:Spectrum_binned_SI_inelastic}, for gypsum in the HE scenario with normalization consistent with the limits reported by LUX--ZEPLIN~\cite{LZ:2023lvz} for $\delta_{m}=50~\mathrm{keV}/c^{2}$, the spectra for the spin-independent NREFT operators are smaller than those for elastic scattering presented in the right panel of Fig.~\ref{fig:Spectrum_binned_SI}. However, for the $\mathcal{O}^{s}_{1}$ and $\mathcal{O}^{s}_{11}$ operators, which do not explicitly depend on $\vec{v}_{\text{inel}}^{\,\perp}$ as defined in Eq.~(\ref{u_inelastic}), the signal-to-background ratio shown in the bottom-left panel of Fig.~\ref{fig:Spectrum_binned_SI_inelastic} remains clearly distinguishable from the relative background uncertainty, leading to better sensitivity than that reported by LUX--ZEPLIN for these operators. In contrast, for the $\mathcal{O}^{s}_{5}$ and $\mathcal{O}^{s}_{8}$ operators, which depend explicitly on $\vec{v}_{\text{inel}}^{\,\perp}$, the signal-to-background ratio is comparable to the relative background uncertainty, resulting in only a modest improvement in sensitivity relative to LUX--ZEPLIN. 

For a mass splitting of $\delta_{m}=100~\mathrm{keV}/c^{2}$—close to the upper limit of $\delta_{m}$ values that paleo-detectors can probe—gypsum is projected to have a sensitivity similar to that reported by LUX--ZEPLIN for the $\mathcal{O}^{s}_{1}$ and $\mathcal{O}^{s}_{11}$ operators, since the signal-to-background ratio, illustrated in the bottom-right panel of Fig.~\ref{fig:Spectrum_binned_SI_inelastic}, is comparable to the relative background uncertainty. Conversely, for the $\mathcal{O}^{s}_{5}$ and $\mathcal{O}^{s}_{8}$ operators, the signal-to-background ratio is clearly smaller than the background uncertainty, resulting in weaker sensitivity than that reported by LUX--ZEPLIN. Finally, paleo-detectors are projected to probe lower DM masses than conventional direct detection experiments for inelastic scattering. In particular, gypsum is projected to set constraints for $m_{\chi} \gtrsim 30~\mathrm{GeV}/c^{2}$ when $\delta_{m}=50~\mathrm{keV}/c^{2}$, and for $m_{\chi} \gtrsim 200~\mathrm{GeV}/c^{2}$ when $\delta_{m}=100~\mathrm{keV}/c^{2}$.

In this section, we have derived the projected 90\% confidence level exclusion limits on isoscalar WIMP--nucleon NREFT coupling constants for both elastic and inelastic scattering, considering the HR and HE scenarios and four representative minerals: gypsum, halite, olivine, and muscovite. Projected sensitivities for additional target minerals---sinjarite, epsomite, phlogopite, and nchwaningite---are presented in the Appendix. We have also compared the projected sensitivity of paleo-detectors with that of conventional direct detection experiments, including XENON100~\cite{XENON:2017fdd}, LUX--ZEPLIN~\cite{LZ:2023lvz}, PandaX--II~\cite{PandaX-II:2018woa}, and SuperCDMS~\cite{SuperCDMS:2022crd}.

For elastic scattering, paleo-detectors with sufficiently high read-out resolution are projected to probe WIMP--nucleon couplings well below current experimental limits for $m_\chi \lesssim 10~\mathrm{GeV}/c^2$. At higher masses ($m_\chi \gtrsim 10~\mathrm{GeV}/c^2$), the high--exposure scenario yields sufficient statistics for paleo-detectors to place constraints on NREFT coupling constants projected to be comparable to or stronger than those from conventional direct detection experiments, particularly for minerals such as gypsum. 

In the case of inelastic scattering, paleo-detectors are projected to provide improved sensitivity relative to conventional direct detection experiments for DM masses $m_\chi \gtrsim 50~\mathrm{GeV}/c^2$ and mass splittings $\delta_m \lesssim 50~\mathrm{keV}/c^2$ for most NREFT operators. For larger mass splittings, $\delta_m \sim 100~\mathrm{keV}/c^2$, conventional direct detection experiments outperform the projected sensitivity of paleo-detectors due to heavier target isotopes, while paleo-detectors lose sensitivity to inelastic interactions with $\delta_m \gtrsim 100~\mathrm{keV}/c^2$ owing to the absence of sufficiently heavy nuclei in typical target minerals.

\section{Conclusion} \label{conclusion}

In this work, we estimated the sensitivity of paleo-detectors—natural minerals that record and retain nuclear damage tracks over geological timescales—to WIMP–nucleon interactions described by a NREFT, considering both elastic and inelastic scattering, and isoscalar couplings. Specifically, considering solar, supernova, and atmospheric neutrinos, as well as radiogenic backgrounds computed following Ref.~\cite{Baum:2021jak}, we derived the projected 90\% confidence-level upper limits on the isoscalar NREFT coupling constants for both scattering types.

We considered two representative read-out scenarios: a high-resolution scenario with a spatial resolution of $\sigma_x = 1\ \mathrm{nm}$ and a sample mass of $M = 10\ \mathrm{mg}$, and a high-exposure scenario with $\sigma_x = 15\ \mathrm{nm}$ and $M = 100\ \mathrm{g}$. The analysis has been performed for the minerals gypsum, halite, olivine, and muscovite (discussed in the main text), and for sinjarite, epsomite, phlogopite, and nchwaningite (included in the Appendix).

The projected paleo-detector sensitivities were compared with the 90$\%$ confidence-level limits from the XENON100 \cite{XENON:2017fdd}, LUX–ZEPLIN \cite{LZ:2023lvz}, and PandaX–II~\cite{PandaX-II:2018woa} experiments, as well as with the 95$\%$ Bayesian credible region of the two-dimensional marginalized posterior distribution from SuperCDMS~\cite{SuperCDMS:2022crd}. For DM masses between $1\ \mathrm{GeV}/c^2$ and $10\ \mathrm{GeV}/c^2$, where the background is dominated by neutrinos and the expected track lengths are relatively short, paleo-detectors exhibit superior sensitivity in the HR scenario compared to conventional direct detection experiments. For heavier DM masses in the range $10\ \mathrm{GeV}/c^2 - 5\ \mathrm{TeV}/c^2$, where the background is dominated by radiogenics and the expected track lengths are relatively long, several paleo-detectors provide sensitivities comparable to or better than those of existing experiments for most NREFT operators in the HE scenario. In the main text, we have presented track length spectra for the spin-independent NREFT operators $\mathcal{O}_{1}^{s}$, $\mathcal{O}_{5}^{s}$, $\mathcal{O}_{8}^{s}$, and $\mathcal{O}_{11}^{s}$, including the corresponding background spectra, while results for the remaining NREFT operators are provided in the Appendix.

Throughout this analysis, we assumed an one-to-one relationship between the observed track length $x_T$ and the recoil energy $E_R$ for DM masses in the range $1\ \mathrm{GeV}/c^2 - 5\ \mathrm{TeV}/c^2$. Within this mass window, this approximation yields accurate results, as demonstrated in Ref.~\cite{Fung:2025cub}. A natural extension of this work is to explore lighter DM masses, for which the one-to-one mapping becomes inaccurate. In that regime, following the method described in Ref.~\cite{Fung:2025cub}, a recoil with energy $E_R$ can produce a distribution of possible track lengths $x_T$, and the probability of track formation for a given recoil energy becomes non-trivial and must be modeled explicitly.

We also identify as an important direction for future work the use of paleo-detectors to distinguish signals arising from different NREFT operators. Such analyses have been performed for conventional direct detection experiments, for example in Ref.~\cite{Kavanagh:2015jma}, which investigates how nuclear recoil direction, as well as energy, can be used to distinguish between different NREFT operators. Rather than directional information, paleo-detectors could leverage relatively large exposures to determine how well recoil track length spectra for different (combinations of) NREFT operators compare to data from a given WIMP signal.

\acknowledgments

This work was supported by the U.S. National Science Foundation Growing Convergence Research award 2428507.
KF is grateful for
support from the Jeff \& Gail Kodosky Endowed Chair in
Physics at the University of Texas. KF acknowledges
support from the Swedish Research Council (Contract
No. 638-2013-8993). The work of CK~is supported in part by the U.S.~Department of Energy, Office of Science, Office of High Energy Physics under Award Number DE-SC0024693.
The work of PS is co-funded by the European Union's Horizon Europe research and innovation program under the Marie Sklodowska-Curie COFUND Postdoctoral Programme grant agreement No. 101081355-SMASH and by the Republic of Slovenia and the European Union from the European Regional Development Fund. Views and opinions expressed are however those of the authors only and do not necessarily reflect those of the European Union or European Research Executive Agency. Neither the European Union nor the granting authority can be held responsible for them.

\appendix* \label{appendix}
\section{}

Within the main body of the text, we explored the predicted spectra in paleo-detector experiments for WIMP--nucleus scattering in the case of the spin-independent NREFT operators \(\mathcal{O}_{1}^{s}\), \(\mathcal{O}_{5}^{s}\), \(\mathcal{O}_{8}^{s}\), and \(\mathcal{O}_{11}^{s}\) for a particular set of minerals that are representative candidates for paleo-detector target minerals. As mentioned previously, \(\mathcal{O}_{1}^{s}\) is the canonical spin-independent operator often used to illustrate cross-section/mass bounds from direct detection experiments. Here in the Appendix, we study instead the predicted spectra for spin-dependent NREFT operators.
We note that one of these, \(\mathcal{O}_{4}^{s}\) is the canonical spin-dependent operator often used to illustrate cross-section/mass bounds from direct detection experiments. Further, the Appendix shows predictions on target minerals different from the ones considered in the main text, for all NREFT operators.

\begin{figure}
    \captionsetup{justification=raggedright,singlelinecheck=false}
    \centering
    \begin{subfigure}{0.33\textwidth}
    \includegraphics[width=\linewidth]{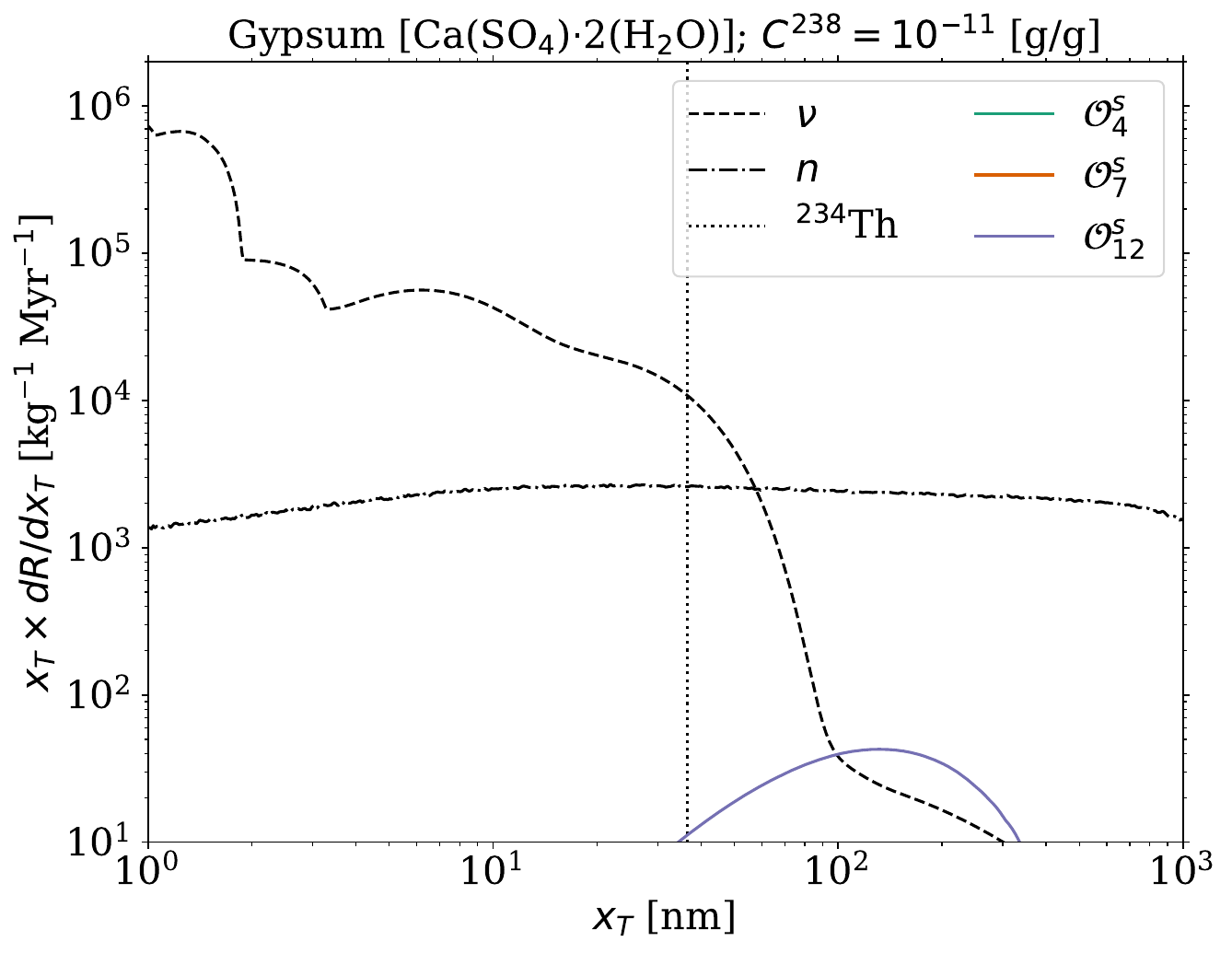}
    \end{subfigure}
    \begin{subfigure}{0.33\textwidth}
    \includegraphics[width=\linewidth]{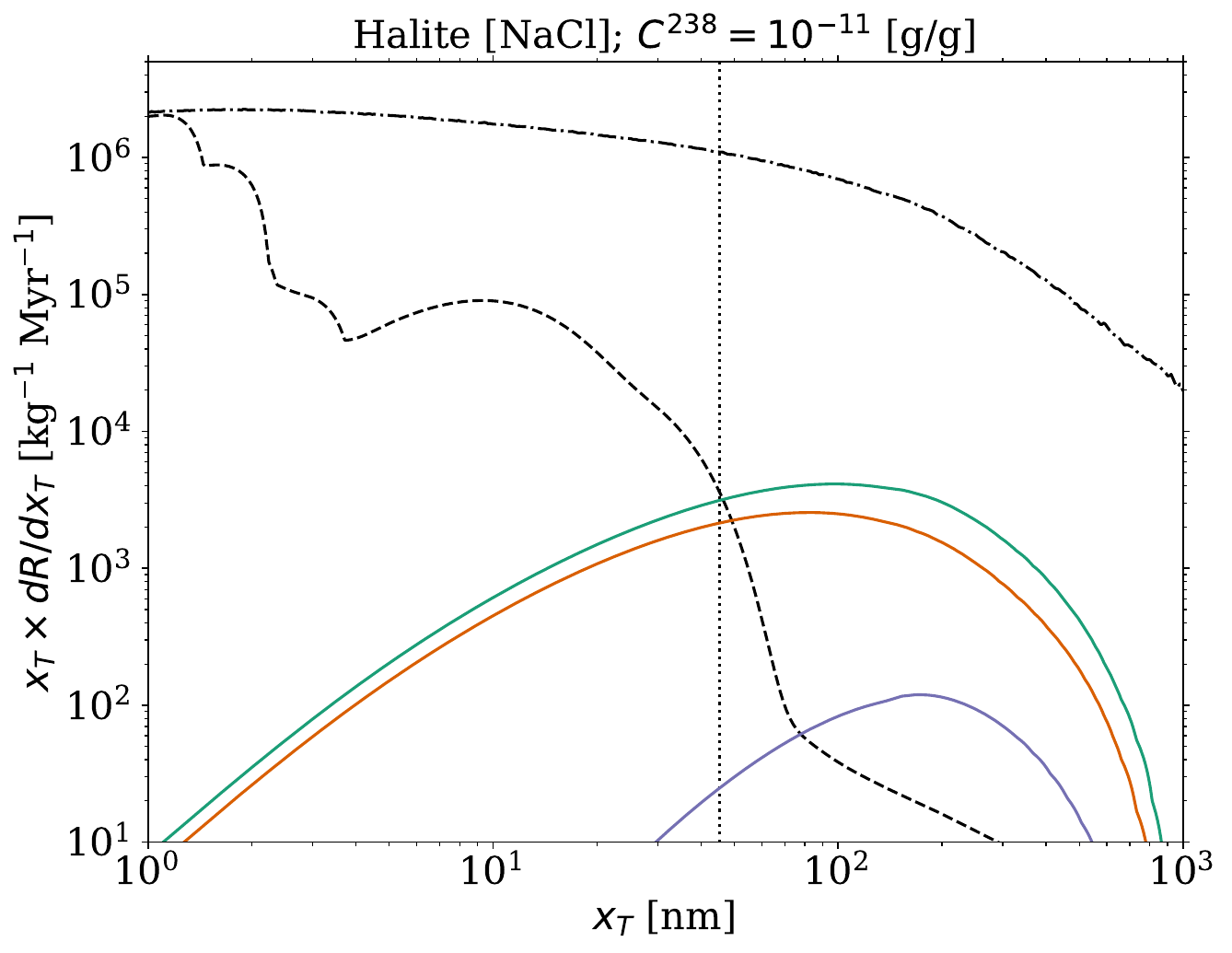}
    \end{subfigure}
    
    \vspace{0.1ex}

    \begin{subfigure}{0.33\textwidth}
    \includegraphics[width=\linewidth]{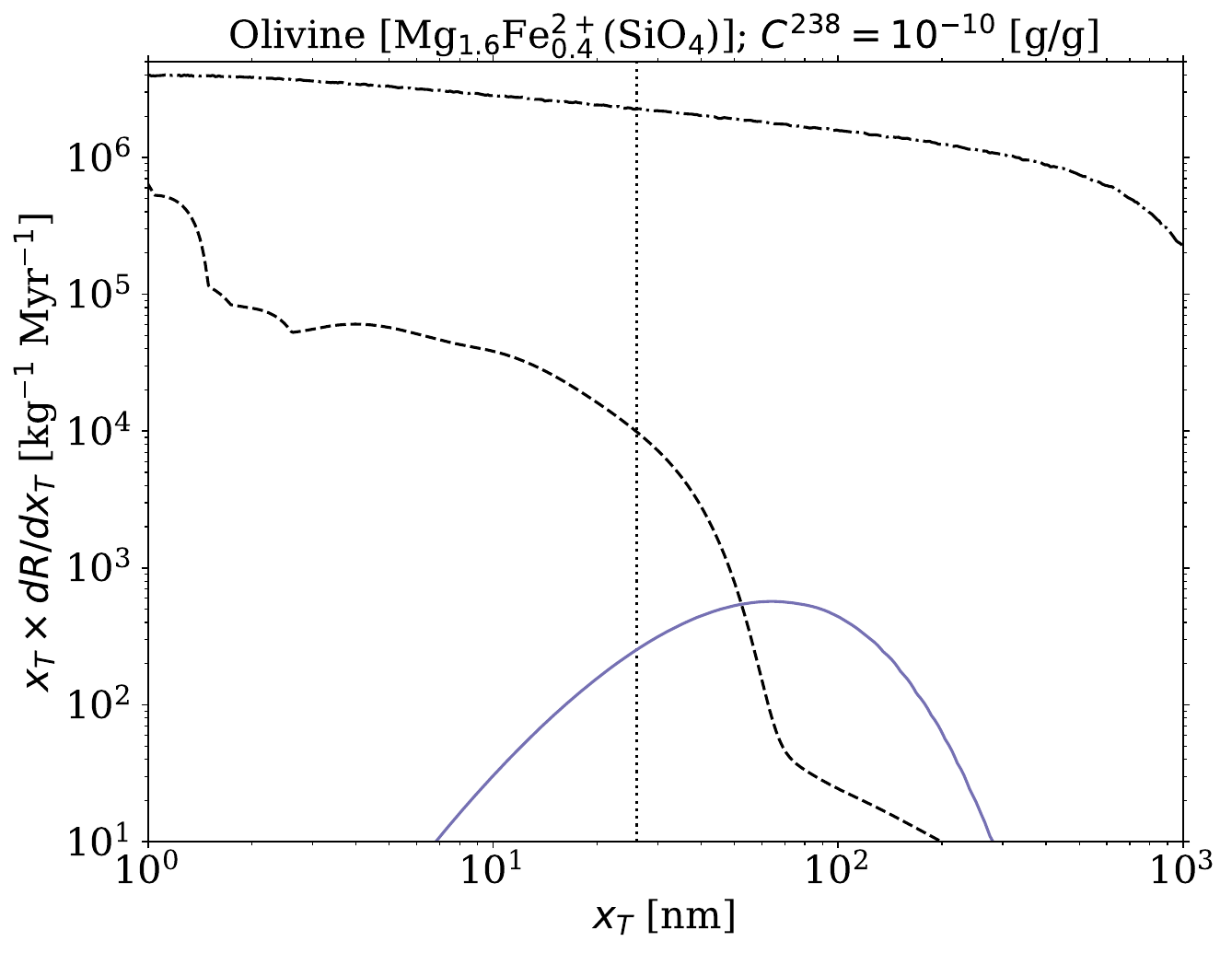}
    \end{subfigure}
    \begin{subfigure}{0.33\textwidth}
    \includegraphics[width=\linewidth]{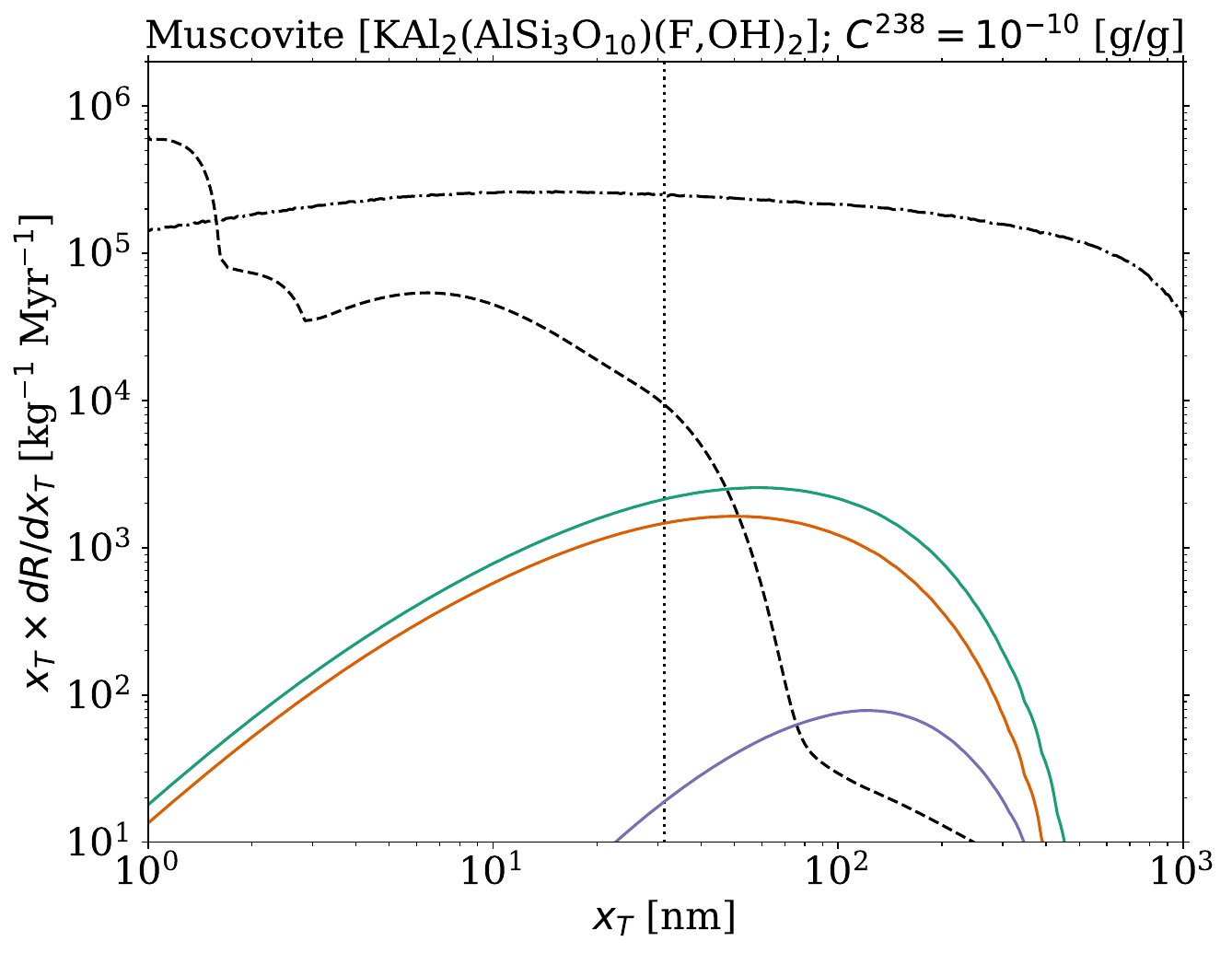}
    \end{subfigure}
    \caption{Differential track production rate $dR/d(\ln x_T)$ for elastic scattering, given per unit exposure and per logarithmic interval in track length, shown as a function of the track length $x_T$ for the spin–dependent NREFT operators $\mathcal{O}^{s}_{4}$, $\mathcal{O}^{s}_{7}$, and $\mathcal{O}^{s}_{12}$, which are independent of the momentum transfer $\vec{q}$. The calculations assume isoscalar interactions ($c^{p}=c^{n}$). Results are shown for gypsum and halite with ${}^{238}$U concentrations of $10^{-11}\ \mathrm{g/g}$, and for olivine and muscovite with ${}^{238}$U concentrations of $10^{-10}\ \mathrm{g/g}$. The DM mass is fixed at $500\ \mathrm{GeV}/c^{2}$. For comparison, we also show background spectra induced by neutrinos ($\nu$), radiogenic neutrons ($n$), and ${}^{238}\text{U}\to{}^{234}\text{Th}+\alpha$ recoils (${}^{234}\text{Th}$); see Sec.~\ref{background}. For the normalization of the DM track production spectra, we set the NREFT coupling constants to $(c^{s}_{4}\ m_{v}^{2})^{2}=2\cdot10^{-3}$, $(c^{s}_{7}\ m_{v}^{2})^{2}=1\cdot10^{3}$, and $(c^{s}_{12}\ m_{v}^{2})^{2}=3\cdot10^{-4}$, where $m_{v} = 246.2\ \mathrm{GeV}$ denotes the electroweak mass scale. These values are chosen to be compatible with the upper limits from the LUX–ZEPLIN experiment \cite{LZ:2023lvz}. The dominant isotopes of gypsum and olivine have spin-zero ground states, and therefore the differential track production rates for the operators $\mathcal{O}_4^s$ and $\mathcal{O}_7^s$ are negligible and are not shown for these two minerals.}
    \label{fig:Spectrum_SDq0}
\end{figure}

\begin{figure}
    \captionsetup{justification=raggedright,singlelinecheck=false}
    \centering
    \begin{subfigure}{0.33\textwidth}
    \includegraphics[width=\linewidth]{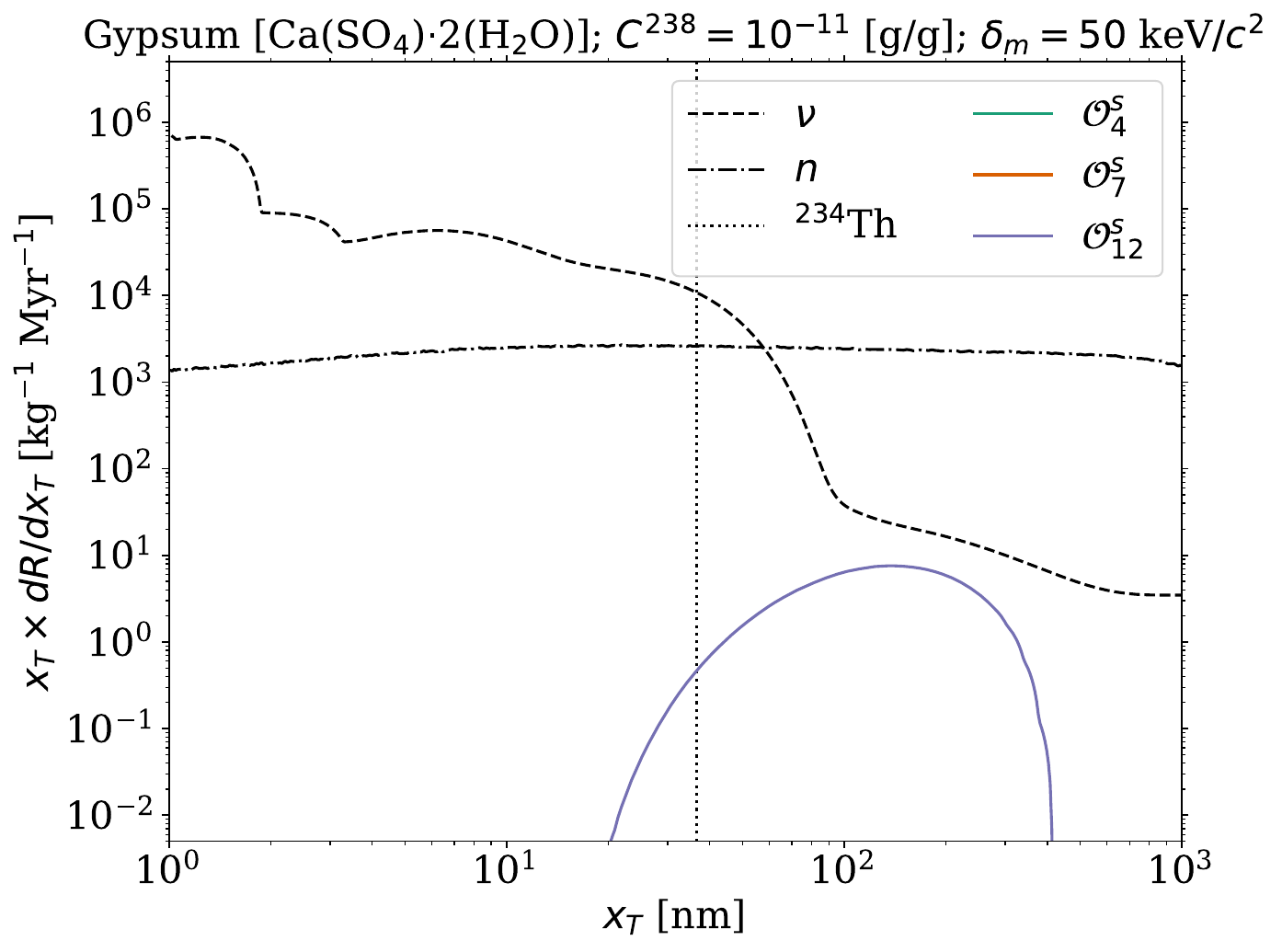}
    \end{subfigure}
    \begin{subfigure}{0.33\textwidth}
    \includegraphics[width=\linewidth]{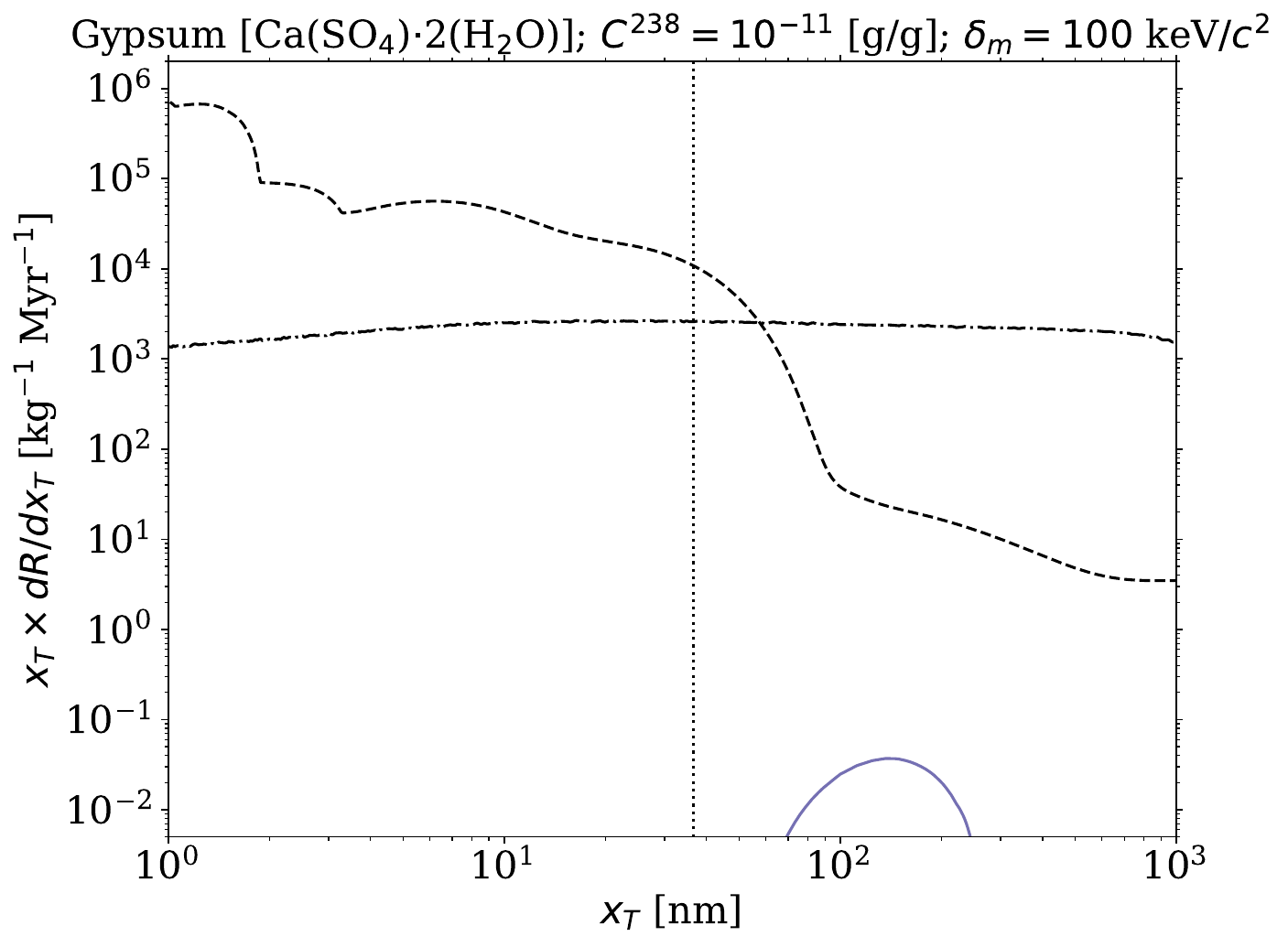}
    \end{subfigure}
    
    \vspace{0.1ex}

    \begin{subfigure}{0.33\textwidth}
    \includegraphics[width=\linewidth]{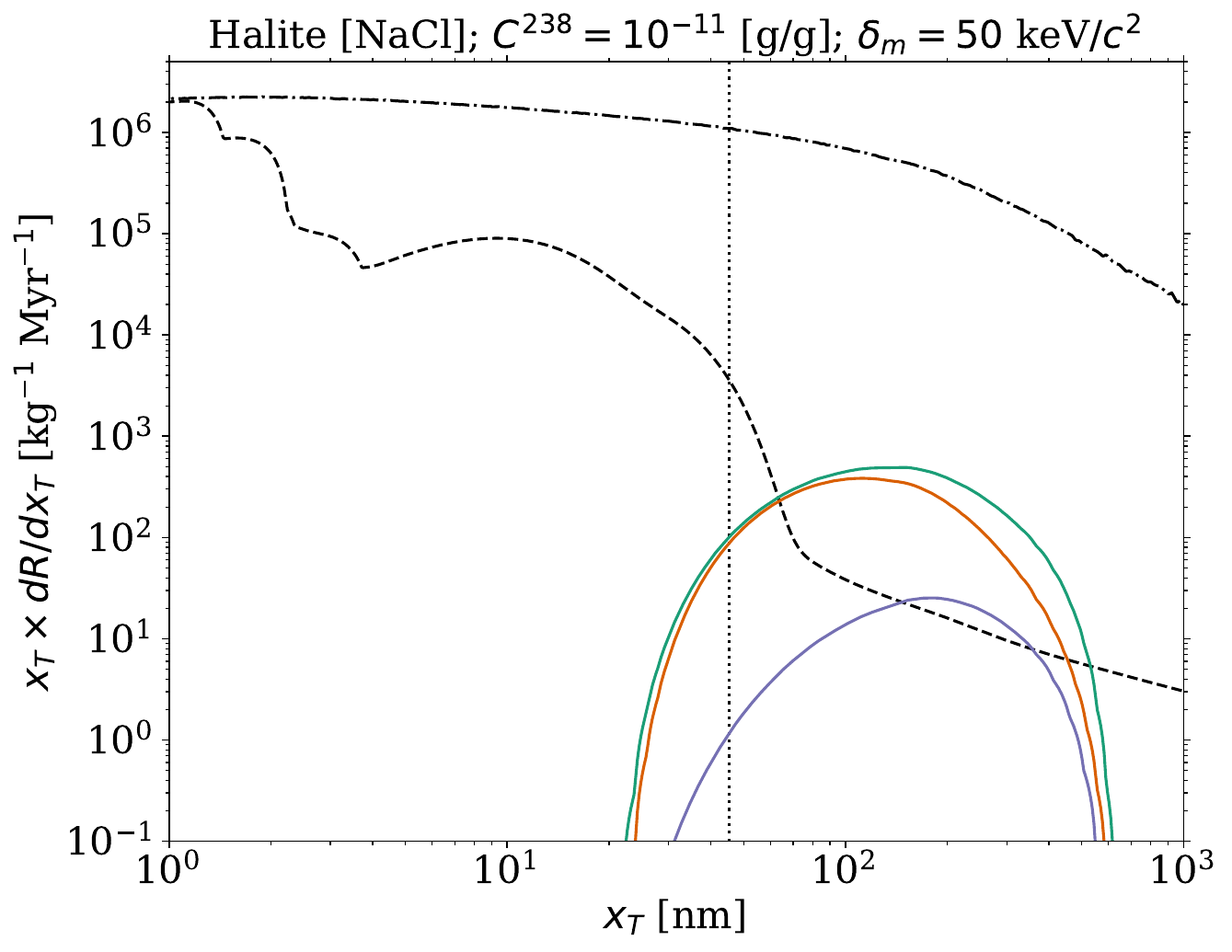}
    \end{subfigure}
    \begin{subfigure}{0.33\textwidth}
    \includegraphics[width=\linewidth]{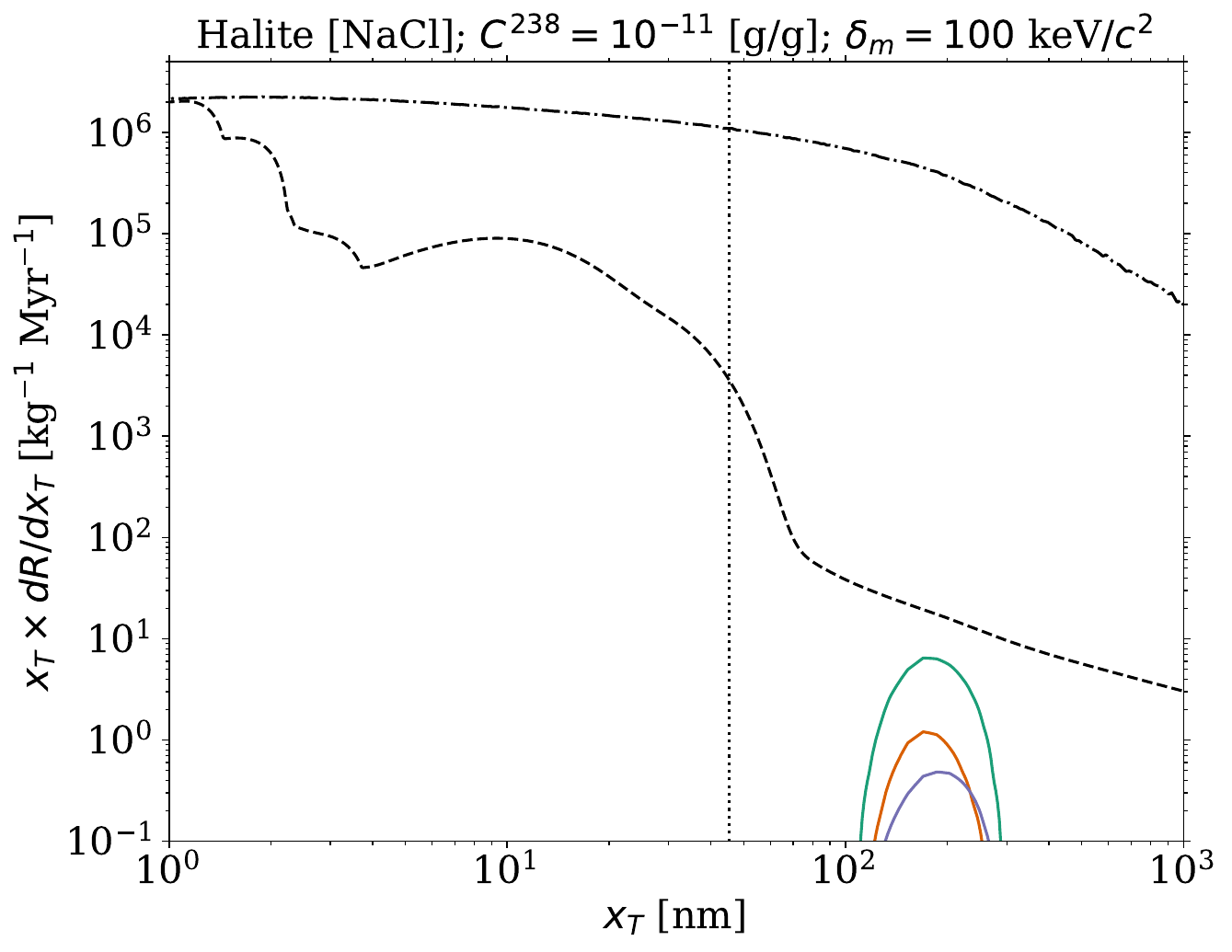}
    \end{subfigure}
    \caption{Differential track production rate $dR/d(\ln x_T)$ for inelastic scattering, given per unit exposure and per logarithmic interval in track length, shown as a function of the track length $x_T$ for the spin–dependent NREFT operators $\mathcal{O}^{s}_{4}$, $\mathcal{O}^{s}_{7}$, and $\mathcal{O}^{s}_{12}$, which are independent of the momentum transfer $\vec{q}$. The calculations assume isoscalar interactions ($c^{p}=c^{n}$). The DM mass is fixed at $500\ \mathrm{GeV}/c^{2}$. In the left panel, the mass splitting is $\delta_{m}=50\ \mathrm{keV}/c^{2}$. The NREFT coupling constants are set to $(c^{s}_{4} m_{v}^{2})^{2}=4.5\cdot10^{-3}$, $(c^{s}_{7} m_{v}^{2})^{2}=4.5\cdot10^{3}$, and $(c^{s}_{12} m_{v}^{2})^{2}=6\cdot10^{-4}$, where $m_{v}=246.2\ \mathrm{GeV}$ denotes the electroweak mass scale. In the right panel, the mass splitting is $\delta_{m}=100\ \mathrm{keV}/c^{2}$. The coupling constants are set to $(c^{s}_{4} m_{v}^{2})^{2}=1.6\cdot10^{-2}$, $(c^{s}_{7} m_{v}^{2})^{2}=2.3\cdot10^{4}$, and $(c^{s}_{12} m_{v}^{2})^{2}=2.3\cdot10^{-3}$. In both panels, the values for the coupling constants are chosen to be compatible with the upper limits for inelastic WIMP--nucleon scattering from the LUX--ZEPLIN experiment \cite{LZ:2023lvz}. For comparison, background spectra induced by neutrinos ($\nu$), radiogenic neutrons ($n$), and ${}^{238}\text{U}\to{}^{234}\text{Th}+\alpha$ recoils (${}^{234}\text{Th}$) are also shown; see Sec.~\ref{background}. Results are presented for gypsum (upper panels) and halite (lower panels) with a ${}^{238}$U concentration of $10^{-11}\,\mathrm{g/g}$. The dominant isotopes of gypsum have spin-zero ground states, and therefore the differential track production rates for the operators $\mathcal{O}_4^s$ and $\mathcal{O}_7^s$ are negligible and are not shown for this mineral.}
    \label{fig:Spectrum_SDq0_inelastic}
\end{figure}

We perform an exploration of spin-dependent NREFT operators. In Fig.~\ref{fig:Spectrum_SDq0}, we show the differential track production rate \(dR/dx_T\) for elastic, isoscalar interactions described by the spin-dependent NREFT operators \(\mathcal{O}_{4}^{s}\), \(\mathcal{O}_{7}^{s}\), and \(\mathcal{O}_{12}^{s}\), which are independent from the momentum transfer $\vec{q}$, assuming a DM mass of \(500~\mathrm{GeV}/c^{2}\). For the normalization of the DM track production spectra, the NREFT coupling constants are chosen to be consistent with the upper limits set by the LUX–ZEPLIN experiment~\cite{LZ:2023lvz}. Results are shown for the minerals: gypsum, halite, olivine, and muscovite, together with the background spectra discussed in Section~\ref{background}. In Fig.~\ref{fig:Spectrum_SDq0_inelastic}, we present the corresponding spectra for inelastic, isoscalar interactions described by the same spin-dependent NREFT operators, for a DM mass of \(500~\mathrm{GeV}/c^{2}\) and mass splittings of \(\delta_{m}=50~\mathrm{keV}/c^{2}\) (left panel) and \(\delta_{m}=100~\mathrm{keV}/c^{2}\) (right panel). Here, the NREFT coupling constants are set to values consistent with the upper limits for inelastic WIMP–nucleon scattering from the LUX–ZEPLIN experiment~\cite{LZ:2023lvz}. Results are shown for gypsum and halite.  

\begin{figure}
    \captionsetup{justification=raggedright,singlelinecheck=false}
    \centering
    
    \begin{subfigure}{0.33\textwidth}
    \includegraphics[width=\linewidth]{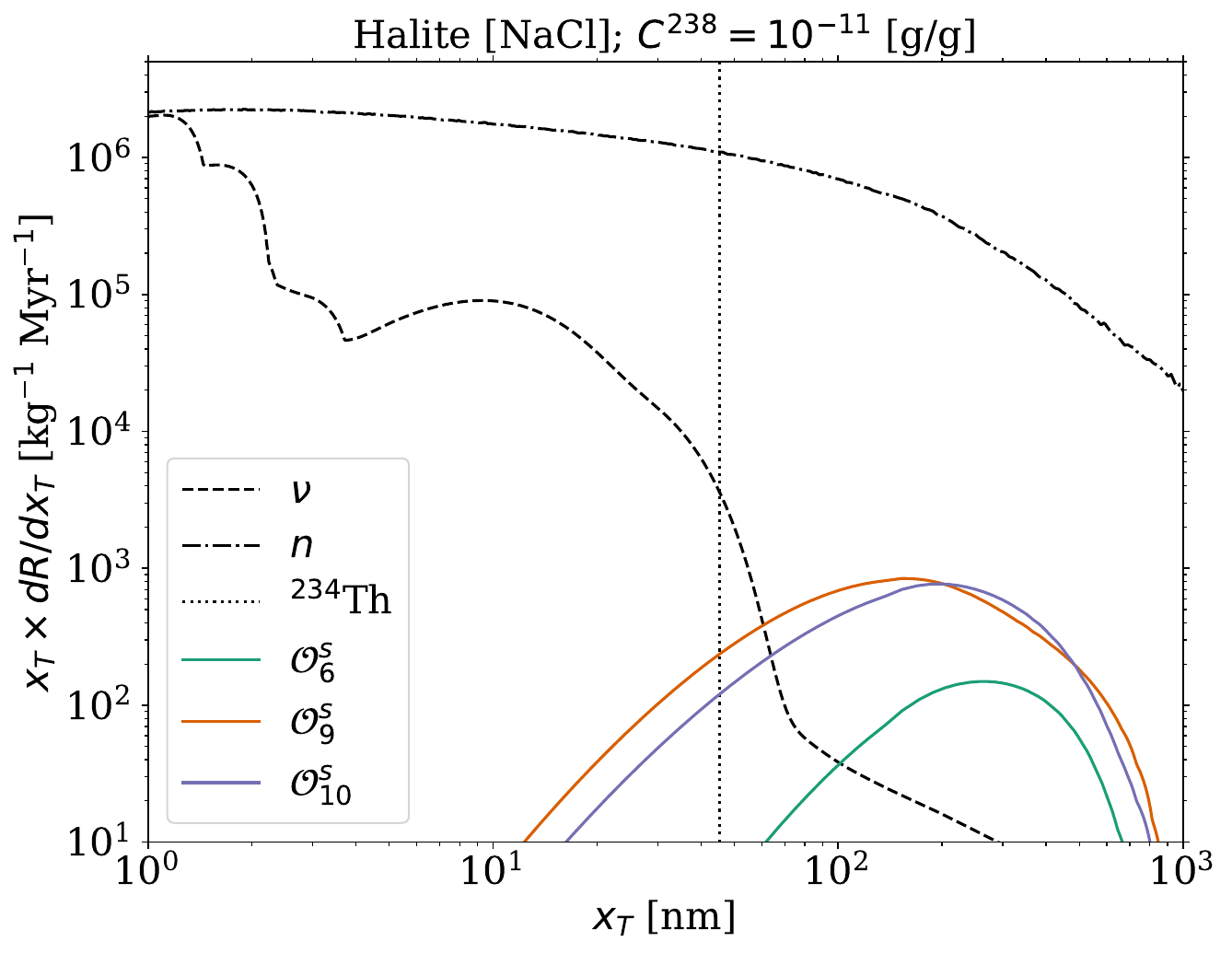}
    \end{subfigure}
    \begin{subfigure}{0.33\textwidth}
    \includegraphics[width=\linewidth]{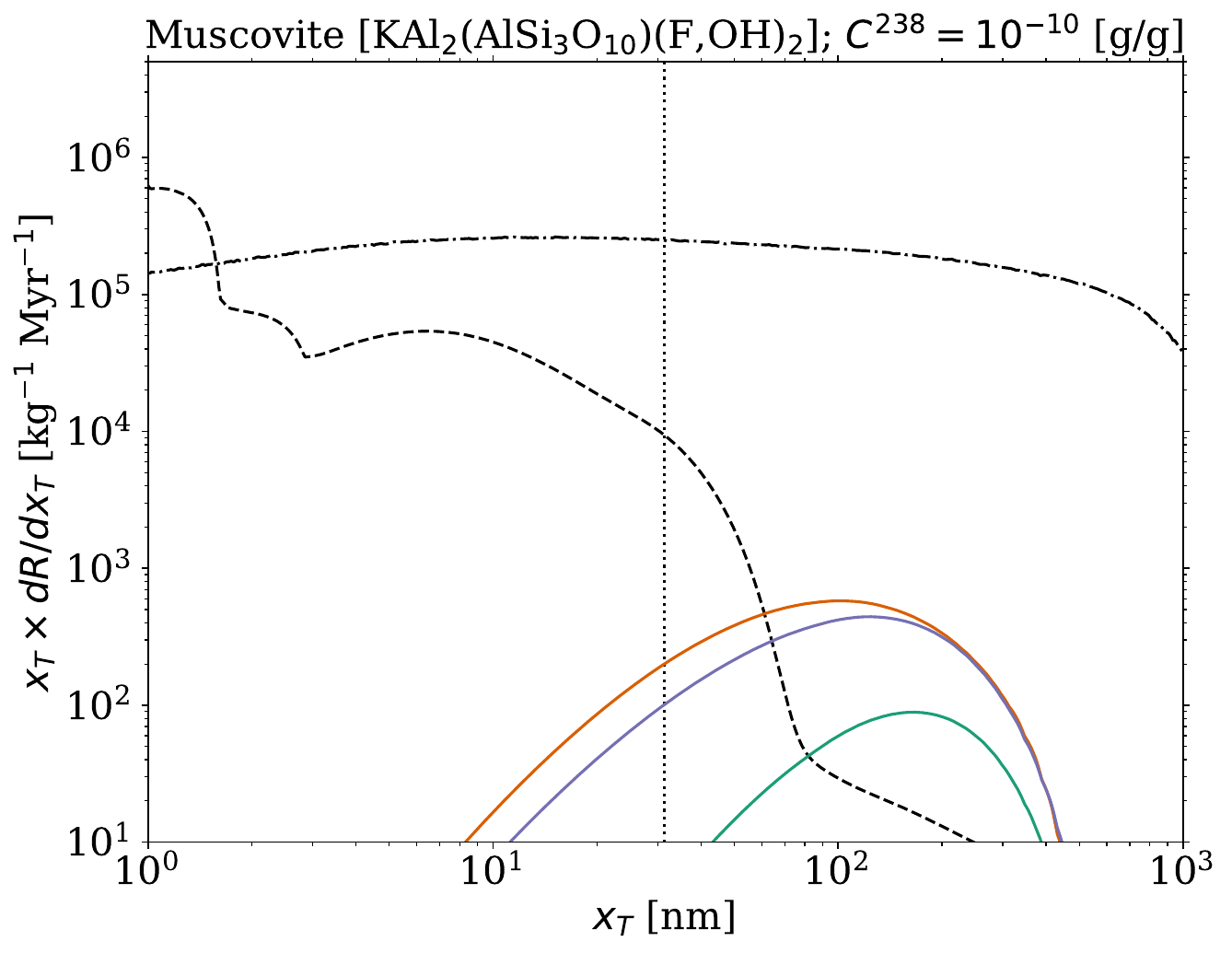}
    \end{subfigure}

    \caption{Differential track production rate $dR/d(\ln x_T)$ for elastic scattering, given per unit exposure and per logarithmic interval in track length, shown as a function of the track length $x_T$ for the spin–dependent NREFT operators $\mathcal{O}^{s}_{6}$, $\mathcal{O}^{s}_{9}$, and $\mathcal{O}^{s}_{10}$, which are independent of the component of the WIMP–nucleon relative velocity perpendicular to the momentum transfer, $\vec{v}^{\perp}$. The calculations assume isoscalar interactions ($c^{p}=c^{n}$). Results are shown for halite with a ${}^{238}$U concentration of $10^{-11}\ \mathrm{g/g}$, and for muscovite with a ${}^{238}$U concentration of $10^{-10}\ \mathrm{g/g}$. The DM mass is fixed at $500\ \mathrm{GeV}/c^{2}$. For comparison, we also show background spectra induced by neutrinos ($\nu$), radiogenic neutrons ($n$), and ${}^{238}\text{U}\to{}^{234}\text{Th}+\alpha$ recoils (${}^{234}\text{Th}$); see Sec.~\ref{background}. For the normalization of the DM track production spectra, we set the NREFT coupling constants to $(c^{s}_{6}\ m_{v}^{2})^{2}=5$, $(c^{s}_{9}\ m_{v}^{2})^{2}=2.3\cdot10^{-1}$, and $(c^{s}_{10}\ m_{v}^{2})^{2}=5\cdot10^{-2}$, where $m_{v} = 246.2\ \mathrm{GeV}$ denotes the electroweak mass scale. These values are chosen to be compatible with the upper limits from the LUX–ZEPLIN experiment \cite{LZ:2023lvz}.}
    \label{fig:Spectrum_SDq1v0}
\end{figure}

\begin{figure}
    \captionsetup{justification=raggedright,singlelinecheck=false}
    \centering
    \begin{subfigure}{0.33\textwidth}
    \includegraphics[width=\linewidth]{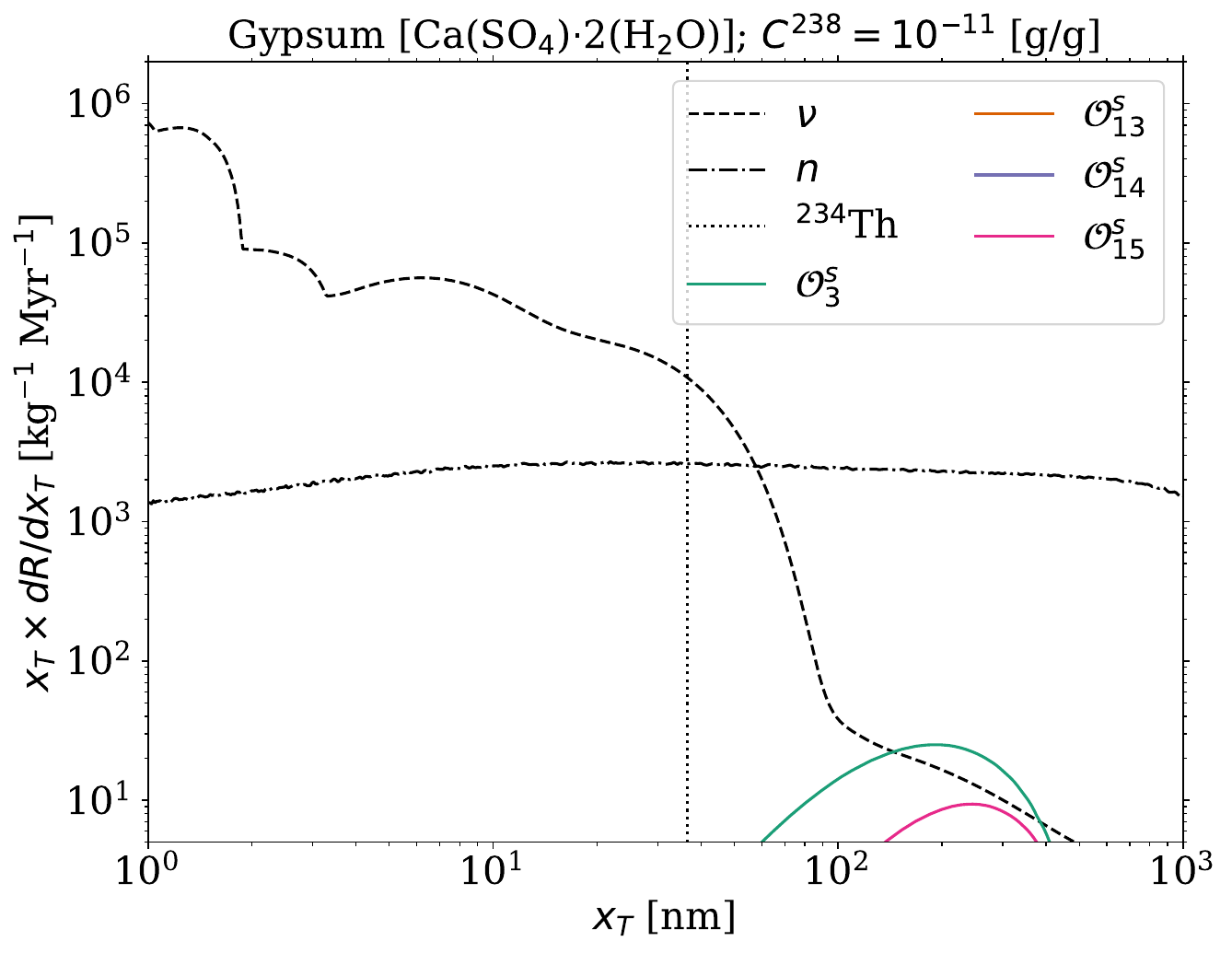}
    \end{subfigure}
    \begin{subfigure}{0.33\textwidth}
    \includegraphics[width=\linewidth]{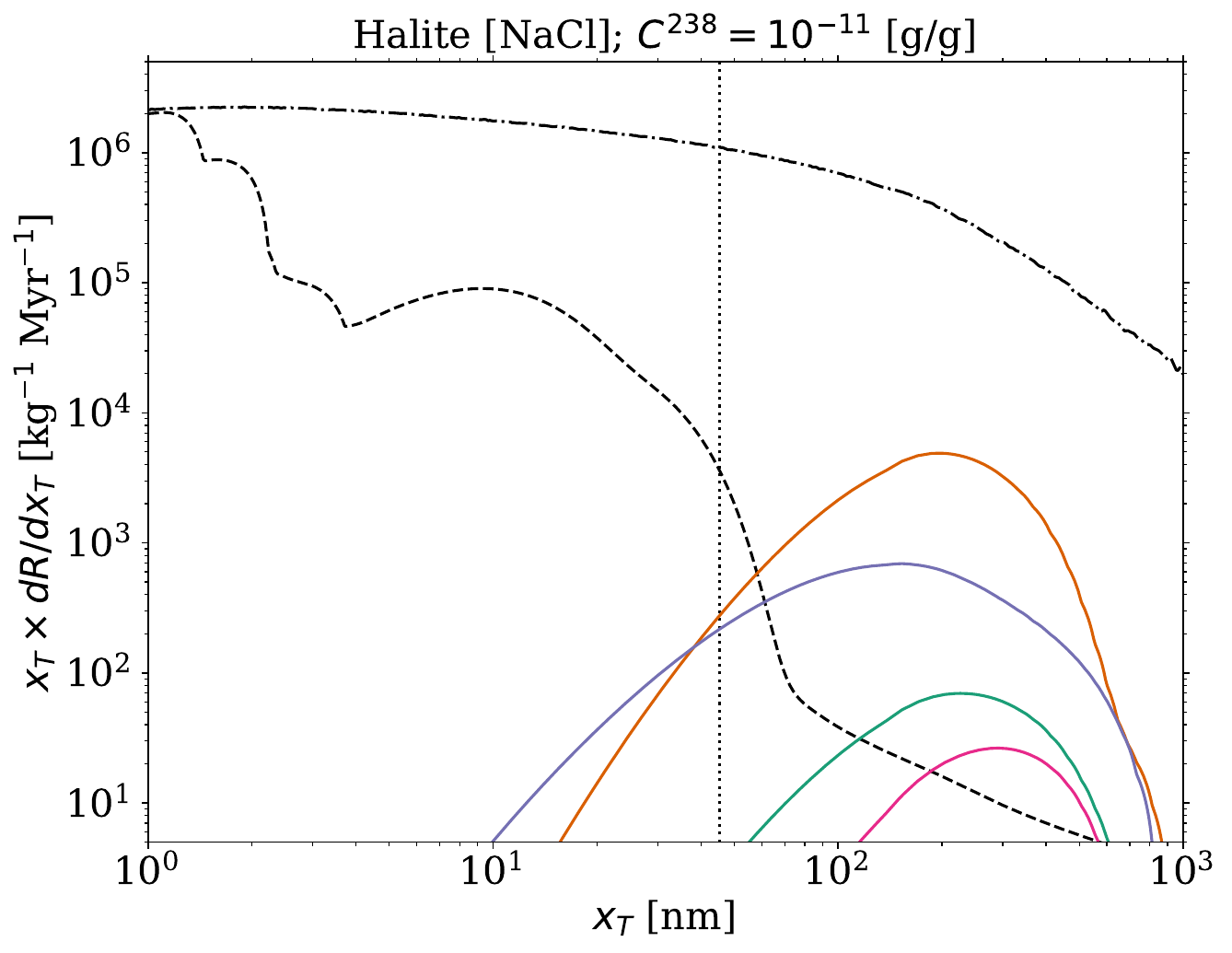}
    \end{subfigure}
    
    \vspace{0.1ex}

    \begin{subfigure}{0.33\textwidth}
    \includegraphics[width=\linewidth]{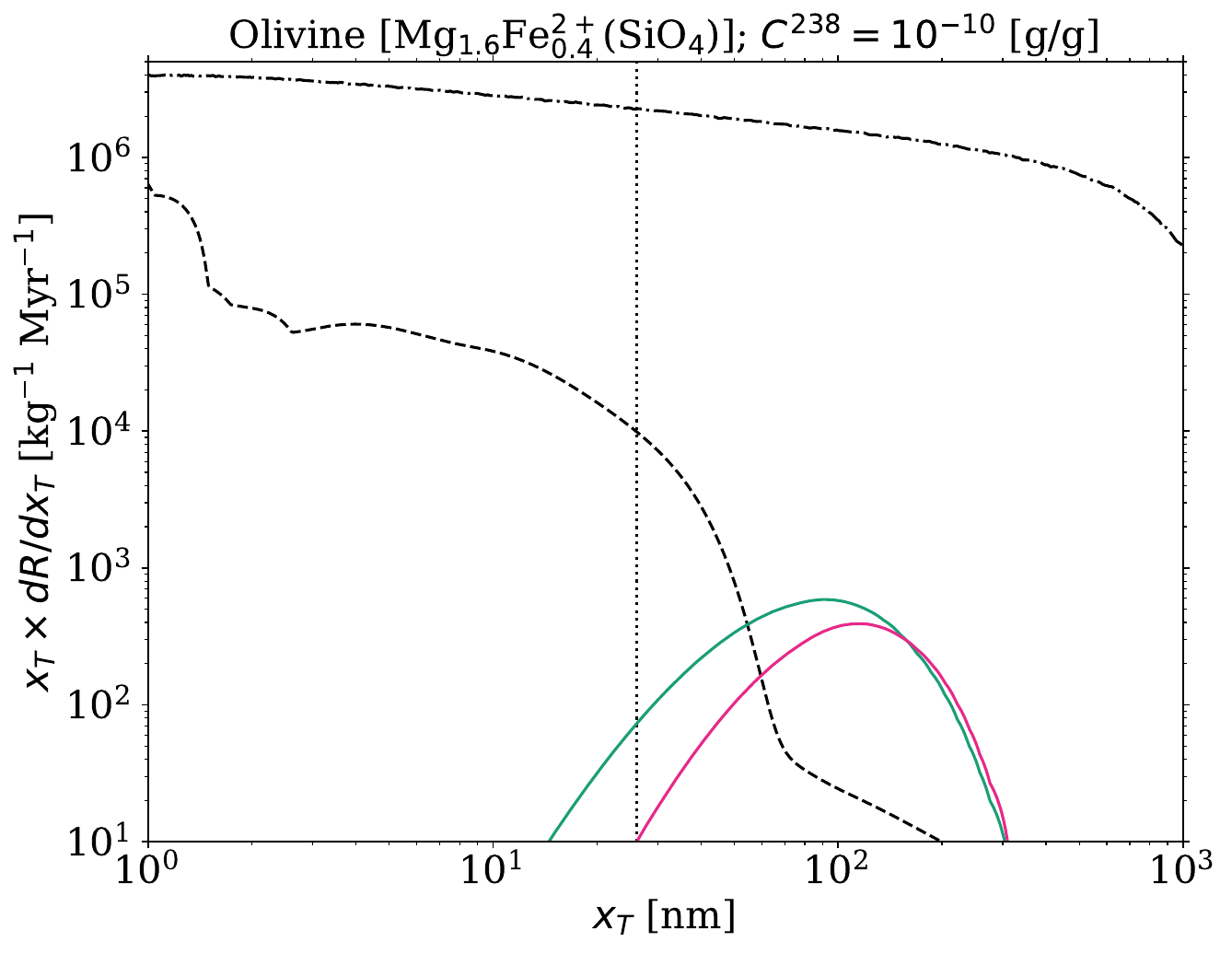}
    \end{subfigure}
    \begin{subfigure}{0.33\textwidth}
    \includegraphics[width=\linewidth]{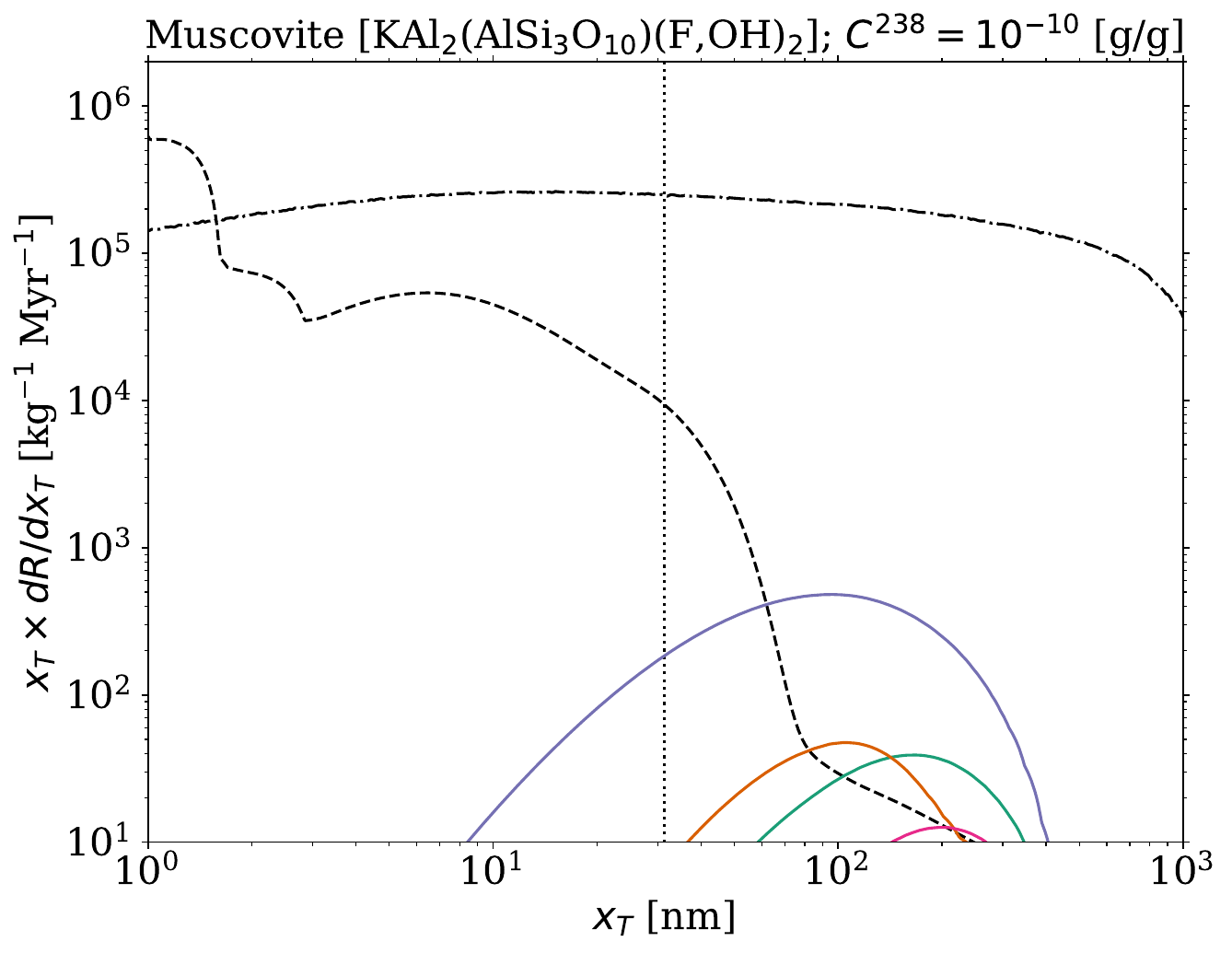}
    \end{subfigure}
    \caption{Differential track production rate $dR/d(\ln x_T)$ for elastic scattering, given per unit exposure and per logarithmic interval in track length, shown as a function of the track length $x_T$ for the spin–dependent NREFT operators $\mathcal{O}^{s}_{3}$, $\mathcal{O}^{s}_{13}$, $\mathcal{O}^{s}_{14}$, and $\mathcal{O}^{s}_{15}$, which depend on both the momentum transfer $\vec{q}$, and the WIMP–nucleon relative velocity perpendicular to the momentum transfer, $\vec{v}^{\perp}$. The calculations assume isoscalar interactions ($c^{p}=c^{n}$). Results are shown for gypsum and halite with ${}^{238}$U concentrations of $10^{-11}\ \mathrm{g/g}$, and for olivine and muscovite with ${}^{238}$U concentrations of $10^{-10}\ \mathrm{g/g}$. The DM mass is fixed at $500\ \mathrm{GeV}/c^{2}$. For comparison, we also show background spectra induced by neutrinos ($\nu$), radiogenic neutrons ($n$), and ${}^{238}\text{U}\to{}^{234}\text{Th}+\alpha$ recoils (${}^{234}\text{Th}$); see Sec.~\ref{background}. For the normalization of the DM track production spectra, we set the NREFT coupling constants to $(c^{s}_{3}\ m_{v}^{2})^{2}=6\cdot10^{-3}$, $(c^{s}_{13}\ m_{v}^{2})^{2}=4\cdot10^{2}$, $(c^{s}_{14}\ m_{v}^{2})^{2}=4\cdot10^{5}$, and $(c^{s}_{15}\ m_{v}^{2})^{2}=9\cdot10^{-1}$, where $m_{v} = 246.2\ \mathrm{GeV}$ denotes the electroweak mass scale. These values are chosen to be compatible with the upper limits from the LUX–ZEPLIN experiment \cite{LZ:2023lvz}. The dominant isotopes of gypsum and olivine have spin-zero ground states, and therefore the differential track production rates for the operators $\mathcal{O}_{13}^s$ and $\mathcal{O}_{14}^s$ are negligible and are not shown for these two minerals.}
    \label{fig:Spectrum_SDq1v1}
\end{figure}

\begin{figure}
    \captionsetup{justification=raggedright,singlelinecheck=false}
    \centering
    \includegraphics[width=0.35\textwidth]{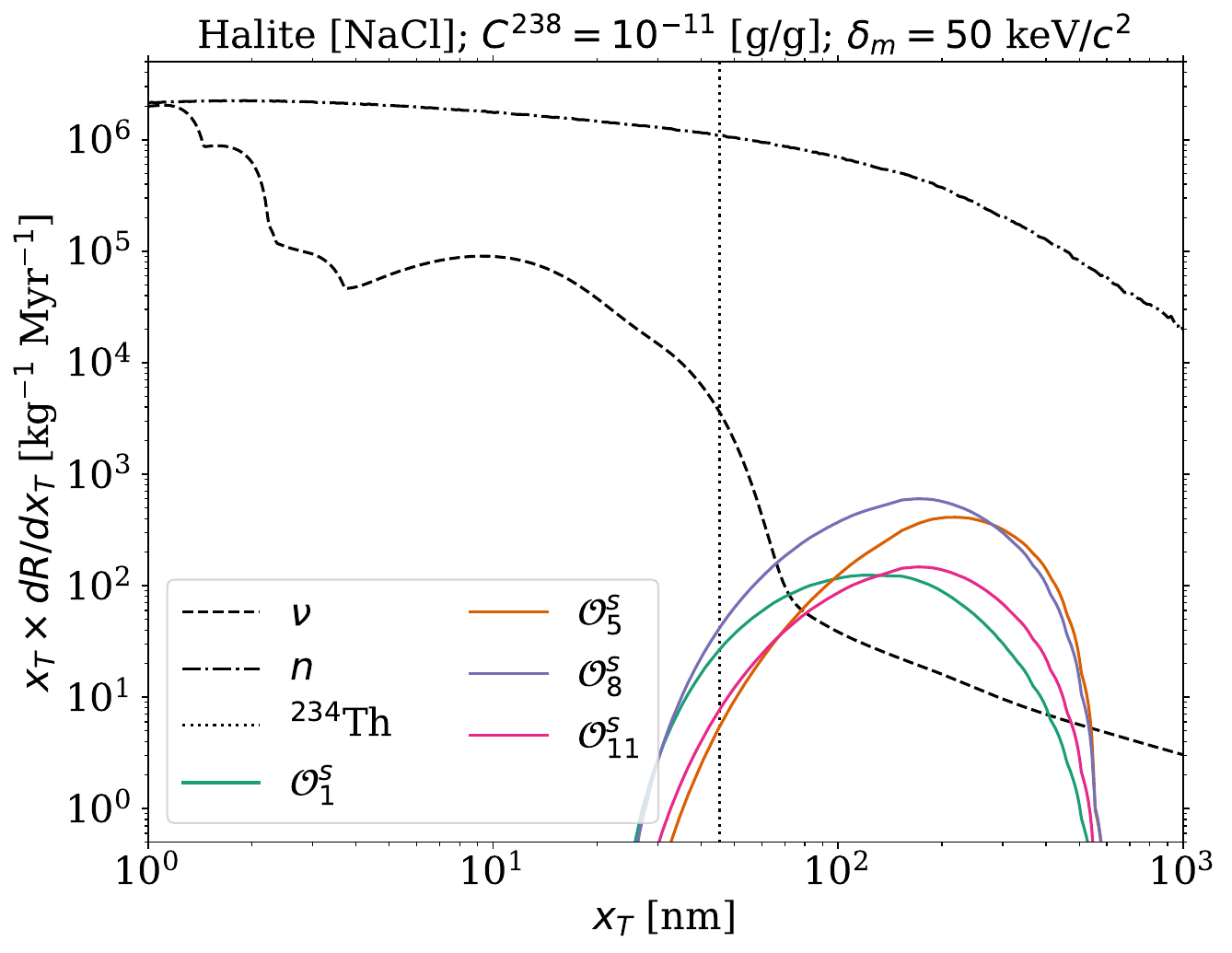}
    \includegraphics[width=0.356\textwidth]{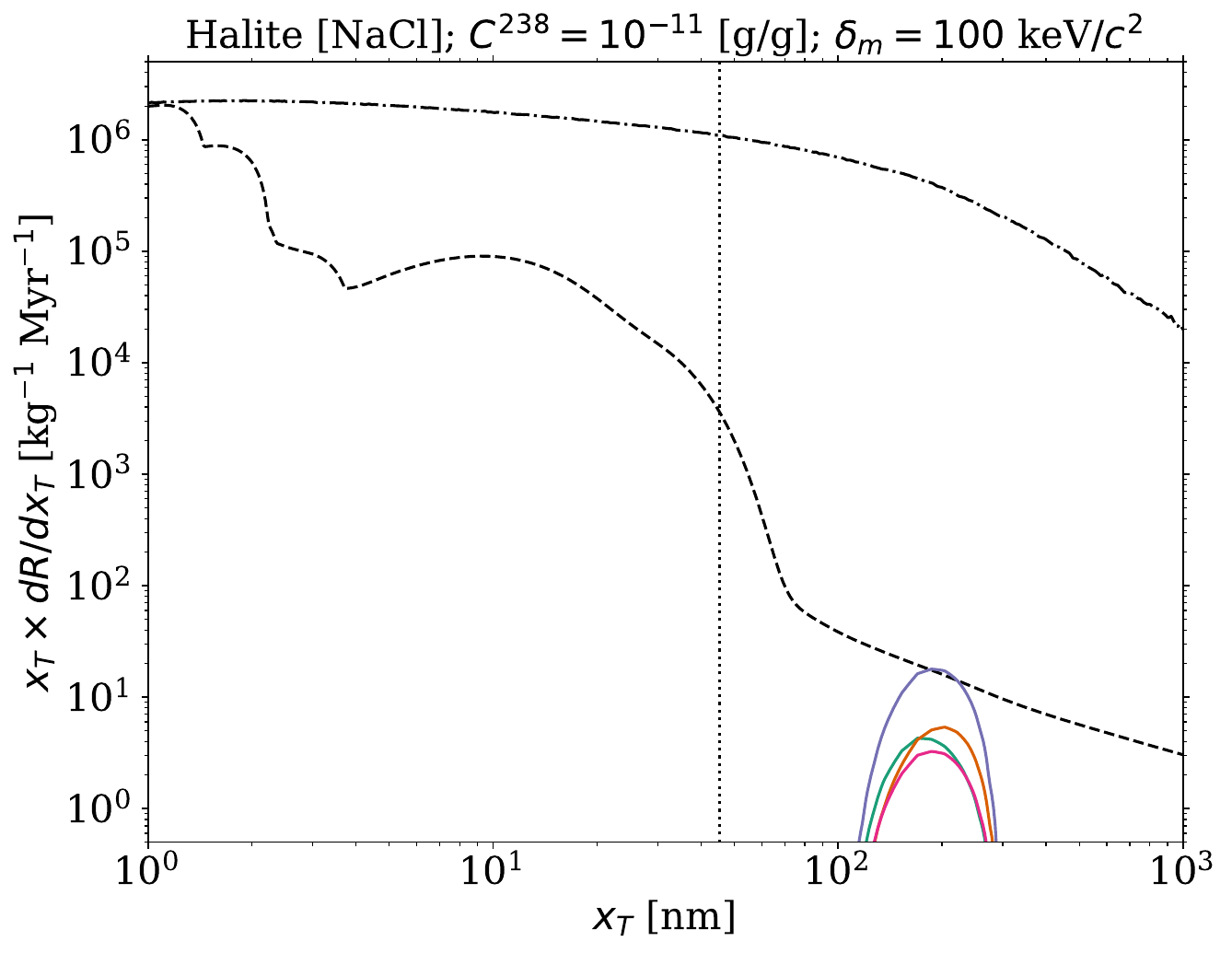}
    \caption{Differential track production rate $dR/d(\ln x_T)$ for inelastic scattering, given per unit exposure and per logarithmic interval in track length, shown as a function of the track length $x_T$ for the spin–independent NREFT operators $\mathcal{O}^{s}_{1}$, $\mathcal{O}^{s}_{5}$, $\mathcal{O}^{s}_{8}$, and $\mathcal{O}^{s}_{11}$. The calculations assume isoscalar interactions ($c^{p}=c^{n}$). The DM mass is fixed at $500\ \mathrm{GeV}/c^{2}$. In the left panel, the mass splitting is $\delta_{m}=50\ \mathrm{keV}/c^{2}$. The NREFT coupling constants are set to $(c^{s}_{1} m_{v}^{2})^{2}=1.5\cdot10^{-8}$, $(c^{s}_{5} m_{v}^{2})^{2}=5$, $(c^{s}_{8} m_{v}^{2})^{2}=5.6\cdot10^{-2}$, and $(c^{s}_{11} m_{v}^{2})^{2}=1.2\cdot10^{-5}$, where $m_{v}=246.2\ \mathrm{GeV}$ denotes the electroweak mass scale. In the right panel, the mass splitting is $\delta_{m}=100\ \mathrm{keV}/c^{2}$. The coupling constants are set to $(c^{s}_{1} m_{v}^{2})^{2}=1.3\cdot10^{-7}$, $(c^{s}_{5} m_{v}^{2})^{2}=1.4\cdot10^{1}$, $(c^{s}_{8} m_{v}^{2})^{2}=3.7\cdot10^{-1}$, and $(c^{s}_{11} m_{v}^{2})^{2}=5.4\cdot10^{-5}$. In both panels, the values for the coupling constants are chosen to be compatible with the upper limits for inelastic WIMP--nucleon scattering from the LUX--ZEPLIN experiment \cite{LZ:2023lvz}. For comparison, background spectra induced by neutrinos ($\nu$), radiogenic neutrons ($n$), and ${}^{238}\text{U}\to{}^{234}\text{Th}+\alpha$ recoils (${}^{234}\text{Th}$) are also shown; see Sec.~\ref{background}. Results are presented for halite with a ${}^{238}$U concentration of $10^{-11}\,\mathrm{g/g}$.}
    \label{fig:Spectrum_SI_Halite_inelastic}
\end{figure}

\begin{figure}
    \captionsetup{justification=raggedright,singlelinecheck=false}
    \centering
    \includegraphics[width=0.35\textwidth]{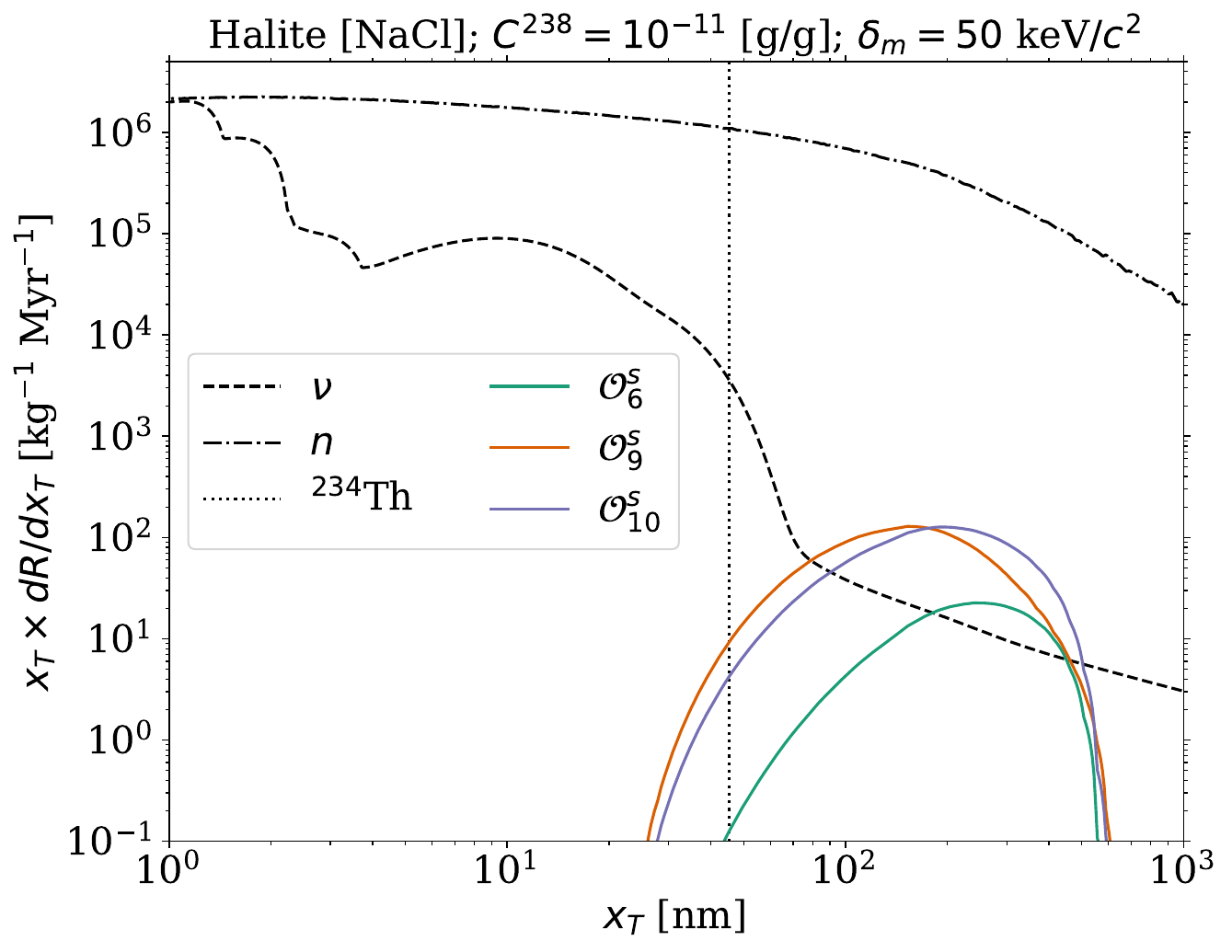}
    \includegraphics[width=0.356\textwidth]{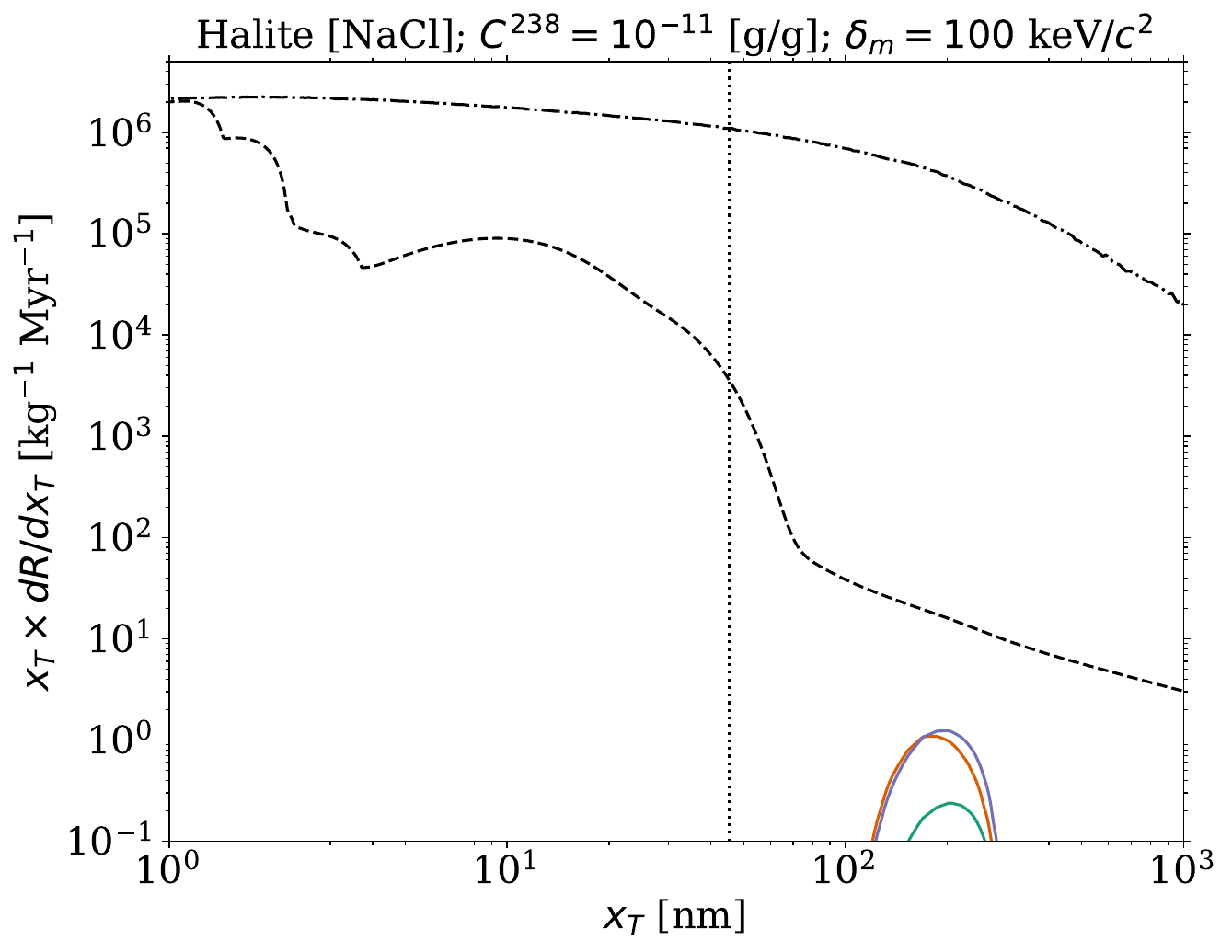}
    \caption{Differential track production rate $dR/d(\ln x_T)$ for inelastic scattering, given per unit exposure and per logarithmic interval in track length, shown as a function of the track length $x_T$ for the spin–dependent NREFT operators $\mathcal{O}^{s}_{6}$, $\mathcal{O}^{s}_{9}$, and $\mathcal{O}^{s}_{10}$, which are independent of the component of the WIMP–nucleon relative velocity perpendicular to the momentum transfer, $\vec{v}^{\perp}$. The calculations assume isoscalar interactions ($c^{p}=c^{n}$). The DM mass is fixed at $500\ \mathrm{GeV}/c^{2}$. In the left panel, the mass splitting is $\delta_{m}=50\ \mathrm{keV}/c^{2}$. The NREFT coupling constants are set to $(c^{s}_{6} m_{v}^{2})^{2}=7$, $(c^{s}_{9} m_{v}^{2})^{2}=4.6\cdot10^{-1}$, and $(c^{s}_{10} m_{v}^{2})^{2}=8.6\cdot10^{-2}$, where $m_{v}=246.2\ \mathrm{GeV}$ denotes the electroweak mass scale. In the right panel, the mass splitting is $\delta_{m}=100\ \mathrm{keV}/c^{2}$. The coupling constants are set to $(c^{s}_{6} m_{v}^{2})^{2}=1.6\cdot10^{1}$, $(c^{s}_{9} m_{v}^{2})^{2}=9.3\cdot10^{-1}$, and $(c^{s}_{10} m_{v}^{2})^{2}=1.7\cdot10^{-1}$. In both panels, the values for the coupling constants are chosen to be compatible with the upper limits for inelastic WIMP--nucleon scattering from the LUX--ZEPLIN experiment \cite{LZ:2023lvz}. For comparison, background spectra induced by neutrinos ($\nu$), radiogenic neutrons ($n$), and ${}^{238}\text{U}\to{}^{234}\text{Th}+\alpha$ recoils (${}^{234}\text{Th}$) are also shown; see Sec.~\ref{background}. Results are presented for halite with a ${}^{238}$U concentration of $10^{-11}\,\mathrm{g/g}$.}
    \label{fig:Spectrum_SDq1v0_inelastic}
\end{figure}

\begin{figure}
    \captionsetup{justification=raggedright,singlelinecheck=false}
    \centering
    \begin{subfigure}{0.33\textwidth}
    \includegraphics[width=\linewidth]{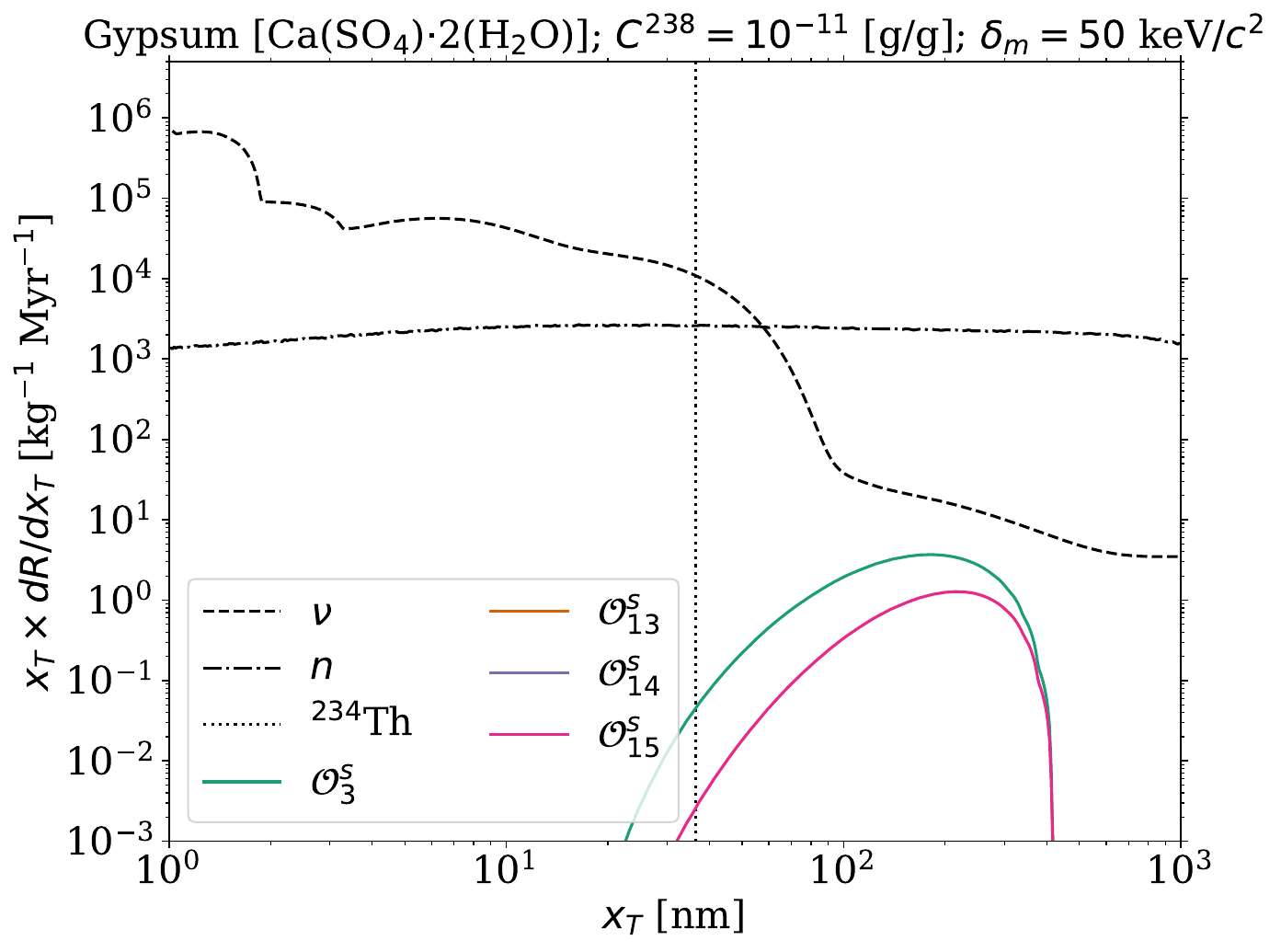}
    \end{subfigure}
    \begin{subfigure}{0.33\textwidth}
    \includegraphics[width=\linewidth]{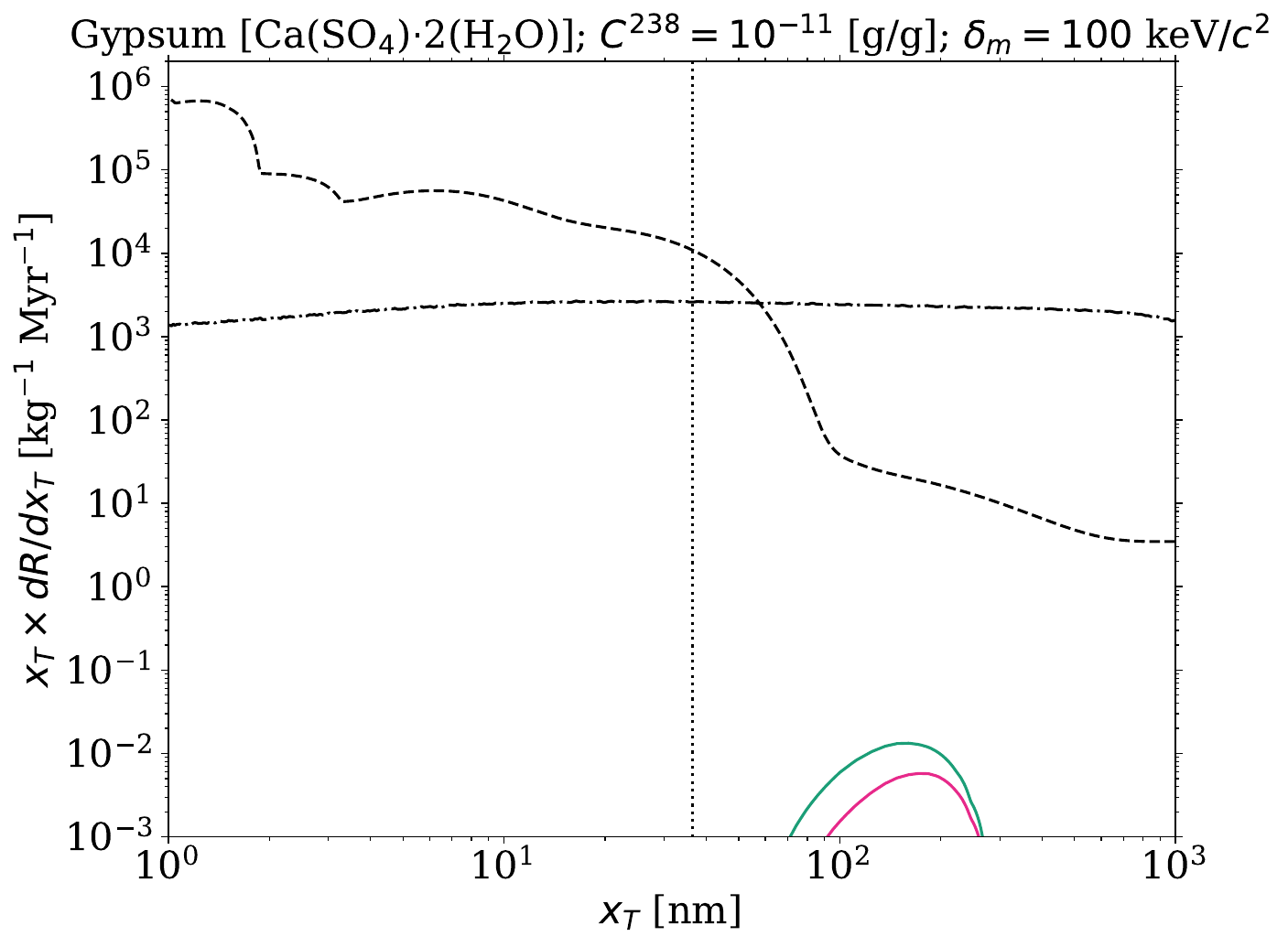}
    \end{subfigure}
    
    \vspace{0.1ex}

    \begin{subfigure}{0.33\textwidth}
    \includegraphics[width=\linewidth]{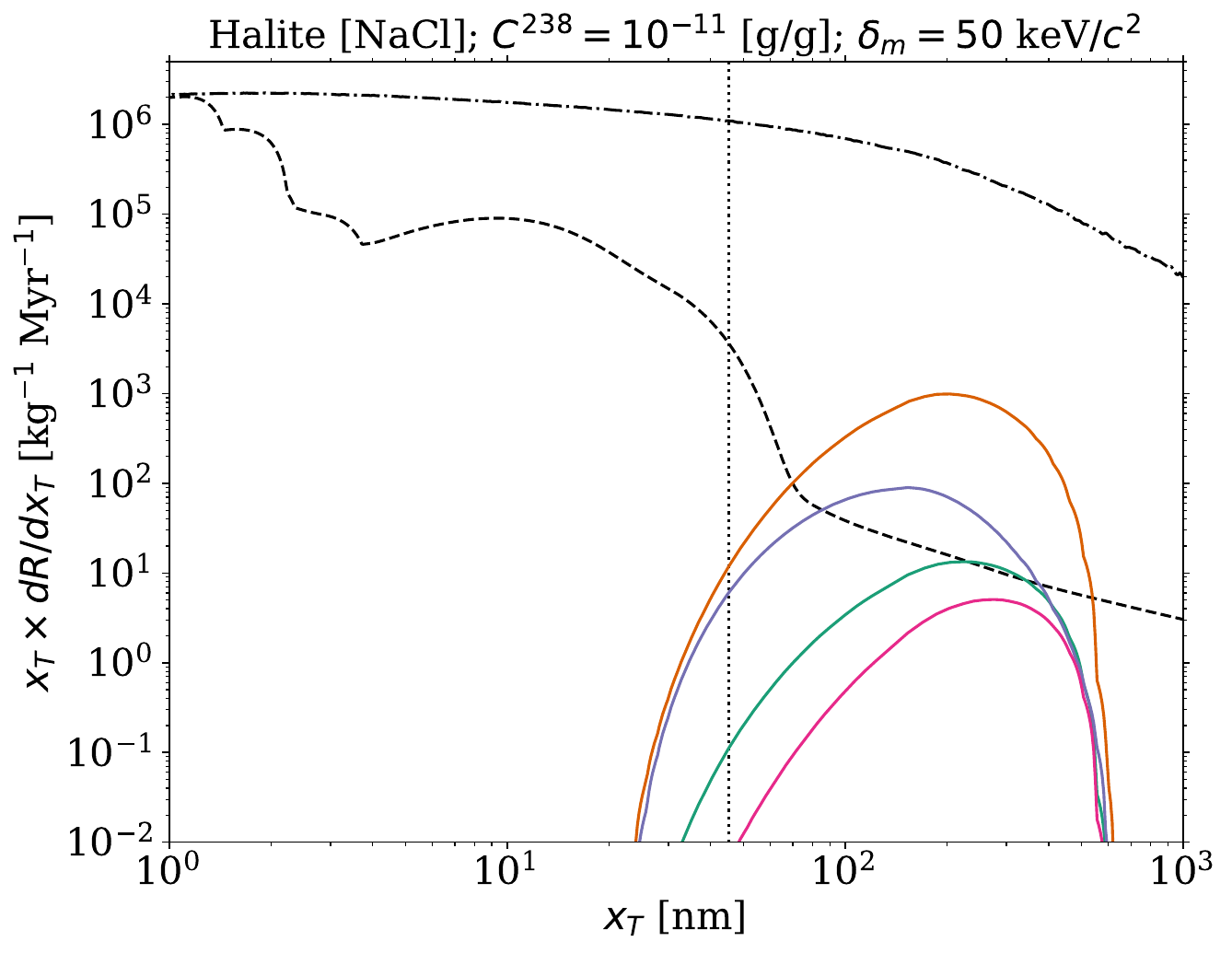}
    \end{subfigure}
    \begin{subfigure}{0.33\textwidth}
    \includegraphics[width=\linewidth]{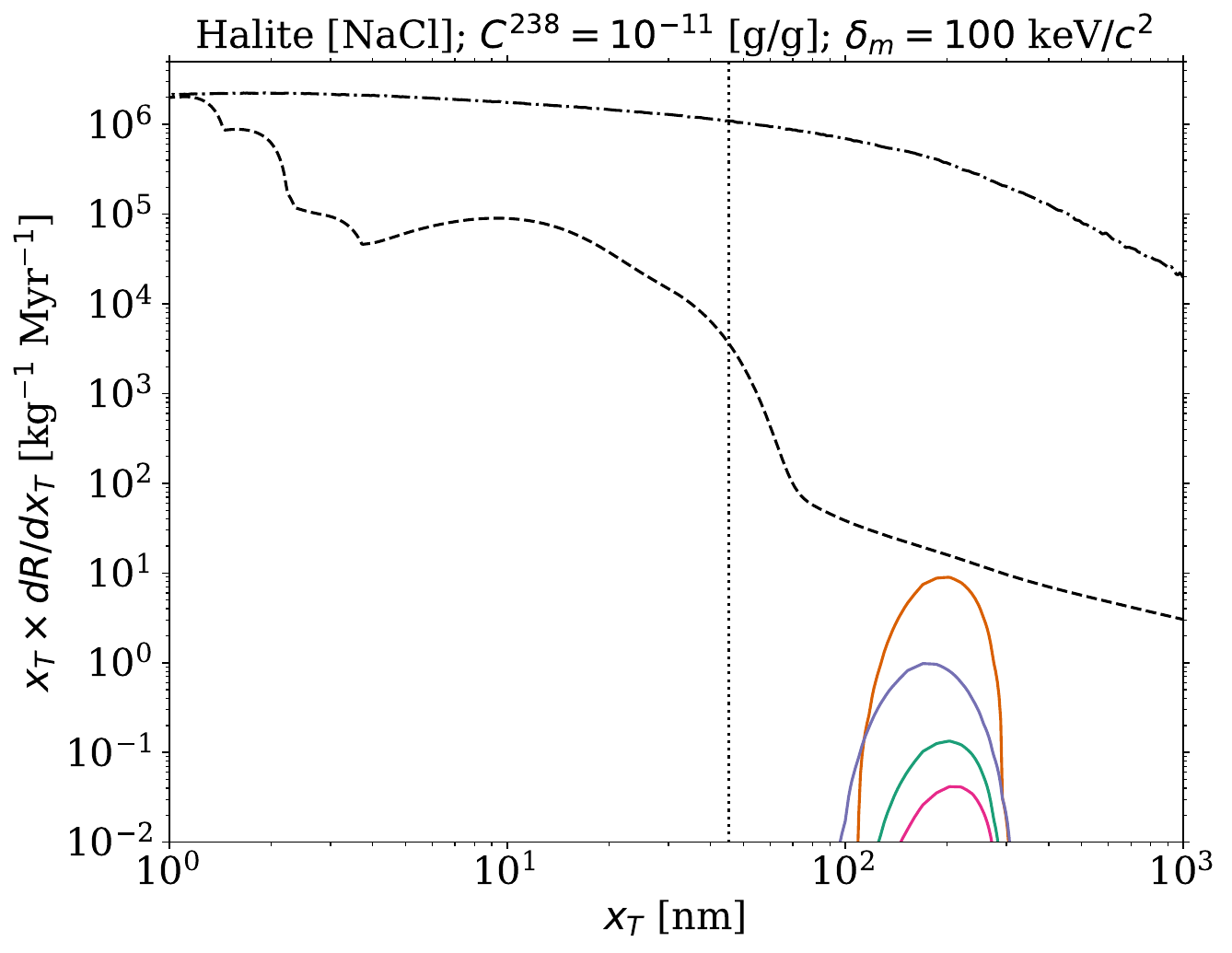}
    \end{subfigure}
    \caption{Differential track production rate $dR/d(\ln x_T)$ for inelastic scattering, given per unit exposure and per logarithmic interval in track length, shown as a function of the track length $x_T$ for the spin–dependent NREFT operators $\mathcal{O}^{s}_{3}$, $\mathcal{O}^{s}_{13}$, $\mathcal{O}^{s}_{14}$, and $\mathcal{O}^{s}_{15}$, which depend on both the momentum transfer $\vec{q}$, and the WIMP–nucleon relative velocity perpendicular to the momentum transfer, $\vec{v}^{\perp}$. The calculations assume isoscalar interactions ($c^{p}=c^{n}$). The DM mass is fixed at $500\ \mathrm{GeV}/c^{2}$. In the left panel, the mass splitting is $\delta_{m}=50\ \mathrm{keV}/c^{2}$. The NREFT coupling constants are set to $(c^{s}_{3} m_{v}^{2})^{2}=1\cdot10^{-2}$, $(c^{s}_{13} m_{v}^{2})^{2}=6.8\cdot10^{2}$, $(c^{s}_{14} m_{v}^{2})^{2}=8.6\cdot10^{5}$, and $(c^{s}_{15} m_{v}^{2})^{2}=1.5$, where $m_{v}=246.2\ \mathrm{GeV}$ denotes the electroweak mass scale. In the right panel, the mass splitting is $\delta_{m}=100\ \mathrm{keV}/c^{2}$. The coupling constants are set to $(c^{s}_{3} m_{v}^{2})^{2}=2\cdot10^{-2}$, $(c^{s}_{13} m_{v}^{2})^{2}=1.2\cdot10^{3}$, $(c^{s}_{14} m_{v}^{2})^{2}=2.1\cdot10^{6}$, and $(c^{s}_{15} m_{v}^{2})^{2}=3$. In both panels, the values for the coupling constants are chosen to be compatible with the upper limits for inelastic WIMP--nucleon scattering from the LUX--ZEPLIN experiment \cite{LZ:2023lvz}. For comparison, background spectra induced by neutrinos ($\nu$), radiogenic neutrons ($n$), and ${}^{238}\text{U}\to{}^{234}\text{Th}+\alpha$ recoils (${}^{234}\text{Th}$) are also shown; see Sec.~\ref{background}. Results are presented for gypsum (upper panels) and halite (lower panels) with a ${}^{238}$U concentration of $10^{-11}\,\mathrm{g/g}$. The dominant isotopes of gypsum have spin-zero ground states, and therefore the differential track production rates for the operators $\mathcal{O}_{13}^s$ and $\mathcal{O}_{14}^s$ are negligible and are not shown for this mineral.}
    \label{fig:Spectrum_SDq1v1_inelastic}
\end{figure}

Similar results can be obtained for the rest of the spin-dependent NREFT operators. In Figs.~\ref{fig:Spectrum_SDq1v0} and \ref{fig:Spectrum_SDq1v0_inelastic}, we present the track production spectra for the spin-dependent NREFT operators \(\mathcal{O}_{6}^{s}\), \(\mathcal{O}_{9}^{s}\), and \(\mathcal{O}_{10}^{s}\), which are independent from the component of the WIMP–nucleon relative velocity perpendicular to the momentum transfer, $\vec{v}^{\perp}$, for elastic and inelastic interactions, respectively. Additionally, the spectra for the spin-dependent NREFT operators \(\mathcal{O}_{3}^{s}\), \(\mathcal{O}_{13}^{s}\), \(\mathcal{O}_{14}^{s}\), and \(\mathcal{O}_{15}^{s}\), which depend on both $\vec{q}$ and $\vec{v}^{\perp}$, for elastic and inelastic scattering, are shown in Figs.~\ref{fig:Spectrum_SDq1v1} and \ref{fig:Spectrum_SDq1v1_inelastic}, respectively. In Fig.~\ref{fig:Spectrum_SI_Halite_inelastic}, we present the spectra for inelastic scattering via the spin-independent operators \(\mathcal{O}_{1}^{s}\), \(\mathcal{O}_{5}^{s}\), \(\mathcal{O}_{8}^{s}\), and \(\mathcal{O}_{11}^{s}\) for halite; the corresponding spectra for inelastic scattering in gypsum are shown in the main text in Fig.~\ref{fig:Spectrum_SI_inelastic}.

\begin{figure}
    \captionsetup{justification=raggedright,singlelinecheck=false}
    \centering
    \begin{subfigure}{0.33\textwidth}
    \includegraphics[width=\linewidth]{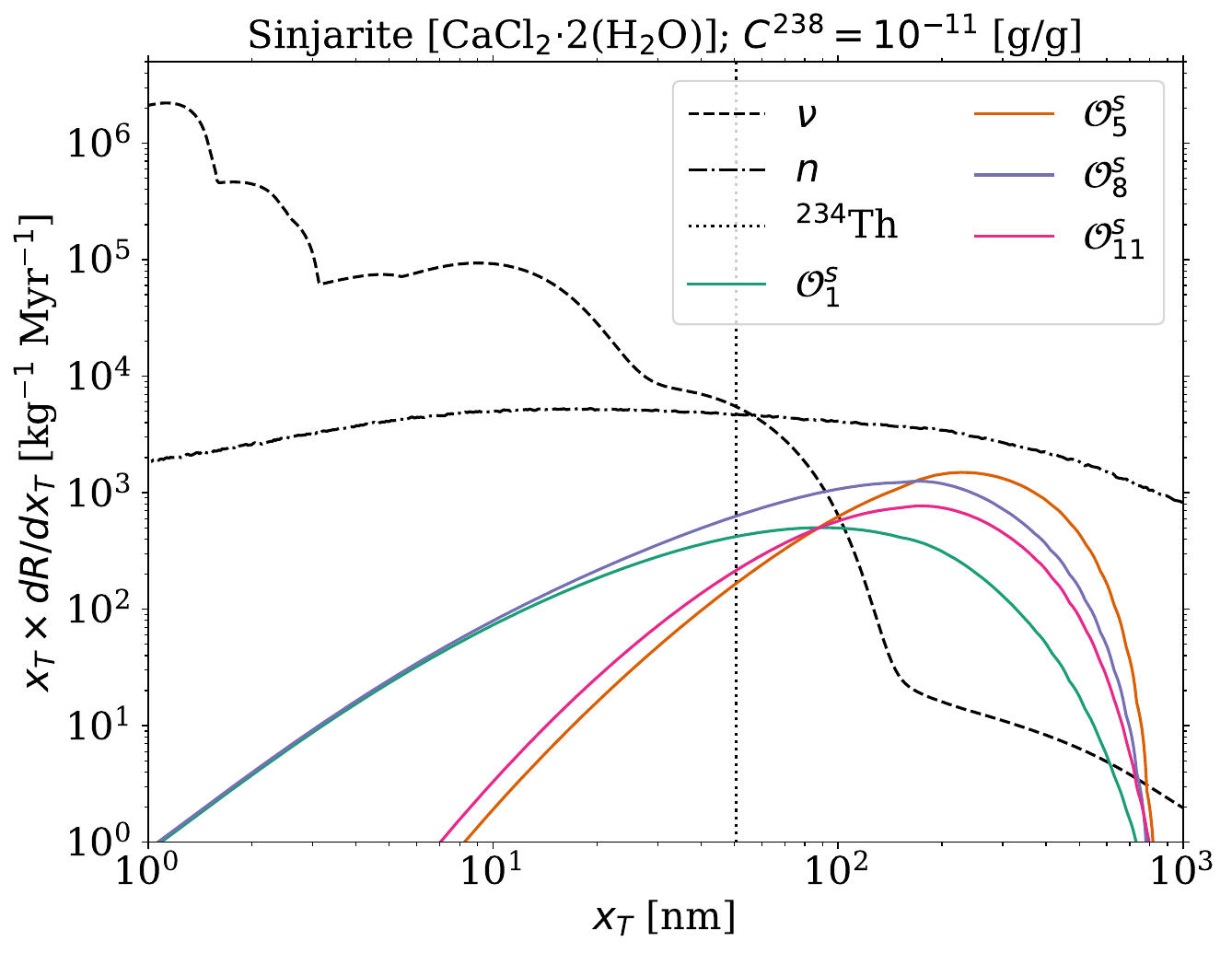}
    \end{subfigure}
    \begin{subfigure}{0.33\textwidth}
    \includegraphics[width=\linewidth]{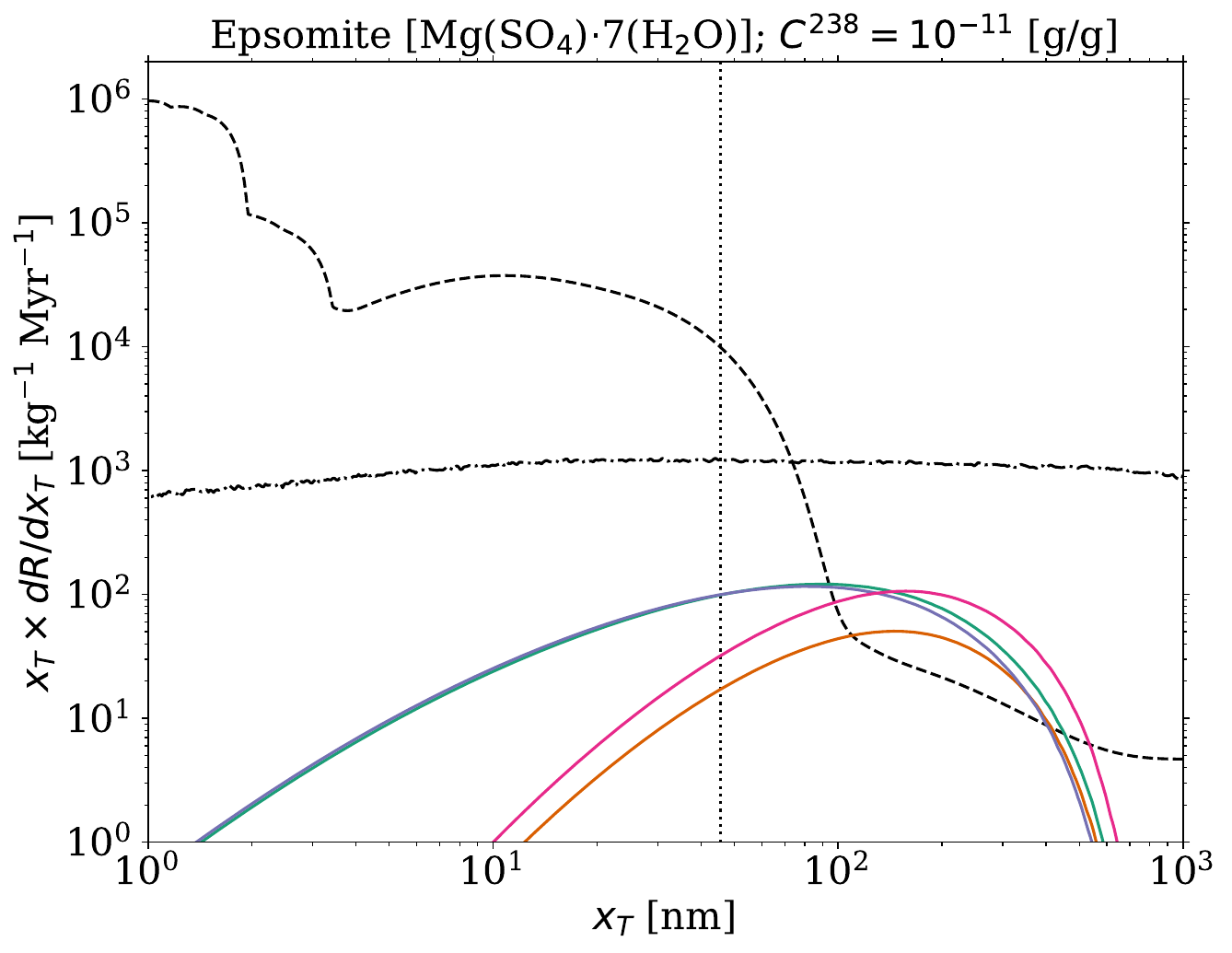}
    \end{subfigure}
    
    \vspace{0.1ex}

    \begin{subfigure}{0.33\textwidth}
    \includegraphics[width=\linewidth]{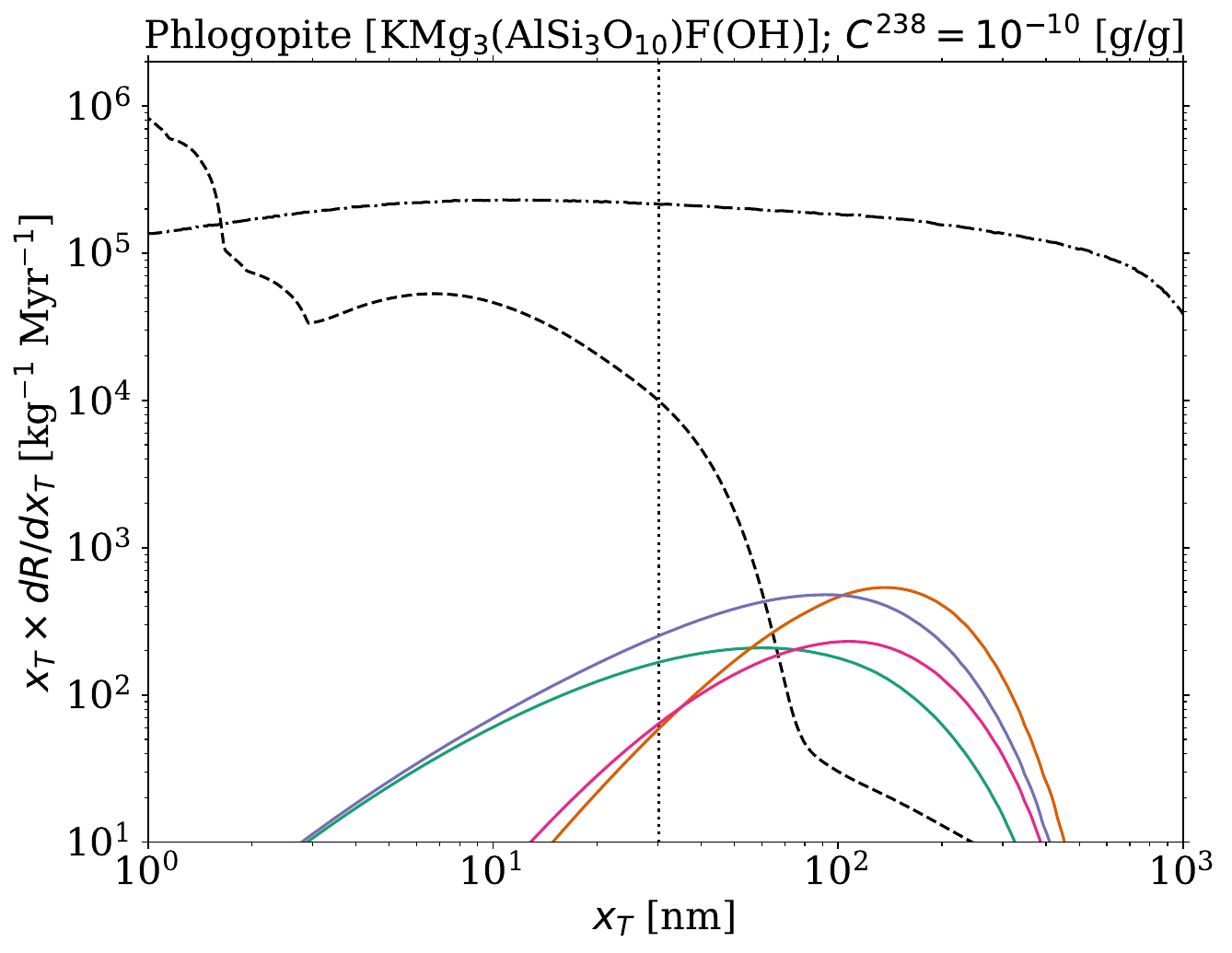}
    \end{subfigure}
    \begin{subfigure}{0.33\textwidth}
    \includegraphics[width=\linewidth]{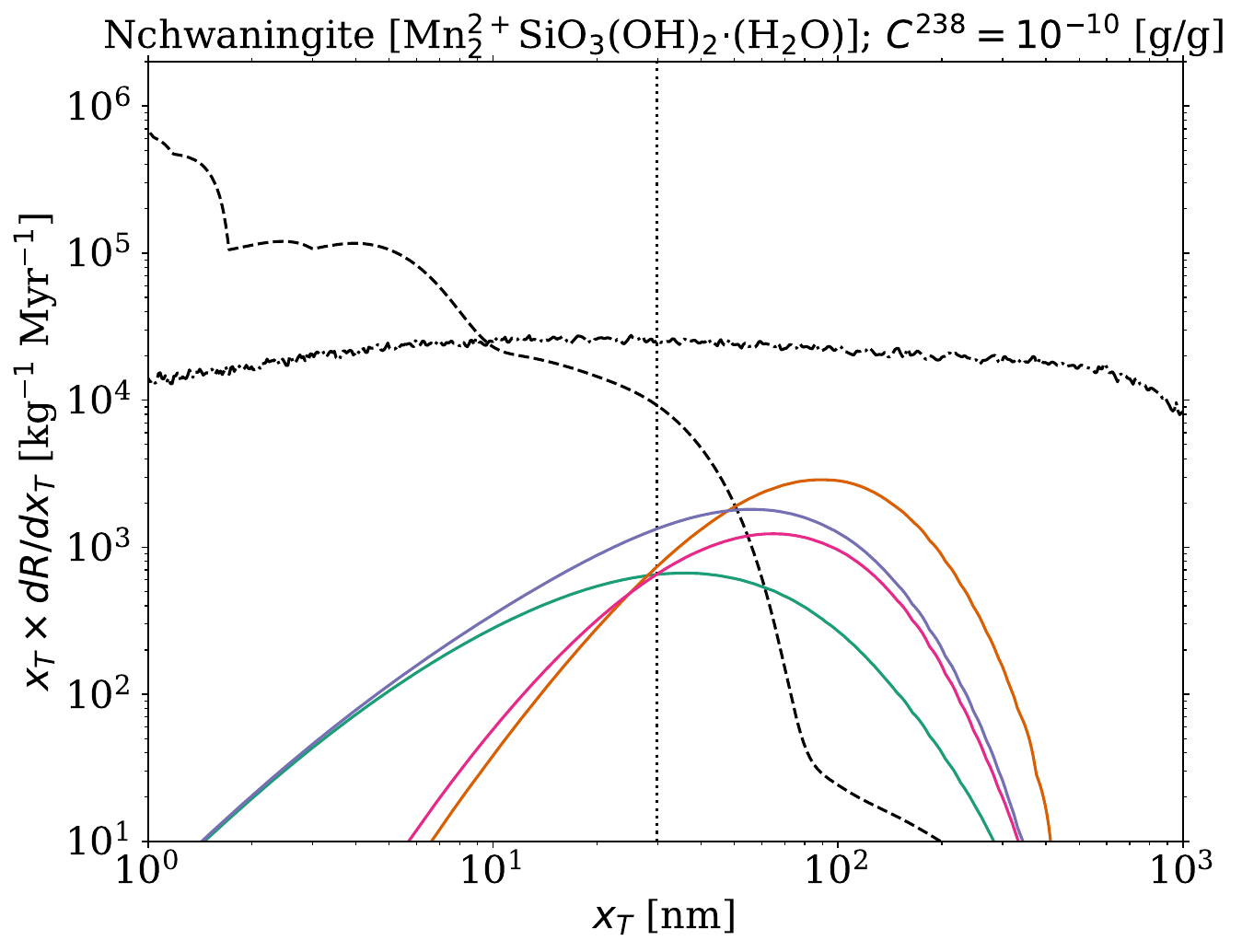}
    \end{subfigure}
    \caption{Differential track production rate $dR/d(\ln x_T)$ for elastic scattering, given per unit exposure and per logarithmic interval in track length, shown as a function of the track length $x_T$ for the spin–independent NREFT operators $\mathcal{O}^{s}_{1}$, $\mathcal{O}^{s}_{5}$, $\mathcal{O}^{s}_{8}$, and $\mathcal{O}^{s}_{11}$. The calculations assume isoscalar interactions ($c^{p}=c^{n}$). Results are shown for sinjarite and epsomite with ${}^{238}$U concentrations of $10^{-11}\ \mathrm{g/g}$, and for phlogopite and nchwaningite with ${}^{238}$U concentrations of $10^{-10}\ \mathrm{g/g}$. The DM mass is fixed at $500\ \mathrm{GeV}/c^{2}$. For comparison, we also show background spectra induced by neutrinos ($\nu$), radiogenic neutrons ($n$), and ${}^{238}\text{U}\to{}^{234}\text{Th}+\alpha$ recoils (${}^{234}\text{Th}$); see Sec.~\ref{background}. For the normalization of the DM track production spectra, we set the NREFT coupling constants to $(c^{s}_{1}\ m_{v}^{2})^{2}=4\cdot10^{-9}$, $(c^{s}_{5}\ m_{v}^{2})^{2}=2.5$, $(c^{s}_{8}\ m_{v}^{2})^{2}=1.5\cdot10^{-2}$, and $(c^{s}_{11}\ m_{v}^{2})^{2}=5\cdot10^{-6}$, where $m_{v} = 246.2\ \mathrm{GeV}$ denotes the electroweak mass scale. These values are chosen to be compatible with the upper limits from the LUX–ZEPLIN experiment \cite{LZ:2023lvz}.}
    \label{fig:Spectrum_SI_ap}
\end{figure}

\begin{figure}
    \captionsetup{justification=raggedright,singlelinecheck=false}
    \centering
    \begin{subfigure}{0.33\textwidth}
    \includegraphics[width=\linewidth]{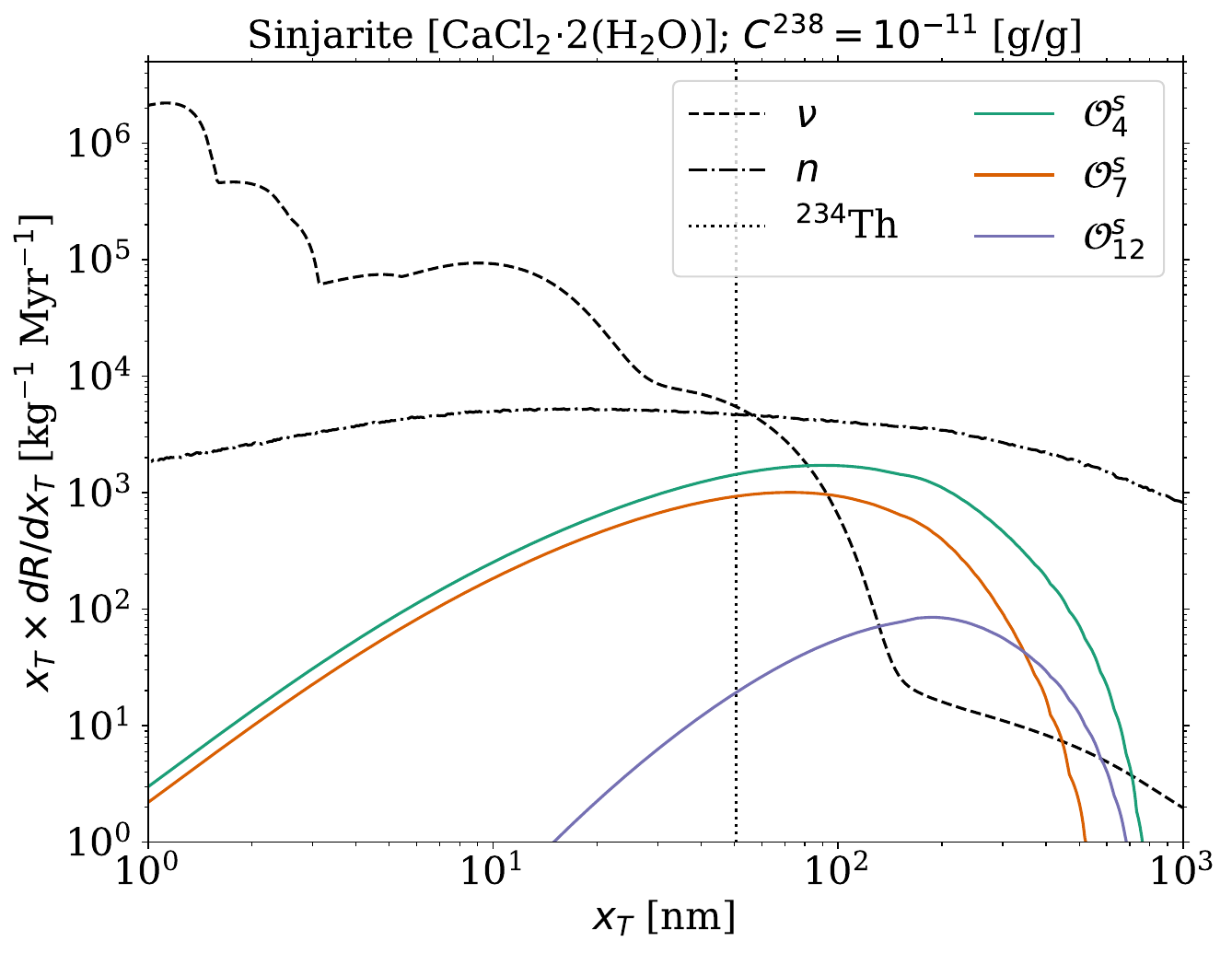}
    \end{subfigure}
    \begin{subfigure}{0.33\textwidth}
    \includegraphics[width=\linewidth]{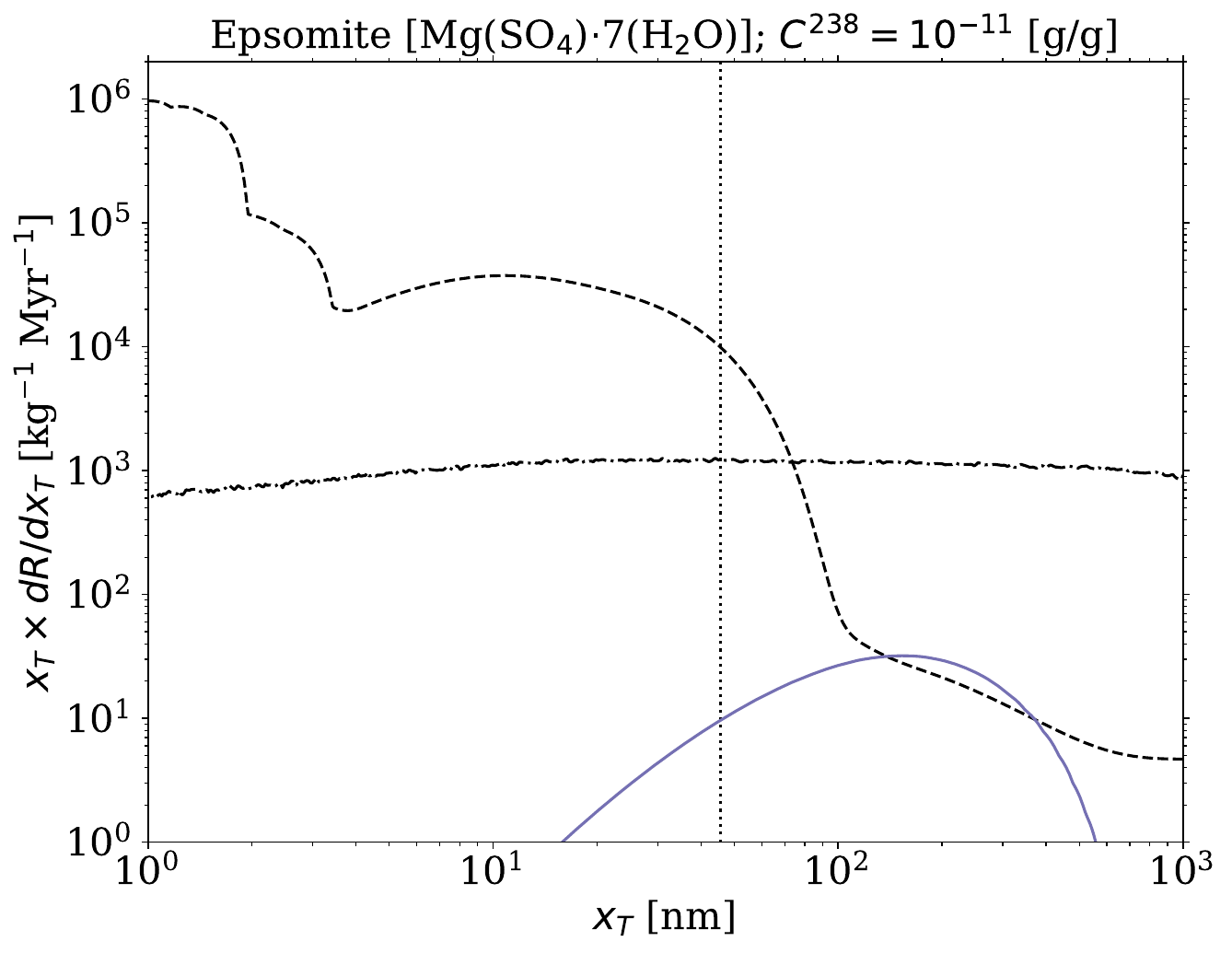}
    \end{subfigure}
    
    \vspace{0.1ex}

    \begin{subfigure}{0.33\textwidth}
    \includegraphics[width=\linewidth]{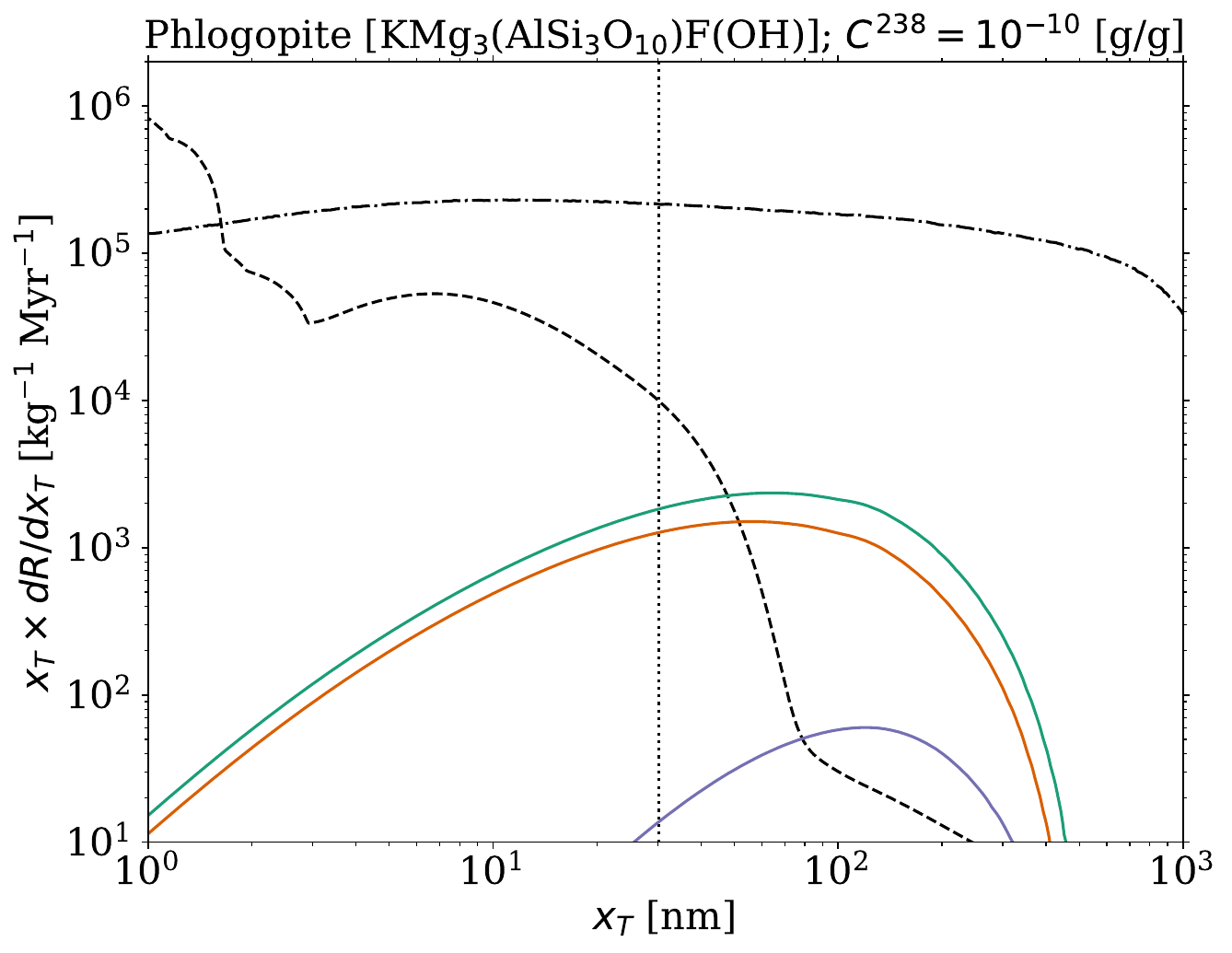}
    \end{subfigure}
    \begin{subfigure}{0.33\textwidth}
    \includegraphics[width=\linewidth]{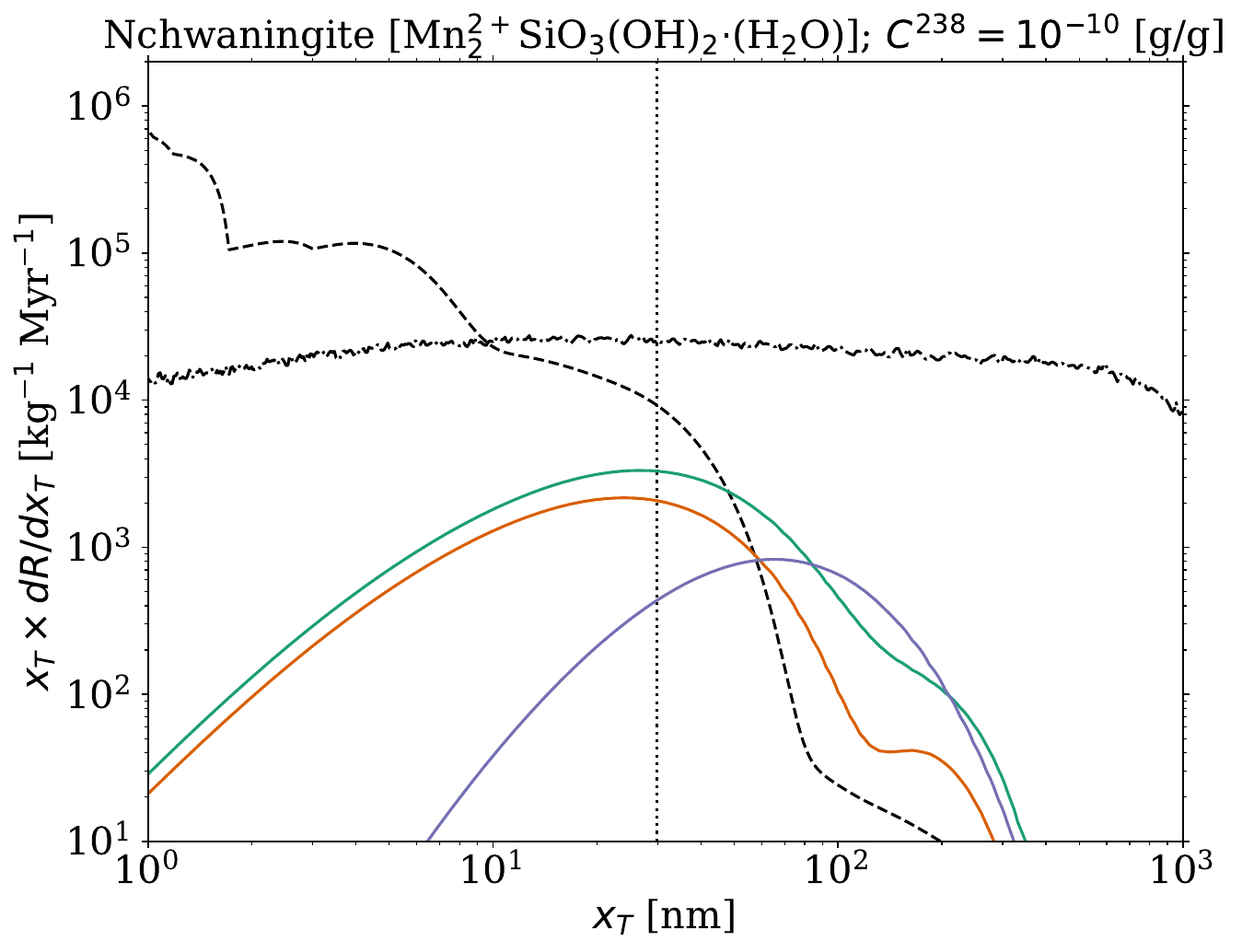}
    \end{subfigure}
    \caption{Differential track production rate $dR/d(\ln x_T)$ for elastic scattering, given per unit exposure and per logarithmic interval in track length, shown as a function of the track length $x_T$ for the spin–dependent NREFT operators $\mathcal{O}^{s}_{4}$, $\mathcal{O}^{s}_{7}$, and $\mathcal{O}^{s}_{12}$, which are independent of the momentum transfer $\vec{q}$. The calculations assume isoscalar interactions ($c^{p}=c^{n}$). Results are shown for sinjarite and epsomite with ${}^{238}$U concentrations of $10^{-11}\ \mathrm{g/g}$, and for phlogopite and nchwaningite with ${}^{238}$U concentrations of $10^{-10}\ \mathrm{g/g}$. The DM mass is fixed at $500\ \mathrm{GeV}/c^{2}$. For comparison, we also show background spectra induced by neutrinos ($\nu$), radiogenic neutrons ($n$), and ${}^{238}\text{U}\to{}^{234}\text{Th}+\alpha$ recoils (${}^{234}\text{Th}$); see Sec.~\ref{background}. For the normalization of the DM track production spectra, we set the NREFT coupling constants to $(c^{s}_{4}\ m_{v}^{2})^{2}=2\cdot10^{-3}$, $(c^{s}_{7}\ m_{v}^{2})^{2}=1\cdot10^{3}$, and $(c^{s}_{12}\ m_{v}^{2})^{2}=3\cdot10^{-4}$, where $m_{v} = 246.2\ \mathrm{GeV}$ denotes the electroweak mass scale. These values are chosen to be compatible with the upper limits from the LUX–ZEPLIN experiment \cite{LZ:2023lvz}. The dominant isotopes of epsomite have spin-zero ground states, and therefore the differential track production rates for the operators $\mathcal{O}_4^s$ and $\mathcal{O}_7^s$ are negligible and are not shown for this mineral.}
    \label{fig:Spectrum_SDq0_ap}
\end{figure}

\begin{figure}
    \captionsetup{justification=raggedright,singlelinecheck=false}
    \centering
    \begin{subfigure}{0.32\textwidth}
    \includegraphics[width=\linewidth]{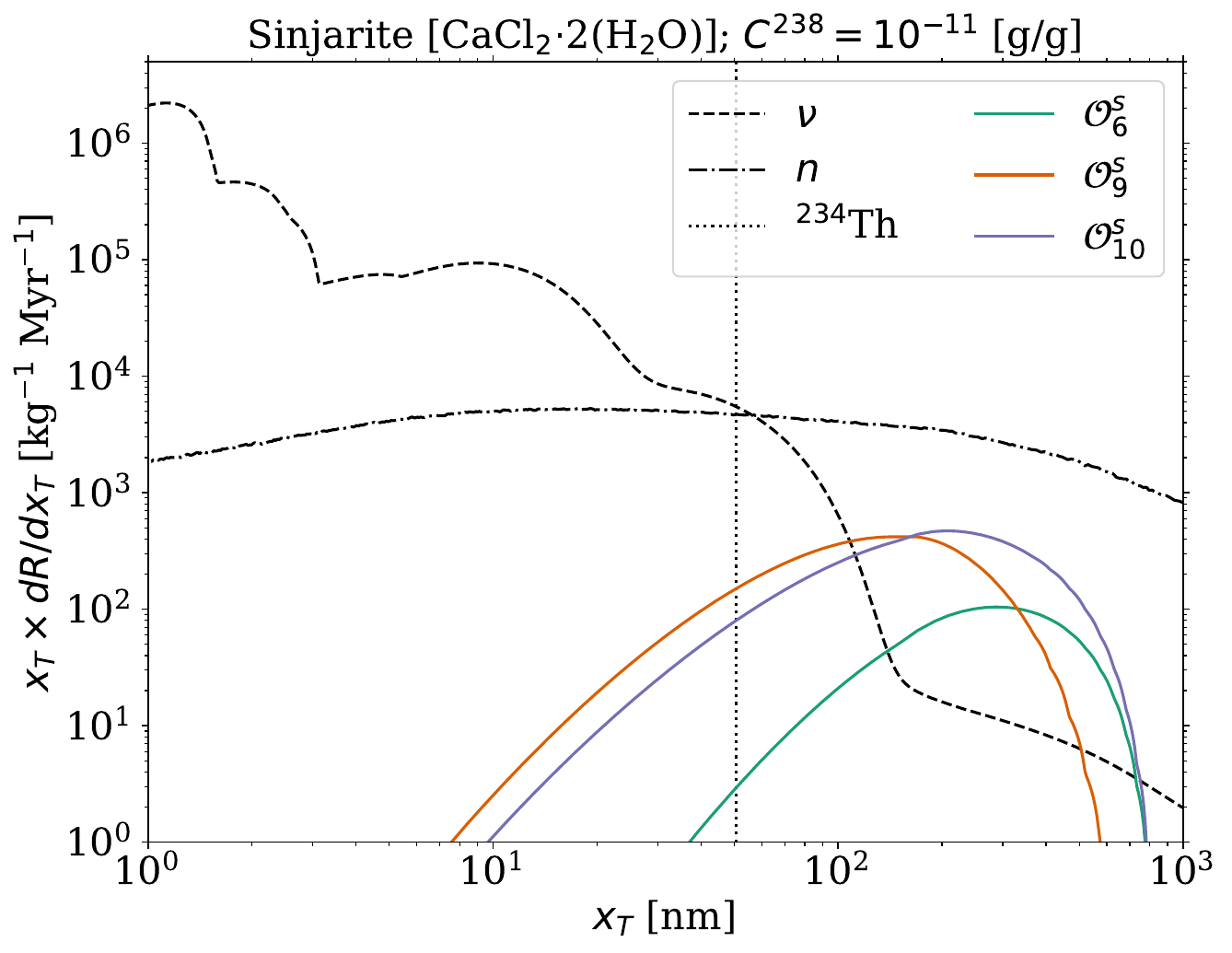}
    \end{subfigure}
    \begin{subfigure}{0.32\textwidth}
    \includegraphics[width=\linewidth]{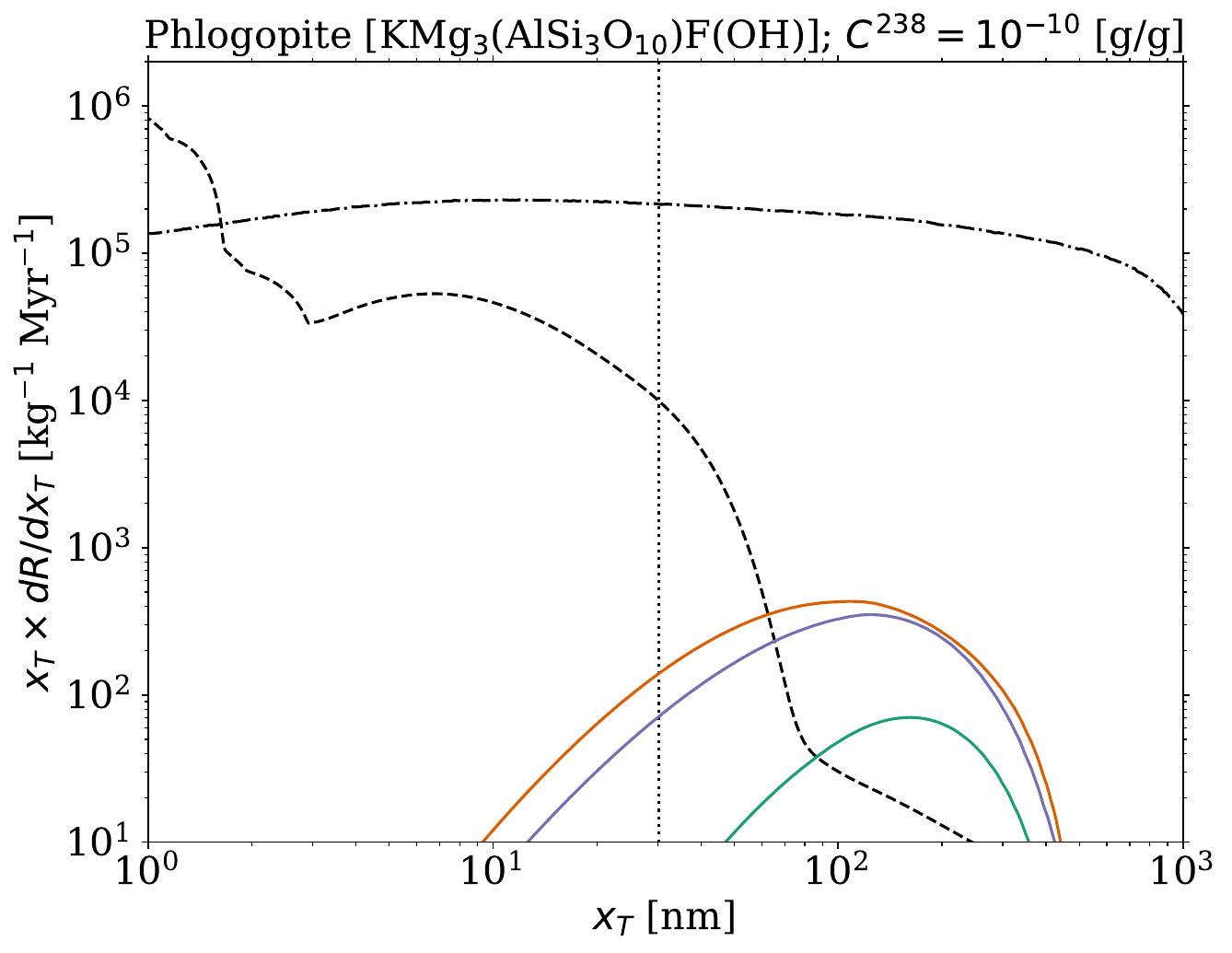}
    \end{subfigure}
    \begin{subfigure}{0.32\textwidth}
    \includegraphics[width=\linewidth]{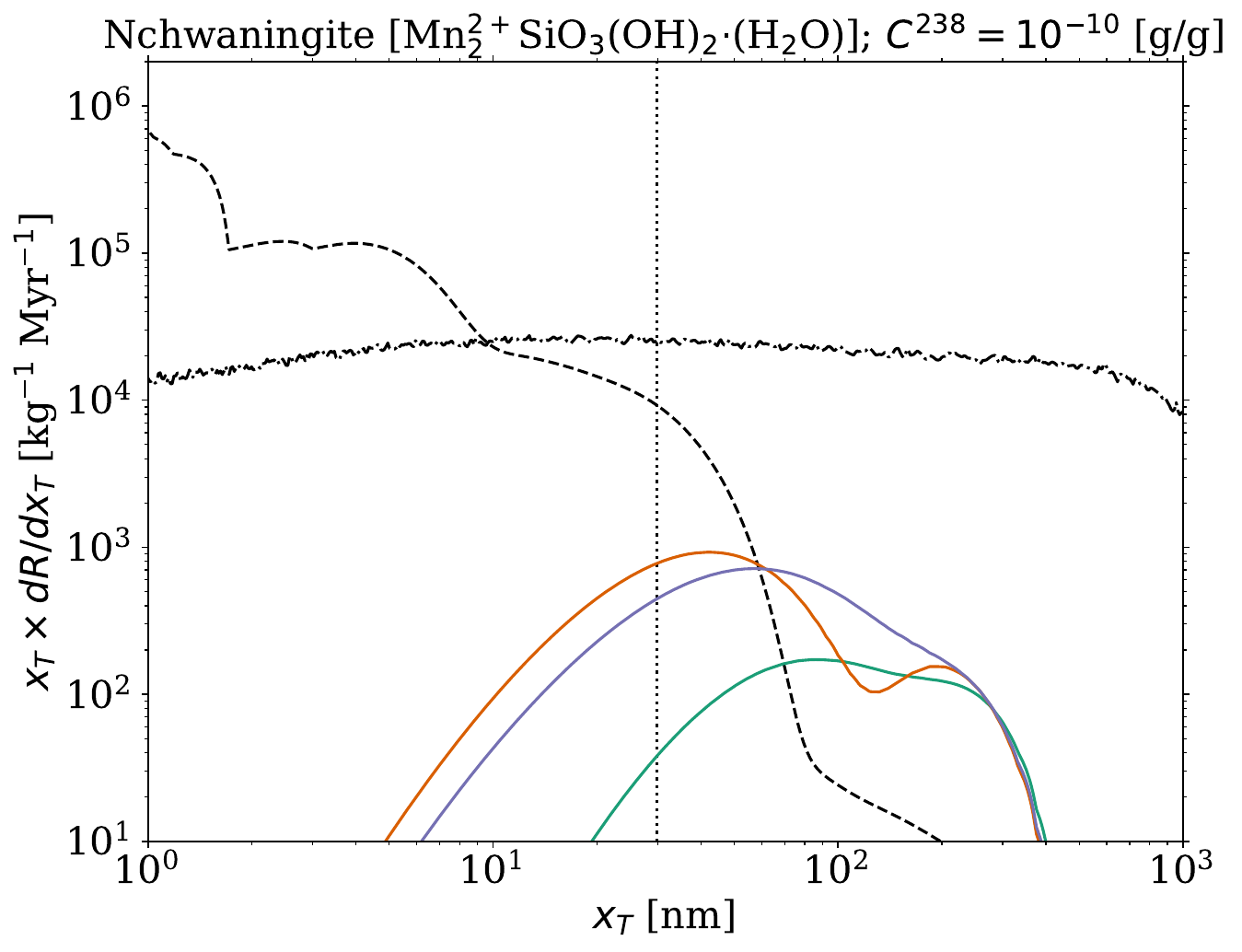}
    \end{subfigure}
    \caption{Differential track production rate $dR/d(\ln x_T)$ for elastic scattering, given per unit exposure and per logarithmic interval in track length, shown as a function of the track length $x_T$ for the spin–dependent NREFT operators $\mathcal{O}^{s}_{6}$, $\mathcal{O}^{s}_{9}$, and $\mathcal{O}^{s}_{10}$, which are independent of the component of the WIMP–nucleon relative velocity perpendicular to the momentum transfer, $\vec{v}^{\perp}$. The calculations assume isoscalar interactions ($c^{p}=c^{n}$). Results are shown for sinjarite with a ${}^{238}$U concentration of $10^{-11}\ \mathrm{g/g}$, and for phlogopite and nchwaningite with ${}^{238}$U concentrations of $10^{-10}\ \mathrm{g/g}$. The DM mass is fixed at $500\ \mathrm{GeV}/c^{2}$. For comparison, we also show background spectra induced by neutrinos ($\nu$), radiogenic neutrons ($n$), and ${}^{238}\text{U}\to{}^{234}\text{Th}+\alpha$ recoils (${}^{234}\text{Th}$); see Sec.~\ref{background}. For the normalization of the DM track production spectra, we set the NREFT coupling constants to $(c^{s}_{6}\ m_{v}^{2})^{2}=5$, $(c^{s}_{9}\ m_{v}^{2})^{2}=2.3\cdot10^{-1}$, and $(c^{s}_{10}\ m_{v}^{2})^{2}=5\cdot10^{-2}$, where $m_{v} = 246.2\ \mathrm{GeV}$ denotes the electroweak mass scale. These values are chosen to be compatible with the upper limits from the LUX–ZEPLIN experiment \cite{LZ:2023lvz}.}
    \label{fig:Spectrum_SDq1v0_ap}
\end{figure}

\begin{figure}
    \captionsetup{justification=raggedright,singlelinecheck=false}
    \centering
    \begin{subfigure}{0.33\textwidth}
    \includegraphics[width=\linewidth]{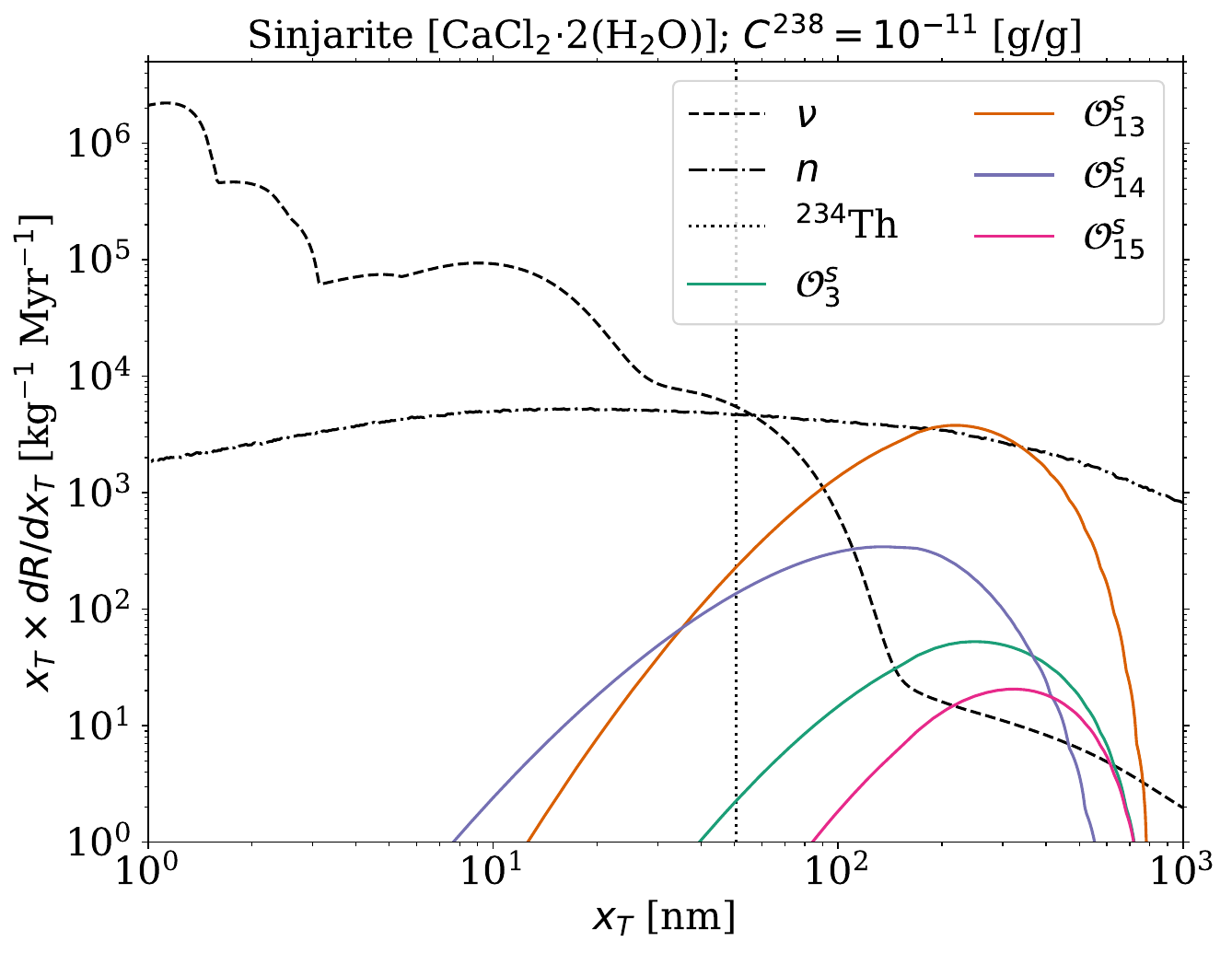}
    \end{subfigure}
    \begin{subfigure}{0.33\textwidth}
    \includegraphics[width=\linewidth]{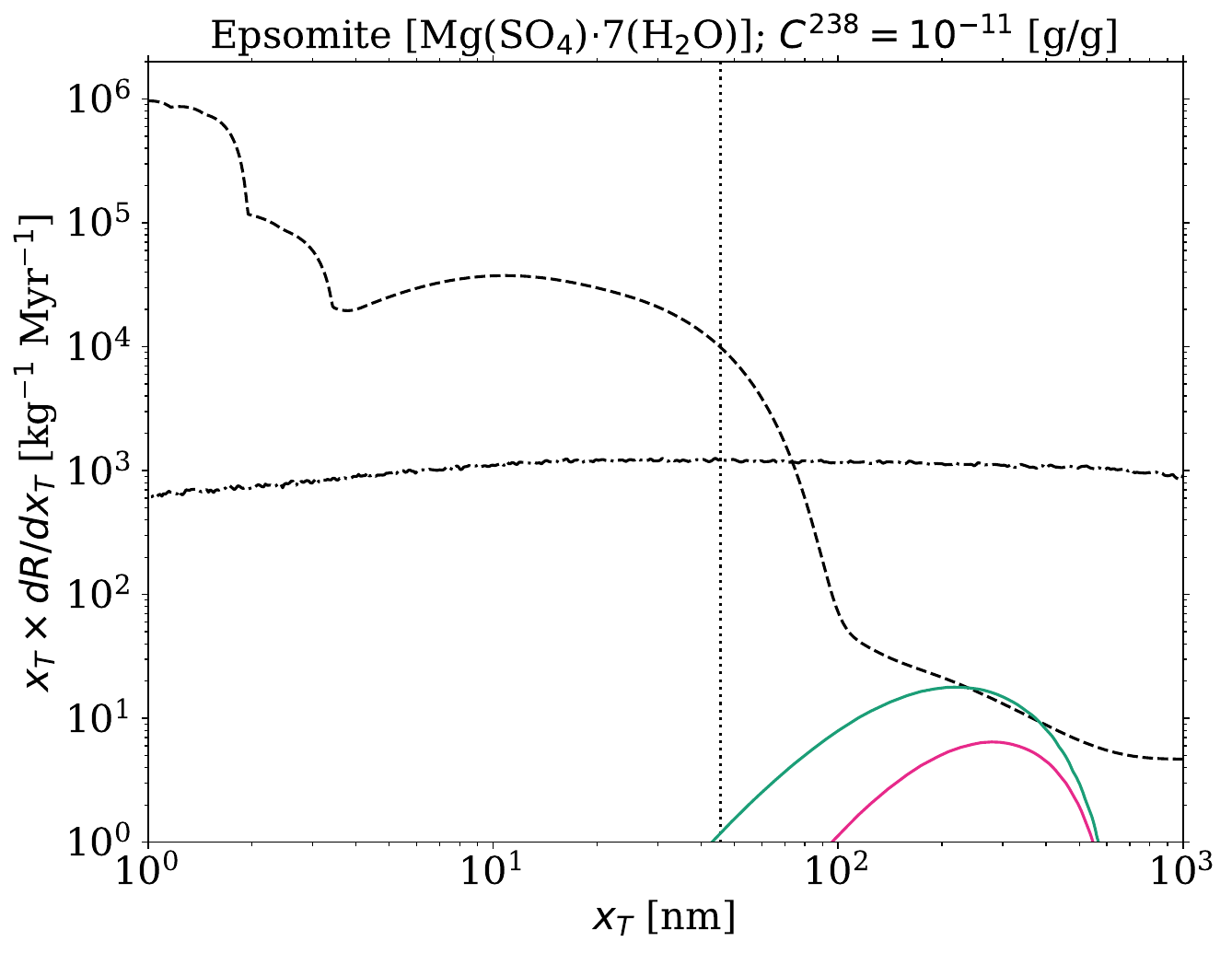}
    \end{subfigure}
    
    \vspace{0.1ex}

    \begin{subfigure}{0.33\textwidth}
    \includegraphics[width=\linewidth]{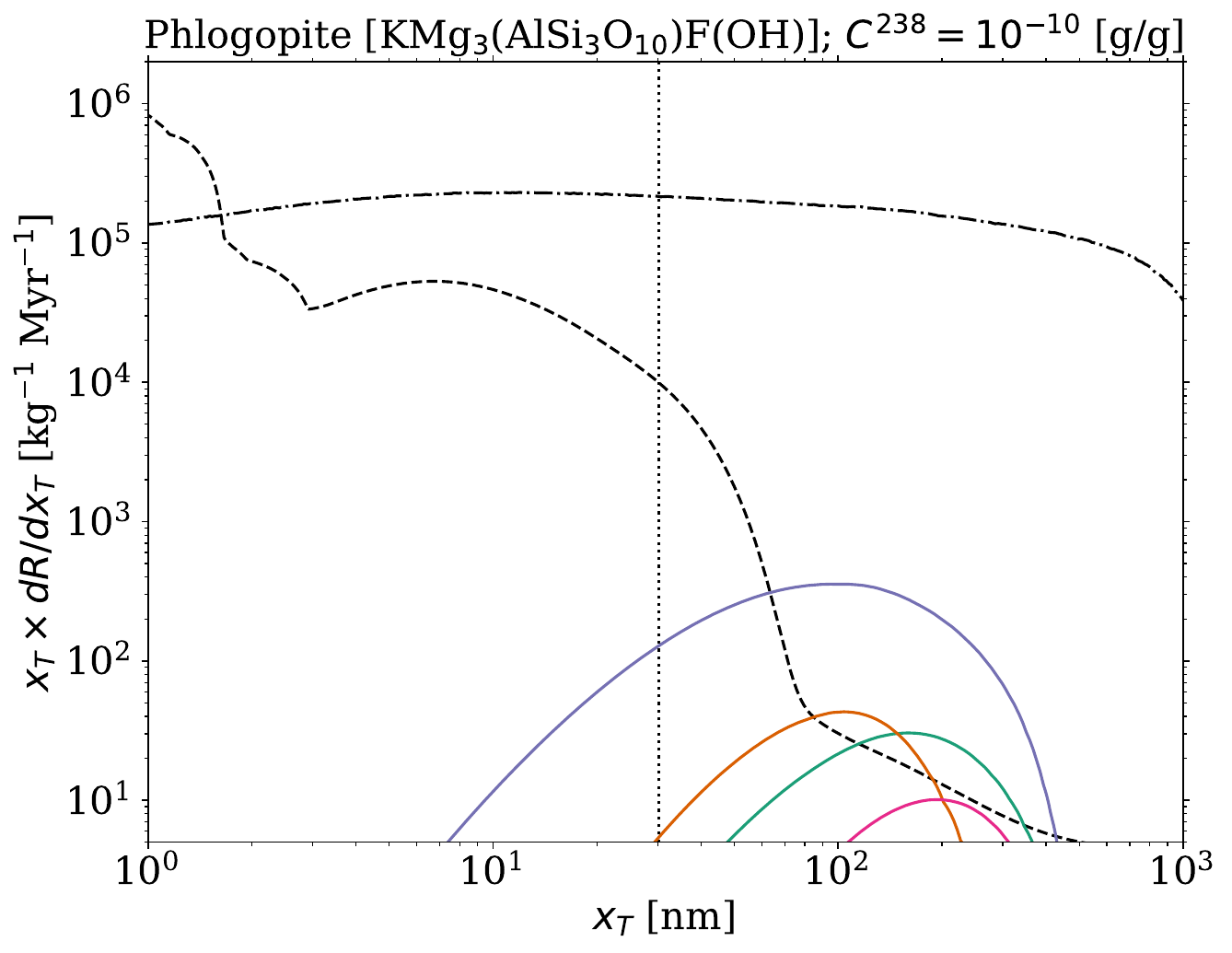}
    \end{subfigure}
    \begin{subfigure}{0.33\textwidth}
    \includegraphics[width=\linewidth]{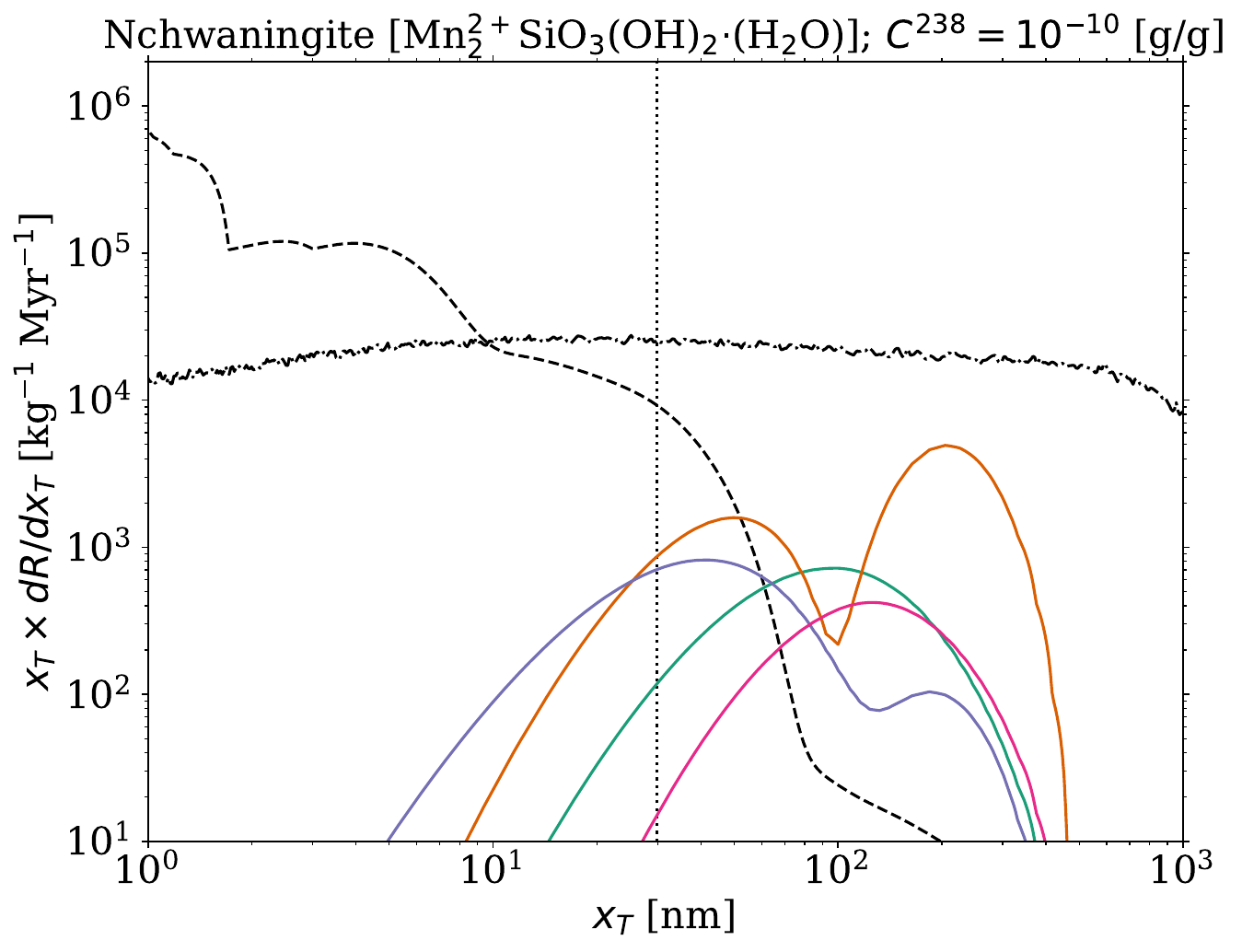}
    \end{subfigure}
    \caption{Differential track production rate $dR/d(\ln x_T)$ for elastic scattering, given per unit exposure and per logarithmic interval in track length, shown as a function of the track length $x_T$ for the spin–dependent NREFT operators $\mathcal{O}^{s}_{3}$, $\mathcal{O}^{s}_{13}$, $\mathcal{O}^{s}_{14}$, and $\mathcal{O}^{s}_{15}$, which depend on both the momentum transfer $\vec{q}$, and the WIMP–nucleon relative velocity perpendicular to the momentum transfer, $\vec{v}^{\perp}$. The calculations assume isoscalar interactions ($c^{p}=c^{n}$). Results are shown for sinjarite and epsomite with ${}^{238}$U concentrations of $10^{-11}\ \mathrm{g/g}$, and for phlogopite and nchwaningite with ${}^{238}$U concentrations of $10^{-10}\ \mathrm{g/g}$. The DM mass is fixed at $500\ \mathrm{GeV}/c^{2}$. For comparison, we also show background spectra induced by neutrinos ($\nu$), radiogenic neutrons ($n$), and ${}^{238}\text{U}\to{}^{234}\text{Th}+\alpha$ recoils (${}^{234}\text{Th}$); see Sec.~\ref{background}. For the normalization of the DM track production spectra, we set the NREFT coupling constants to $(c^{s}_{3}\ m_{v}^{2})^{2}=6\cdot10^{-3}$, $(c^{s}_{13}\ m_{v}^{2})^{2}=4\cdot10^{2}$, $(c^{s}_{14}\ m_{v}^{2})^{2}=4\cdot10^{5}$, and $(c^{s}_{15}\ m_{v}^{2})^{2}=9\cdot10^{-1}$, where $m_{v} = 246.2\ \mathrm{GeV}$ denotes the electroweak mass scale. These values are chosen to be compatible with the upper limits from the LUX–ZEPLIN experiment \cite{LZ:2023lvz}. The dominant isotopes of epsomite have spin-zero ground states, and therefore the differential track production rates for the operators $\mathcal{O}_{13}^s$ and $\mathcal{O}_{14}^s$ are negligible and are not shown for this mineral.}
    \label{fig:Spectrum_SDq1v1_ap}
\end{figure}

In addition to the minerals gypsum, halite, olivine, and muscovite, which are common in nature, we consider some additional minerals, namely sinjarite, epsomite, phlogopite, and nchwaningite, for which chemical compositions and $^{238}$U concentrations are also included in Table \ref{tab:U238_concentration}. These additional minerals are chosen partly for comparisons to previous studies \cite{Drukier:2018pdy,Baum:2021jak,Baum:2019fqm,Edwards:2018hcf,Baum:2018tfw,Tapia-Arellano:2021cml}. Additionally, sinjarite and nchwaningite, despite being rare in nature, are illustrative examples of how the mineral's composition can enhance the sensitivity of paleo-detectors. The track length spectra for these four minerals, for elastic interactions via all the NREFT operators, are presented in Figs.~\ref{fig:Spectrum_SI_ap}, \ref{fig:Spectrum_SDq0_ap}, \ref{fig:Spectrum_SDq1v0_ap}, and \ref{fig:Spectrum_SDq1v1_ap}.

\begin{figure}
    \captionsetup{justification=raggedright,singlelinecheck=false}
    \centering
    \begin{subfigure}{0.30\textwidth}
    \includegraphics[width=\linewidth]{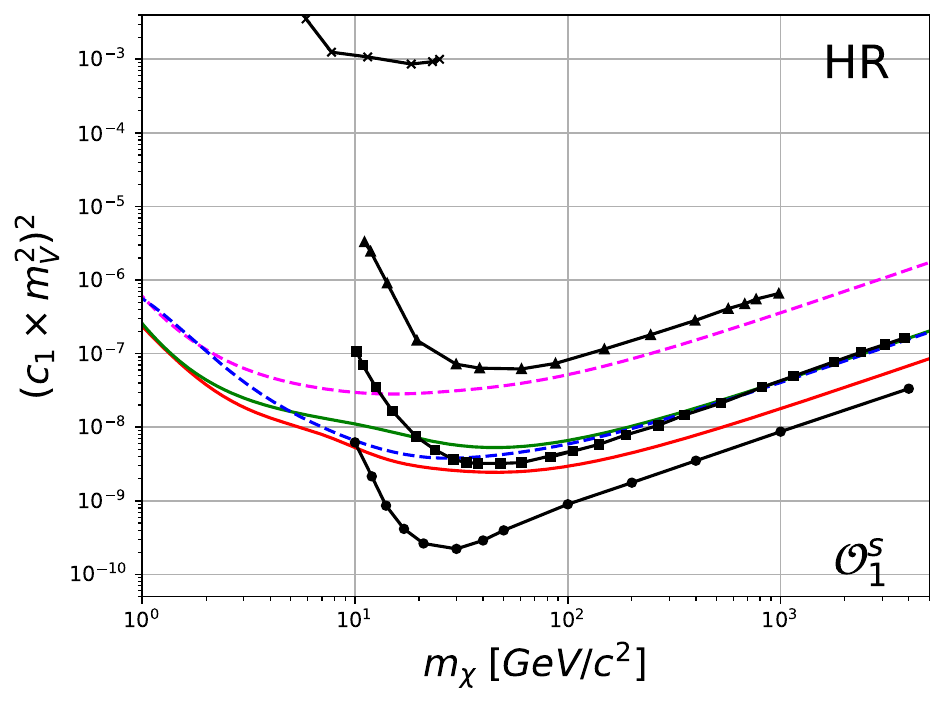}
    \end{subfigure}
    \begin{subfigure}{0.30\textwidth}
    \includegraphics[width=\linewidth]{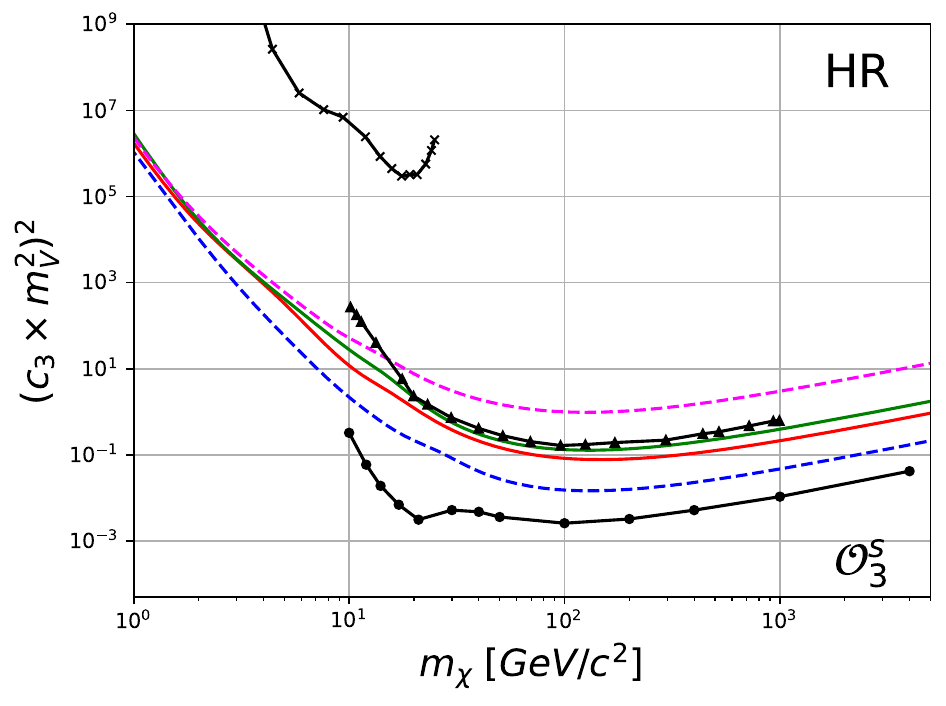}
    \end{subfigure}
    \begin{subfigure}{0.30\textwidth}
    \includegraphics[width=\linewidth]{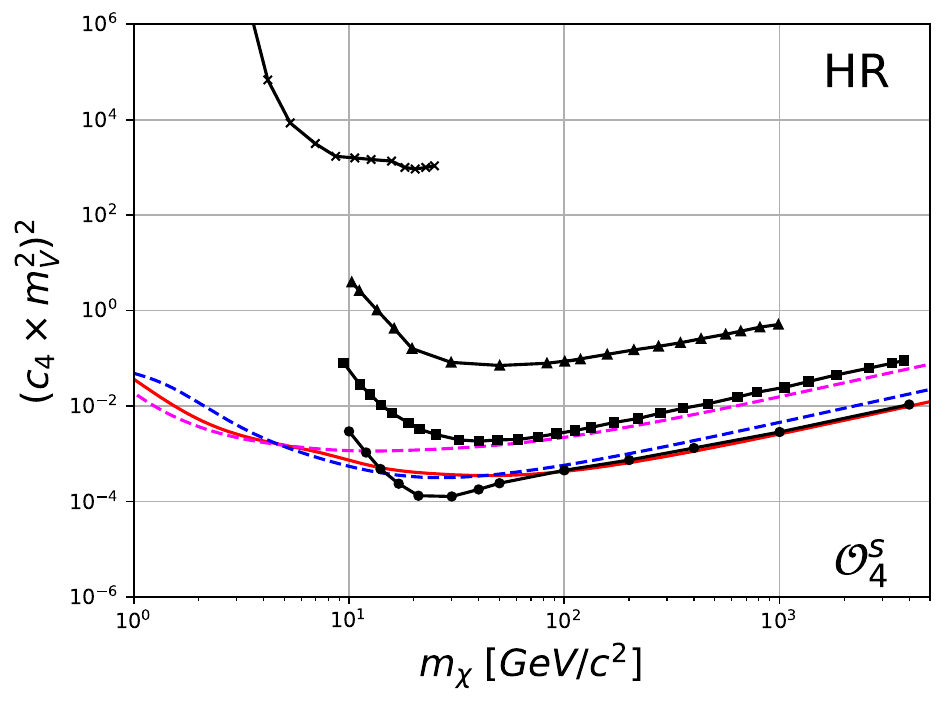}
    \end{subfigure}

    \vspace{0.1ex}

    \begin{subfigure}{0.30\textwidth}
    \includegraphics[width=\linewidth]{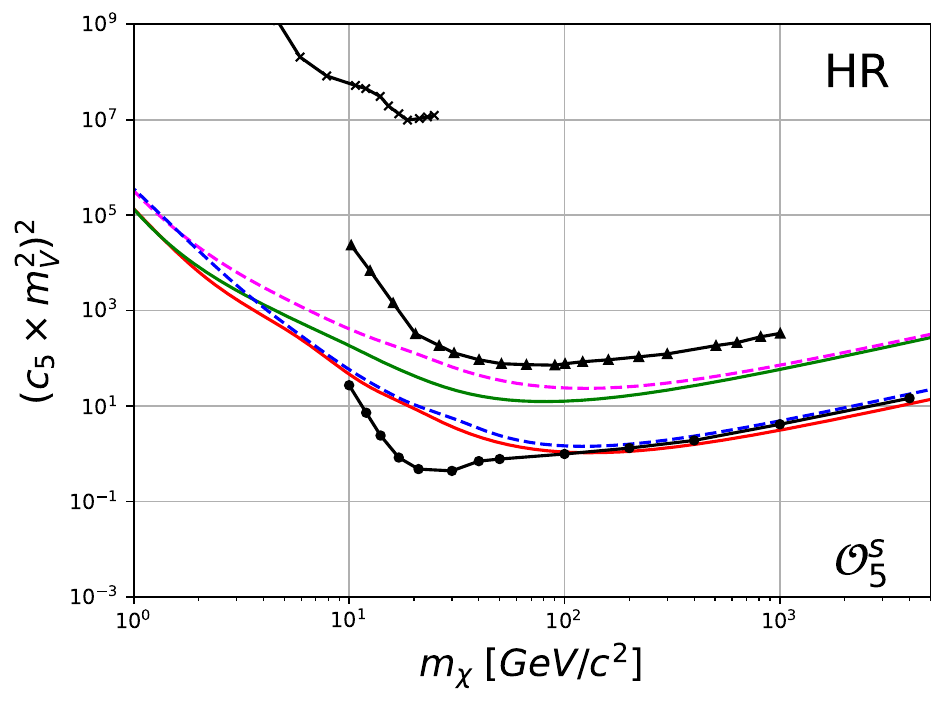}
    \end{subfigure}
    \begin{subfigure}{0.30\textwidth}
    \includegraphics[width=\linewidth]{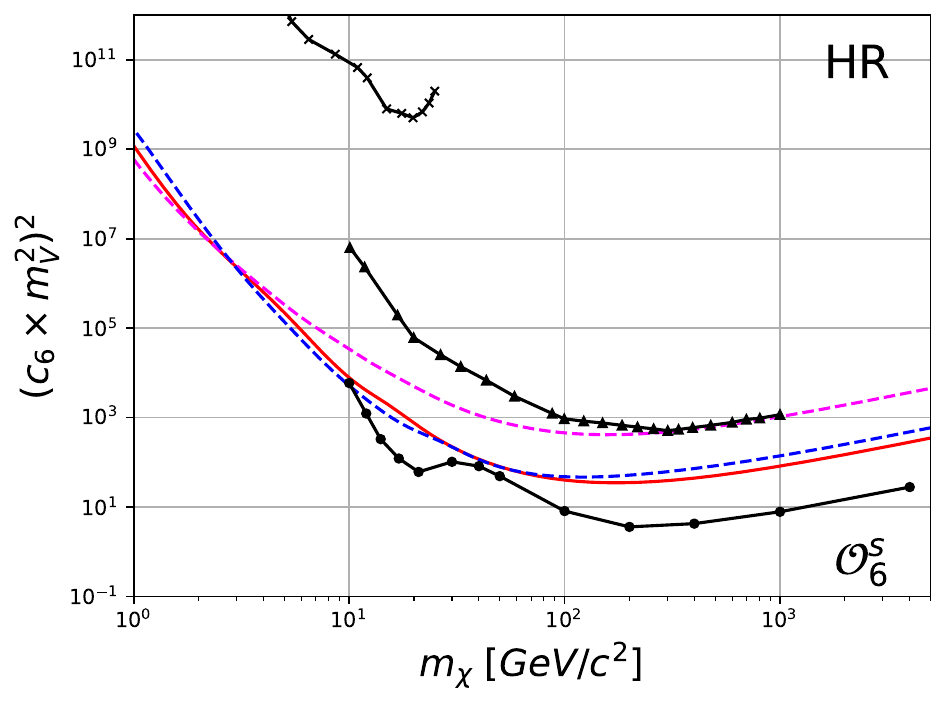}
    \end{subfigure}
    \begin{subfigure}{0.30\textwidth}
    \includegraphics[width=\linewidth]{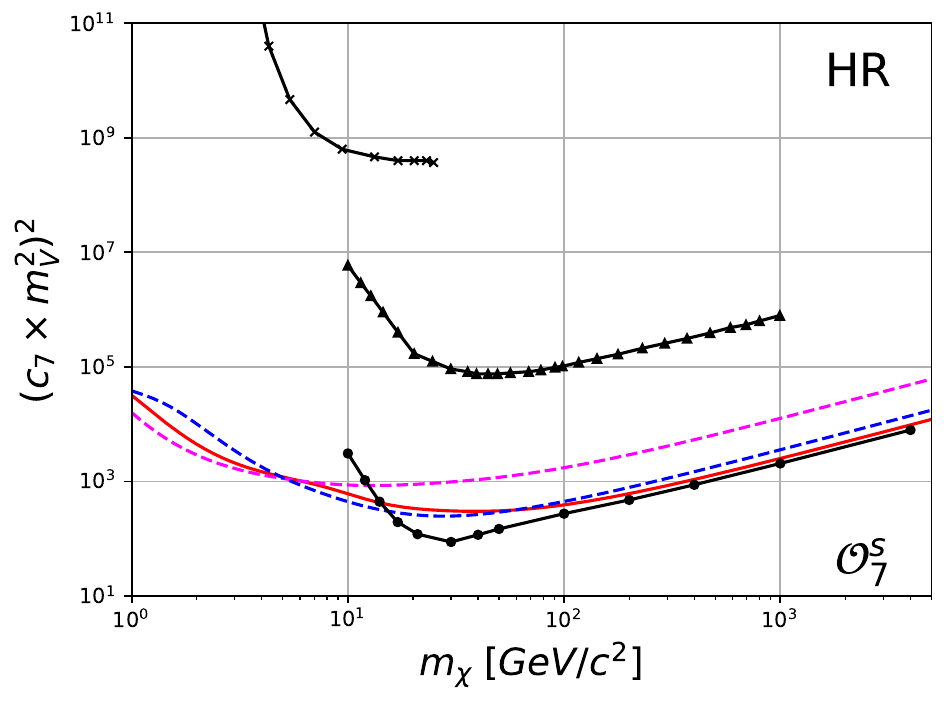}
    \end{subfigure}

    \vspace{0.1ex}

    \begin{subfigure}{0.30\textwidth}
    \includegraphics[width=\linewidth]{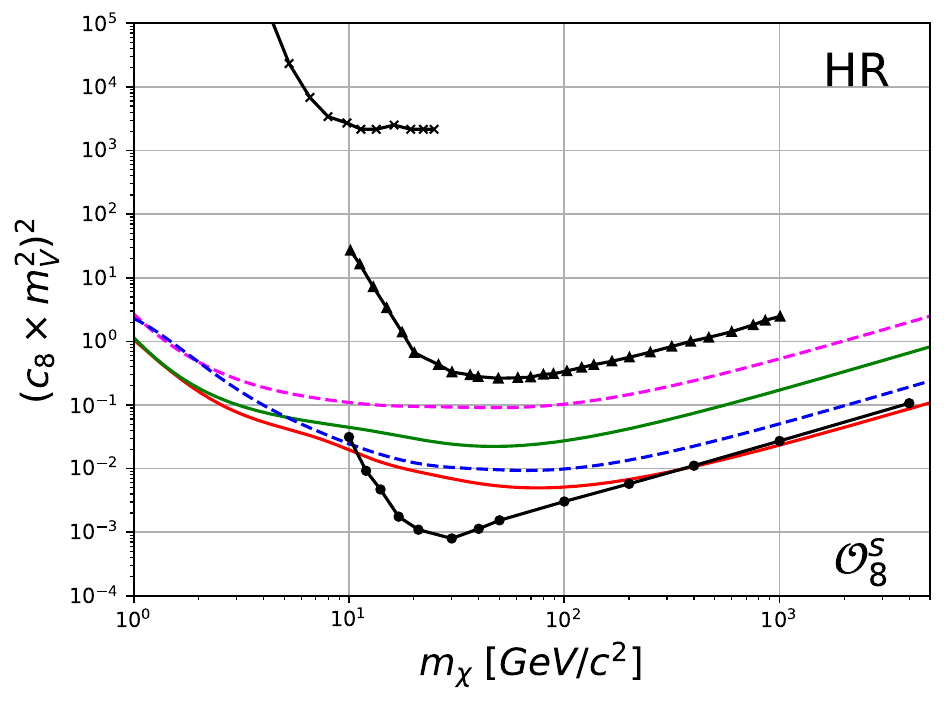}
    \end{subfigure}
    \begin{subfigure}{0.30\textwidth}
    \includegraphics[width=\linewidth]{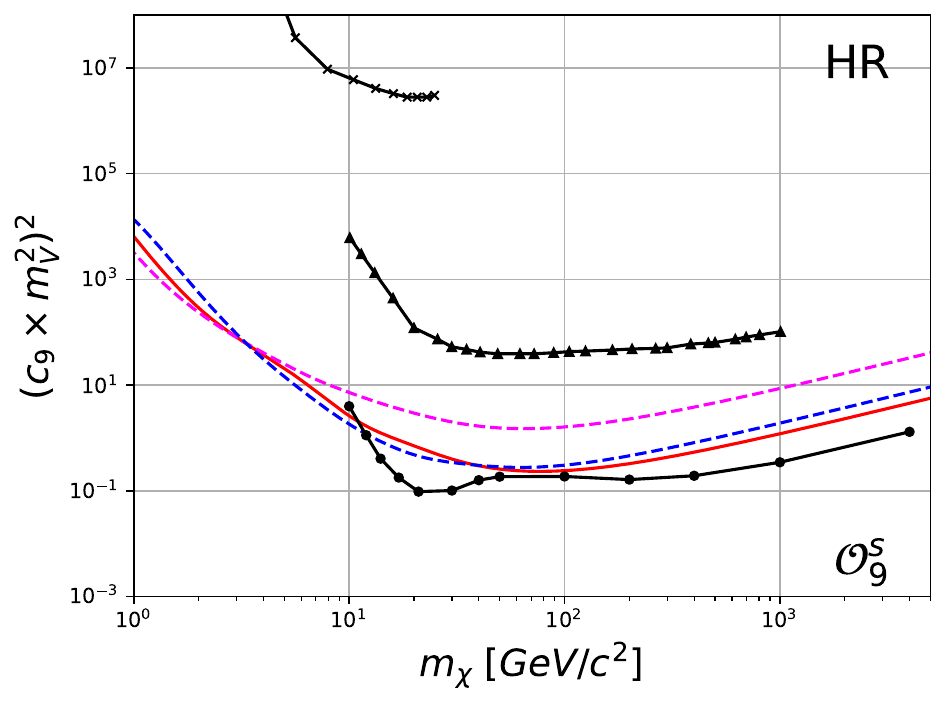}
    \end{subfigure}
    \begin{subfigure}{0.30\textwidth}
    \includegraphics[width=\linewidth]{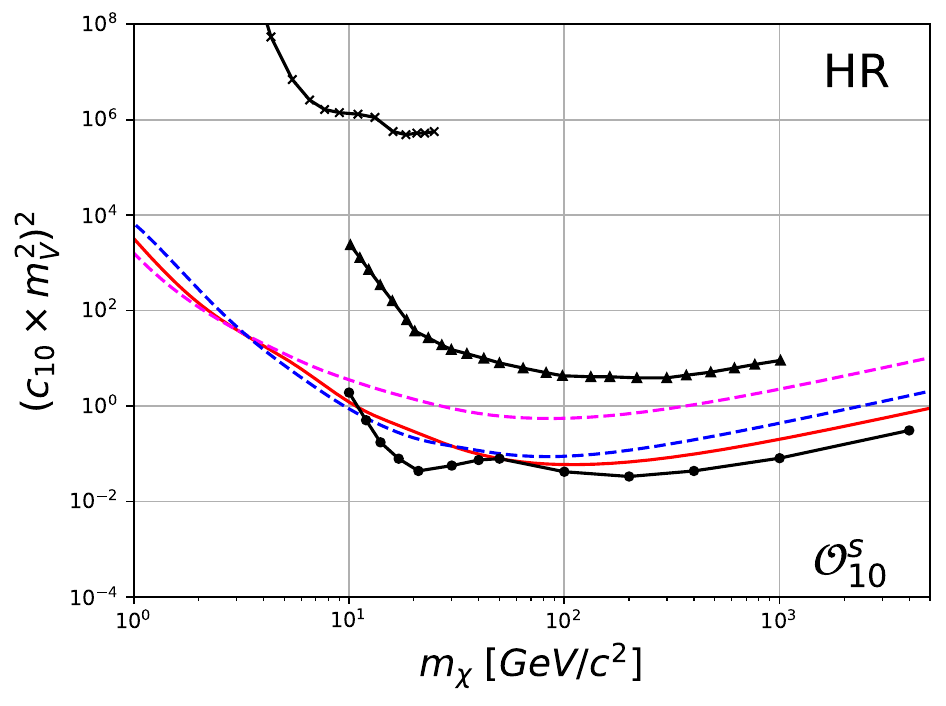}
    \end{subfigure}

    \vspace{0.1ex}

    \begin{subfigure}{0.30\textwidth}
    \includegraphics[width=\linewidth]{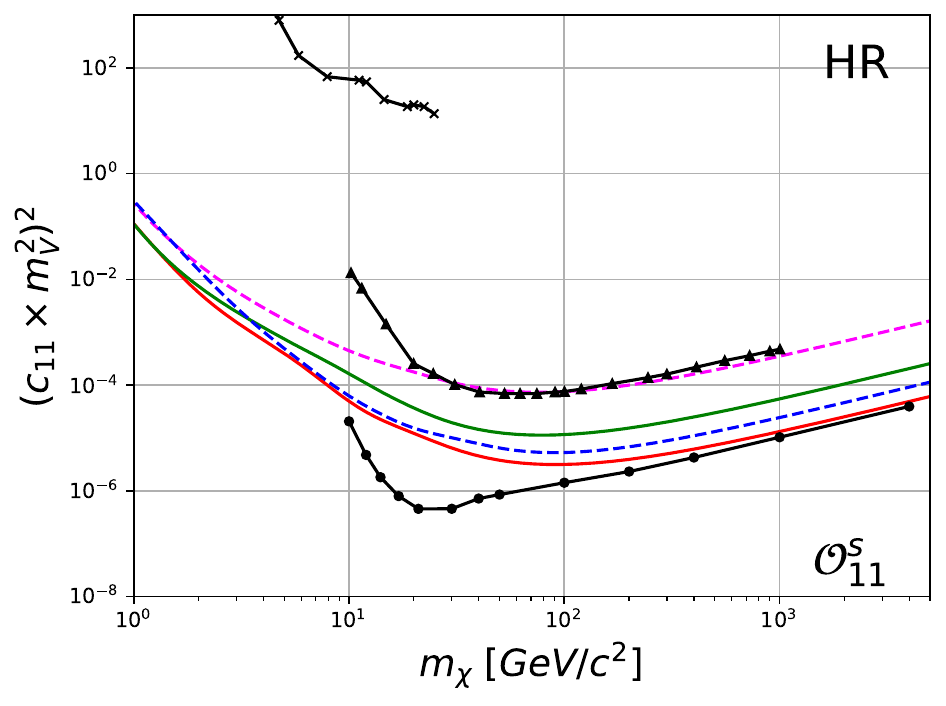}
    \end{subfigure}
    \begin{subfigure}{0.30\textwidth}
    \includegraphics[width=\linewidth]{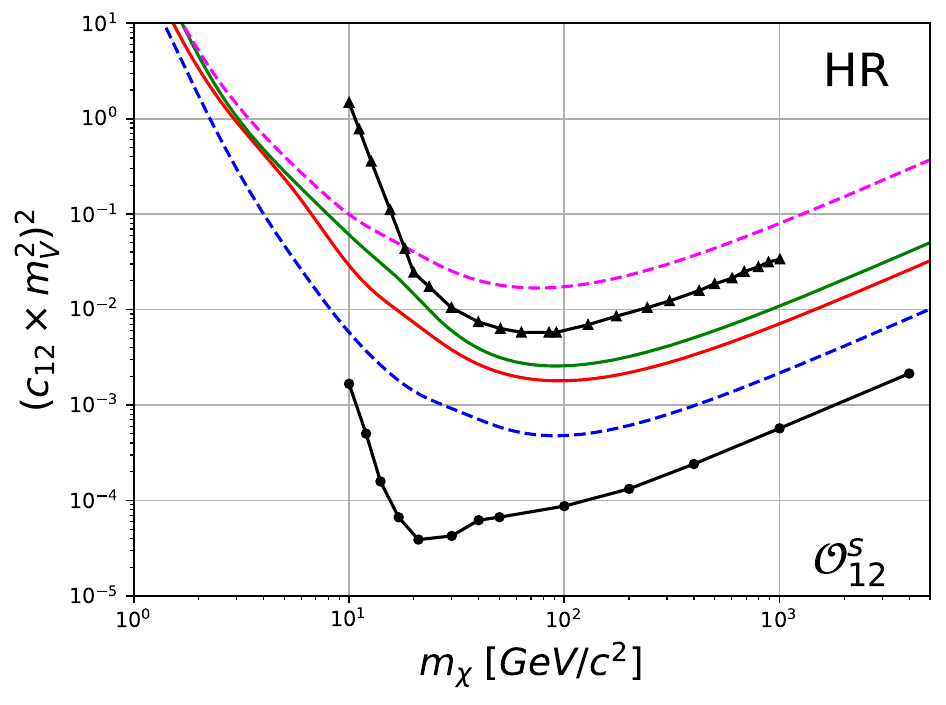}
    \end{subfigure}
    \begin{subfigure}{0.30\textwidth}
    \includegraphics[width=\linewidth]{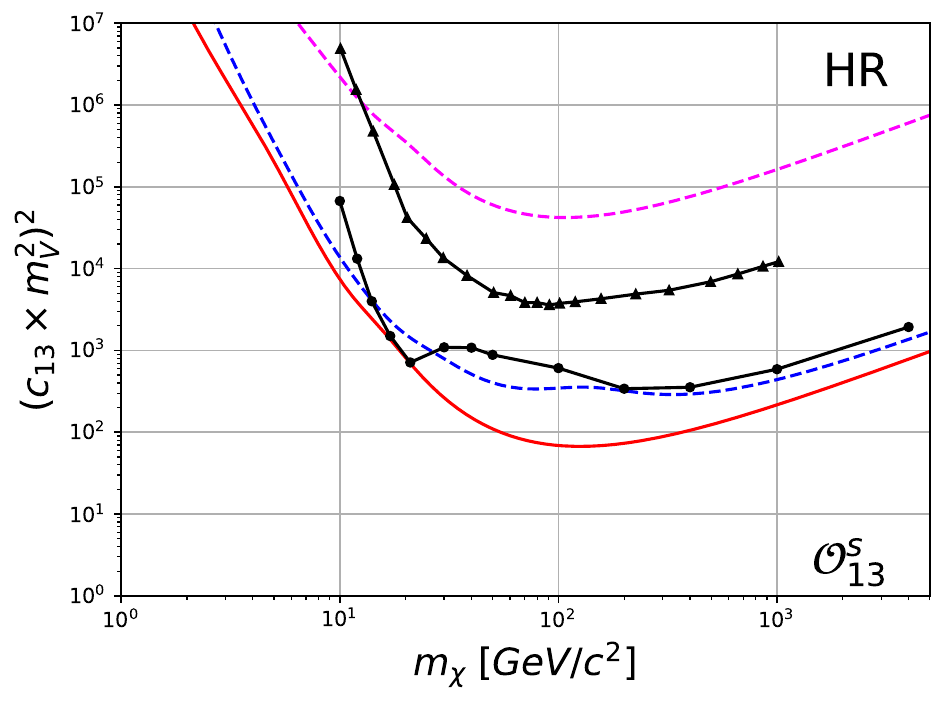}
    \end{subfigure}

    \vspace{0.1ex}

    \begin{subfigure}{0.30\textwidth}
    \includegraphics[width=\linewidth]{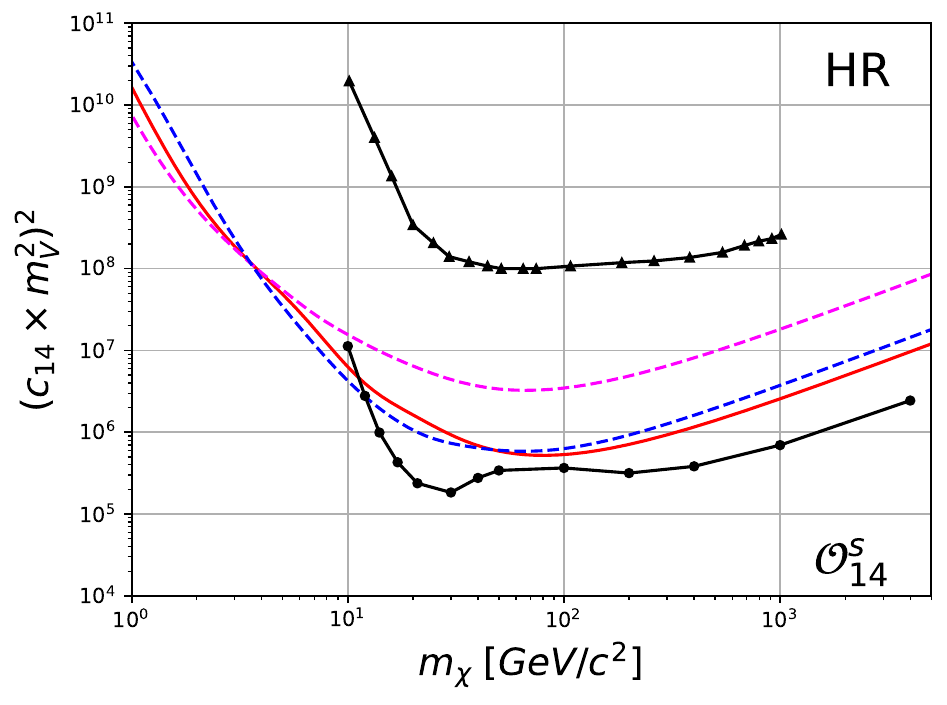}
    \end{subfigure}
    \begin{subfigure}{0.30\textwidth}
    \includegraphics[width=\linewidth]{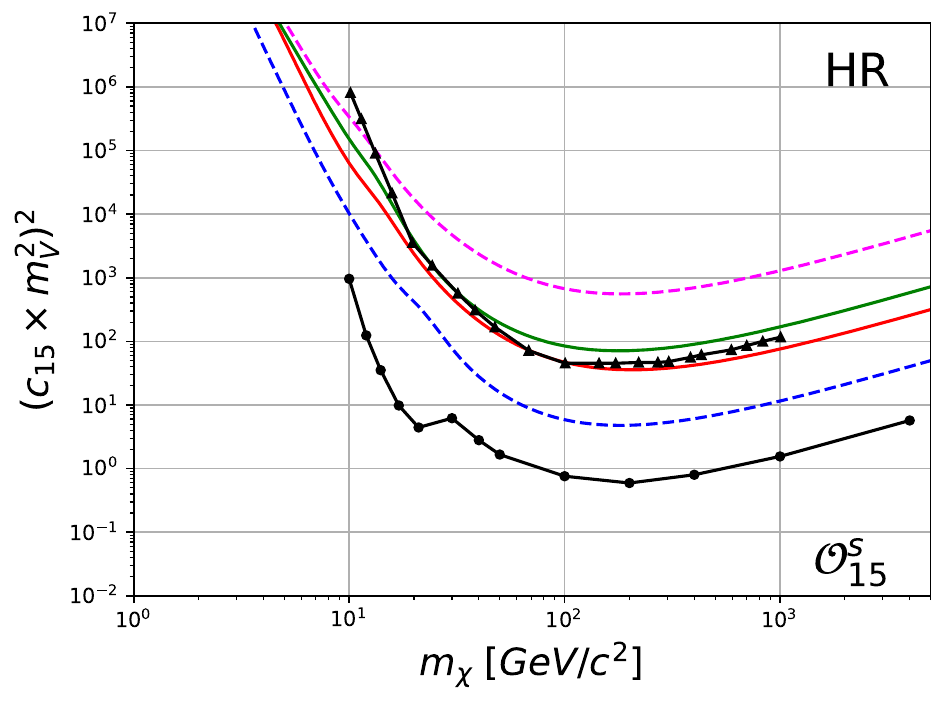}
    \end{subfigure}\hspace{4.85em}
    \begin{subfigure}{0.21\textwidth}
    \includegraphics[width=\linewidth]{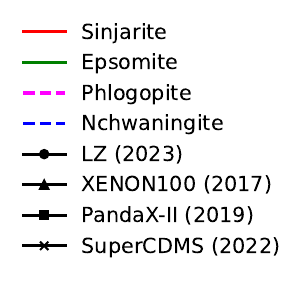} 
    \end{subfigure}
    \caption{Sensitivity, in the HR scenario, for additional minerals (not considered in the main body of the text): Projected 90\% confidence level upper limits on the dimensionless isoscalar WIMP--nucleon NREFT coupling constants for elastic scattering, assuming a small exposure $M=10$ mg and $t_{ \text{age}} = 1$ Gyr, and a high read-out resolution $\sigma_{x}=1$ nm. The solid (dashed) lines indicate minerals with $C^{238}=10^{-11}$ g/g ($C^{238}=10^{-10}$ g/g), see Table \ref{tab:U238_concentration}. With black lines shown are the NREFT results from other experiments: the 90\% confidence level upper limits from XENON100 \cite{XENON:2017fdd}, LUX--ZEPLIN \cite{LZ:2023lvz}, PandaX--II \cite{PandaX-II:2018woa}, as well as the 95\% Bayesian credible region of the two-dimensional marginalized posterior distribution from SuperCDMS \cite{SuperCDMS:2022crd}. For elastic scattering, the HR scenario illustrates sensitivity to WIMP--nucleon cross sections below existing experimental bounds for $m_\chi \lesssim 10~\mathrm{GeV}/c^2$.}
    \label{fig:HR_ap}
\end{figure}

\begin{figure}
    \captionsetup{justification=raggedright,singlelinecheck=false}
    \centering
    \begin{subfigure}{0.30\textwidth}
    \includegraphics[width=\linewidth]{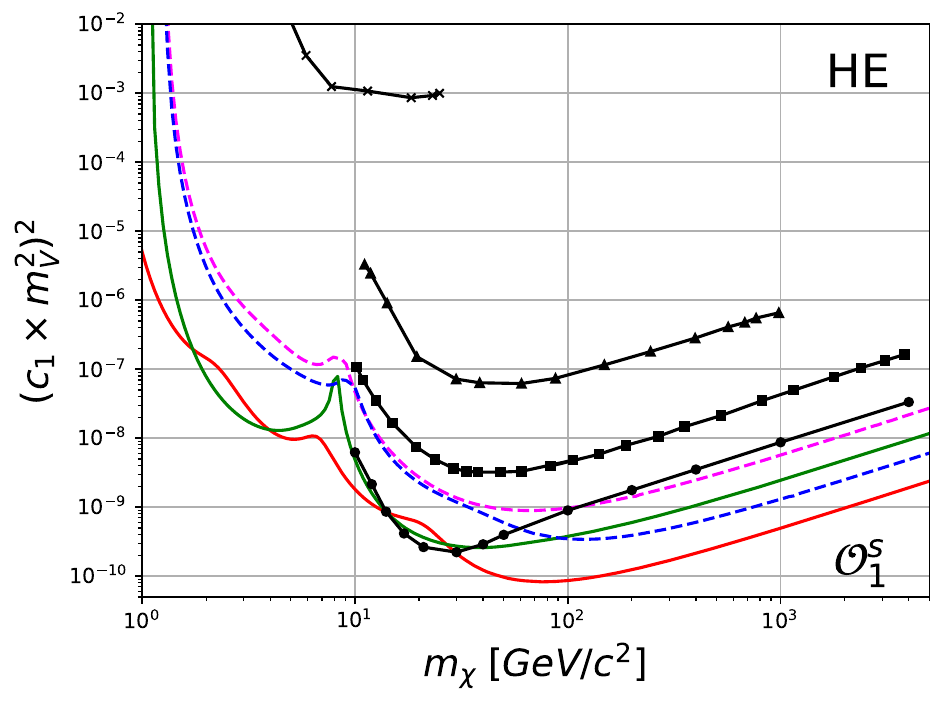}
    \end{subfigure}
    \begin{subfigure}{0.30\textwidth}
    \includegraphics[width=\linewidth]{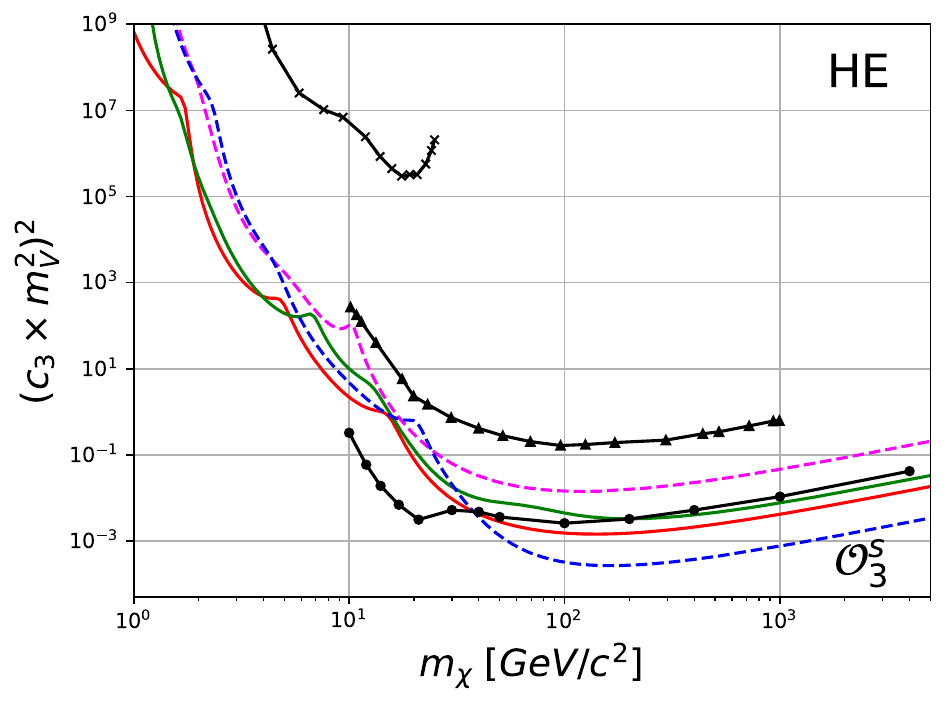}
    \end{subfigure}
    \begin{subfigure}{0.30\textwidth}
    \includegraphics[width=\linewidth]{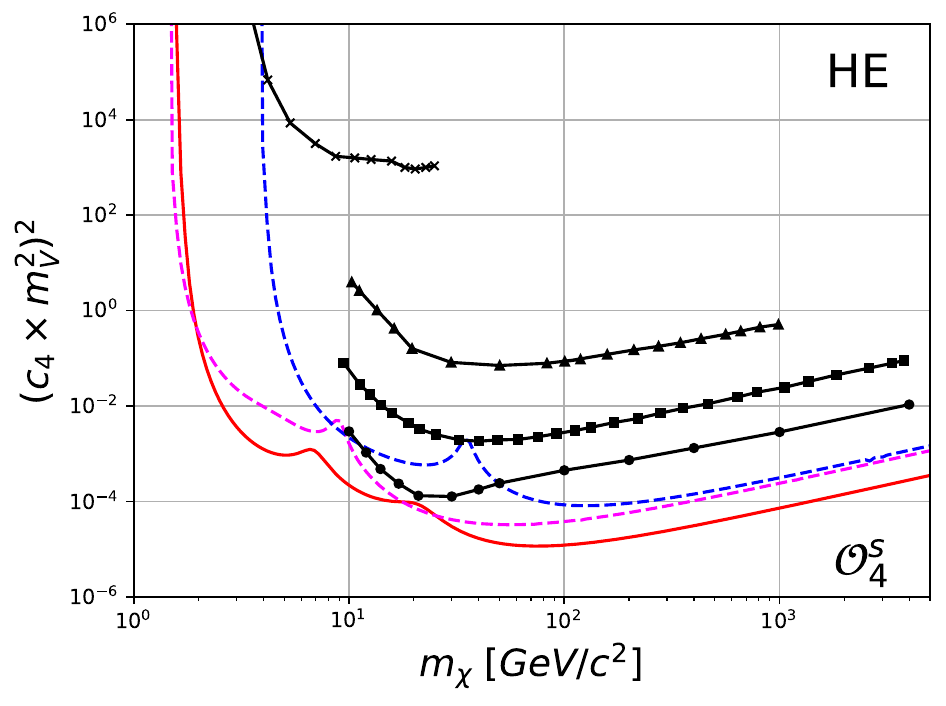}
    \end{subfigure}

    \vspace{0.1ex}

    \begin{subfigure}{0.30\textwidth}
    \includegraphics[width=\linewidth]{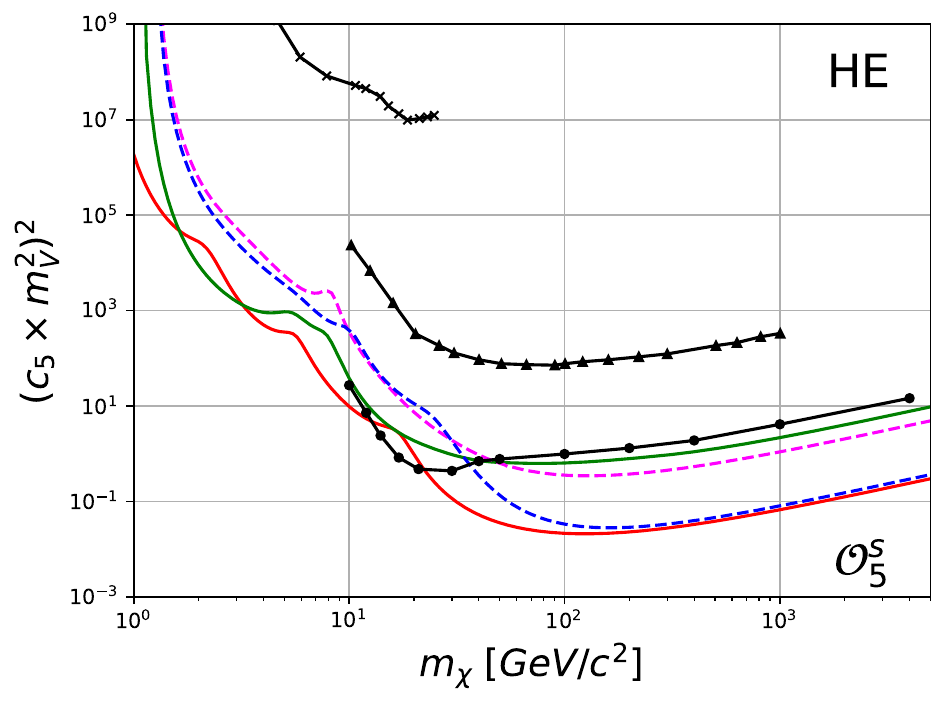}
    \end{subfigure}
    \begin{subfigure}{0.30\textwidth}
    \includegraphics[width=\linewidth]{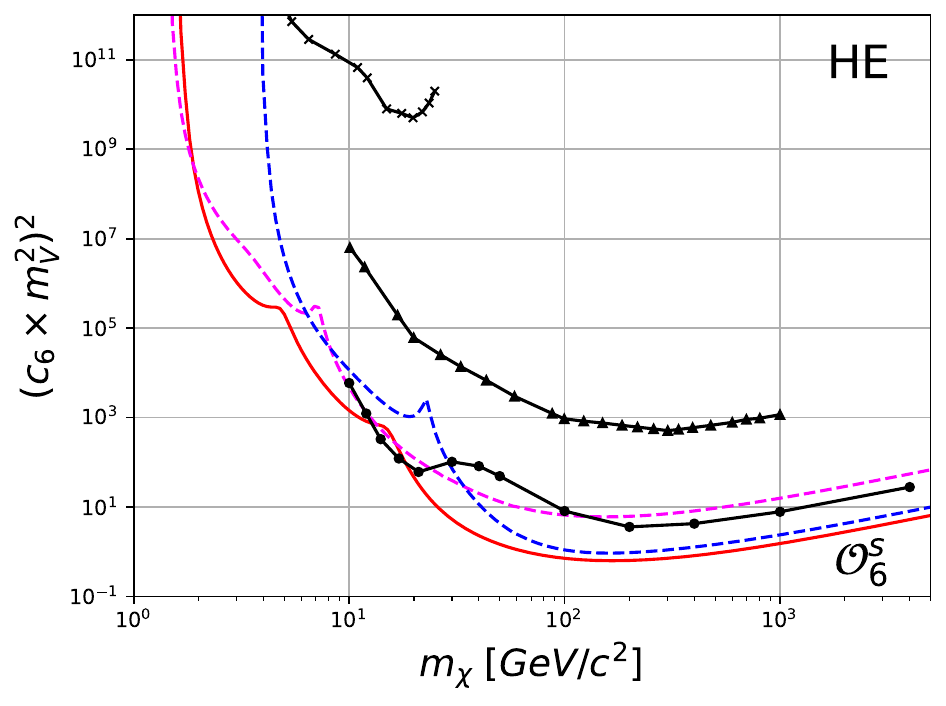}
    \end{subfigure}
    \begin{subfigure}{0.30\textwidth}
    \includegraphics[width=\linewidth]{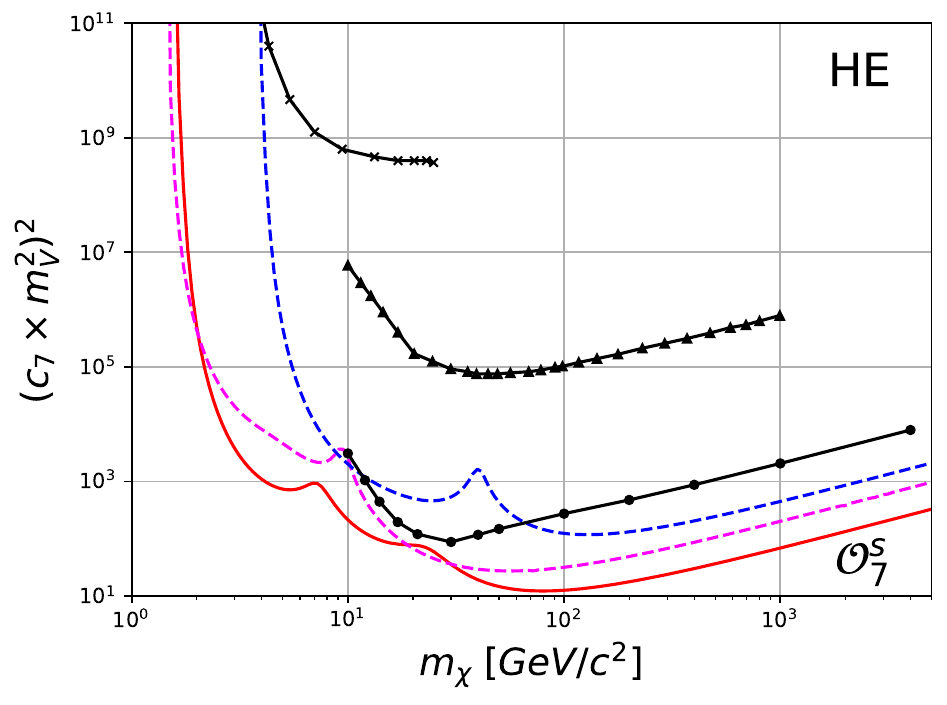}
    \end{subfigure}

    \vspace{0.1ex}

    \begin{subfigure}{0.30\textwidth}
    \includegraphics[width=\linewidth]{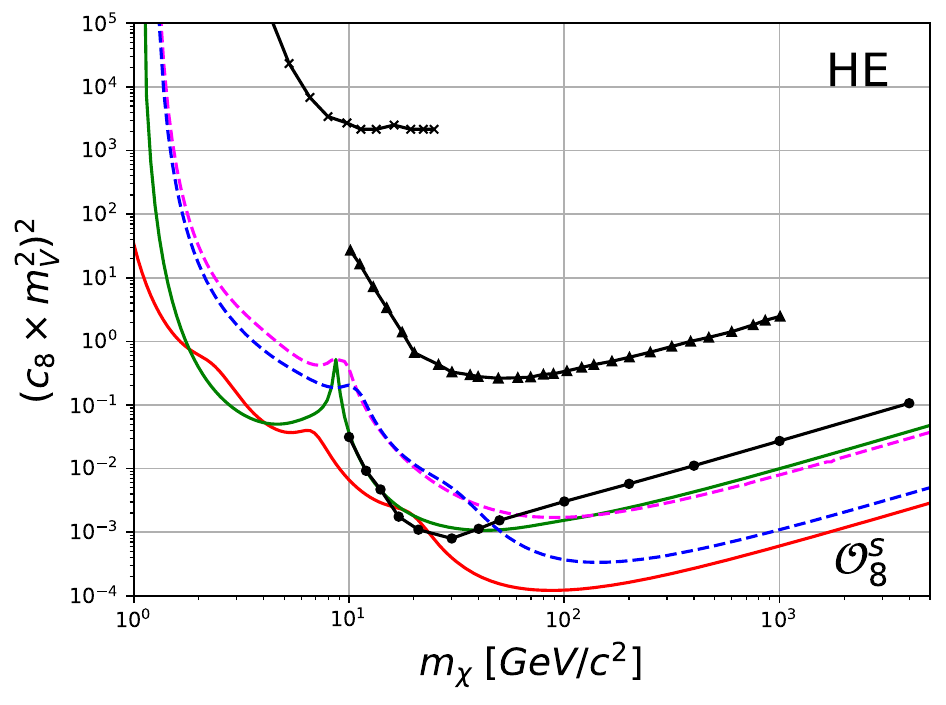}
    \end{subfigure}
    \begin{subfigure}{0.30\textwidth}
    \includegraphics[width=\linewidth]{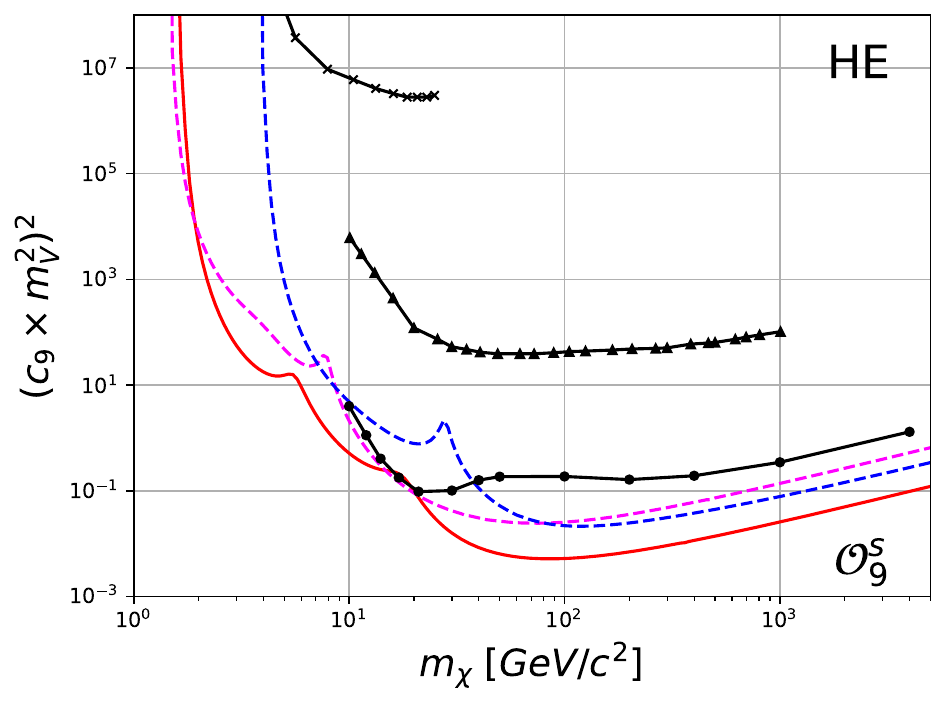}
    \end{subfigure}
    \begin{subfigure}{0.30\textwidth}
    \includegraphics[width=\linewidth]{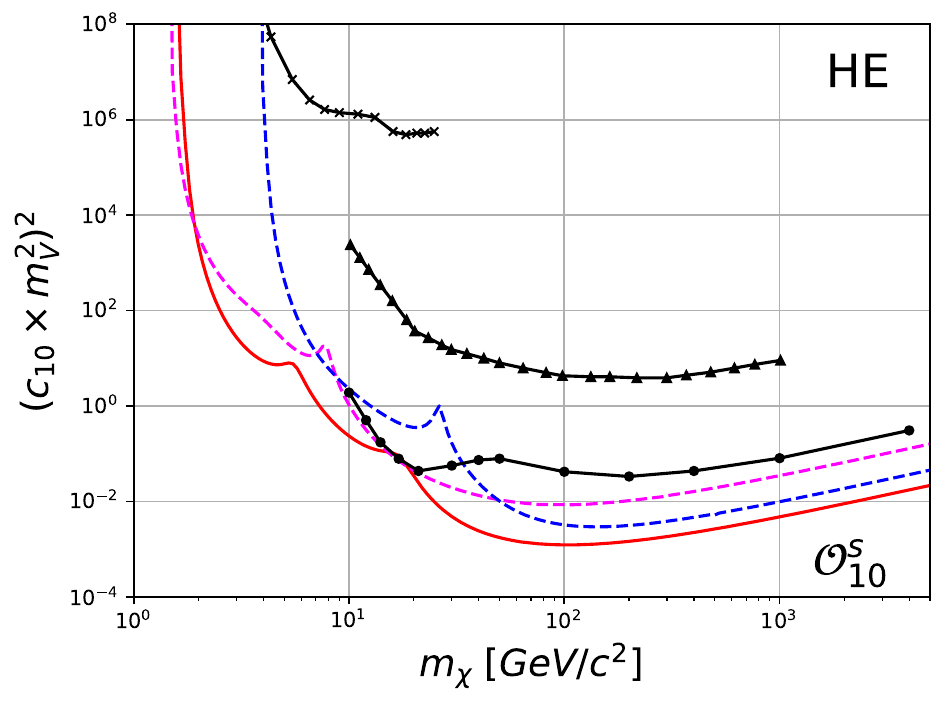}
    \end{subfigure}

    \vspace{0.1ex}

    \begin{subfigure}{0.30\textwidth}
    \includegraphics[width=\linewidth]{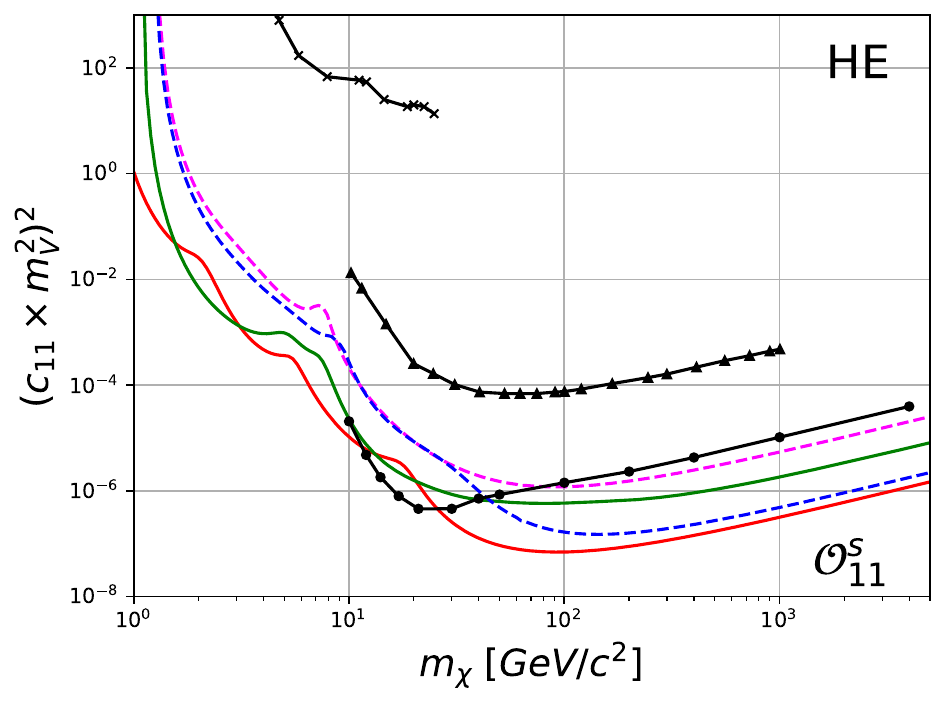}
    \end{subfigure}
    \begin{subfigure}{0.30\textwidth}
    \includegraphics[width=\linewidth]{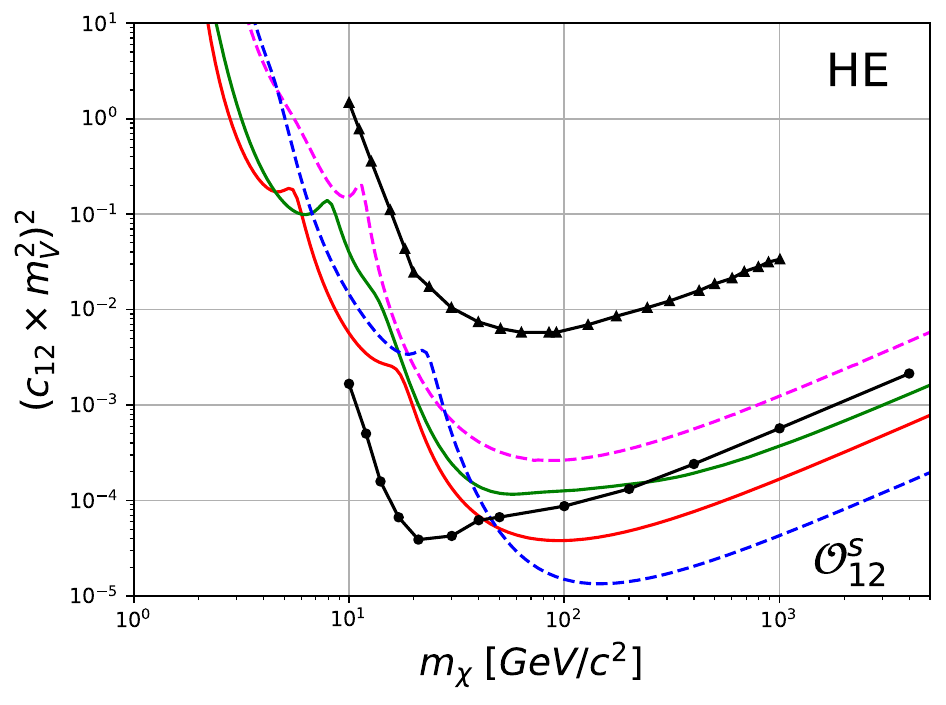}
    \end{subfigure}
    \begin{subfigure}{0.30\textwidth}
    \includegraphics[width=\linewidth]{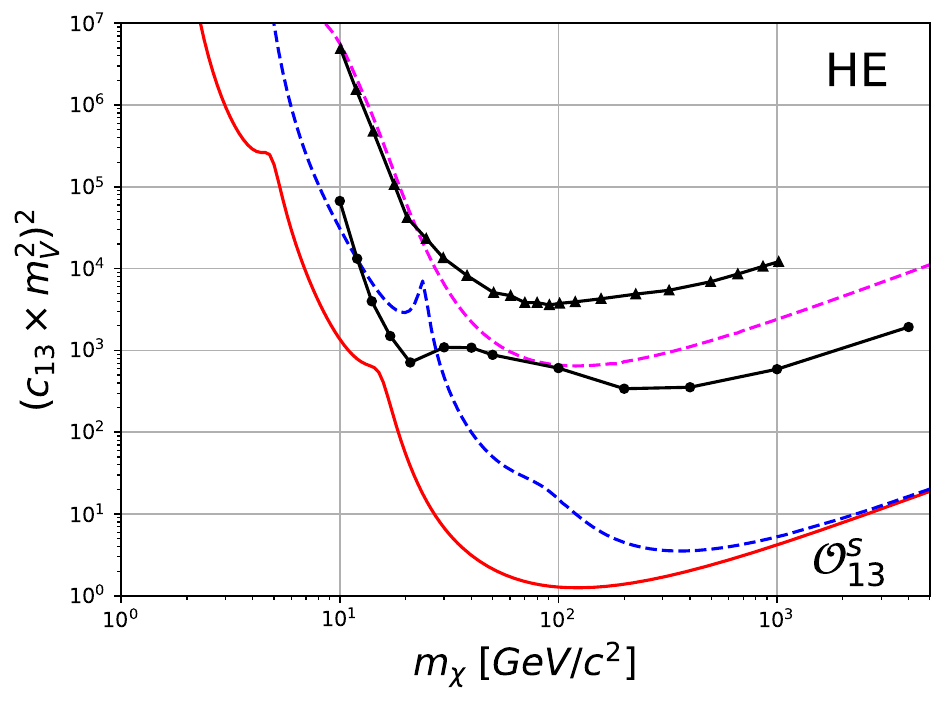}
    \end{subfigure}

    \vspace{0.1ex}

    \begin{subfigure}{0.30\textwidth}
    \includegraphics[width=\linewidth]{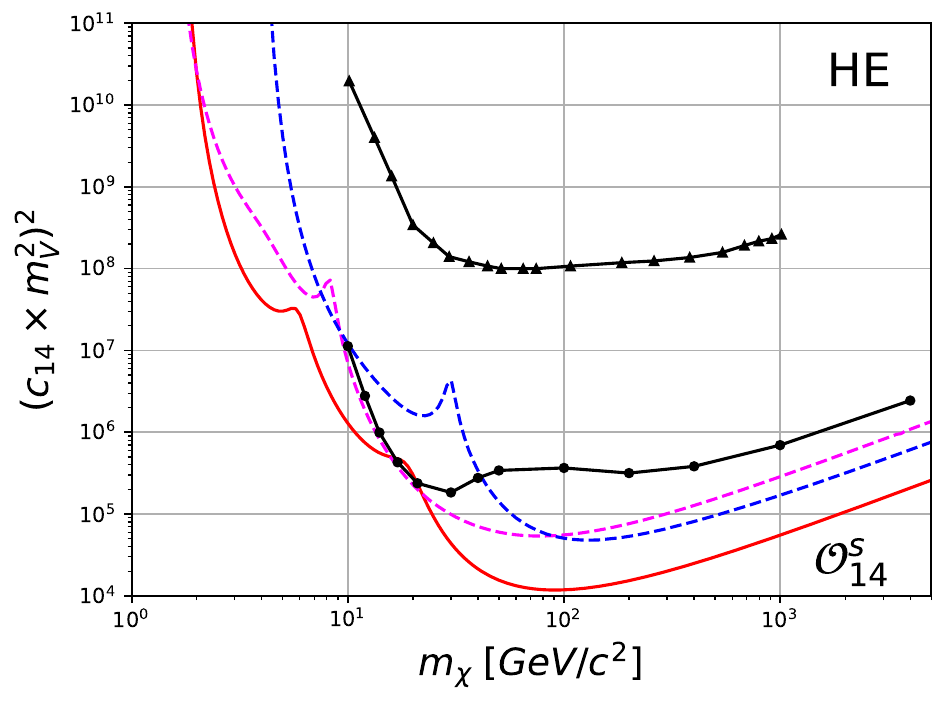}
    \end{subfigure}
    \begin{subfigure}{0.30\textwidth}
    \includegraphics[width=\linewidth]{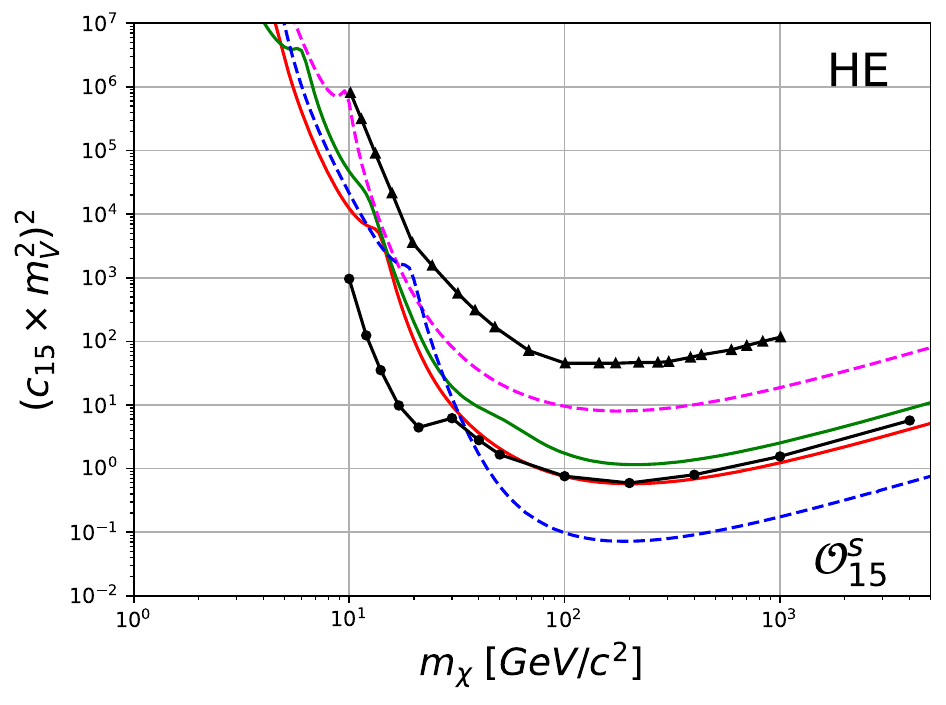}
    \end{subfigure}\hspace{4.85em}
    \begin{subfigure}{0.21\textwidth}
    \includegraphics[width=\linewidth]{figures/legend_ap.pdf} 
    \end{subfigure}
    \caption{Sensitivity, in the HE scenario, for additional minerals (not considered in the main body of the text): Projected 90\% confidence level upper limits on the dimensionless isoscalar WIMP--nucleon NREFT coupling constants for elastic scattering, assuming a large exposure $M=100$ g and $t_{ \text{age}} = 1$ Gyr, and a low read-out resolution $\sigma_{x}=15$ nm. The solid (dashed) lines indicate minerals with $C^{238}=10^{-11}$ g/g ($C^{238}=10^{-10}$ g/g), see Table \ref{tab:U238_concentration}. With black lines shown are the NREFT results from other experiments: the 90\% confidence level upper limits from XENON100 \cite{XENON:2017fdd}, LUX--ZEPLIN \cite{LZ:2023lvz}, PandaX--II \cite{PandaX-II:2018woa}, as well as the 95\% Bayesian credible region of the two-dimensional marginalized posterior distribution from SuperCDMS \cite{SuperCDMS:2022crd}. For elastic scattering, the HE scenario illustrates enhanced sensitivity to NREFT coupling constants relative to conventional direct detection experiments for $m_\chi \gtrsim 100~\mathrm{GeV}/c^2$, particularly for minerals such as sinjarite.}
    \label{fig:HE_ap}
\end{figure}

In Figs.~\ref{fig:HR_ap} and \ref{fig:HE_ap}, we present the projected 90\% confidence-level exclusion limits on the isoscalar WIMP–nucleon NREFT coupling constants ($c^{p}=c^{n}$) for elastic scattering in the minerals sinjarite, epsomite, phlogopite, and nchwaningite, for the HR scenario ($M=10~\mathrm{mg}$, $\sigma_{x}=1~\mathrm{nm}$) and the HE scenario ($M=100~\mathrm{g}$, $\sigma_{x}=15~\mathrm{nm}$), respectively. Among the minerals considered in this work, sinjarite provides the best projected sensitivity in both read-out scenarios for most of the NREFT operators. This is primarily due to both its hydrogen content and small $^{238}$U concentrations, which consequently yield low $^{238}$U–induced backgrounds. Additionally, sinjarite's chemical composition enhances its sensitivity, as it contains isotopes of Ca and Cl also found in gypsum and halite, respectively. In contrast to gypsum, sinjarite possesses isotopes with non-zero spin ground states yielding non-zero spectra for the spin-dependent operators. Additionally the presence of Cl implies boosted spectra for the operators $\mathcal{O}_{5}^{s}$, $\mathcal{O}_{8}^{s}$, and $\mathcal{O}_{13}^{s}$, as explained for halite in the main text. Moreover, the isotope $^{55}$Mn in nchwaningite, with a nuclear ground state of spin and parity $J^{\pi} = 5/2^{-}$, has a partially filled angular-momentum orbital that enhances the $\Phi''$ response, thereby increasing the mineral's sensitivity to the operators $\mathcal{O}_{3}$, $\mathcal{O}_{12}$, and $\mathcal{O}_{15}$, and yielding the most stringent projected limits among all the minerals considered.


\bibliography{refs}

\end{document}